%% file: main.tex
\documentclass[10pt,onecolumn,english]{book}

\include{TexFiles/settings}

\def\phdThesis{1}

\newcommand{\currentyear}{2023}
\newcommand{\authorname}{Weixing Zhang}
\newcommand{\mytitle}{Towards Automated Support for the Co-Evolution of Meta-Models and Grammars}
\newcommand{\division}{Interaction Design \& Software Engineering}
\newcommand{\techReportNumber}{xxxx}

\newcommand{\phdISBNNumber}{978-91-7905-833-3}

\begin{document}
\frontmatter

\include{TexFiles/firstpage}

\include{TexFiles/dedication}

\cleardoublepage \addcontentsline{toc}{chapter}{Abstract}
\include{TexFiles/abstract}

\cleardoublepage \addcontentsline{toc}{chapter}{Acknowledgement}
\include{TexFiles/acknowledgment}

\cleardoublepage \addcontentsline{toc}{chapter}{List of Publications}
\include{TexFiles/listPub}

\cleardoublepage \addcontentsline{toc}{chapter}{Personal Contribution}
\include{TexFiles/contribution}


\tableofcontents 

 
\mainmatter

\include{TexFiles/introduction}

\cleardoublepage 
\include{TexFiles/appendedPapers}

\cleardoublepage\addcontentsline{toc}{chapter}{Bibliography}
\bibliographystyle{IEEEtran}

\bibliography{lit}

%
\cleardoublepage
\end{document}

%% file: TexFiles/settings.tex
\usepackage{zref-abspage}
\usepackage{geometry}
\usepackage{fancyhdr}
\usepackage{graphicx}
\usepackage{url}
\usepackage{csquotes}
\usepackage[english]{babel}
\usepackage{amsmath}
\usepackage{cite}
\usepackage{textcomp}
\usepackage{psfrag}
\usepackage{rotating}
\usepackage{pdflscape}
\usepackage[final]{pdfpages}
\usepackage{xspace}
\usepackage{relsize}
\usepackage{url}
\usepackage{IEEEtrantools}
\usepackage{microtype}
\usepackage{todonotes}

\usepackage{cite}
\usepackage{amsmath,amssymb,amsfonts}
\usepackage{algorithmic}
\usepackage{graphicx}
\usepackage{textcomp}
\usepackage{xcolor}
\usepackage{lipsum}
\usepackage{caption}
\def\BibTeX{{\rm B\kern-.05em{\sc i\kern-.025em b}\kern-.08em
    T\kern-.1667em\lower.7ex\hbox{E}\kern-.125emX}}
\usepackage{placeins}
\usepackage{tabularx}
\usepackage{longtable}
\usepackage{fontawesome}
\usepackage{soul}
\usepackage{tcolorbox}
\usepackage{colortbl}
\usepackage{changepage}
\usepackage{setspace}
\usepackage{tcolorbox}
\usepackage{wrapfig}
\usepackage{hyperref}
\newcommand{\interviewquote}[2]{
 \def\FrameCommand{%
    \hspace{0pt}%
    {\color{blue}\vrule width 1.5pt}%
    {\color{white}\vrule width 4pt}%
    \colorbox{white}
  }%
  \MakeFramed{\advance\hsize-\width\FrameRestore}%
  \noindent\hspace{-4.55pt}
  \begin{adjustwidth}{}{7pt}
  {\footnotesize ``\emph{#1}'' - {#2}}\vspace{0.5pt}\end{adjustwidth}\endMakeFramed%
}

\usepackage{cite}
\usepackage{algorithmic}
\usepackage{graphicx}
\usepackage{textcomp}
\usepackage{xcolor}
\usepackage{lipsum}
\usepackage{caption}
\def\BibTeX{{\rm B\kern-.05em{\sc i\kern-.025em b}\kern-.08em
    T\kern-.1667em\lower.7ex\hbox{E}\kern-.125emX}}
\usepackage{placeins}
\usepackage{tabularx}
\usepackage{longtable}
\usepackage{fontawesome}
\usepackage{soul}
\usepackage{tcolorbox}
\usepackage{colortbl}
\usepackage{changepage}
\usepackage{setspace}
 \usepackage{hyperref}

 \usepackage{multirow}
 \usepackage{orcidlink}
 \usepackage{rotating}
 \usepackage{environ}
\usepackage{pdflscape}
\usepackage{stmaryrd}

\usepackage{cite}
\usepackage{algorithmic}
\usepackage{graphicx}
\usepackage{textcomp}
\usepackage{xcolor}
\usepackage{soul}        
\usepackage{array}
\usepackage{multirow}
\usepackage{tabularx}
\usepackage{adjustbox}
\usepackage{amssymb}
\usepackage{hyperref}
\hypersetup{
hidelinks
}
\usepackage{booktabs}

\usepackage{pifont}
\usepackage[T1]{fontenc}

\def\BibTeX{{\rm B\kern-.05em{\sc i\kern-.025em b}\kern-.08em
T\kern-.1667em\lower.7ex\hbox{E}\kern-.125emX}}
\usepackage[hyphenbreaks]{breakurl}
\usepackage{changepage}
\usepackage{float}
\usepackage{makecell}
\usepackage{xcolor,pifont}

\usepackage[numbers]{natbib}
\usepackage{enumerate}
\usepackage{listings}
\usepackage{tikz}
\usepackage{harveyballs}
\usepackage{afterpage}
\usepackage{caption,booktabs}
\usepackage{subfig}
\usepackage{footmisc}
\usepackage{ifthen}
\usepackage{fancyvrb}
\usepackage{arydshln}
\usepackage[nomessages]{fp}
\usepackage{mdframed}
\usepackage{nameref}
\usepackage[outline]{contour}
\usepackage{colortbl,multirow,hhline}
\usepackage{threeparttable}

\newcommand{\fe}[1]{\textsf{\small #1}}

\newcommand{\figref}[1]{Fig.~\ref{#1}}

\newcommand{\secref}[1]{Section~\ref{#1}}

\definecolor{yellow}{RGB}{255,255,0}

\newboolean{showannotations}
\setboolean{showannotations}{true} 

\newboolean{showcomments}
\setboolean{showcomments}{true} 

\newcommand{\newtextcolor}{blue}
\newcommand{\oldtextcolor}{red}

\newcommand{\rem}[1]{%
    \ifthenelse{\boolean{showannotations}}%
    {\textcolor{\oldtextcolor}{\st{#1}}}%
    {}%
}

\newcommand\add[1]{%
    \ifthenelse{\boolean{showannotations}}%
    {\textcolor{\newtextcolor}{{\textbf{#1}}}}%
    {#1}%
}

\newcommand\xofy[3]{%
    \FPeval{perc}{round(100.0 / #2 * #1, 0)}%
    \ifthenelse{\equal{#3}{}}{%
        {#1} of {#2} (\perc\%)%
    }%
    {%
        {#1} of {#2} {#3} (\perc\%)%
    }%
}

\newcommand\xofyp[2]{%
    \FPeval{perc}{round(100.0 / #2 * #1, 0)}%
    {#1} of {#2} -- \perc\%%
}

\setul{0.5ex}{0.3ex}
\definecolor{purple}{RGB}{235, 52, 225}
\setulcolor{purple}

\newcommand{\ie}{i.e.,\xspace}
\newcommand{\eg}{e.g.,\xspace}
\newcommand{\etc}{etc.\xspace}

\newcolumntype{C}[1]{>{\centering\arraybackslash}p{#1}}

\makeatletter
\newcommand{\phasecolor}[1]{
\ifnum\pdf@strcmp{#1}{1}=0 red%
\else orange%
\fi
}

{\newcommand{\nb}[2]{
    \fcolorbox{gray}{yellow}{\bfseries\sffamily\scriptsize#1}
    {$\blacktriangleright$#2$\blacktriangleleft$}
  }
  
}
{\newcommand{\nb}[2]{}
  
}

\newboolean{shownames}
\setboolean{shownames}{true} 

\definecolor{gray50}{gray}{.5}
\definecolor{gray40}{gray}{.6}
\definecolor{gray30}{gray}{.7}
\definecolor{gray20}{gray}{.8}
\definecolor{gray10}{gray}{.9}
\definecolor{gray05}{gray}{.95}

\definecolor{green_sms}{RGB}{69,191,71}
\definecolor{barcolor}{RGB}{0, 245, 255}
\definecolor{barbackground}{RGB}{160, 160, 160}

\newcommand{\maxnum}{36}
\newcommand{\cellvspace}{-3.5mm}
\newcommand{\colheight}{-.3}
\newlength{\maxlen}

\newcommand{\databar}[2][]{%
	\settowidth{\maxlen}{\maxnum}%
	\addtolength{\maxlen}{\tabcolsep}%
	\FPeval\result{round(#2/\maxnum:4)}%
	\rlap{\color{barcolor}\hspace*{-.5\tabcolsep}\rule[\colheight\ht\strutbox]{\result\maxlen}{1.2\ht\strutbox}}%
	\vspace{\cellvspace}
	\makebox[\dimexpr\maxlen-\tabcolsep][l]{#1}~({#2\%})%
}

\newcommand{\relativeDatabar}[2]{%
\FPeval{relativeShare}{round( 100.0 / #2 * #1, 0)}
\databar[#1]{\relativeShare}
}

\newcommand{\numberAcademicScreenedApprox}{5,000}
\newcommand{\numberAcademicScreened}{4,975}
\newcommand{\numberAcademicIncluded}{77}
\newcommand{\numberAcademicTools}{68}

\newcommand{\numberGreyScreenedApprox}{1,500}
\newcommand{\numberGreyScreened}{1,494}
\newcommand{\numberGreyTools}{68}

\newcommand{\numberToolCandidates}{133}

\newcommand{\numberToolsIdentified}{26}

\newcommand{\eatxt}{\textsc{Eatxt}\xspace}
\newcommand{\eastadl}{\textsc{East-Adl}\xspace}
\newcommand{\grammaroptimizer}{\textsc{GrammarOptimizer}\xspace}


%% file: TexFiles/firstpage.tex
\ifx\phdThesis\undefined
\newcommand{\degreetitle}{Licentiate of Engineering}
\newcommand{\reportno}{TODO}
\newcommand{\reportNoText}{Technical Report No \techReportNumber \\
 ISSN 1652-876X\\ }
\else
\newcommand{\degreetitle}{Licentiate of Engineering}
\newcommand{\reportNoText}{ISBN \phdISBNNumber\\
ISSN 0346-718X\\ 
\vspace{1cm}

\noindent
Technical Report No \techReportNumber \\
}
\fi


\thispagestyle{empty} 
\begin{center}
  \textsc{Thesis for The Degree of \degreetitle}\\
\end{center}

\vspace{6cm}
\begin{center} \Large \mytitle
\end{center}

\vspace{1cm}
\begin{center}
\textsc{\authorname} \\
\end{center}

\vspace{2cm}
\begin{figure}[h]
  \begin{center}
     \includegraphics[width=30mm]{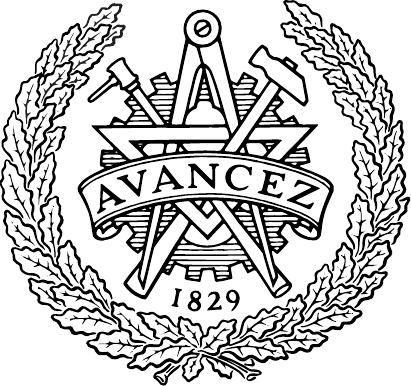}
     \hspace{1cm}
     \includegraphics[width=30mm]{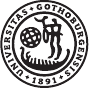}
  \end{center}
\end{figure}
\vspace{2cm}

\begin{center}
Division of \division \\
Department of Computer Science \& Engineering\\
Chalmers University of Technology and University of Gothenburg \\
Gothenburg, Sweden, \currentyear \\
\end{center}

\newpage
\thispagestyle{plain}

\vspace{2cm} \noindent \textbf
\mytitle\\

\noindent
\textsc{\authorname}\\

\vfill

\noindent
Copyright \copyright \currentyear \space \authorname \\
except where otherwise stated. \\
All rights reserved. \vspace{1cm}

\noindent Department of Computer Science \& Engineering\\
Division of \division \\
Chalmers University of Technology and University of Gothenburg\\
Gothenburg, Sweden\\
\vspace{1cm}

\noindent This thesis has been prepared using \LaTeX.

\noindent
Printed by Chalmers Reproservice,\\
Gothenburg, Sweden \currentyear.

%% file: TexFiles/dedication.tex

\newpage
\begin{table}
  \begin{flushright}
	\large{\emph{``What is life? Life is a process of continuous self-improvement through perception.'' \\- Kazuo Inamori (Renowned Japanese Entrepreneur)}}  
\end{flushright}
\end{table}

\newpage
\thispagestyle{empty}

%% file: TexFiles/abstract.tex
\thispagestyle{empty}
\section*{Abstract}

Blended modeling is an emerging paradigm involving seamless interaction between multiple notations for the same underlying modeling language. We focus on a model-driven engineering (MDE) approach based on meta-models to develop textual languages to improve the blended modeling capabilities of modeling tools. In this thesis, we propose an approach that can support the co-evolution of meta-models and grammars as language engineers develop textual languages in a meta-model-based MDE setting. Firstly, we comprehensively report on the challenges and limitations of modeling tools that support blended modeling, as well as opportunities to improve them. Second, we demonstrate how language engineers can extend Xtext's generator capabilities according to their needs. Third, we propose a semi-automatic method to transform a language with a generated grammar into a Python-style language. Finally, we provide a solution (i.e., GrammarOptimizer) that can support rapid prototyping of languages in different styles and the co-evolution of meta-models and grammars of evolving languages.

\vspace{5 mm} \noindent{\textbf{Keywords}}

\vspace{3 mm} \noindent{Blended Modeling, Systematic Literature Review, Xtext, Grammar Optimization, Co-Evolution}

\newpage
\thispagestyle{empty}

%% file: TexFiles/acknowledgment.tex
\chapter*{Acknowledgment}
\vspace{5 mm}

I want to express my sincere gratitude to my three supervisors: Dr. Jan-Philipp Steghöfer, Dr. Regina Hebig, and Dr. Daniel Strüber (in the order by time). They provided me with ample assistance, support, enthusiasm, patience, and tolerance. They guided me with their extensive expertise and rigorous scholarly attitude, helping me acquire knowledge, skills, and methodology in my research field, and successfully transitioning my mindset from an almost rigid engineer to a researcher. I have also learned an important thing from them: Do things in the right way. All of those have laid a solid foundation for my future academic career. I would also like to thank my former colleague, independent researcher Dr. Jörg Holtmann. During his time in the CSE department, he offered me a lot of help, both in research and engineering.

Furthermore, I want to express my gratitude to my examiner, Prof. Johan Karlsson, and other members of the doctoral school for their constructive feedback and administrative assistance. Special thanks to the post-doc Dr. Shiliang Zhang at the University of Oslo, senior researcher Dr. Yemao Man at ABB company, and my colleague Wardah Mahmood. Shiliang and Yemao have answered many of my questions and provided valuable and helpful advice. In the second year of my Ph.D., I felt a lot of pressure and lacked confidence during a certain period, and Wardah provided many effective suggestions on how to adjust my mentality, and that helped me get through the hard times. Additionally, I would like to thank all my colleagues in the IDSE division and the wonderful atmosphere they collectively created. Working and studying in the IDSE division is the luckiest thing. In addition, I would also like to thank Dawen Liang, a senior software engineer from the United States, and Jiawen Wu, a doctoral student at the University of Jyväskylä, who have always supported me with their best wishes and encouragement.

Finally, I am deeply thankful to my wife Jinying Li, and my whole family. Without their support and blessings, I would not have been able to study abroad, let alone live abroad for my doctoral studies. In particular, I would like to express my gratitude to my wife for her incredible and endless love, support, and patience. Without her sacrifices and positive energy, I would not have been able to come this far.

\newpage
\thispagestyle{empty}

%% file: TexFiles/listPub.tex
\chapter*{List of Publications}
\label{A:Papers}

\section*{Appended publications}
This thesis is based on the following publications:
\renewcommand{\labelenumi}{[\Alph{enumi}]}
\begin{enumerate}




\item I.~David, M.~Latifaj, J.~Pietron, W.~Zhang, F.~Ciccozzi, I.~Malavolta, A.~Raschke, J.~Steghöfer, R.~Hebig\\
\newblock ``Blended Modeling in Commercial and Open-source Model-Driven Software Engineering Tools: A Systematic Study''\\
\newblock {\em Software and Systems Modeling (SoSyM), 2023, 22(1), pp. 415-447}.

\item J. Holtmann, J.~Steghöfer, W.~Zhang\\
\newblock ``Exploiting Meta-Model Structures in the Generation of Xtext Editors''\\
\newblock {\em 11th International Conference on Model-Based Software and Systems Engineering, SciTePress, 2023, pp. 218-225}.

\item W.~Zhang, R.~Hebig, J.~Steghöfer, J. Holtmann\\
\newblock ``Creating Python-style Domain Specific Languages: A Semi-automated Approach and Intermediate Results''\\
\newblock {\em 11th International Conference on Model-Based Software and Systems Engineering, SciTePress, 2023, pp. 210-217}.

\item W.~Zhang, J. Holtmann, D. Strüber, R.~Hebig, J.~Steghöfer\\
 \newblock ``Supporting Meta-model-based Language Evolution and Rapid Prototyping with Automated Grammar Optimization''\\
 \newblock {\em Revised and Re-submitted to Journal of Systems and Software, 2023}

\end{enumerate}

\newpage
\section*{Other publications}
The following publications were published during my PhD studies.
However, they are not appended to this thesis, due to contents overlapping that of appended publications or contents not related to this licentiate thesis.


\renewcommand{\labelenumi}{[\alph{enumi}]}
\begin{enumerate}
\item Wenli~Zhang, Weixing~Zhang, D. Strüber, R.~Hebig\\
 \newblock ``Manual Abstraction in the Wild: A Multiple-Case Study on OSS Systems’ Class Diagrams and Implementations''\\
 \newblock {\em Accepted in 26th International Conference on Model Driven Engineering Languages and Systems (MODELS). ACM. 2023.}
\item W.~Zhang, R.~Hebig, D. Strüber, J.~Steghöfer\\
 \newblock ``Automated Extraction of Grammar Optimization Rule Configurations for Metamodel-Grammar Co-evolution''\\
 \newblock {\em 16th ACM SIGPLAN International Conference on Software Language Engineering (SLE). ACM. 2023, pp. 84–96.}
 \end{enumerate}

%% file: TexFiles/contribution.tex
\thispagestyle{empty}
\section*{Research Contribution}
I was the main driver and contributor of papers C and D. Also, I have significant contributions to papers A and B. My contributions in these papers are classified according to the Contributor Roles Taxonomy (CRediT).

In Paper A, I did an entire gray literature review with Dr. Steghöfer and Dr. Hebig. I participated in developing the methodological design of the gray literature review and jointly performed the investigation of about 1,500 web pages. I jointly identified modeling tools in the gray literature and participated in data collection based on the identified tools. This data collection requires downloading and trying out the tool and completing the data collection by verifying the functionality and features of the tool. I reported the results of the gray literature review to the paper team and participated in writing the content of the gray literature.

In paper B, I worked with Dr. Steghöfer and Dr. Holtmann on the implementation of the project involved in the paper. To address the four technical challenges in this paper, I investigated technical manuals and open-source web pages. I provided technical solutions for two of the four challenges, i.e. ``content-assist for new model elements with unique names'' and ``scoping for cross-references''. I also investigated materials for the writing in this paper, i.e., finding related work. I also participated in the validation work of the paper, i.e., validating whether the engineering implementation we have completed conforms to the design of the technical solutions for solving the four technical challenges. I once provided a presentation on the academic work of the EATXT editor at the SE division seminar, which included the research content of paper B.

In paper C, I completed the conceptualization of engineering. During the engineering, I developed a script to automate grammar modification work and applied this script to multiple languages as a validation. I wrote the initial full draft of this paper and then combined it with the review comments of Dr. Hebig and Dr. Steghöfer to make it perfect. Dr. Hebig suggested I add the validation part which is one of the key steps to make the paper perfect. Before this paper, Dr. Holtmann, Dr. Steghöfer, and I jointly implemented the development of the EATXT editor which the content is involved in paper B. The concept of paper C originally originated from the EATXT technical discussion, but it was not implemented in EATXT in the end. Therefore, I extracted the concept separately and searched for the language background of the case to complete paper C.

The concept of paper D was first proposed by my co-author Dr. Hebig, while the concept originated from a script I developed, so I further refined and determined the concept. I, together with Dr. Holtmann and Dr. Hebig, performed a large number of systematic analyses in this paper, most of which were performed by me, resulting in many documents and data. The development of the paper also involved a significant amount of Java programming, most of which was performed by me. The model-driven architecture idea of the software developed in this paper came from Dr. Holtmann. Dr. Steghöfer implemented the initial steps, and I fully developed the architecture. Early in the paper, I made contributions to the investigation by checking whether initial sample languages identified by Dr. Steghöfer were appropriate for our research context and purpose. The entire process of the paper was managed as a project, and the initial administration plan was developed by Dr. Steghöfer, later, the plan was maintained by me, in particular, during the revision in the journal revision process. Supported by Dr. Holtmann and Dr. Hebig, I performed the validation of the research results of the paper (i.e., the software we developed), taking on the majority of the effort. I made major contributions to the manuscript in both its initial and the revised version. For the revised version, I addressed the majority of the reviewer comments, supported by Dr. Strüber, who joined the team of authors for the revision.
Additionally, I gave presentations on the paper on several occasions.

\begin{table*}[hb]
\scriptsize
\centering
\label{tab:CRediT}
\begin{tabular}{@{}lllll@{}}
\toprule
\textbf{Role}   & \textbf{Paper A}     & \textbf{Paper B} & \textbf{Paper C} & \textbf{Paper D} \\
\midrule
    Conceptualization &  &  & X & X \\  
    Data curation & X & X & X & X \\
    Formal analysis &  &  &  &   \\
    Funding acquisition &  &  &  &   \\
    Investigation & X & X & X & X  \\
    Methodology & X &  & X &   \\
    Project administration &  &  & X & X  \\
    Resources &  &  &  &   \\
    Software &  & X & X & X  \\
    Supervision &  &  &  &   \\
    Validation &  & X & X & X  \\
    Visualization &  & X & X & X  \\
    Writing – original draft & X & X & X & X  \\
    Writing – Review and Editing &  &  & X & X  \\
\bottomrule
\end{tabular}
\end{table*}

%% file: TexFiles/introduction.tex
\chapter{Introduction}
\label{sec:intro_lic}
Blended modeling is a rapidly emerging modeling technology that involves seamless interactions between multiple notations (i.e., concrete syntax) and a single model (i.e., abstract syntax), allowing for a certain degree of temporary inconsistency~\cite{Ciccozzi.2019}. Different modeling notations have their respective advantages. For example, the visual connections between elements representing domain concepts in graphical notations make it easier for users to understand the relationships between concepts, the tree-based notation of the model makes its hierarchical structure clearer for users, and textual notations have unique advantages in fast editing and global replacement. By blending different modeling notations in the same modeling tool and making them adhere to the same abstract syntax, respective advantages of different modeling notations will be available at the same time. Modifications made to the model in one notation can be synchronized to the other notations. Blended modeling increases modeling flexibility and efficiency as well as productivity. Additionally, engineers can choose their preferred notation for modeling based on their preferences.

Adding textual notations to a language through development is a way to improve a language's blended modeling capability.
There are many existing tools for developing textual domain-specific languages (DSLs), and Xtext~\cite{Xtext} is one of the popular textual DSL development tools~\cite{erdweg2015evaluating}. 
Xtext replies on the Eclipse Modeling Framework (EMF)~\cite{EMF} and uses its Ecore (meta-)modeling facilities as a basis. Developing a textual DSL in Xtext involves two main artifacts: a grammar, which defines the concrete syntax of the language, and a meta-model, which defines the abstract syntax. Xtext allows either the grammar or the meta-model to be created first, and then automatically generates the one from the other (or alternatively, writing both manually and aligning them). 
We use the term ``model-driven engineering (MDE) approach'' to refer to the strategy of first creating a meta-model and then generating a grammar from that meta-model.
Applying the MDE approach, language engineers
can generate a textual grammar from the meta-model instead of designing the grammar from scratch. 
Additionally, the MDE approach is most suited for the scenario where multiple concrete syntaxes adhere to a single abstract syntax, 
which coincides with the goal of blended modeling.

However, in the MDE approach, grammars generated from metamodels (hereinafter referred to as \textbf{generated grammars}) often require adaptation before being put into use. The grammar generated by Xtext from the meta-model is composed of grammar rules. These grammar rules and their contained attributes, keywords, and other elements always follow a fixed format. In a real DSL application, some of these elements may be unnecessary or not expressive enough. For example, to express the semantics of ``assign \texttt{0} to the variable \texttt{value}'', the default output generated by Xtext ``\texttt{value 0}'' does not agree with how variable assignment is typically represented. One solution is for the language engineer to modify the grammar definition, i.e., add an equal sign ``='' after the keyword ``value''. Moreover, the curly braces included by default in the generated grammar often lead to deep nesting in the program, which makes the grammar cumbersome to use.

\begin{figure*}[tb]
  \centering
  \includegraphics[width=\linewidth]{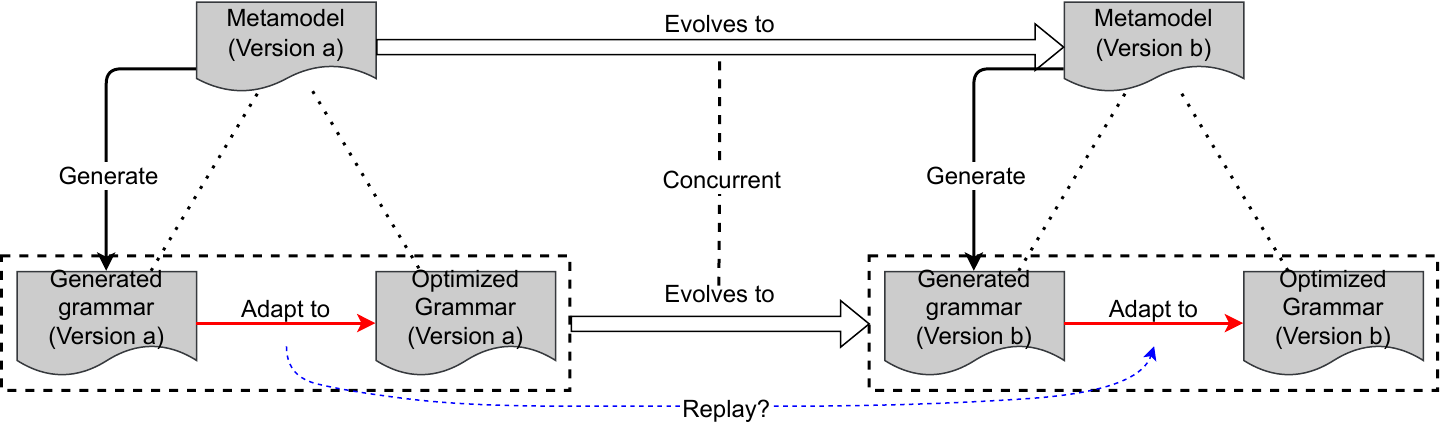}
  \caption{In a MDE approach, the grammar is adapted before being put into use, and when the language evolves, the grammar generated from the evolved meta-model needs to be adapted again.}
  \label{fig:core_problem}
\end{figure*}

Because of that generated grammars require manual adaptation before the can be used. However, a further problem faced when adapting grammar is related to the workload of adaptation. When manually adapting a generated grammar, language engineers are faced with repeating the same operation across many grammar rules. For example, moving an attribute to the outside of the container braces in many grammar rules would be time-consuming. Another problem of grammar adaptation comes from the evolution of language. There is a practical scenario showing the evolution of a language (as shown in Figure~\ref{fig:core_problem}). When the meta-model evolves, the original grammar would be out of date, so then the grammar needs to be regenerated from the evolved meta-model. 
However, the grammar generated from the evolved version of the meta-model does not contain the manual improvements in the previous version.
In this case, language engineers have to manually adapt the newly generated grammar once again and thus leading to repetitive work. Note that this is an issue of the MDE approach to language development. We will later discuss an alternative, specifically, a grammar-driven approach, and the respective advantages and benefits.

Finally, after the grammar is ready, a complete infrastructure including a parser, compiler, etc. can be obtained based on the grammar using Xtext. 
However, some comment features of modern editors, such as template proposals, are still not supported in the editor composed of this infrastructure which is based on Xtext out-of-the-box. 
We identified here another issue that needed to be addressed, i.e., the default Xtext's generator capabilities are limited.


In this thesis, we propose an approach that can support the co-evolution of meta-models and grammars as language engineers develop textual languages in a meta-model-based MDE setting.
To this end, first, we studied the state-of-the-art and practice of modeling tools that support blended modeling by conducting a systematic literature review. After studying the limitations and opportunities of existing modeling tools that support blended modeling, we decided to explore and improve blended modeling technologies on a specific case language. As a start, we developed a textual language (i.e., EATXT) for EAST-ADL based on the MDE approach. In this process, to address the limitations of the default Xtext generators' capability and its implementation, we proposed ways in which language engineers can extend the Xtext generators' capability and its implementation according to their own needs. Meanwhile, to solve the inherent problems of grammars generated from meta-models, we first proposed a semi-automatic method that can change the language with the generated grammar to a Python-style language. To solve the problem that manual improvements to the generated grammar cannot be replayed in evolved versions, we proposed a more general grammar adaptation method, i.e., GrammarOptimizer (we name grammar adaptation \textbf{grammar optimization}), which can optimize the generated grammar of languages in different styles. Moreover, its optimization on the generated grammar of the previous version is saved in the form of configurations. These configurations can be reused in the generated grammar of the evolved version, thereby supporting the co-evolution of meta-model and grammar.


To guide our research, we address the following research questions in this thesis:

\emph{\textbf{RQ1:} What are the user-oriented characteristics of modeling tools most suitable for supporting blended modeling?}

By answering this research question, we aim to identify the external characteristics of modeling tools that are relevant to their adoption and use, e.g., the notations (types) they support, etc.

\emph{\textbf{RQ2:} What are the realization-oriented characteristics of modeling tools most suitable for supporting blended modeling?}

By answering this research question, we aim to identify the internal characteristics of modeling tools, as well as the technologies used to implement these characteristics, such as the implementation platforms they use, etc. Answering the above two questions is crucial because practitioners can learn from the answers about the limitations of current blended modeling tools and how they can improve the tools, and researchers, including us, can learn from the answers the state of the practice in blended modeling tools, including the gaps that need to be filled.

\emph{\textbf{RQ3:} How can we build a solution to adapt generated grammars to produce the same language as available expert-created grammars?}

We developed a textual language for EAST-ADL to explore and improve blended modeling technologies after studying the state-of-the-art and practice of blended modeling tools and we proposed a method that optimizes the generated grammar and we propose a generalized grammar optimization method. By answering this research question, we aimed to evaluate the generalizability of this optimization method, i.e. whether it can adapt the generated grammar to languages to produce the same language as available expert-created grammars.

\emph{\textbf{RQ4:} Can our solution support the co-evolution of generated grammars when the meta-model evolves?}

By answering this research question, we aim to evaluate whether the proposed grammar optimization method supports the co-evolution of meta-models and grammars as the language evolves.





\section{Background and Related Work}

In this section, I will first introduce the background and related work such as blended modeling, Xtext, and metamodel-based DSL engineering, co-evolution in MDE contexts, and then other concepts such as EAST-ADL will be introduced in subsequent chapters.

\subsection{Blended Modeling}
The integration of graphical modeling and textual modeling was not a new thing~\cite{addazi2017towards,atkinson2013harmonizing,perez2008domain,charfi2009hybrid,scheidgen2008textual,maro2015integrating}, however, Ciccozzi et al. formally conceptualized \textbf{blended modeling} for the first time in~\cite{ciccozzi2019blended}, defining it as follows:

\emph{``Blended modeling is the activity of interacting seamlessly with a single model (i.e., abstract syntax) through multiple notations (i.e., concrete syntaxes), allowing a certain degree of temporary inconsistencies.''}

It distinguishes itself from Multi-View Modeling (MVM)~\cite{von2012multi} and Multi-Paradigm Modeling (MPM)~\cite{mosterman2004computer} by possessing the following three characteristics: Multiple notations, Seamless interaction, and Flexible consistency management.

There are existing many modeling tools, e.g., SequenceDiagramOrg~\cite{SequenceDiagramOrg} which blends both graphical and textual notations. However, which tools support blended modeling and how they support it remains largely unknown. 

\subsection{EAST-ADL}
EAST-ADL~\cite{cuenot2007managing} is an architectural description language used in the domain of automotive embedded systems and is based on a metamodel with approximately 300 meta-classes and a hierarchy in which nested elements describe different aspects of the electronic vehicle system. As mentioned before, we use EAST-ADL as a case language to explore and improve blended modeling technology. EAST-ADL can be edited by EATOP~\cite{eatop-bitbucket} which is an eclipse-based modeling tool specifically for EAST-ADL, which provides tree-based and table-based editing capabilities but does not provide text symbol-based editing capabilities.

\subsection{Xtext and Meta-Model-Based DSL Engineering}
Eclipse Xtext is a framework used for the development of programming languages and domain-specific languages (DSLs)~\cite{behrens2008xtext}. It is one of the many popular DSL development tools~\cite{erdweg2015evaluating}. Xtext supports both grammar-based and meta-model-based DSL development approaches. In the grammar-based approach, users directly define the grammar of the DSL, and the meta-model can be generated from this grammar. Xtext artifacts are then generated from the grammar, providing the foundation for textual editor infrastructure. Xtext also supports the meta-model-based approach, where users first represent domain concepts and their relationships by creating a meta-model. From this meta-model, grammar is generated, and subsequently, a textual editor is generated from this grammar. The generation process is controlled by the Modeling Workflow Engine (MWE2). Running MWE2 allows for the generation of a full infrastructure, including a parser, linker, type checker, compiler, and editing support for Eclipse, any editor that supports the Language Server Protocol, and web browsers~\cite{Xtext}.

At the beginning of the language creation process, it does not need to be clear whether the new language is text-based, graphical, or both~\cite{kleppe2007towards}. In fact, following the philosophy of blended modeling~\cite{ciccozzi2019blended}, there are good reasons to support multiple syntaxes, as different developers are likely to benefit from different syntax paradigms. The MDE approach is most suited for this philosophy. Additionally, generating grammar directly from the meta-model avoids designing grammar from scratch. Therefore, in this thesis, we adopted the MDE approach for developing the DSL.

The full infrastructure used for generating text editors, known as the language generator (in the form of Java code), is available out of the box and can be extended or replaced as needed. For large languages such as EAST-ADL, the default content assistant features may be insufficient, allowing users to modify and extend the existing generator. Furthermore, when Xtext generates grammar from the meta-model, it generates a grammar rule for each meta-class and also generates an attribute in the grammar rule for each attribute in a meta-class, resulting in grammar closely adhering to the meta-model. The generated grammar has a default format, where each grammar rule includes a keyword with the same name as the rule and is enclosed in curly braces, containing multiple attributes. Each attribute includes a keyword followed by an attribute string. This inherent format introduces certain challenges, e.g., redundant keywords and curly braces, or the default keywords not effectively conveying semantics, etc. These are inherent characteristics and limitations of grammar generated from the meta-model.

\subsection{Co-Evolution in MDE Contexts}
In model-driven engineering, it is well-known that evolutionary changes to an artifact may affect other artifacts, which leads to several co-evolution scenarios.
The most prominent one is \textit{meta-model/model} co-evolution, in which a meta-model is evolved and corresponding instances have to be updated to stay in sync with the meta-model.  This scenario has inspired a substantial body of work.  Hebig et al.~\cite{hebig2016approaches} survey 31 relevant approaches, classifying them according to their support for change collection, change identification, and model resolution.
Beyond meta-model/model co-evolution, co-evolution between meta-models and other MDE artifacts have received attention as well, including associated OCL constraints \cite{khelladi2017semi}, model transformations \cite{kusel2015consistent,khelladi2017exploratory},  code generators \cite{lombardi2021co}, and graphical editor models \cite{di2011automated}.
Inconsistencies between evolved meta-models and general MDE artifacts have also been addressed in the context of \textit{technical debt management}, with an approach that assists the modeler with the aid of interactive visualization tools \cite{di2023modeling}.
However, except for GrammarOptimizer described in Chapter~\ref{chap_paper_d}, on which we build and improve with our contribution, we are not aware of previous work on meta-model/grammar co-evolution.

Model federation~\cite{guychard2013conceptual,golra2016addressing,drouot2019model} deals with the challenges of keeping several models synchronized, which is related to our addressed co-evolution scenario. However, to the best of our knowledge, there is no previous work that applies model federation techniques to grammars. Previous work is often focused on establishing links between the different involved artifacts, which, in our scenario, is a non-issue. However, the actual modification for keeping several artifacts synchronized is often simpler if only models are involved, than in our case deals with concrete textual syntaxes. For example, the order of attributes in the grammar does not have to be consistent with the corresponding meta-model attributes but can be changed freely according to the developer’s design intention.
In fact, the approach enabled by our contribution could be used to augment available model federation frameworks to make them applicable to grammars as well. 

\section{Methodology}
\begin{figure*}[htp]
  \centering
  \includegraphics[width=\linewidth, angle=-90, width=4in]{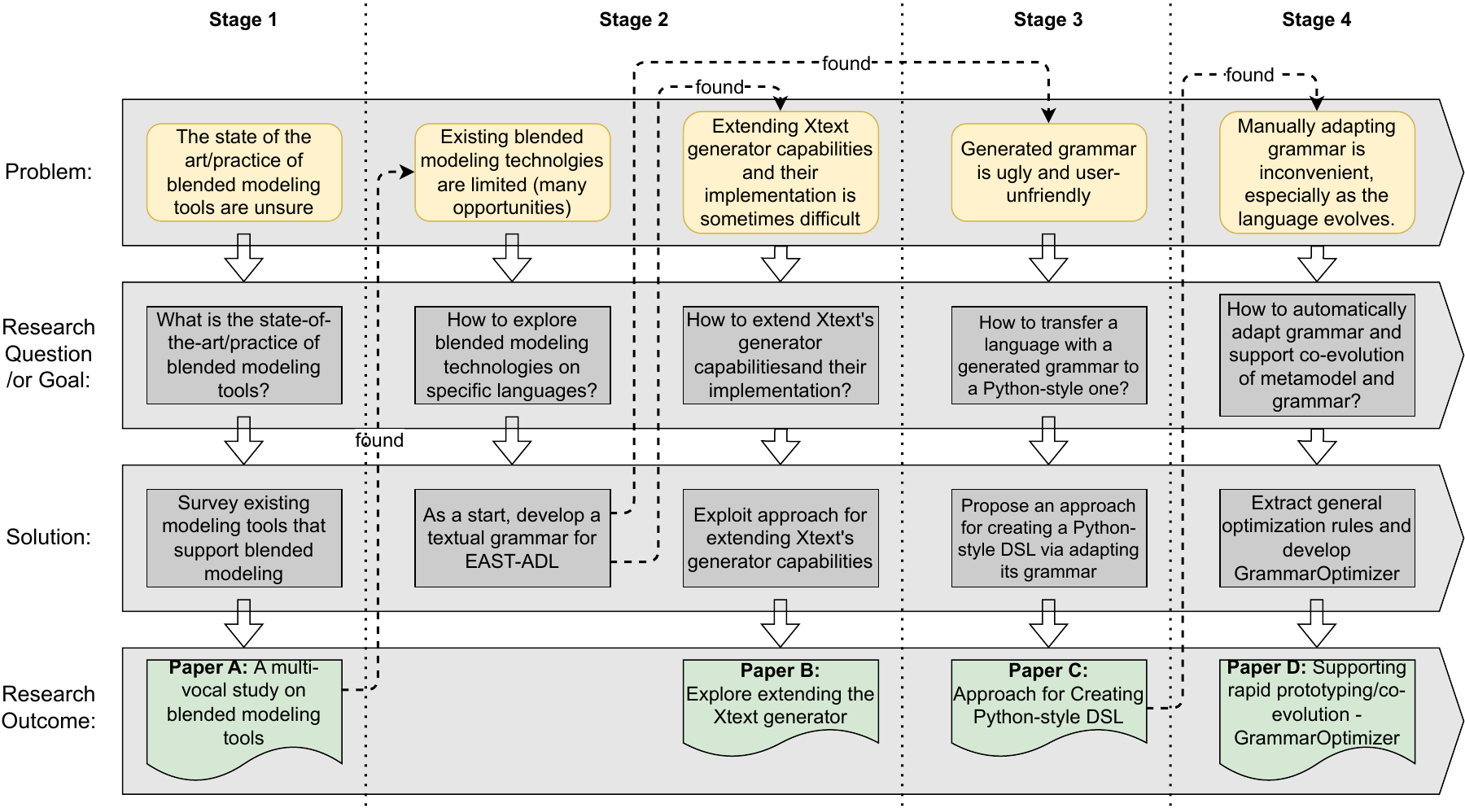}
  \caption{The research included in this thesis involves multiple problems. We have implemented different research activities and obtained some research results for these different problems.}
  \label{fig:overview_of_research}
\end{figure*}

Figure~\ref{fig:overview_of_research} depicts all the studies included in this thesis and the problems and methodologies involved. 
We went through four stages to complete these studies. 
In \textbf{Stage 1}, to understand the potential of current commercial and open-source modeling tools to support blended modeling, we have designed and carried out a systematic literature review.
Through this study, we discovered the shortcomings of existing blended modeling technology and the opportunities of improving blended modeling technologies, which motivated us to explore and improve blended modeling technology. In \textbf{Stage 2}, we developed a textual language for EAST-ADL as a starting point for exploring and improving blended modeling technologies. We introduced ways to extend the Xtext's generator capabilities. 
In \textbf{Stage 3}, we proposed a semi-automated method for turning a language with generated grammar into a Python-style language, thereby making the language more concise and user-friendly. 
In \textbf{Stage 4}, we extracted grammar optimization rules from the analysis on seven case languages, and developed a grammar optimization tool GrammarOptimizer to include these general optimization rules, and finally evaluated our approach through multiple exemplar languages. 
The above methodologies aimed to address the identified problems described in Section~\ref{sec:intro_lic}, and will be introduced in detail in the following subsections.

\subsection{Stage 1: Multi-Vocal Literature Review}

\label{sec:problem_survey}
To understand the state-of-the-art and practice of modeling tools that support blended modeling, we conducted a Multi-vocal Literature Review (MLR) in this stage. An MLR is a form of Systematic Literature Review (SLR) encompassing formally published literature (e.g., journal and conference papers) and Gray Literature (GL), such as blog posts and whitepapers~\cite{garousi2016and}. The specific focus of the MLR in this thesis is blended modeling, making the SLR an academic synthesis of evidence of blended modeling. As a supplement, we also conduct a Grey Literature Review (GLR). 

Software engineering practitioners typically lack the time, channels, or expertise to review formally peer-reviewed papers. They are more likely to turn to sources like online blogs, technical reports, and other GL. Furthermore, they are willing to contribute their thoughts and experiences in the form of GL, which can serve as a valuable source of data for research~\cite{garousi2020benefitting}. 
Moreover, the research objective of the investigation was modeling tools, which often provide specific information about the tools in their user manuals, such as whether they support textual modeling notations. Based on the above advantages of GL, we conducted GLR as the supplement for this investigation.

At the same time, GL also have their inherent limitations. I.e., GL have not gone through conventional publication channels, which means that they do not undergo rigorous and formal peer review~\cite{petticrew2008systematic}, and gray literature can be difficult to search and retrieve for evidence synthesis. Therefore, in this investigation, we combine SLR with GLR instead of adopting one of them alone.

We followed the common process for conducting a GLR, which involves the following steps~\cite{kamei2021grey}:
1) Select where to search, such as Google. 2) Set search strings. The search engine will search based on them. 3) Source selection. There may be many search results, so it is necessary to set a selection criterion, such as what sources should be excluded, and perform the selection according to this criterion. 4) Data extraction. Each source may provide a variety of different information, so in this step, it is needed to set the extraction goal, i.e., what data to extract. 5) Data synthesis. This step is for summarizing and analyzing the extracted data to draw meaningful insights and conclusions.

MLR, which combines SLR and GLR, provides several advantages for our investigation in this thesis. SLR ensures rigorous analysis of published academic literature to provide a solid foundation for existing research. On the other hand, GLR expands the scope and captures information lacking in research papers, such as descriptions of product features in user manuals in modeling tools. This combination allows our investigation to access a wider range of data sources, resulting in a more comprehensive investigation of blended modeling tools and reducing publication bias. Practitioners and researchers, including ourselves, can benefit from such a comprehensive investigation to enhance decision-making and facilitate the identification of gaps and future research directions in academic and practical contexts.

\subsection{Stage 2: Extending Xtext's Generator Capabilities}
\label{sec:explore}
The results of MLR in Stage 1 showed that textual notation is a popular type of notation. However, there are still modeling tools that do not support textual notations, including EAST-ADL. We used EAST-ADL as a case language and improved its blended modeling capabilities by developing textual notations for it. EAST-ADL is a language based on a large meta-model. The ground fact that Xtext can generate a textual grammar directly from the meta-model allows us to avoid designing a grammar from scratch that involves a large number of domain concepts. However, due to limitations of Xtext's default generator capabilities, part of the common features of modern editors (e.g., template proposals) are not supported by DSLs based on Xtext out-of-the-box.
Therefore in this stage, we take EATXT as an example to describe how language engineers can extend Xtext's generator capabilities according to their own needs. 

Our research methodology was based on \textit{design science}~\cite{wieringa2014design} and consisted of four iterations. We extended one capability of the Xtext generator in each of the first three iterations and improved the scoping feature in the last iteration. We added new features incrementally, i.e., for each iteration, we provide the editor with added features from that iteration to our industrial partners and receive feedback from them.

The work is limited in the scope of using Xtext to create textual languages for DSLs. Xtext generates Xtext artifacts such as parser from the grammar to build the editor.
The generation process is controlled by a workflow for the Modeling Workflow Engine (MWE2)~\cite{MWE2}. Xtext provides a language generator that can be customized and extended with custom fragments. A fragment generates code based on the generator’s configuration, the grammar, and the corresponding meta-model. Xtext out-of-the-box provides a number of such fragments, which can be extended or replaced. These fragments add a number of features to the generated editors. We use Xtext’s ability to change the standard configuration to add custom fragments that provide better formatting, content-assist, and template proposals. These custom fragments are written in Xtend~\cite{Xtend}.

For the scoping feature, our approach is to generate a cross-reference lookup map from the plug-in's activator. This generation traverses the metamodel exactly once with a complexity of \textit{O(n)}. This lookup map contains the corresponding type of the cross-reference target for any source context type, and we compute it by iterating over all cross-references in the metamodel. We generate the lookup map during the first activation of the plugin. After that, the scoping/\texttt{(language name)ScopeProvider}.java accesses it via an interface but does not need to perform the same computation on every cross-reference content-assist keystroke. The lookup map is implemented as a Java HashMap whose \texttt{get()} method has a complexity of \textit{O(1)} in most cases.


\subsection{Stage 3: Python-Style Prototyping}
\label{sec:prototyping_lic}
As mentioned previously, when we explored and improved blended modeling technologies based on the case language EAST-ADL, we identified a problem with the grammar generated from the meta-model, i.e., the generated grammar was not concise and user-friendly. Our industrial partner provided some requirements of appearance for the language. For example, they requested that curly braces and keywords be reduced in the grammar to make the language more concise. 
Python is a language renowned for conciseness and provides a very clean coding style, and it is considered easy to learn~\cite{gmys2020comparative}. Therefore, in this stage, we proposed a method that turns a DSL with a generated grammar into a Python-style language.

Our methodology at this stage was based on \textit{constructive research}~\cite{crnkovic2010constructive}. First, we took a small architecture description language DemoADL as a case language,
and generated a grammar from the meta-model of it and wrote example code that conforms to the generated grammar. We then wrote pseudocode expressing the same content in Python style. We compared the two pieces of source codes and summarized the gaps between languages in those two styles. To address these gaps, we proposed a series of steps for adapting the text of the grammar and developed a script to semi-automate these adaptation steps. 

To evaluate the usability and generality of the proposed method, we apply it to two other DSLs, i.e., Xenia and ACME. With the help of the script, we completed the adaptation of the generated grammars for these two languages. We observe whether the adapted grammar is with a Python-style feature, i.e., using spaces and indents to express hierarchy. For each of the two languages, we successively compared the adapted grammar with the generated grammar, and compared the adapted grammar with the expert-created grammar, to observe the improvement of conciseness, and being as compact as the expert-created grammar.

\subsection{Stage 4: Generalize Grammar Optimization Approach}
In the last stage, we learned that some generalization (for the case of Python styles) is possible, which can be re-applied to other languages. Therefore, in this stage, we aimed to develop a more complete system that can adapt the generated grammar, which is rule-based and can be applied to different styles of languages. Also, our work on generating infrastructure in stage 2 inspired our ideas for generating adaptations. Therefore, the proposed rule-based system in this stage will include rule configurations that support the generation of grammar adaptations in the evolved version. In this section, we will introduce the methodology we adopted in \textbf{Stage 4}, including the extraction of candidate optimization rules, two iterations, and how to evaluate the proposed methodology.

We selected seven case languages, i.e., ATL\footnote{\url{https://eclipse.dev/atl/}}, Bibtex\footnote{\url{https://www.bibtex.com/}}, DOT\footnote{\url{https://graphviz.org/doc/info/lang.html}}, SML\footnote{\url{http://scenariotools.org/scenario-modeling-language/}}, Spectra\footnote{\url{http://smlab.cs.tau.ac.il/syntech/spectra/index.html}}, Xcore\footnote{\url{https://wiki.eclipse.org/Xcore}}, and Xenia\footnote{\url{https://github.com/rodchenk/xenia}}, and obtained their meta-models and expert-created grammars. To smoothly generate grammar from the meta-model, we slightly process some meta-models, e.g., adding values to the namespace \texttt{URI} and \texttt{prefix} in Bibtex. We took two iterations to analyze the grammars of these languages and extracted candidate grammar optimization rules from the analysis. To reduce the complexity of the analysis, we excluded OCL expression language parts from the meta-models and grammars of both ATL and SML.

Regarding grammar analysis, we drew on ideas from \textit{design science}. We analyzed different languages in different iterations and incrementally added extracted candidate optimization rules.
We analyzed four of the seven languages in the first iteration. This analysis identifies differences between the generated grammar and the expert-created grammar, and we extracted candidate optimization rules from this analysis. These candidate rules can optimize the generated grammar, and the optimized grammar produces the same language as the expert-created grammar. We obtained a set of optimization rules by excluding the duplicate candidate rules. In the second iteration, we applied the optimization rules extracted from the first iteration to optimize the generated grammars for the other three languages and analyzed any remaining differences between the optimized grammars and the expert-created grammars. If there are differences, we derived more optimization rules. We developed the grammar optimization tool GrammarOptimizer. In this development, we implemented the optimization rules.

We evaluated the method proposed in this stage (i.e., GrammarOptimizer) from two aspects. On the one hand, we applied it to the seven case languages to evaluate its usability and generalization. For each language, we configured the GrammarOptimizer to optimize the grammar generated by that language, aiming to adapt it to a grammar that produces the same language as the expert-created grammar. On the other hand, we applied GrammarOptimizer to multiple versions of two languages (i.e., EAST-ADL and QVTo\footnote{\url{https://wiki.eclipse.org/QVTo}}) to evaluate its support for language evolution. For the first version of each language (e.g., QVTo's V1.0), we configured GrammarOptimizer to optimize the generated grammar, and the optimized grammar can produce the same language as the expert-created grammar. We reused these optimization rule configurations so that they optimize the generated grammar of evolved versions (e.g., QVTo's V1.1). We then compared this optimized grammar and the expert-created grammar and modified the configurations based on the difference(s). The goal was that GrammarOptimizer could completely optimize the grammar driven by the modified configurations, i.e., the optimized grammar could produce the same language as the expert-created grammar. We evaluated GrammarOptimizer's support for language evolution by counting the number of modifications to optimization rule configurations in the evolved version.

\section{Results and Evaluation}
\label{sec:result}
As mentioned before, the work of this thesis consists of four stages, and the four stages of work are highly connected. Our research results in the first stage comprehensively reported the state-of-the-art and practices of blended modeling tools, including the limitations of existing tools. This motivates us to explore and improve existing blended modeling techniques. As a start, we developed a textual language EATXT for the case language EAST-ADL in Stage 2. During this development, we used EATXT as a case to demonstrate how language engineers can extend the Xtext generator capabilities and its implementation according to their own needs. While designing EATXT's grammar, we identified another problem, i.e., the grammar generated from the meta-model was cumbersome and non-user-friendly. To this end, in stage 3 we proposed a semi-automated method for converting a language with a generated grammar into a Python-style language, thus improving its conciseness and user-friendliness. We learned from the results of Stage 3 that the generalization of some rules about grammar adaptation is possible, and we aimed to support the co-evolution of meta-models and grammars, which motivated us to propose a new solution in Stage 4, the solution is a rule-based system consisting of general optimization rules. It can support the optimization of different languages and support the co-evolution of meta-models and grammars. In this section, we will elaborate on the results of the four stages respectively.

\subsection{Results and Evaluation in Stage 1}
\label{sec:result_paper_a}
To understand the potential of existing modeling tools to support blended modeling, we conducted a GLR in \textbf{Stage 1} in which we reviewed nearly 5,000 academic papers and nearly 1,500 gray literature. Based on these, we identified 133 candidate modeling tools that involve blended modeling technologies and finally identified 26 of them as the most advanced and practical modeling tools that represent the current range of modeling tools. We investigated these 26 modeling tools for their support of various blended aspects, such as inconsistency tolerance, and then obtained many results.

The obtained results were divided into results on user-oriented characteristics and results on realization-oriented characteristics. From the perspective of user-oriented characteristics, the results showed that more than half of the 26 tools supported two modeling notations, and about one-third of the tools supported three modeling notations. Among them, Boston Professional~\cite{boston} and TopBraid Composer~\cite{topbraid} supported four modeling notations. Graphical notation and textual notations were the most available notations, i.e., all of the 26 tools supported graphic notations, and most of them supported textual notations.
Usability aspects are generally difficult to measure, so we evaluated a tool's usability by evaluating whether it supports multi-notations visualization and provides seamless navigation between notations. The results showed that all 26 tools supported the visualization of two or more notations. Our evaluation of the navigation across notations was mainly divided into two points, i.e., whether the navigation between multiple notations is synchronized, and the speed of navigation between different notations. The results showed that more than half of the tools provided simultaneous navigation facilities, while the majority of the tools provided immediate navigation from one notation to another.
Flexibility is the tolerance of inconsistent user-related embodiment at various levels of abstraction across the modeling stack and various modeling facilities. We considered three categories of flexibility, i.e., model, language, and persistence. The results showed that most of the 26 tools did not support deviations between different notations describing the same model, while the other six supported model-level flexibility. Second, most tools did not allow inconsistencies between the model and the language, and four tools allowed such inconsistencies. The tool Umple~\cite{umple} supported both model-level and language-level flexibility. Third, most tools did not support persisting inconsistency models.

We elaborate on our findings related to the realization characteristics of sampled tools from the following three aspects, i.e., mapping and platforms, change propagation and traceability, and inconsistency management. 
The mapping between abstract syntax and notations is usually implemented in a parser- or projection-based fashion. 
In the parser-based approach, the user modifies the model through different notations and the parser generates an abstract syntax tree. 
However, in the projection method, the abstract syntax tree is modified directly. 
The results showed that 22 of the 26 tools implement parser-based editors, while four tools have projection tools. 
The platforms used by these 26 sampled tools are almost all Eclipse, and only mbdeddr~\cite{mbeddr} is the only one of these tools based on MPS~\cite{mps}. 
In addition, MagicDraw~\cite{magicdraw} supports multiple platforms.
During the data extraction phase, we failed to obtain any useful information in the categories ``change propagation'' and ``traceability''.
In the context of inconsistency management for modeling tools, a majority of the tools (58\%) lack visualization for inconsistencies (Table 19). In terms of inconsistency management types, there are two fundamental approaches: prevention and allow-and-resolve. Half of the tools (50\%) focus on prevention, while others either manage inconsistencies on the fly (42\%) or on-demand (8\%) (Table 20). When it comes to inconsistency management automation, 50\% of the tools follow a preventive approach and do not offer inconsistency resolution, while the remaining 13 tools provide varying degrees of automation for resolving inconsistencies, with only two relying on manual resolution (Table 21).

\subsection{Results and Evaluation in Stage 2}
\label{sec:result_B}

As we mentioned in Section~\ref{sec:explore}, we developed a textual language (i.e., EATXT) for EAST-ADL as a starting point for exploring and improving blended modeling technology, and during this period demonstrated how language engineers could extend Xtext's generator capabilities according to their own needs. In this section, we present the research results of Stage 2.

Firstly, in the case of EATXT, the generated template file contains 194 code templates with a total of more than 1,000 XML lines and covers all meta-classes of the metamodel and their mandatory sub-elements. 
Secondly, for each meta-class with the mandatory attribute, our approach generates a proposal provider that includes a corresponding overriding content-assist method which proposes a unique name. Overall, in the case of EATXT, the generated EatxtProposalProvider encompasses 188 such methods. 
Thirdly, for the formatter feature, we implemented the generator fragment formatting2/EatxtFormatter2Fragment.xtend as part of the plugin. This fragment is executed by the MWE2 workflow and automatically generates the formatter class formatting2/EatxtFormatter.xtend as part of the same plugin. The generated formatter class in the EATXT case encompasses 51 dispatch methods~\cite{lang_impl} and 141 calls of the formatting methods for the nested sub-elements.
Fourthly, we generated a cross-reference lookup map in the activator of the plugin (i.e., org.bumble.eatxt in the case of EATXT). In the EATXT case, the map encompasses the source context meta-classes and corresponding target metaclasses for 261 cross-references of the EAST-ADL metamodel.

\subsection{Results and Evaluation in Stage 3}
\label{sec:result_paper_c}
In Stage 2, we developed a textual language (i.e., EATXT) for EAST-ADL. When designing the grammar for EATXT, we identified the problem that the grammar generated from the metamodel was usually cumbersome and not user-friendly. To improve the conciseness and user-friendliness of the language with a generated grammar, in this stage, we proposed a method that can transform the language into a Python-style language. In this section, we present the results and our evaluation in Stage 3.

Firstly, we compared the program conforming to the generated grammar and the program in Python style, and observed that the program conforming to the generated grammar has issues in the following aspects: 1) inappropriate positioning of identifiers, 2) heavy separation of code blocks; 3) duplicate keywords, and 4) nested curly braces.
In this regard, we proposed a method to address these problems, which consists of steps that directly modify the text of grammar definition. The steps are: 1) introduce the white-space-awareness feature and remove curly braces, 2) reposition the identifier, 3) remove commas, and 4) refine keywords (especially remove duplicate keywords).
As stated in Section~\ref{sec:prototyping_lic}, we semi-automated these adaptation steps by developing a script.

We applied the proposed method along with the script to two other DSLs, i.e., Xenia and ACME. For each of them, we compared the adapted grammar to the generated grammar and compared the adapted grammar to the expert-created grammar. The comparison showed that the adapted grammar was more concise and user-friendly than the generated grammar, and like Python, the program used whitespace and indents to express hierarchy. Moreover, the adapted grammar was closer to the expert-created grammar in terms of compactness.
Our case languages Xenia and ACME were different languages, which showed that the proposed method and its script could be adapted to different DSLs. Language engineers could use it to quickly reach a Python-like grammar, which could then be used as a basis for further refinement of the grammar.

\subsection{Results and Evaluation in Stage 4}

In Stage 3, we proposed a method that makes the grammar of a language more concise and user-friendly by adapting the text of the generated grammar. We learned that some generalization (for the case of Python style) is possible, which can be re-applied to other languages. Therefore, we provide a solution (i.e. GrammarOptimizer) in stage 4. GrammarOptimizer provides general grammar optimization rules, can support the adaptation of different styles of languages, and supports the co-evolution of meta-models and grammars. We present the results of stage 4 in this section.

Before providing the solution GrammarOptimizer, we completed the following two works. First, through two iterations of comparative analysis of generated grammars and expert-created grammars for seven case languages, we extracted 56 general grammar optimization rules. Secondly, we developed an Eclipse-based grammar optimization tool, i.e., GrammarOptimizer. In this development, we implemented 56 general grammar optimization rules.

To evaluate the usability and generalizability of the proposed method, we applied GrammarOptimizer to the seven case languages, i.e. we used it to optimize the generated grammars of these case languages. The results showed that the generated grammars of all DSLs except Spectra can be fully optimized by GrammarOptimizer until they produce languages as same as the expert-created grammars, i.e., the optimized grammar produces the same language as the expert-created grammar. For Spectra, we were able to use GrammarOptimizer to optimize it to be very close to the expert-created grammar of Spectra, i.e., 96.30\% (54/56) of the grammar rules could be optimized to be equivalent to the corresponding grammar rules in the expert-created grammar. This was a limitation of this method, which will explained in detail in Chapter~\ref{chap_paper_d}.

To evaluate the proposed approach's support for language evolution, we applied GrammarOptimizer to two versions of EAST-ADL and four versions of QVTo. The grammar generated by the simplified version of EAST-ADL contains 755 lines of text. We configured 22 optimization rule configurations which completed the optimization of the generated grammar. The grammar generated by the full version 2.2 of EAST-ADL has 2839 lines of text. We only modified the configuration of 10 optimization rules to complete the optimization of the generated grammar of the full version. Similarly, for the 1026 lines of text in the generated grammar of version 1.0 of QVTo, we configured 733 grammar optimization rule configurations to complete the optimization of the generated grammar. However, facing the generated grammars of QVTo versions 1.1, 1.2, and 1.3 with similar amounts of text, we only modified two, zero, and one optimization rule configurations respectively to complete the optimization of the generated grammars respectively.

\section{Answers to the RQs}
We will answer all the RQs in this section.


\textbf{RQ1: } What are the user-oriented characteristics of modeling tools most suitable for supporting blended modeling?

From a user-oriented perspective, results indicated that over half of the 26 tools supported two modeling notations, while about one-third supported three. Notably, Boston Professional and TopBraid Composer each accommodated four modeling notations. Graphical and textual notations were the most common, with all 26 tools offering graphical notations and the majority providing textual options. Assessing usability, a generally challenging task, involved evaluating whether tools support multi-notations visualization and enable seamless navigation between notations. All 26 tools facilitated the visualization of two or more notations. Evaluating cross-notation navigation mainly considered synchronization and speed; more than half of the tools allowed simultaneous navigation, with most providing immediate transitions between notations. Flexibility, pertaining to inconsistent user-related aspects across modeling levels and facilities, was divided into three categories: model, language, and persistence. Most tools did not support deviations between different notations describing the same model; only six offered model-level flexibility. For language and model inconsistencies, the majority of tools didn't allow them, except for four. Notably, Umple supported both model-level and language-level flexibility. Concerning persisting inconsistency models, most tools did not offer support for this. For more details, please refer to Section~\ref{sec:rq1} in Chapter~\ref{chap_paper_a}.

\textbf{RQ2: } What are the realization-oriented characteristics of modeling tools most suitable for supporting blended modeling?
From a realization-oriented perspective, there are three aspects: mapping and platforms, change propagation and traceability, and inconsistency management. The results showed that most sampled tools (22 out of 26) use parser-based editors while four employ projection tools. Eclipse is the primary platform for these tools, except for mbdeddr (based on MPS), and MagicDraw supports multiple platforms. In terms of inconsistency management, the majority of tools (58\%) lack visualization. They mainly fall into two categories: prevention-focused (50\%) and real-time/on-demand resolution (42\% and 8\%). Automation-wise, 50\% follow a preventive approach, while 13 offer varying levels of automation for inconsistency resolution, with only 2 relying on manual resolution. For more details, please refer to Section~\ref{sec:rq2} in Chapter~\ref{chap_paper_a}.



\textbf{RQ3: } How can we build a solution to adapt generated grammars to produce the same language as available expert-created grammars?

The result and evaluations showed that the solution (i.e., GrammarOptimizer) developed in \textbf{Stage 4} could adapt the generated grammar to produce the same language as an expert-created grammar.
GrammarOptimizer consists of grammar optimization rules, which can be applied to different grammar adaptation needs. We extracted grammar optimization rules from seven different case languages to maximize their usability and generalization.
For the grammar comparison, we used a manual comparison to confirm whether the adapted grammar (i.e., the optimized grammar) and the expert-created grammar produce the same language. For example, for the language DOT, in the expert-created EBNF grammar, the grammar rule \texttt{node\_id} does not contain any curly braces, so in the optimized grammar (in Xtext), the corresponding grammar rule \texttt{NodeId} should not contain any curly braces either.

\textbf{RQ4: } Can our solution support the co-evolution of generated grammars when the meta-model evolves?

When using GrammarOptimizer to optimize a grammar, language engineers need to configure the grammar optimization rules in the form of configurations. These configurations drive GrammarOptimizer which rules to apply to adapt the grammar. The language engineers reuse these configurations when the language evolves and new grammar is generated from the evolved meta-model. They only need to adjust the configurations accordingly for the changes between meta-models to execute the optimization on the new version of the generated grammar, thereby supporting the co-evolution of the meta-model and grammar.

\section{Threats to Validity}
In this section, we will discuss threats to both internal validity and external validity. There will be also additional threats to validity discussed in the subsequent chapters.


\subsection{External Validity}
As mentioned in the methodology section, we developed a textual language called EATXT for EAST-ADL. We used EATXT as a case to illustrate how language engineers can extend Xtext generators' capabilities and its implementation according to their own needs, which was the completed work in Paper B. There is a potential threat, i.e. whether the method proposed in Paper B is also applied to other languages. First, we have explicitly stated in Paper B that the discussed approach is for the Xtext editor. Xtext provides a mechanism for extending functionality, and our method demonstrates how to use this extension mechanism with EATXT as an example. Although EATXT as a language has its metamodel distinct from other languages. However, this extension mechanism is generic and is not limited to a specific metamodel.

\subsection{Internal Validity}
In the evaluation part of Stage 4, the number of optimization rule configurations required to optimize the grammar generated by different versions of QVTo differs very little, i.e., the maximum difference is two rule configurations. This data serves as one of the evidences that our solution GrammarOptimizer supports language evolution. However, the number of rule configurations required to optimize the generated grammar of any version of QVTo exceeds 700, which threatens the causal because the effort required to configure GrammarOptimizer does not reflect any less effort than manually adapting the grammar.
However, we argue that it is more efficient to configure GrammarOptimizer once than to manually rewrite grammar rules every time the language changes – under the assumption that the configuration can be reused for new versions of the grammar. In that case, the effort invested in configuring GrammarOptimizer would quickly pay off when a language is going through changes, e.g., while rapidly prototyping modifications or when the language is evolving. We evaluated this assumption in stage 4.
Furthermore, the effort required to configure optimization rules for optimizing the generated grammar can technically be reduced, one solution being to extract the rule configurations by comparing the generated grammar and the target grammar rather than manually writing the configuration from scratch~\cite{zhang2023automated}.

\section{Summary of Contributions}

\begin{figure*}[htp]
  \centering
  \includegraphics[width=\linewidth]{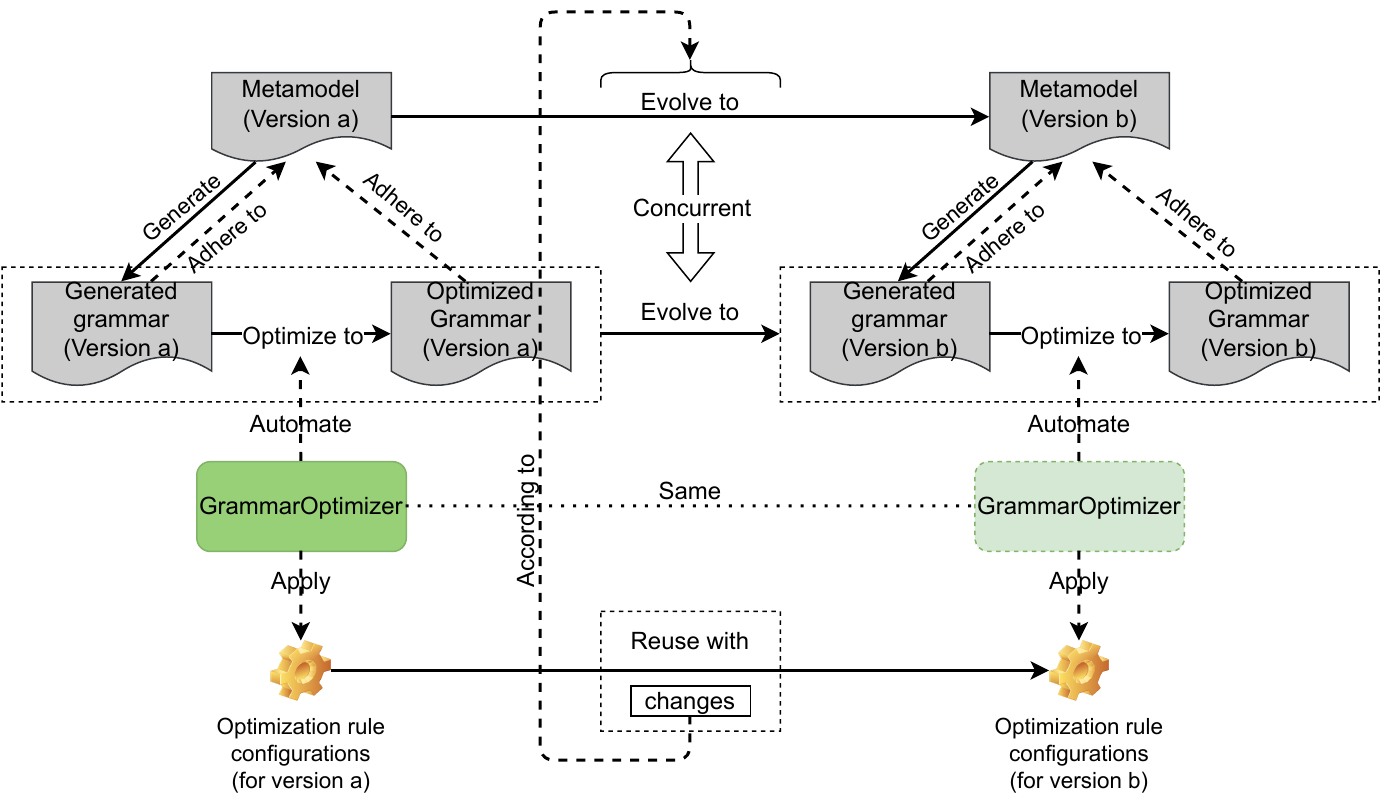}
  \caption{Schematic diagram of the proposed solution in \textbf{Stage 4} to support the co-evolution of meta-models and grammars.}
  \label{fig:tech_co_evolution}
\end{figure*}

In this thesis, there is a clear contribution in each of our four stages of work.

\paragraph{The state-of-the-art of blended modeling tools.} First, prior to our study of Stage 1, a clear overview of the state of the art and practices in modeling tools that support blended modeling was missing. Our study of Stage 1 filled this gap, including indicating limitations of current tools and opportunities for future research or industrial practice.

\paragraph{Extension of Xtext's generator capabilities.} In Stage 1, we comprehensively reported on the state-of-the-art and practices of blended modeling tools. We learned from this report that textual notations are a very popular type of notation. However, not all modeling tools support textual notations, including EAST-ADL. In Stage 2, we developed textual notations for EAST-ADL to explore and improve blended modeling techniques. We identified issues with the limitations of Xtext's generator capabilities, therefore we used EAST-ADL as a case to explore and show how to extend it and its implementation. Our contribution in Stage 2 is the method of how language engineers can extend Xtext's generator capabilities and its implementation according to their own needs. Language engineers can get a reference from our specific implementation on the EATXT case.

\paragraph{Python-style prototyping.} In Stage 2, we developed a textual language (i.e., EATXT) for EAST-ADL to explore and improve blended modeling techniques. When designing the grammar for EATXT, we identified the problem that the generated grammar was cumbersome and user-unfriendly. Our contribution in Stage 3 is that we proposed a general method for converting a language with a generated grammar into a Python-style language, including providing a script that can semi-automate the execution of this method.



\paragraph{Tool GrammarOptimizer.} Fourth, manually optimizing a grammar generated from a metamodel can be cumbersome and time-consuming, and as the language evolves, manual improvements made in the generated grammar of the previous version of the language can not be replayed in the generated grammar of the evolved version. This causes language engineers to have to perform similar optimizations on the generated grammar once again. Our work in \textbf{Stage 4} provided a solution, i.e., the tool GrammarOptimizer, which includes 60 general optimization rules (four added during the evaluation). By configuring these optimization rules, the solution can be applied to different languages for rapid prototyping and co-evolution of meta-models and grammars. 

Figure~\ref{fig:tech_co_evolution} depicts how GrammarOptimizer supports the co-evolution of meta-model and grammar from version a to version b. By configuring GrammarOptimizer, language engineers can automate the optimization of the generated grammar of version a. In the same version, generated grammar and optimized grammar adhere to the same meta-model. As the meta-model evolves, language engineers regenerate the grammar from the meta-model (of version b). To optimize the generated grammar (of version b), the language engineers reuse the configurations of version a to replay the previous optimization in the generated grammar (of version b). In addition, language engineers adjust configurations based on differences between versions to optimize the changed parts.

\section{Conclusion and Future Work}
In this thesis, we comprehensively report on the challenges and limitations of modeling tools that support blended modeling, and opportunities for improving them. Our report helps tool providers identify the limitations of their tools in supporting blended modeling, and researchers can use this work to better contextualize their research and better position their work in terms of applicability. To contribute to the exploration and improvement of blended modeling technology, in stage 2, this thesis takes EAST-ADL and its textual language EATXT as an example to demonstrate how language engineers can extend the Xtext generator capabilities and its implementation according to their needs. Users and our peers who use Xtext to develop DSLs using the MDE approach will benefit from this. In stage 3, this thesis proposes a semi-automated method that can transform the language with a generated grammar into a Python-style language. Language engineers can apply this method to quickly improve the conciseness and user-friendliness of the language. We provide a systematic rule-based solution in Stage 4 that can generate adaptations in evolved versions' generated grammar of the evolving language, enabling semi-automated co-evolution of metamodels and grammars.

In future work, we plan to expand GrammarOptimizer into a more mature language workbench, supporting advanced features such as automatic extraction of configuration, a ``what you see is what you get'' view of the optimization of the grammar, optional language style library, and co-evolution of model instances and grammar~\cite{zhang2023rapid}. We will also explore integration with workflows that generate graphical editors for blended modeling, which will further improve existing blended modeling techniques.

%% file: TexFiles/appendedPapers.tex

\chapter{Paper A}
\label{chap_paper_a}
\thispagestyle{empty}
\subsection*{Blended Modeling in Commercial and Open-source Model-Driven Software Engineering Tools: A Systematic Study}
\subsubsection*{I.~David, M.~Latifaj, J.~Pietron, W.~Zhang, F.~Ciccozzi, I.~Malavolta, A.~Raschke, J.~Steghöfer, R.~Hebig}
\subsubsection*{{\em Software and Systems Modeling (SoSyM), 2023, 22(1), pp. 415-447}.}
\newpage
\thispagestyle{empty}
\mbox{}
\newpage
\addtocounter{page}{-2}

\input{TexFiles/paperA/Abstract}
\input{TexFiles/paperA/Introduction}
\input{TexFiles/paperA/Background}

\input{TexFiles/paperA/Study_Design}
\input{TexFiles/paperA/Results}
\input{TexFiles/paperA/Orthogonal_Finding}
\input{TexFiles/paperA/Discussion}

\input{TexFiles/paperA/Threats_to_Validity}
\input{TexFiles/paperA/Conclusions}
\input{TexFiles/paperA/appendix-tools}

\chapter{Paper B}
\label{chap_paper_b}
\thispagestyle{empty}
\subsection*{Exploiting Meta-Model Structures in the Generation of Xtext Editors}
\subsubsection*{J. Holtmann, J.~Steghöfer, W.~Zhang}
\subsubsection*{{\em 11th International Conference on Model-Based Software and Systems Engineering, SciTePress, 2023, pp. 218-225.}}
\newpage
\thispagestyle{empty}
\mbox{}
\newpage
\addtocounter{page}{-2}

\input{TexFiles/paperB/abstract}
\input{TexFiles/paperB/main}

\chapter{Paper C}
\label{chap_paper_c}
\thispagestyle{empty}
\subsection*{Creating Python-style Domain Specific Languages: A Semi-automated Approach and Intermediate Results}
\subsubsection*{W.~Zhang, R.~Hebig, J.~Steghöfer, J. Holtmann}
\subsubsection*{{\em 11th International Conference on Model-Based Software and Systems Engineering, SciTePress, 2023, pp. 210-217.}}
\newpage
\thispagestyle{empty}
\mbox{}
\newpage
\addtocounter{page}{-2}

\input{TexFiles/paperC/abstract}
\input{TexFiles/paperC/main}

\chapter{Paper D}
\label{chap_paper_d}
\thispagestyle{empty}
\subsection*{Supporting Meta-model-based Language Evolution and Rapid Prototyping with Automated Grammar Optimization}
\subsubsection*{W.~Zhang, J. Holtmann, D. Strüber, R.~Hebig, J.~Steghöfer}
\subsubsection*{{\em Revised and Re-submitted to Journal of Systems and Software, 2023}.}
\newpage
\thispagestyle{empty}
\mbox{}
\newpage
\addtocounter{page}{-2}

\input{TexFiles/paperD/Abstract}
\input{TexFiles/paperD/main}

%% file: TexFiles/paperA/Abstract.tex
\newpage
\section*{Abstract}
Blended modeling aims to improve the user experience of modeling activities by prioritizing the seamless interaction with models through multiple notations over the consistency of the models. Inconsistency tolerance, thus, becomes an important aspect in such settings. To understand the potential of current commercial and open-source modeling tools to support blended modeling, we have designed and carried out a systematic study. We identify challenges and opportunities in the tooling aspect of blended modeling. Specifically, we investigate the user-facing and implementation-related characteristics of existing modeling tools that already support multiple types of notations and map their support for other blended aspects, such as inconsistency tolerance, and elevated user experience.
For the sake of completeness, we have conducted a multivocal study, encompassing an academic review, and grey literature review. We have reviewed nearly \numberAcademicScreenedApprox{} academic papers and nearly \numberGreyScreenedApprox{} entries of grey literature. We have identified \numberToolCandidates{} candidate tools, and eventually selected \numberToolsIdentified{} of them to represent the current spectrum of modeling tools.

\newpage

%% file: TexFiles/paperA/Introduction.tex
\section{Introduction}\label{sec:intro}

Model-driven engineering (MDE) advocates modeling the engineered system at high levels of abstraction before it gets realized. The resulting models serve crucial roles in ensuring the appropriateness (e.g., correctness, safety, optimality) of the system.
To keep the cognitive flow of modeling effective and efficient, stakeholders shall be equipped with proper formalisms, notations, and supporting computer-aided mechanisms. This is especially important in the design of modern systems, as their complexity has been increasing exponentially over the past years~\cite{persson2013characterization}.
Modeling does not remove complexity from the engineering process, but rather, it replaces the accidental complexity of complex systems with essential complexity that is easier to manage~\cite{atkinson2008reducing}. Nonetheless, as a consequence of the increasing complexity of modern systems, modeling itself is becoming more complex.

In this paper, we focus on a specific manifestation of this added complexity stemming from the need for an orchestrated ensemble of modeling notations, aiming to enable seamless interaction with models through any of the notations. Such a need has been reported in multiple academic~\cite{broy2012software} and industrial domains, e.g., automotive~\cite{huning2021seamless}, avionics~\cite{gu2003end}, cyber-physical systems~\cite{yang2017constraint}, and product lines~\cite{schulze2012automotive}.
In such an approach, user experience may also be (temporarily) prioritized over the correctness of the described system, in an effort to enable a smooth process of expressing the stakeholder's cognitive models in terms of the modeling language. This approach is referred to as \textit{blended modeling}~\cite{addazi2021blended}.

\subsection{What is blended modeling?}
Blended modeling was first introduced by Ciccozzi et al.~\cite{ciccozzi2019blended} as follows:

\begin{center}
\makebox{\centering\parbox{0.9\linewidth}{\textit{Blended modeling is the activity of interacting seamlessly with a single model (i.e., abstract syntax) through multiple notations (i.e., concrete syntaxes), allowing a certain degree of temporary inconsistencies.}}}\end{center}\hfill

That is, blended modeling is characterized by the following three features.

\begin{description}
    \item[\textbf{Multiple notations}.] This is not to be confused with multiple \textit{languages}. In our terminology, a language is composed of (i) a metamodel (abstract syntax), and (ii) a set of notations (concrete syntax). Blended modeling does not impose different metamodels.
    \item[\textbf{Seamless interaction}.] Different notations have to be carefully integrated and orchestrated to allow for using the most appropriate notation for specific modeling tasks. This requires intuitive navigation between notations, proper change propagation between them, and in many cases, traceability.
    \item[\textbf{Flexible consistency management}.] This aspect entails both vertical inconsistencies~\cite{vanherpen2016ontological} (e.g., inconsistencies between the instance model and its metamodel); and horizontal inconsistencies (e.g., inconsistencies between two notations used to manipulate instances of the same metamodel).
\end{description}

\subsection{What is \textit{not} blended modeling?}

\begin{figure}[!h]
	\centering
	\includegraphics[width=\linewidth]{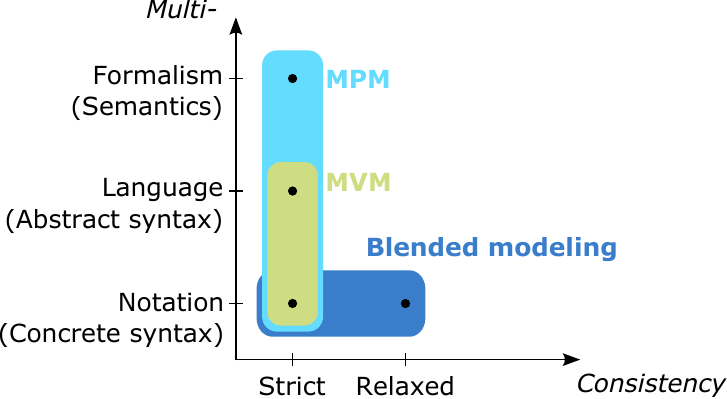}
	\caption{Blended modeling in the context of MVM and MPM.}
	\label{fig:positioning}
\end{figure}

\paragraph{Multi-view modeling is not blended modeling.}~As shown in~\figref{fig:positioning}, Multi-View Modeling (MVM)~\cite{von2012multi} and blended modeling share the trait of \textit{multi-notation}. The main differences are, that (i) MVM further assumes \textit{multiple languages}, while (ii) blended modeling assumes relaxed consistency rules instead. These differences stem from the different aims of the two approaches. MVM is concerned with constructing the appropriate views for stakeholders with varying backgrounds. Blended modeling focuses on the elevated UX with respect to an ensemble of notations, assuming a single underlying model. Prior work has reported challenges in relaxed consistency in multi-language settings such as MVM~\cite{reineke2019basic}. Blended modeling enables relaxed consistency by restricting the number of languages to one.

For example, the SCADE\footnote{\url{https://www.ansys.com/products/embedded-software/ansys-scade-suite}} tool suite provides the user with different languages for different purposes within the same model development environment. These languages facilitate multi-view modeling of the overall system and necessitate different abstract syntaxes. Therefore, working with SCADE cannot be considered blended modeling.

\paragraph{Multi-paradigm modeling is not blended modeling.}~In addition to assuming multiple languages, Multi-Paradigm Modeling (MPM)~\cite{mosterman2004computer} further assumes potentially different semantics behind the languages, giving rise to \textit{multi-formalism} (\figref{fig:positioning}). This added complexity positions MPM even further from blended modeling, and vastly exacerbates consistency management, as reported in prior work~\cite{david2019foundation}.

For example, Matlab/Simulink is a typical combination of formalisms for system design, in which the overall system is graphically designed in Simulink\footnote{\url{https://se.mathworks.com/products/simulink.html}}, which follows causal block diagrams (CBD) semantics; and the low-level functions in the system are textually described in Matlab\footnote{\url{https://www.mathworks.com/products/matlab.html}}, which relies on matrix semantics for complex computations.
While some level of navigation is provided between the two formalisms within the Matlab modeling and development environment, relaxed consistency is completely missing. Therefore, working with Matlab/Simulink cannot be considered blended modeling.

\subsection{Motivation and aim}

Blended modeling is an emerging new concept, thus, a map of current commercial and open-source tools is needed to properly position it in the research-and-development landscape.

In this article, we report the design, execution, and results of our mapping study on tools that are prime candidates to support blended modeling. Our study shows that these are typically tools with multiple notations for a single underlying abstract syntax, but they lack proper inconsistency tolerance mechanisms or fail to leverage such features for an improved user experience. The aim of our study was to identify, classify, and analyze (i) the user-oriented, and (ii) the realization-oriented characteristics of these tools.
To infer this information while ensuring external validity, we surveyed both the academic (peer-reviewed) literature and the grey literature~\cite{rothstein2009grey}, consisting of websites, blogs, and user manuals of engineering tools, following the guidelines for multi-vocal reviews in software engineering~\cite{garousi2018guidelines}. To be able to treat both types of literature uniformly, we made \emph{tools} the primary units of our study, instead of papers. This is motivated by the inherent limitations of grey literature in terms of providing high-fidelity research data. Websites and end-user documentation do not aim to provide such information.
We formulated a surveying protocol based on well-established guidelines and we have meticulously followed this protocol in the execution of our study. Eventually, we screened \numberAcademicScreened{} academic papers and included \numberAcademicIncluded{} of them. Additionally, \numberGreyScreened{} grey literature entries were processed. Out of the academic papers, \numberAcademicTools{} distinct tools were extracted and complemented by \numberGreyTools{} tools extracted from the grey literature. After removing duplicates, the set of \numberToolCandidates{} tools was reviewed according to the tool selection criteria (see \secref{sec:study_design}) and we eventually identified \numberToolsIdentified{} tools to be analyzed in detail. Although this list of tools is not exhaustive, we are reasonably confident about its representativeness of the domain of interest.

The results of this study provide a clear overview of the state of the art and practice of the domain of modeling tools closest to blended modeling. The tool characteristics reported in this paper can be particularly useful for tool providers in identifying the limitations of their tools in supporting blended modeling. Researchers of the three main dimensions of blended modeling (multi-notation, seamless integration of languages, inconsistency tolerance) could use this work to better contextualize their research, and position their work better in terms of applicability.

\subsection{Structure}

The rest of this paper is structured as follows. First, in~\secref{sec:background}, we give an overview of the background concepts of blended modeling and review the related work. In~\secref{sec:study_design}, we define the methodological framework for carrying out this study. In Section~\ref{sec:vertical}, we elaborate on the findings of this study, in particular on the results pertaining to the research questions of this study: the user-oriented and realization-oriented characteristics of the tools of interest. In~\secref{sec:orthogonal}, we provide orthogonal insights on the aggregated data. We discuss the results in~\secref{sec:discussion}, and the threats to validity in~\secref{sec:threats}. Finally, we summarize this paper by drawing the conclusions in~\secref{sec:conclusions}.

%% file: TexFiles/paperA/Background.tex
\section{Background}\label{sec:background}

In this section, we provide the foundational background concepts to contextualize our study. More specifically, we describe the core ingredients of blended modeling: multiple notations (\secref{sec:background-multinotation}), seamless interaction (\secref{sec:background-usability}), and flexibility in managing inconsistencies (\secref{sec:background-inconsistency}). Additionally, we discuss the secondary literature related to our study (\secref{sec:related-secondary}).

\subsection{Multiple notations}\label{sec:background-multinotation}

Interacting with the (abstract) model through multiple notations (concrete syntaxes) is one of the three distinguishing features of blended modeling. A vast body of knowledge on the topic has been produced, especially in relation to multi-view modeling, and multi-paradigm modeling.

\subsubsection{Multi-view modeling}
\label{sec:background-multinotation-multiview}
Multi-view modeling (MVM) tackles the complexity of modeling heterogeneous systems by decomposing the models into multiple views, that are concerned with specific aspects of the system~\cite{von2012multi}. The ISO/IEC/IEEE 42010:2011 standard~\cite{iso42010} defines a view as a set of concerns of specific stakeholders and viewpoints as the specification of conventions utilized to construct a view. The five mutually non-exclusive enabling mechanisms of multi-view modeling are (i) \textit{synthetic}, where views are specified by means of different domain specific modeling languages and synthesized together; (ii) \textit{separate}, a stricter version of synthetic, where synthesis does not take place; (iii) \textit{projective}, where a single metamodel allows for the definition of multiple virtual views; (iv) \textit{orthographic}, where views are orthographic projections of a single underlying model; or (v) \textit{hybrid}, where views represent only a portion of the common metamodel~\cite{cicchetti2019multi}.
MVM has been shown to be an effective approach in several complex domains, such as cyber-physical systems~\cite{persson2013characterization}, and cloud-based software-intensive systems~\cite{corley2016cloud}.
The principles of MVM are similar to those of blended modeling. However, its goal is different. While MVM is oriented towards the identification of multiple views and the management of consistency between them, blended modeling focuses on enabling an elevated user experience while working with multiple notations at the same time.

\subsubsection{Multi-Paradigm Modeling}
Multi-paradigm modeling (MPM) advocates modeling every aspect of the system explicitly, at the most appropriate level of abstraction, and using the most appropriate formalism~\cite{mosterman2004computer,carreira2020foundations}. As such, MPM facilitates the modeling of complex systems that could not be described through a single formalism and at a common level of abstraction due to the heterogeneity of the different components. It combines three research areas: (i) meta-modeling used for the specification of formalisms, (ii) multi-formalism used for the coupling of models specified in different formalisms and their transformations, and (iii) model abstraction used for the relationships among models described in different formalisms \cite{vangheluwe2002introduction}.
The principles of MPM are similar to those of blended modeling, as both approaches promote employing a variety of notations to model the problem at hand. However, MPM achieves this by employing a variety of separate formalisms, i.e., multiple notations with possibly different semantics. Blended modeling assumes a single abstract syntax, and therefore, single semantics. This simplification allows for greater flexibility in terms of temporarily inconsistent designs.

\subsection{Seamless interaction}\label{sec:background-usability}
Usability in terms of the ability to seamlessly interact with models through multiple different notations is one of the three distinguishing features of blended modeling.
In this section, we review how state-of-the-art approaches typically support seamless interaction.
We focus on UML tools here since they have received significant attention from research and tool providers of the software engineering domain in the past. We also mention examples for other modeling languages where appropriate.

\subsubsection{Text-based modeling with graphical visualizations}
Umple~\cite{umple} is a modeling tool that supports the creation of UML models using both textual and graphical notations, where the synchronization between the two notations is automated and on the fly. However, the graphical editor does not offer full editing capabilities, and the existing editing capabilities are only available on class diagrams but not on state machines, composite structures, or feature diagrams. 
\mbox{FXDiagram}\footnote{\url{https://jankoehnlein.github.io/FXDiagram}} is a JavaFX-based framework that can be integrated into Eclipse as well as intelliJ IDEA. It supports the creation of graph diagrams (nodes and edges) and it is typically used for graphical visualization of textual DSLs but does not provide editing functions.
MetaUML\footnote{\url{https://github.com/ogheorghies/MetaUML}} is a GNU GPL library for typesetting UML diagrams, using a textual notation. This notation is used for rendering read-only graphical UML diagrams. Plant\-UML\footnote{\url{https://plantuml.com}} is very similar but supports also non-UML diagrams. ZenUML\footnote{\url{https://www.zenuml.com}} supports sequence diagrams and flowcharts, again defined using a textual notation that is translated into read-only graphical views. The generation of the sequence diagrams is automatic, as the conversion happens on the browser. Excalibur~\cite{ries2018messir} is a tool that relies on Xtext for textual specification and Sirius for graphical views of the textual specification. The model elements are defined using Messir textual DSL and the generated graphical visualization is read-only. Chart Mage\footnote{\url{http://chartmage.com/index.html}} is a web-based tool that supports automatic and on-the-fly generation of sequence diagrams and flowcharts using a textual notation. DotUML\footnote{\url{https://dotuml.com}} is a javascript application that supports the generation of a subset of UML diagrams (i.e., use-case, sequence, class, state, and deployment) from a textual notation.  For all of the aforementioned tools, concrete syntaxes are predefined and not customizable, and the graphical notation is read-only, generated using the textual notation.

\subsubsection{Mixed textual and graphical modeling}

Addazi and Ciccozzi~\cite{addazi2021blended} present a proof-of-concept implementation for UML and UML profiles modeling using blended textual and graphical notations. The stack of technologies used includes Eclipse Modeling Framework (EMF)\footnote{\url{https://www.eclipse.org/modeling/emf}}, Xtext\footnote{\url{https://www.eclipse.org/Xtext}}, and Papyrus~\cite{papyrus}. Their solution includes a single underlying abstract syntax,
two notations (i.e., graphical and textual), and one single persistent resource that is the UML resource. This architecture enables synchronization by means of serialization/deserialization operations across Xtext and UML models. In addition, the authors conduct an experiment to demonstrate that their solution on blended modeling increases user performance compared to single notation modeling. 

Maro et al.~\cite{maro2015integrating} introduce a solution that integrates graphical and textual editors for a specific UML profile-based DSL. Being that the graphical editor is already provided, this work focuses on obtaining the textual editor and switching between views (i.e., graphical and textual). To obtain the textual editor, the UML profile-based DSL is first transformed into an Ecore model using an ATL transformation, and then this Ecore model is consumed by the Xtext plugin to generate the textual editor. Switching between views is achieved by employing ATL transformations.
Scheidgen~\cite{scheidgen2008textual} provides embedded textual editors for graphical editors as an add-on feature. For each selected model element that needs to be edited, the embedded textual editor creates an initial representation that can be changed by the user and using parsing operations, new edited model elements are created. However, the synchronization is on-demand as the changes in the underlying model are not carried out until they are committed by the user and the textual editor is closed.

Laz{\u{a}}r~\cite{lazuar2011integrating} makes use of the Eclipse modeling environment to integrate the existing UML tree-based editor with the textual editor for Alf language\footnote{\url{https://www.omg.org/spec/ALF}} and to create fUML\footnote{\url{https://www.omg.org/spec/FUML}} models. However, the synchronization is on-demand as the changes are carried out upon the occurrence of a save action by the user. 

Charfi et al.~\cite{charfi2009hybrid} define a hybrid language that integrates textual and graphical notations in one concrete syntax. The contribution consists of a visual notation for the most used UML actions and an editor that supports the proposed notation. The hypothesis that the hybrid notation can perform better than the textual notation is backed by an experiment that takes into consideration the learnability of the hybrid notation, the prevented errors, and the circumstances in which the hybrid notation is a better fit than the textual notation. However, this approach is restricted to UML actions only.

Van Rest et al. \cite{vanrest2013robust} implement an approach for the robust synchronization of graphical editors generated with the Graphical Modeling Framework (GMF)\footnote{\url{https://www.eclipse.org/modeling/gmp}} and textual editors generated with Spoofax\footnote{\url{http://strategoxt.org/Spoofax}}. This approach allows error recovery during synchronization and preserves the textual and graphical layout in case of errors. However, layout preservation is not supported at all times, as during cut-paste operations, the elements and their associated layouts are deleted and then recreated, therefore losing the original layout.

\subsubsection{Projectional editing}

Projectional editing is an approach where the abstract syntax tree (AST) is modified directly upon every editing action and bypasses the stages of the parser-based approach, where the parser must first check the correctness of the syntactic aspects, and then construct the AST based on the changes in the notation~\cite{voelter2014towards}. This course of action allows the definition of multiple notations (e.g., tables, diagrams, formulas) that cannot be supported by parser-based approaches, and supports multiple views of the same program, simultaneously. Moreover, a considerable amount of the ambiguities caused during the parsing process are tackled. Projectional editing is a realization of the intentional programming paradigm~\cite{simonyi1995the}, and as such, it encourages the combination of a variety of different notations. Some of the state-of-the-art language workbenches that adopted this principle for providing domain-specific tool engineers with efficient tools~\cite{berger2016efficiency} are JetBrains MPS\footnote{\url{https://www.jetbrains.com/mps}} and MelanEE\footnote{\url{http://www.melanee.org}}. However, even though they provide a greater amount of notations, their support for textual notations is limited compared to parser-based approaches, as it is only a projection that resembles text. In particular, no possibly inconsistent intermediate states are allowed, which consequently restricts the user accustomed to classical text editors and their corresponding free editing features.

\subsection{Inconsistency management}\label{sec:background-inconsistency}

Approaches, such as multi-view modeling (MVM) and multi-paradigm modeling (MPM) advocate modeling the engineered system using the most appropriate notations, formalisms, and abstractions. This allows multiple users to be involved in the modeling of the system, and thus, introduces parallelism, which is beneficial for the overall efficiency of the engineering endeavor. Parallelism, however, gives rise to inconsistencies between the design artifacts, compromising the ultimate correctness of the system. Inconsistency has been shown to be an effective heuristic for managing the ultimate correctness of the system~\cite{david2019foundation}. Techniques, such as blended modeling, make use of this assertion by focusing on the early detection of inconsistencies~\cite{david2018process} and establishing the proper tolerance mechanisms.

The notion of consistency models and their various alternatives have been well-researched already in early distributed systems. Lamport~\cite{lamport1979make} is the first to describe how multi-processor systems should be constructed to ensure proper execution of programs. His notion of \emph{sequential consistency} allows a relaxation of the locking model by assuming a total order of modifications that distributed nodes are guaranteed to observe. Adve and Gharachorloo~\cite{adve1996shared} describe various relaxations of the sequential consistency model, based on architectural choices on the hardware and software level.
\emph{Eventual consistency} has been suggested by Vogel et al.~\cite{vogels2009eventually} to enable a weaker notion of consistency between distributed participants, by embracing that real consistency can never be achieved. In such settings, distributed participants are characterized by the BASE properties: basic availability, soft state, and eventual consistency.
Lately, \emph{strong eventual consistency} (SEC) has been suggested~\cite{balegas2015putting} to combine the liveliness guarantees of eventual consistency with the safety guarantees of strong consistency. Conflict-free replicated data types~\cite{shapiro2011conflict} are the prime examples of their applications.

Inconsistencies are a well-researched area in software engineering~\cite{spanoudakis2001inconsistency}, too. Consistency between models can be categorized into two orthogonal dimensions~\cite{engels2001methodology}: horizontal and vertical consistency; and syntactic and semantic consistency. Horizontal consistency is concerned with models on the same level of abstraction, whereas vertical consistency is defined between models on different levels of abstraction (typically in model-meta\-model contexts)~\cite{vanherpen2018contract}. The majority of inconsistency management techniques rely on syntactic concepts, e.g. synchronization by bi-directional model transformations~\cite{stevens2020maintaining}, triple-graph grammars~\cite{giese2006incremental}, and by version control systems and related mechanisms~\cite{kehrer2013consistency,kelly2017collaborative}. However, semantic techniques have been shown to be beneficial in heterogeneous engineering settings~\cite{vanherpen2016ontological}.
View consistency has been researched in the context of MVM, e.g., in the Vitruvius approach~\cite{klare2021enabling}, which provides languages for consistency preservation, and defines a model-driven development process for enacting consistency rules.

Finkelstein et al.~\cite{finkelstein2000foolish} suggest that inconsistencies are organic elements of any engineering process, and instead of simply removing them from the system, one should apply proper inconsistency management techniques~\cite{finkelstein1994inconsistency}. Such inconsistency management techniques typically entail the activities of detecting, resolving, preventing, and tolerating inconsistencies~\cite{nuseibeh2001making}. Blended modeling heavily relies on the tolerance of inconsistencies. Balzer et al.~\cite{balzer1991tolerating} suggest augmenting inconsistency instances with a state. Inconsistency rules are first deconstructed into appearance and disappearance rules spanning a temporal interval; then, tolerance rules are put in place to trigger repair actions based on temporal constructs. Easterbrook et al.~\cite{easterbrook1994coordinating} propose a similar technique for temporal inconsistency tolerance in the context of MVM. Inconsistency tolerance is achieved via pairs of pre- and post-conditions relying on a user-defined consistency metric. David et al.~\cite{david2016towards} introduce various patterns of inconsistency tolerance for implementing such systems.

\subsection{Related secondary literature}\label{sec:related-secondary}

This paper reports on the first systematic study on blended modeling. There are, however, secondary studies close to our work that are similar in topic, but differ in terms of motivation and objectives, and are generally limited to a narrower scope.

Torres et al.~\cite{torres2020systematic} conduct a  systematic literature review with the aim to identify a list of available tools to support model management and provide a categorization of these tools into (i) tools that can provide consistency checking on models of different domains, (ii) tools that can provide consistency checking on models of the same domain, and (iii) tools that do not provide any consistency checking. Furthermore, the authors identify the inconsistency types, strategies to keep the consistency between models of different domains, and the challenges to manage models of different domains. 
The information retrieved from the primary studies is also complemented with additional data sources (e.g., the official website of the tool). Our study focuses on a broader scope, especially multi-notation and seamless interaction. 
Torres et al. observe that 35\% of their analyzed tools do not provide any consistency checking features, whereas in our study we observe that 64\% of the analyzed tools do not support models inconsistencies. Moreover, Torres et al. identify different strategies that have been used to keep models consistent, \eg by using standard file formats for the models, explicitly modeling dependencies among model elements, mapping model elements to a shared ontology, etc. Our study complements such results by highlighting which inconsistency management strategies involve a manual effort (like keeping a dependency matrix always up to date), a semi-automated procedure (\eg by specifying a priori consistency constraints and checking them during development), or a completely automated one (\eg via the automated application of inconsistency resolution procedures).

Iung et al.~\cite{iung2020systematic} conduct a systematic mapping study with the aim to identify tools, language workbenches, or frameworks for DSL development.
The authors identify 59 tools and they use the feature model proposed by Erdweg et al.~\cite{erdweg2015evaluating} for their comparison. The study focuses on the technologies/tools used for DSL development, their license types, the application domains, and the features of the DSL creation process that these tools support. 48 tools support only one notation  (graphical or textual), seven tools support two notations (graphical and textual), two tools support three notations, and two tools support four notations. Our study focuses on a broader scope, by extending the set of features on which the comparison is based with features such as synchronization mechanisms, collaborative features, or conformance relaxation. We also contextualize our work on a broader timeline, while the authors focus on the period between 2012--2019.
In line with the results of our study, Iung et al. observed that the notations that were more frequently used in combination are \textit{textual} and \textit{graphical}, with the tabular one complementing them. In \cite{iung2020systematic} two language workbenches are identified as particularly relevant for blended modeling: (i) GEMOC Studio, which provides real-time bidirectional synchronization in their generated editors, and (ii) the Whole Platform, which allows language engineers to choose among four different types of notation (\ie textual, graphical, tabular, and symbolic), and to visualize the different translations among them at the model level.

Franzago et al.~\cite{franzago2017collaborative} and David et al.~\cite{david2021collaborative} map the state-of-the-practice of collaborative model-based software engineering. The authors identify and classify collaborative MDSE approaches based on the different categories such as characteristics of the collaborative model editing environments, model versioning mechanisms, model repositories, support for communication and decision making, and more. Additionally, the authors identify limitations and challenges with respect to the state of the art in collaborative MDSE approaches. Regarding model management, they provide a taxonomy for the management support of collaborative MDSE approaches, collaboration support, and communication support. This study covers some of the aspects that we cover in our systematic mapping study (e.g., conflict detection). However, while this study is mostly focused on the characteristics of the collaborative approaches, we aim toward a classification of tools based on a broader set of features such as synchronization mechanisms and their generation, or conformance relaxation.
The results of Franzago et al. and David et al. for collaborative modeling that are confirmed in this study are about: (i) the types of notations, with graphical as the most supported one, followed by textual, (ii) the prevalence of custom/other modeling platforms with respect to Eclipse EMF, (iii) the growth of web-based approaches, (iv) the growth of preventive conflict management, and (v) the prevalence of mechanisms for (semi-)automatically resolving conflicts. We anticipate that 15 out of the 26 tools analyzed in this study support collaborative modeling, with the majority of tools providing off-line collaboration (\ie a la Git), rather than real-time collaboration (\ie a la Google Docs); this result is different for academia where, according to Franzago et al. and David et al., researchers focus primarily on real-time collaboration. Another difference with respect to the state of the art in collaborative modeling is that blended modeling tools are primarily parser-based, whereas collaborative modeling approaches tend to be equally distributed between parser-based and projectional approaches. 
Interestingly, while researchers are recently investigating more on eventual consistency for collaborative modeling~\cite{david2021collaborative}, in our study we observe that blended modeling tools provide limited support for consistency tolerance that would allow deviations between different notations describing the same model. 

Granada et al.~\cite{granada2017model} map model-based language workbenches that can be used to generate editors for visual DSLs and point out their features and functionalities. 
The authors identify eight language workbenches for the generation of editors for visual DSL. The features taken into consideration for their analysis are the following: scope, framework, the distinction between abstract and concrete syntax, abstract syntax, concrete syntax, editing capabilities, use of models, automation, usability, and methodological basis. The conclusions point out that the most complete commercial language workbenches are MetaEdit+\footnote{\url{https://www.metacase.com/products.html}} and ObeoDesigner\footnote{\url{https://www.obeodesigner.com/en}}, while the most complete open-source ones are Eugenia\footnote{\url{https://www.eclipse.org/epsilon/doc/eugenia}}, GMF\footnote{\url{https://www.eclipse.org/modeling/gmp}},\\ Graphiti\footnote{\url{https://www.eclipse.org/graphiti}}, and Sirius\footnote{\url{https://www.eclipse.org/sirius}}. Our study differs in scope, as we focus on tools that provide multiple notations, not only on tools that can be used to develop editors for a single visual DSL. Indeed, none of the tools identified by Granada et al. support the definition of more than one (visual) concrete syntax for the same abstract syntax; this means that language engineers willing to develop blended modeling environments should either use a dedicated language workbench for blended modeling or suitably combine the languages produced by two or more of the language workbenches mentioned above.

Do Nascimento et al.~\cite{donascimento2012systematic} perform a large-scale systematic mapping study on DSLs and their related tools. The tools are categorized into (i) tools for using DSLs, (ii) tools for creating DSLs, and (iii) language workbenches. Our study differs from this work, as we focus on DSLs tool comparison, while the authors provide a brief categorization of DSL tools, and do not go into the details of conducting a comparison of the technical features.
It is interesting to note that Do Nascimento et al. observed that tool support for a single DSL is well-studied in the literature, but at that time (2012) there was little knowledge about how to support multiple DSLs and notations in a single modeling environment. They claim that supporting multiple DSLs and multiple notations is fundamental when describing large-scale industrial systems and that methods and tool support are needed for the success of multi-DSL development. Based on the results of our study and the ones on multi-notation modeling (see Section~\ref{sec:background-multinotation}, we can confirm that in the last years, the MDE scientific community actively worked and contributed to filling this research gap.

There are additional studies related to our research that are not systematic in nature, but their takeaways are still relevant.
Negm et al.~\cite{negm2019survey} compare 14 language workbenches based on (i) structure (grammar-driven or model-driven), (ii) editor (parser-based or projectional), (iii) language notations (textual, tabular, symbols, or graphical), (iv) semantics  (translational or interpretive), and (v) composability language aspects. However, this study is limited to language workbenches and does not cover aspects such as synchronization mechanisms and their generation, or collaborative features.
Some of the results obtained by Negm et al. are relevant for blended modeling as well. Firstly, out of nine analyzed parser-based language workbenches, only one (\ie Ens\~{o}) supports both textual and graphical concrete syntaxes; this capability is achieved by having a bidirectional mapping between tokens in the textual representation of the model and elements in the object graph. Moreover, all four considered projection-based language workbenches support multiple concrete syntaxes, with the \textit{Whole} platform and MPS supporting four different syntaxes: textual, graphical, tabular, and symbolic. The main advantage of projection-based workbenches is that they can rely on a shared common representation of all modeling elements (\eg the AST in MPS), whereas parser-based workbenches have a dedicated parser for each concrete syntax.
One of the claimed advantages of parser-based language workbenches (especially the textual ones) is the flexibility with respect to the models' conformance; the textual representation of parser-based models can still be opened and inspected, whereas projectional editors work directly on the abstract representation of the model. Similarly, according to Negm et al., textual parser-based workbenches avoid tool lock-in since the modeler is not limited to using any specific editor and can be easily integrated with other tools.

Erdweg et al.~\cite{erdweg2015evaluating} conduct a comparison study of 10 language workbenches participating in the 2013 edition of the Language Workbench Challenge (LWC). The comparison of the language workbenches is based on a feature model that includes: notation, semantics, editor support, validation, testing, and composability, where some of them support multiple notations (fully or partially). The conclusions state that no language workbench realizes all features. However, this study is limited to the language workbenches presented in LWC'13. 
For what concerns blended modeling, the results obtained by Erdweg et al. are in line with the ones reported by Negm et al.~\cite{negm2019survey}, where projectional language workbenches are better supporting the combination of different concrete syntaxes, with Enz\~{o} and MPS again as the ones supporting all types of concrete syntaxes. Erdweg et al. also highlight the need for integrating ``different notational styles'', which is at the core of blended modeling.

Merkle~\cite{merkle2010textual} conduct a comparison study of textual language workbenches categorizing them into pure text-based and projectional-based with a textual projection. The language workbenches compared in this study are Xtext\footnote{\url{https://www.eclipse.org/Xtext}}, TEF\footnote{\url{https://www2.informatik.hu-berlin.de/sam/meta-tools/tef/tool.html}}, EMFText\footnote{\url{https://github.com/DevBoost/EMFText}}, and MPS\footnote{\url{https://www.jetbrains.com/mps}}. The language workbenches are compared based on workflow, abstract/concrete syntax, and editor. However, this study is limited to textual language workbenches, while our focus is on tools that provide multiple notations.
In the study by Merkle, the only language workbench supporting a combination of concrete syntaxes is TEF (Textual Editing Framework\footnote{\url{http://www2.informatik.hu-berlin.de/sam/meta-tools/tef/tool.html}}), an Eclipse-based language workbench focusing primarily on textual editors, but with the possibility of embedding them into other editors supporting other concrete syntaxes~\cite{scheidgen2008textual}. Internally, TEF follows the \textit{background parsing} strategy for the textual concrete syntax, where textual models are always represented and edited as plain text, and their parsing is demanded by a background process. TEF also provides some basic form of blending, where modelers can bring up a textual editor from either a graphical or a tree-based editor (\eg by opening a small overlay window); however, TEF-based modeling tools cannot be considered as blended since model updates the embedded textual editor is not seamlessly integrated into its host editor, and model updates are propagated on-demand to the host editor only when the modeler closes the textual editor.

%% file: TexFiles/paperA/Study_Design.tex
\section{Study design}\label{sec:study_design}

The goal of this study is to characterize the state of the art and the state of the practice of modeling tools in relation to blended modeling. More specifically, we formulate such high-level goal by using the Goal-Question-Metric perspectives~\cite{gqm}, shown in Table~\ref{tab:gqm}. 
\begin{center}
    \begin{table}[!htbp]
    \small
    \begin{tabular}{ p{1.3cm} | p{6.5cm} }
    \textit{Purpose} & Identify, classify, and analyze \\ 
    \textit{Issue} & the user-oriented and implementation-oriented characteristics of \\ 
    \textit{Object} & existing modeling tools
    \\ 
    \textit{Context} & in relation to the principles of blended modeling,\\
    \textit{Viewpoint} & from a researcher's and practitioner's point of view.\\
    \end{tabular}
    \caption{Goal of this study.}
    \label{tab:gqm}
    \end{table}
\end{center}

\subsection{Process}

\begin{figure*}[!htbp]
	\centering
	\includegraphics[width=\linewidth]{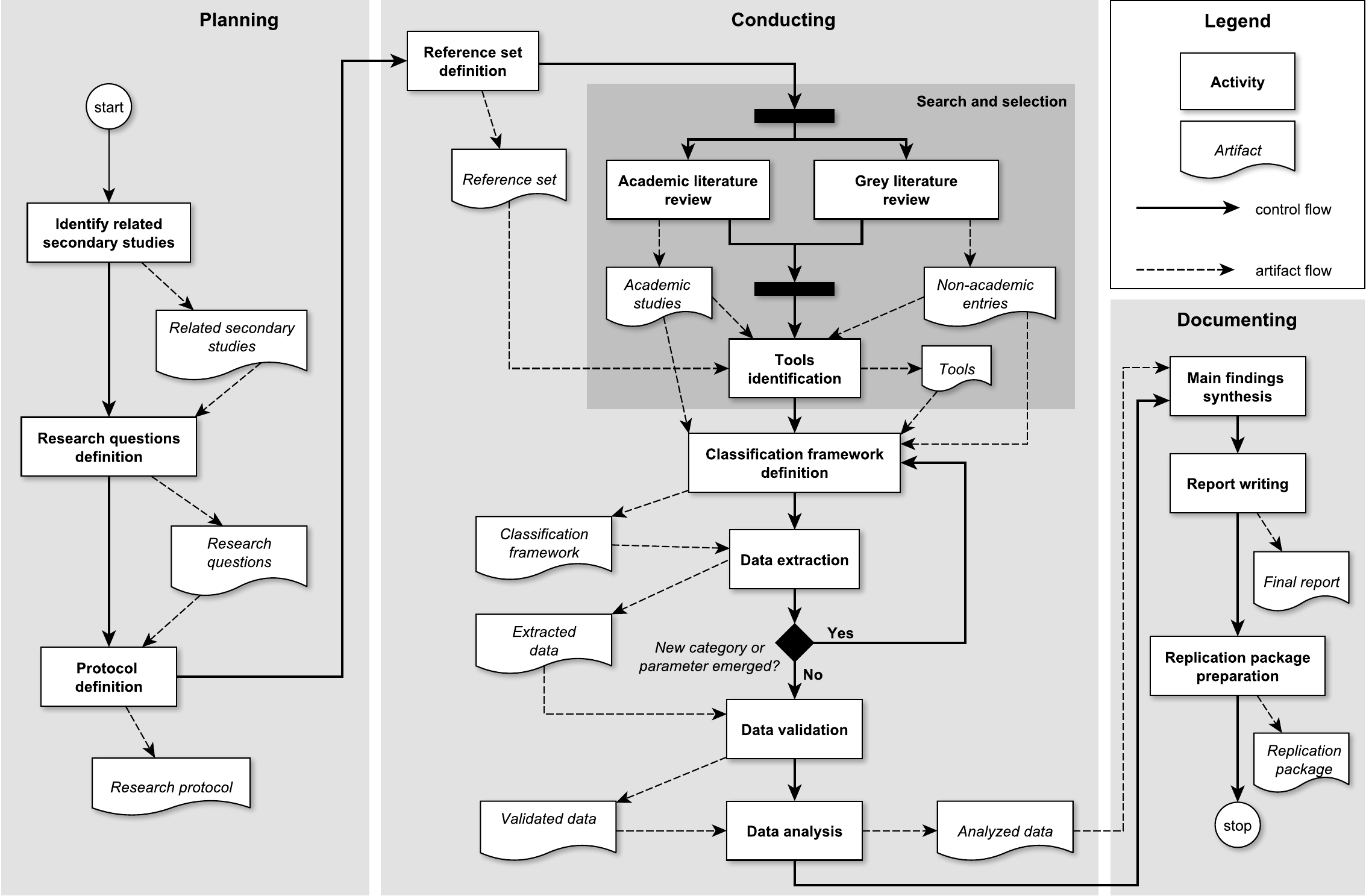}
	\caption{Overview of the whole review process}
	\label{fig:process}
\end{figure*}

This research was carried out by following the process shown in Figure~\ref{fig:process}. Our process can be divided into three main phases, all well-established in systematic secondary studies~\cite{kitchenham2013systematic}\cite{wohlin2012experimentation}\cite{petersen2015guidelines}: planning, conducting and documenting. In the following, we present the three phases of the process.

\subsubsection{Planning}\label{sec:planning}
This phase aims at defining the plan for carrying out all the activities of this study. More specifically, we first identified \emph{related secondary studies}, \ie surveys and literature reviews with a scope similar to the current review's scope (\secref{sec:related-secondary}). Subsequently, we formulated the \emph{research questions} (\secref{sec:rq}), and compiled the \emph{research protocol}.

The research protocol is a document reporting the methodological details of this study. Specifically, the research protocol contains a detailed description of all the steps we followed in the subsequent \emph{Conducting} and \emph{Documenting} phases.
To mitigate potential threats to validity and any bias, the research protocol was defined \textit{prior to} conducting the study, and it was reviewed by two experts. The experts were asked to provide feedback on the protocol, particularly on possible unidentified threats to validity, problems in the overall construction of the review, and the appropriateness of the proposed research protocol and final reports for the aim of this study. Both experts are well-established professors of Computer Science, with substantial experience in empirical research.

\subsubsection{Conducting}\label{sec:conducting}
In this phase, the mapping study is carried out according to the research protocol. More specifically, we carry out the following activities.

\begin{description}
  \item[\textit{Reference set definition.}] The goal of this activity is to identify the modeling tools that could be part of the final set of modeling tools. This set will serve as a guideline for the subsequent steps of the study design, especially formulating the inclusion and exclusion criteria. The inclusion and exclusion criteria will be tested against this set, and thus, the reference set is subject to change until the criteria are not final.
  We identify the initial reference set based on (i) the modeling tools mentioned in related secondary studies (\secref{sec:related-secondary}); (ii) the authors' experiences with tools partially supporting blended modeling (\eg \cite{ciccozzi2019blended,addazi2021blended}); (iii) searches in generic web search engines; and (iv) knowledge garnered from existing networks of experts, \eg by accessing forums and mailing lists (\eg the Eclipse EMF community forum\footnote{\url{https://www.eclipse.org/forums/index.php?t=thread&frm_id=108}}). The results of the subsequent \textit{Search and selection} activity are eventually compared to the reference set for validation purposes.
  The eventual reference set is composed of the following tools:
  MagicDraw~\cite{magicdraw}, Eclipse Papyrus~\cite{papyrus},\\ MetaEdit+\footnote{\url{https://www.metacase.com/products.html}}, Umple~\cite{umple}, and the Open Source AADL Tool Environment (OSATE)~\cite{osate}.

  \item[\textit{Search and selection.} (\secref{sec:selection})] The goal of this activity is to identify as many (possibly blended) modeling tools as possible. Two parallel activities are carried out: the \emph{Academic literature review}, and the \emph{Grey literature review}. 
  In both search activities, we perform a combination of automated search, manual search, and backward-forward snowballing \cite{wohlin2014guidelines}.
  These activities yield two types of artifacts: (i) \textit{Academic studies} (\eg articles published in scientific journals, and proceedings of scientific conferences)
  and (ii) \textit{Non-academic entries} (\eg blog posts, technical reports).
  Because the subject of this study are the \emph{tools} these artifacts describe, both types of artifacts are screened for a specific \textit{Tool} in the \textit{Tools identification} activity. 
  Here, we manually analyze all academic studies and non-academic entries and identify every modeling tool mentioned in their contents. Moreover, in this activity, we keep track of pointers and links referring to the relevant documentation about each tool (\eg its official documentation, its wiki-based knowledge base, \etc).

   \item[\textit{Classification framework definition.} (\secref{sec:classificationFramework})] The\\ goal of this activity is to define the set of categories and their possible values to classify the identified modeling tools.
   
   \item[\textit{Data extraction.} (\secref{sec:extraction})] The goal of this activity is to collect relevant information about each modeling tool. In this activity, multiple researchers collaboratively (i) read the full text of the relevant documentation of each modeling tool, and (ii) populate the data extraction form with the collected data. Upon the emergence of a new category or new possible value in the domain of previously defined categories, the classification framework can be dynamically adapted. In such cases, the previously extracted data entries are updated in accordance with the new framework.
 
  \item[\textit{Data validation.} (\secref{sec:validation})] To ensure the validity of the extracted data, the tool vendors and knowledgeable experts are contacted to review the data extracted in the previous step.
 
  \item[\textit{Data analysis.} (\secref{sec:analysis})] The goal of this activity is to analyze the extracted data in accordance with the research questions. The activity involves both quantitative and qualitative analyses.
  
\end{description}

\subsubsection{Documenting}\label{sec:documenting}
The main activities performed in this phase are: (i) a thorough elaboration on the data analyzed in the previous phase with the aim of discovering the main findings of the study;
(ii) reporting the possible threats to validity, especially the ones identified during the definition of the review protocol;
and (iii) producing the final report. The final report is evaluated by external reviewers and forms the basis of this article. 

The complete \textit{replication package} is available online\footnote{\url{https://zenodo.org/record/6402743}\label{fn:replication-package}} to allow independent researchers to replicate and verify our study, and to reuse our data for other purposes. The replication package includes the research protocol, the list of all academic and non-academic entries considered in the search and selection phase, the complete list of all identified tools, raw data, the scripts for data analysis, and the details on technical requirements.

\subsection{Research questions}\label{sec:rq}

The research questions of this study are reported below. 
\begin{enumerate}[\bfseries{RQ}1.]
    
    \item \emph{What are the \textbf{user-oriented characteristics} of modeling tools most suitable for supporting blended modeling?}
    
    Modeling tools are designed and developed to be adopted by specific users, application domains, and usage scenarios.
    
    By answering this research question, we aim to identify the external characteristics of modeling tools, pertaining to their adoption and usage~\cite{ciccozzi2019blended}. Typical examples include: supported (types of) notations, human-computer interfaces, application domains, and addressed user groups.
    
    Practitioners can benefit from the answer to this research question by understanding how specific state-of-the-art tools address their problems, what are their limitations in terms of blended modeling, and how they can be improved.
    
    \item \emph{What are the \textbf{realization-oriented characteristics} of modeling tools most suitable for supporting blended modeling?}
    
    With the advent of model-based approaches and domain-specific modeling, in particular, several modeling tools are being developed to support certain levels of blending, formalisms, and semantics. Moreover, until the recent spread of mainstream language workbenches (\eg Xtext\footnote{\url{https://www.eclipse.org/Xtext}}, Sirius\footnote{\url{https://www.eclipse.org/sirius}}, MPS\footnote{\url{https://www.jetbrains.com/mps}}, etc.), the development of such modeling tools had been relatively ad-hoc.
    
    By answering this research question, we aim to identify the internal characteristics of modeling tools, and that, in terms of (i) their features and (ii) the techniques employed to implement those features. Typical examples include: implementation platforms, consistency mechanisms, change propagation, traceability, and the linguistic level of model-to-model correspondence are investigated.
    
    Researchers can benefit from the answer to this research question by understanding the state of the practice on the techniques of blended modeling tools, including the gaps to fill.
    
\end{enumerate}

The identified research questions drive the whole study, with a special influence on (i) the search and selection, (ii) data extraction (including the definition of the classification framework), and (iii) data and main findings synthesis.

\subsection{Search and selection}\label{sec:selection}

The goal of the search and selection phase is to retrieve a representative set of modeling tools supporting multiple modeling notations, as demanded by the principles of blended modeling.
First, we perform a \textit{systematic review} of both the academic (\ie  scientific articles published at peer-reviewed academic venues) and grey literature (\ie websites, online blogs, \etc), and discuss the results in~\secref{sec:systematic_reviews}. The output of these two activities (\ie academic studies and non-academic entries) is then further analyzed in order to identify the modeling tools either considered, mentioned, or discussed in them (\secref{sec:toolsIdentification}). 

\subsubsection{Systematic Reviews}\label{sec:systematic_reviews}

We follow the same overall process when reviewing both the academic and grey literature. In this phase, it is \mbox{fundamental} to achieve a good trade-off between the coverage of existing results on the considered topic and having a manageable number of studies to be analyzed \cite{kitchenham2013systematic,garousi2018guidelines}.
To achieve the above-mentioned trade-off, our search and selection process has been designed as a multi-stage process; this gives us full control over the number and characteristics of the entries being either selected or excluded during the various stages. In the following, we present each step of our systematic review process.
In the remainder of this report, we refer to both academic studies and non-academic entries as \textit{primary studies}, unless specifically noted otherwise.
The systematic review is divided into three subsequent and complementary steps of (i) automatic search, (ii) application of selection criteria, and (iii) snowballing. 

\paragraph{Automatic search.}\label{sec:automatic_search}
In this step, we automatically inspect all the results returned from a query execution on (i) Google Scholar for academic studies and (ii) the Google Search engine for grey literature. The automatic searches for both academic and non-academic literature are executed in November 2020.

For the \textit{academic literature}, we use \textit{Google Scholar}. We use Google Scholar as the data source for the following main reasons: (i) it is one of the largest and most complete databases and indexing systems for scientific literature; (ii) as reported in \cite{wohlin2014guidelines}, the adoption of this data source has proved to be a sound choice to identify the initial set of literature studies for the snowballing process (\secref{sec:snowballing}), producing a reasonable number of false positives, but no false negatives (thus, no information is lost); (iii) the query results can be automatically processed via already existing tools.

Below we report the search string used in this study. 
In order to cover as many potentially relevant studies as possible, we defined the search string so that it includes academic studies on blended modeling. The search string can be divided into three main components: the first component captures the model-driven paradigm, the second one captures the focus on multiple entities (\eg multiple notations) and blending, and the third one is used for ensuring that our results focus on software aspects. 
To keep the results of this initial search as focused as possible, the query has been applied to the title of the targeted studies.

\begin{verbatim}
("modeling" OR "modelling" OR "model based"
  OR "model driven")
    AND
("multi*" OR "blended")
    AND
("notation*" OR "syntax*" OR "editor"
  OR "tool" OR "software")
\end{verbatim}

The search string has been tested by executing pilot searches on Google Scholar. At the time of writing, Google Scholar produced a total of 280 hits when searching with the reported search string.

For the \textit{grey literature}, we target the regular \textit{Google Search Engine}. The search engine is selected in accordance with the recommendations for including grey literature in software engineering multi-vocal reviews \cite{garousi2018guidelines}. The search string used for the academic literature yields mostly academic results even in a general web search. We have, therefore, adapted our search strategy to find non-academic sources. In particular, we identified a number of relevant hits through manual searches early on. These manual hits could be classified as either \emph{lists} (\eg Wikipedia's ``List of Unified Modeling Language tools'') or \emph{tool-specific pages} (\eg tool vendor pages or blog posts about how specific tools are used).

We experimented with several search strings to ensure that we find all relevant hits. In particular, we tried to combine different modeling languages and diagram types into one large all-encompassing search string to simplify our search and make it easier to extract results. However, on prototyping this approach, we realized that the \texttt{OR} clauses that we used did not have the desired effect and we did not find the tools we expected, and in particular, not the lists that we expected. In comparison, a search string such as \texttt{(MARTE) AND (tool OR editor OR notation OR modelling)} yields 162 results on Google, whereas our combined search string that included MARTE\footnote{\url{https://www.omg.org/omgmarte}} and many other languages only yielded 150 results.

Therefore, we decided to carry out an independent search for popular modeling languages. We ran the different searches independently and merged the results later on. We selected the relevant modeling languages using a mixture of expert knowledge, browsing the web pages of well-known modeling tools from the reference set and beyond (\eg Eclipse Capella\footnote{\url{https://www.eclipse.org/capella}} and Enterprise Architect\footnote{\url{https://sparxsystems.com/products/ea}}), using lists such as Wikipedia's page on "Modeling Languages". We narrowed down the resulting list of around 40 potential modeling languages by searching for \texttt{(Language Name) AND (tool OR editor OR notation OR modelling)} in Google, and analyzing the first ten non-academic hits (\ie search results that were not academic papers). Since the search term explicitly contains the terms "tool" and "editor", we expected that the Google search engine would include such a tool within the first ten non-academic hits if it exists, and has any practical relevance. Experiments where we checked later result pages for selected searches confirmed this expectation. We thus only included modeling languages for which Google does report a link to a modeling tool. Otherwise, we disregarded it.

To address the large number of hits we would get this way, we limited the search results for each included search string to the first 50 unique results  (if less than 50 hits are reported, we collect all of them), which is based on the suggestion from Garousi et al.~\cite{garousi2018guidelines}. The eventual result set included \numberGreyScreened{} hits, typically containing blog posts, user manuals, websites, technical reports, white papers, academic articles, etc. 

\paragraph{Application of Selection criteria.}\label{sec:selectioncriteria}
In this step, the identified potentially relevant entries undergo rigorous filtering based on the application of a set of selection criteria.
Following the guidelines for systematic literature review for software engineering~\cite{kitchenham2013systematic}, we define the set of inclusion and exclusion criteria \textit{a priori}, in order to reduce the likelihood of bias. 
The potentially relevant entries are rigorously examined by adopting multiple selection rounds in an adaptive reading depth fashion~\cite{petersen2008systematic}.
Specifically, in the first round, the title of the entry is examined. This first step enables us to discard all those papers or web pages that clearly do not fall within the scope of this study. In the second exclusion round, the introduction and conclusion sections are inspected (if present). Finally, the entries are further inspected by considering their full text, in order to ensure that only the ones relevant to answering the research questions are selected.
While processing the full text of a paper/web page, we also keep track of all the mentioned modeling tools and consider them in the tools' identification phase (\secref{sec:toolsIdentification}).

In the following, we detail the set of inclusion and exclusion criteria that guide the selection of the academic and non-academic entries for our systematic review.\footnote{The identifiers used in this section are consistent with those used in the replication package to enable better traceability.} A potentially relevant entry is selected if it (i) satisfies \textit{all} inclusion criteria and (ii) does not satisfy \textit{any} of the exclusion criteria. The selection criteria are divided into three categories, namely: \textit{generic} (\ie they apply for both academic and non-academic studies), \textit{academic-specific}, and \textit{grey-specific}. The decision of adopting three categories of criteria originates from the different nature of the sources we considered (\ie Google Scholar and the Google Search Engine). By defining three different sets, it is possible to design selection criteria specifically tailored to the specific characteristics of academic and non-academic entries, and hence, improve the overall quality of the selection process.

\noindent Generic inclusion criteria:
\begin{enumerate}[{GEN-I}1)]
    \item Entries on modeling tools, \ie where models are used as first-class entities and used as a substantial abstraction from the problem domain (\eg OSATE~\cite{osate} for modeling hardware/software systems according to the AADL modeling language).
    
    \item Entries discussing at least two different notations (possibly for the same abstract syntax). The notations can be of the same type (\eg both textual). 
    
\end{enumerate}

\noindent Generic exclusion criteria:
\begin{enumerate}[{GEN-E}1)]

    \item Entries on non-modeling tools. For example, articles on IDEs, programming tools, drawing tools, \etc
    
    \item Entries that are not in English.
    
    \item Duplicates of already included entries.

    \item Entries that are not available, and hence not analyzable (\eg the full text of a scientific article is not accessible or the link to a web page is broken).

\end{enumerate}

\noindent Exclusion criteria specific to academic sources:

\begin{enumerate}[{A-E}1)]

    \item Studies in the form of full proceedings and books since they are too broad for being thoroughly analyzed in this phase of the study.

    \item Studies that have not been peer-reviewed, as peer-reviewing is the \textit{de facto} standard of quality assurance for scientific literature.

\end{enumerate}

\noindent Exclusion criteria specific to grey literature:

\begin{enumerate}[{G-E}1)]

    \item Web pages reporting exclusively the basic principles of modeling techniques, without mentioning any modeling tool.

    \item Web pages reporting exclusively abstract best practices while applying modeling techniques.

    \item Web pages reporting an implementation without a discussion of its benefits and/or drawbacks.

    \item Academic literature, since such type of studies is considered by a different process in our protocol.

    \item Videos, podcasts, and webinars since they are too time-consuming to be considered for this phase of the study.

\end{enumerate}

\paragraph{Snowballing.}\label{sec:snowballing}
In this step, we complement the preliminary set of academic studies by applying the snowballing 
procedure~\cite{wohlin2014guidelines}.
To mitigate a potential bias with respect to the construct validity of the study, backward and forward snowballing is used to complement the automatic search of the academic literature~\cite{Greenhalgh:2005}.  
In particular, this process is carried out by considering the scientific publications selected in the initial automatic search, and subsequently selecting relevant studies among those cited by one of the initially selected ones (backward snowballing). 
Then, we also perform forward snowballing, \ie selecting relevant studies among those citing one of the initially selected academic studies~\cite{wohlin2014guidelines}. In this context, the \textit{Google Scholar}\footnote{\url{https://scholar.google.com}}
bibliographic database is adopted to retrieve the studies citing the ones selected through the initial search phase.
The final decision about the inclusion of the newly considered publications in the study is based on the application of the selection criteria presented in Section~\ref{sec:selection}.

\subsubsection{Tool identification}\label{sec:toolsIdentification}

In the tool identification activity, each primary study is manually analyzed and the mentioned modeling tools are identified. This is achieved by investigating the full text of each primary study, and collecting every modeling tool mentioned in it, independently of whether it is blended or not. Then, the set of identified modeling tools is filtered for duplicates, which are subsequently merged, regardless of whether the tool originates from an academic or a non-academic source.
After the merge, we obtained a total of \numberToolCandidates{} modeling tools.
For each tool, we have collected the following information: (i) name, (ii) link/reference to official documentation, (ii) organization(s) implementing, maintaining, and supporting the tool, and (iii) tracing information towards all primary studies mentioning the tool.

In order to ensure that the identified tools support us in answering the research questions of this study, we further filter the list of all modeling tools according to a set of selection criteria.
Below we report the inclusion and exclusion criteria.

\begin{enumerate}[{TI}1)]
    \item The tool allows its users to edit the same model in multiple notations. The user can switch between these notations easily and without an extra processing step  (\ie the tool supports some level of blended modeling). The tool allows a certain degree of temporary inconsistencies. Notations like an overview tree for navigation purposes or any textual representation used for file persistency purposes only are not considered (\eg XMI).

    \item The tool is publicly available (either as an open-source or commercial product).

    \item The documentation of the tool is publicly available.
\end{enumerate}

\begin{enumerate}[{TE}1)]
    \item The tool is a language workbench. (Our study focuses on modeling tools themselves.)
    \item The tool is not available for download as a binary that can be run on current operating systems from an official website or an affiliated platform supporting it (\eg a GitHub repository).
    
    \item The documentation of the tool is not in English.
\end{enumerate}

\begin{figure*}[!htbp]
	\centering
	\includegraphics[width=\linewidth]{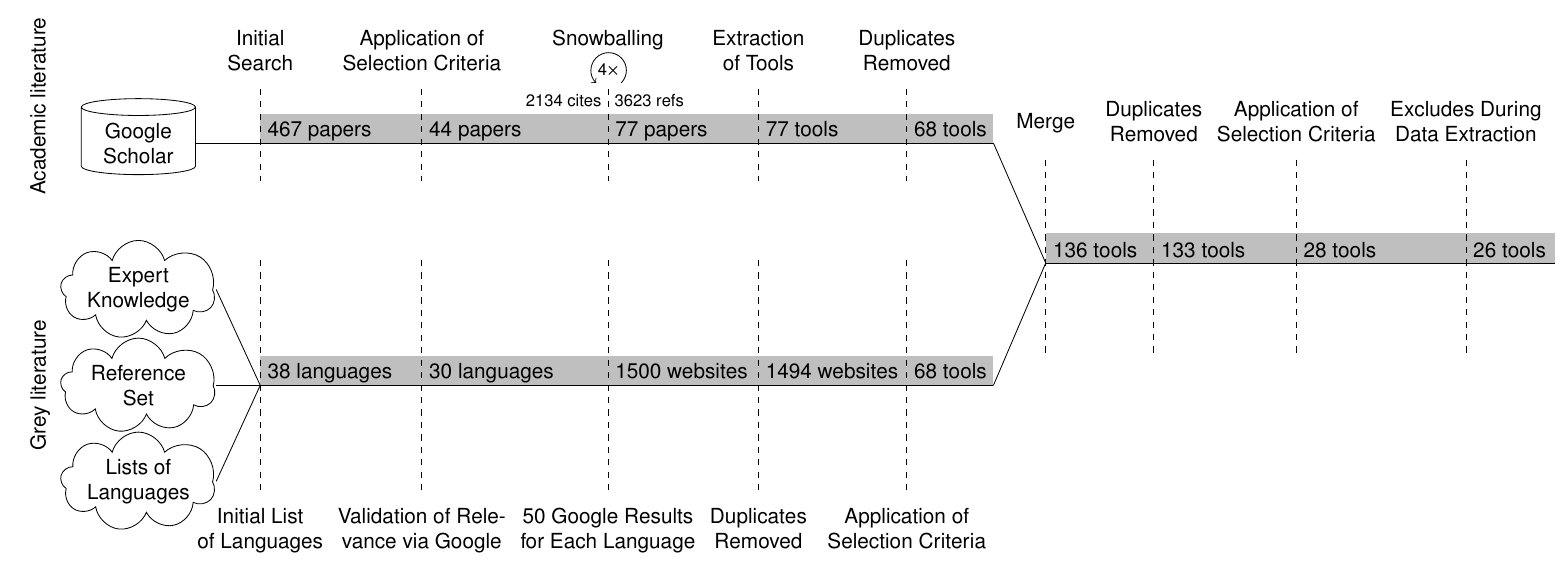}
	\caption{Overview of the conducted search and selection steps}
	\label{fig:numbers}
\end{figure*}

A potentially relevant modeling tool is included if it satisfies \textit{all} inclusion criteria (TI1-TI3), and discarded if it satisfies \textit{any} exclusion criterion (TE1-TE2).

To minimize bias, this activity is performed by five researchers and organized as follows. First, two researchers are randomly assigned to each of the potentially relevant tools. Then, the researchers independently apply the tool selection criteria to their assigned tools; each researcher could mark a tool as \texttt{included}, \texttt{excluded}, \texttt{maybe}. For the 12 of \numberToolCandidates{} tools where at least one researcher indicates an uncertainty (\texttt{maybe}), the conflicts are resolved with the intervention of a randomly--assigned third researcher and, when needed, discuss plenary among all researchers involved in this study.

After the final set of modeling tools has been established, we check whether each tool in the reference set is also included in this final set of tools. If all tools in the reference set are indeed included in the final set of tools, we continue with the subsequent phases of the protocol (\ie data extraction). Otherwise, a dedicated meeting is set up, and a refinement of the systematic review process is designed and conducted again.\\
Eventually, the final list of modeling tools contains all tools of the reference set.

Figure \ref{fig:numbers} shows the different steps performed in the search and selection phase. Out of the 467 papers in the initial scientific search, 44 papers were included in the snowballing process. The snowballing was performed four times before no more new papers were included. During this process, a total of 2,134 cited and 3,623 referenced papers were reviewed. In summary, \numberAcademicTools{} distinct tools were extracted from the included papers. 
For the grey literature part, 30 relevant languages were identified as described above, for which the different search terms yielded \numberGreyScreened{} distinct websites. After applying the selection criteria, \numberGreyTools{} tools were included in the tools set. 
Merging the academic and grey literature parts resulted in \numberToolCandidates{} distinct tools, of which 30 tools were selected according to the tool selection criteria. Two tools had to be excluded during the data extraction process due to lack of availability or semantically out of scope (see \secref{sec:classificationFramework}). Eventually, \numberToolsIdentified{} modeling tools were sampled, shown in the Referred Tools section at the end of this paper.

\input{TexFiles/paperA/classificationfw}

\subsection{Classification framework definition}\label{sec:classificationFramework}

Table~\ref{tab:classificationfw} shows the classification framework of this study. The classification framework is composed of three distinct facets; the first facet is about generic characteristics of modeling tools (\eg release dates, vendor, main motivation for blending notations); the second and third facets directly address research questions RQ1 and RQ2.

We partially reuse the results of previous work~\cite{ciccozzi2019blended} related to blended modeling for defining the initial version of the classification framework.
Then, as suggested in \cite{wohlin2012experimentation}, the customization of the classification framework is performed as follows: (i) firstly we select a random sample of 10 modeling tools, (ii) then two researchers independently extract the data from the 10 modeling tools by using the initial version of the classification framework, (iii) the two researchers then discuss the results of the data extraction with a third researcher, with a special focus on too generic/abstract parameters, parameters which did not fully fit with the characteristics of the tools, parameters with redundant values, and recurrent missing concepts, (iv) the classification framework is customized according to the discussion, and lastly (v) the final version of the classification framework is applied to all remaining modeling tools.
It is important to note that when analyzing the remaining 26 tools, the classification framework can still be enriched/updated based on the characteristics of the currently analyzed tool.
The details about how we extracted data for each modeling tool are provided in the next section.

\subsection{Data extraction}\label{sec:extraction}

The main goal of this activity is to extract relevant data about each modeling tool for answering the research questions. The inputs to this activity are: (i) the set of 28 modeling tools, out of which 26 remained after excluding two additional tools during this phase; and (ii) the textual contents of the academic studies and non-academic entries referring to the tools, and the tools' official documentation (when publicly available). Moreover, when we are not able to collect all relevant data for some specific aspects of a tool (\eg the internal consistency mechanisms of a proprietary tool) we perform a series of ad-hoc Web searches and contact the support team of the tool for collecting the missing data. For the sake of external verifiability, full tracing information is kept between the extracted data and the considered data sources and it is included in the replication package of the study.\footref{fn:replication-package}

To carry out a rigorous data extraction process, and to ease the control and the subsequent analysis of the extracted data, a predefined data extraction form is designed prior to the data extraction process.
The structure of the data extraction form is based on the various categories of the classification framework.

\subsection{Data validation}\label{sec:validation}
To ensure the validity of the extracted data, the tool vendors are contacted and the data and the explanation of the reference framework are made available to them. If a tool does not have a clearly identified vendor, we identify knowledgeable experts who published scientific papers related to the tool.
The vendors and experts are asked to identify any invalid data related to their tool. The contact is initiated via email with the vendors and experts having an option to ask and discuss the details with our research team. The majority of interactions happened in email. Some vendors and experts preferred a live discussion during a video call, which we also accommodated.

Eventually, we have contacted vendors and experts of 24 tools. The authors of this paper have developed or extensively contributed to the remaining 2 tools, and validated them internally.
The validation phase ran for three weeks, between February 28 and March 22, 2022. 69\% of tool vendors or experts replied either with minor change suggestions or with the approval of the extracted data. Based on their responses, 3.8\% of the data (20 of 520 records) has been updated. The most changes, five, were observed in the model-level flexibility category.

\subsection{Data analysis}\label{sec:analysis}

The data analysis activity involves collating and summarizing the data, aiming at understanding, analyzing, and classifying the state of the art of modeling tools \cite[$\S$~6.5]{kitchenham2007guidelines}.
The data synthesis is divided into two main phases: vertical analysis and horizontal analysis.
In both cases, we perform a combination of content
analysis~\cite{franzosi2010quantitative} (mainly for categorizing and coding
tools under broad thematic categories) and narrative
synthesis~\cite{rodgers2009testing} (mainly for detailed explanation and
interpretation of the findings coming from the content analysis).
When performing \textit{vertical analysis}, we analyze the extracted data to find trends and collect information about \emph{each category} of the classification framework.
When performing \textit{horizontal analysis}, we analyze the extracted data to explore possible relations \emph{across different categories} of the classification framework.

\subsubsection{Vertical analysis}\label{sec:horizontal}
Depending on the parameters of the classification framework, in this research, we apply both quantitative and qualitative synthesis methods, separately.
When considering quantitative data, depending on the specific data to be analyzed, we apply descriptive statistics for a better understanding of the data.
When considering qualitative data, we apply the \textit{line of argument} synthesis~\cite{wohlin2012experimentation}, that is: firstly we analyze each tool individually to document it and tabulate its main features with respect to each specific parameter of the classification framework, then we analyze the set of tools as a whole, to reason on potential patterns and trends.
When both quantitative and qualitative analyses are completed, we integrate their results to explain quantitative results by using qualitative results~\cite[$\S$~6.5]{kitchenham2007guidelines}. The results are discussed in~\secref{sec:vertical}.

\subsubsection{Horizontal analysis}\label{sec:horizontal}
Following the best practice of previous secondary studies \cite{david2021collaborative,franzago2017collaborative,difrancesco2019architecting,ciccozzi2018execution}, we explore significant phenomena across pairs of categories as well.
We use contingency tables annotated with the Chi-square statistic at ${\alpha=0.05}$, for identifying statistically significant cases. Following the directions of Haviland~\cite{haviland1990yates}, we report the p-values of the conventional Chi-square test without Yates's correction for continuity. The results are discussed in~\secref{sec:orthogonal}.

%% file: TexFiles/paperA/classificationfw.tex
\begin{table*}[htb]
\sf
\centering
\caption{Categories of the classification framework, and their domain.}
\label{tab:classificationfw}
\scriptsize
\resizebox{\textwidth}{!}{%
\begin{tabular}{lll}
\hline
\rowcolor[HTML]{C0C0C0} 
\multicolumn{1}{|c|}{\cellcolor[HTML]{C0C0C0}\textbf{Category}} & \multicolumn{1}{c|}{\cellcolor[HTML]{C0C0C0}\textbf{Definition}}                                            & \multicolumn{1}{c|}{\cellcolor[HTML]{C0C0C0}\textbf{Type/Domain}}                         \\ \hline
                                                                &                                                                                                             &                                                                                           \\ \hline
\rowcolor[HTML]{C0C0C0} 
\multicolumn{3}{|c|}{\cellcolor[HTML]{C0C0C0}\textbf{GENERIC}}                                                                                                                                                                                                              \\ \hline
\rowcolor[HTML]{EFEFEF} 
\multicolumn{3}{|c|}{\cellcolor[HTML]{EFEFEF}\textbf{META}}                                                                                                                                                                                                               \\ \hline
\multicolumn{1}{|l|}{Tool ID}                                   & \multicolumn{1}{l|}{The internally used ID of the tool.}                                                    & \multicolumn{1}{l|}{"T"+{[}numeric{]}}                                                    \\ \hline
\multicolumn{1}{|l|}{Name}                                      & \multicolumn{1}{l|}{The name of the tool.}                                                                  & \multicolumn{1}{l|}{Free text}                                                            \\ \hline
\multicolumn{1}{|l|}{Analyzed release}                          & \multicolumn{1}{l|}{The version of the release the analysis was carried out on.}                            & \multicolumn{1}{l|}{Free text}                                                            \\ \hline
\rowcolor[HTML]{EFEFEF} 
\multicolumn{3}{|c|}{\cellcolor[HTML]{EFEFEF}\textbf{TOOL}}                                                                                                                                                                                                               \\ \hline
\multicolumn{1}{|l|}{First release}                             & \multicolumn{1}{l|}{Date of the first available release.}                                                   & \multicolumn{1}{l|}{Date}                                                                 \\ \hline
\multicolumn{1}{|l|}{Latest release}                            & \multicolumn{1}{l|}{Date of the latest available release.}                                                  & \multicolumn{1}{l|}{Date}                                                                 \\ \hline
\multicolumn{1}{|l|}{Motivation}                                & \multicolumn{1}{l|}{The self-declared motivation of the tool.}                                              & \multicolumn{1}{l|}{Free text}                                                            \\ \hline
\multicolumn{1}{|l|}{Open-source}                               & \multicolumn{1}{l|}{Whether the tool's sources are available openly.}                                       & \multicolumn{1}{l|}{\{Yes, No\}}                                                          \\ \hline
\multicolumn{1}{|l|}{Web-based}                                 & \multicolumn{1}{l|}{Whether the tool is web-based.}                                                         & \multicolumn{1}{l|}{\{Yes, No\}}                                                          \\ \hline
\multicolumn{1}{|l|}{Collaboration}                             & \multicolumn{1}{l|}{The degree and type of support for collaboration.}                                      & \multicolumn{1}{l|}{\{No, Asynchronous, Synchronous\}}                                    \\ \hline
                                                                &                                                                                                             &                                                                                           \\ \hline
\rowcolor[HTML]{C0C0C0} 
\multicolumn{3}{|c|}{\cellcolor[HTML]{C0C0C0}\textbf{RQ1: USER-ORIENTED CHARACTERISTICS}}                                                                                                                                                                                 \\ \hline
\rowcolor[HTML]{EFEFEF} 
\multicolumn{3}{|c|}{\cellcolor[HTML]{EFEFEF}NOTATIONS}                                                                                                                                                                                                                   \\ \hline
\multicolumn{1}{|l|}{Notation types}                            & \multicolumn{1}{l|}{Types of notations supported by the tool.}                                              & \multicolumn{1}{l|}{\shortstack[l]{\{Textual, Graphical, Tabular,\\ Tree-based, Mixed textual-graphical\}}} \\ \hline
\multicolumn{1}{|l|}{Notation instances (number of)}            & \multicolumn{1}{l|}{Sum number of instances of notation types.}                                             & \multicolumn{1}{l|}{Numeric}                                                              \\ \hline
\multicolumn{1}{|l|}{Embedded notations}                        & \multicolumn{1}{l|}{\shortstack[l]{Whether there are notations\\ that are embedded into each other.}}                         & \multicolumn{1}{l|}{\{Yes, No\}}                                                          \\ \hline
\multicolumn{1}{|l|}{Overlap}                                   & \multicolumn{1}{l|}{The degree of overlap between notations.}                                               & \multicolumn{1}{l|}{\{None, Partial, Complete\}}                                          \\ \hline
\rowcolor[HTML]{EFEFEF} 
\multicolumn{3}{|c|}{\cellcolor[HTML]{EFEFEF}VISUALIZATION AND NAVIGATION}                                                                                                                                                                                                  \\ \hline
\multicolumn{1}{|l|}{Visualize multiple notations}              & \multicolumn{1}{l|}{The ability to visualize more than one notations.}                                      & \multicolumn{1}{l|}{\{Yes, No\}}                                                          \\ \hline
\multicolumn{1}{|l|}{Synchronous navigation}                    & \multicolumn{1}{l|}{\shortstack[l]{Whether the tool supports a synchronous navigation\\ of multiple visualized notations.}}   & \multicolumn{1}{l|}{\{Yes, No\}}                                                          \\ \hline
\multicolumn{1}{|l|}{Navigation among notations}                & \multicolumn{1}{l|}{The dynamics of navigation between different notations.}                                & \multicolumn{1}{l|}{\{Immediate, Complex\}}                                               \\ \hline
\rowcolor[HTML]{EFEFEF} 
\multicolumn{3}{|c|}{\cellcolor[HTML]{EFEFEF}FLEXIBILITY}                                                                                                                                                                                                                 \\ \hline
\multicolumn{1}{|l|}{Flexibility - models}                      & \multicolumn{1}{l|}{\shortstack[l]{Whether the tool supports temporary inconsistency\\ at the level of the instance models.}} & \multicolumn{1}{l|}{\{Yes, No\}}                                                          \\ \hline
\multicolumn{1}{|l|}{Flexibility - language}                    & \multicolumn{1}{l|}{\shortstack[l]{Whether the tool supports temporary inconsistency\\ at the level of the language.}}        & \multicolumn{1}{l|}{\{Yes, No\}}                                                          \\ \hline
\multicolumn{1}{|l|}{Flexibility - persistence}                 & \multicolumn{1}{l|}{Whether the tools can persist inconsistent models.}                                     & \multicolumn{1}{l|}{\{Yes, No\}}                                                          \\ \hline
                                                                &                                                                                                             &                                                                                           \\ \hline
\rowcolor[HTML]{C0C0C0} 
\multicolumn{3}{|c|}{\cellcolor[HTML]{C0C0C0}\textbf{RQ2: REALIZATION-ORIENTED CHARACTERISTICS}}                                                                                                                                                                          \\ \hline
\rowcolor[HTML]{EFEFEF} 
\multicolumn{3}{|c|}{\cellcolor[HTML]{EFEFEF}MAPPING AND PLATFORMS}                                                                                                                                                                                                       \\ \hline
\multicolumn{1}{|l|}{Mapping}                                   & \multicolumn{1}{l|}{The way concrete and abstract syntax are mapped.}                                       & \multicolumn{1}{l|}{\{Parser-based, Projectional\}}                                       \\ \hline
\multicolumn{1}{|l|}{Platform}                                  & \multicolumn{1}{l|}{The platform the tool is built on.}                                                     & \multicolumn{1}{l|}{\{Eclipse, Other\}}                                              \\ \hline
\rowcolor[HTML]{EFEFEF} 
\multicolumn{3}{|c|}{\cellcolor[HTML]{EFEFEF}CHANGE PROPAGATION AND TRACEABILITY}                                                                                                                                                                                         \\ \hline
\multicolumn{1}{|l|}{Change propagation}                        & \multicolumn{1}{l|}{The dynamics of propagating changes across notations.}                                  & \multicolumn{1}{l|}{\{Sequential, Concurrent\}}                                           \\ \hline
\multicolumn{1}{|l|}{Traceability}                              & \multicolumn{1}{l|}{\shortstack[l]{Whether the tool supports explicit\\ traceability between notations.}}                     & \multicolumn{1}{l|}{\{Yes, No\}}                                                          \\ \hline
\rowcolor[HTML]{EFEFEF} 
\multicolumn{3}{|c|}{\cellcolor[HTML]{EFEFEF}INCONSISTENCY MANAGEMENT}                                                                                                                                                                                                    \\ \hline
\multicolumn{1}{|l|}{Inconsistency visualization}               & \multicolumn{1}{l|}{The degree and way the tool visualizes inconsistencies.}                                & \multicolumn{1}{l|}{\{No, Internal, External\}}                                           \\ \hline
\multicolumn{1}{|l|}{Inconsistency management type}             & \multicolumn{1}{l|}{The way the tool manages inconsistencies.}                                              & \multicolumn{1}{l|}{\{On-the-fly, On-demand, Preventive\}}                                \\ \hline
\multicolumn{1}{|l|}{Inconsistency management automation}       & \multicolumn{1}{l|}{\shortstack[l]{The degree of automation of\\ inconsistency management activities.}}                       & \multicolumn{1}{l|}{\shortstack[l]{\{Manual, Partial,\\ Automated, Not applicable\}}}                       \\ \hline
\end{tabular}%
}
\end{table*}

%% file: TexFiles/paperA/Results.tex
\section{Results}\label{sec:vertical}

\lipsum[1][1-4]

\begin{table*}[]
\centering
\rotatebox{-90}{
    \begin{minipage}[t]{1.5\textwidth}
        \centering
        \caption{Relationships between blended aspects (BA) and the research questions (RQ) of this study.}
        \label{tab:blended-rqs}
        \footnotesize
        \begin{tabular}{@{}lm{5.5cm}m{6.5cm}@{}}
        \toprule
         & {\textbf{RQ1: User-oriented characteristics\newline(\secref{sec:rq1})}} & {\textbf{RQ2: Realization-oriented characteristics\newline(\secref{sec:rq2})}}        \\ 
         \midrule
        {\textbf{BA1: Multi-notation}} & Notations\newline(\secref{sec:rq1-blended1})               & Mapping and platforms\newline(\secref{sec:rq2-blended1})                                \\
        {\textbf{BA2: Seamless interaction}}      & Visualization and navigation\newline(\secref{sec:rq1-blended2})    & Change propagation, traceability\newline(\secref{sec:rq2-blended2})       \\ 
        {\textbf{BA3: Flexibility}}    & Model/language/persistence flexibility\newline(\secref{sec:rq1-blended3})                     & Inconsistency management and tolerance\newline(\secref{sec:rq2-blended3}) \\ \bottomrule
        \end{tabular}
        \vspace{10px}
    \end{minipage}}
\end{table*}

\lipsum[2][2-4]

\begin{table*}[]
\sf
\centering
\caption{The list of included tools.}
\label{tab:selectedtools}
\scriptsize
\resizebox{\textwidth}{!}{%
\begin{tabular}{|p{0.4cm}|p{3.2cm}|p{3cm}|p{0.4cm}|p{0.4cm}|p{1.75cm}|p{0.5cm}|p{5.5cm}|}\hline
\rowcolor[HTML]{9B9B9B} 
\multicolumn{3}{|c|}{\cellcolor[HTML]{9B9B9B}\textbf{Tool}}                                                                                                                                      & \multicolumn{3}{c|}{\cellcolor[HTML]{9B9B9B}\textbf{Releases}}                                                                                                                              & \multicolumn{2}{c|}{\cellcolor[HTML]{9B9B9B}\textbf{Info}}                                                                                                                                                                                                                                                                                                                                                                                      \\ \hline
\rowcolor[HTML]{C0C0C0} 
\multicolumn{1}{|c|}{\cellcolor[HTML]{C0C0C0}\textbf{ID}} & \multicolumn{1}{c|}{\cellcolor[HTML]{C0C0C0}\textbf{Name}} & \multicolumn{1}{c|}{\cellcolor[HTML]{C0C0C0}\textbf{Vendor/Maintainer}} & \multicolumn{1}{c|}{\cellcolor[HTML]{C0C0C0}\textbf{First}} & \multicolumn{1}{c|}{\cellcolor[HTML]{C0C0C0}\textbf{Latest}} & \multicolumn{1}{c|}{\cellcolor[HTML]{C0C0C0}\textbf{Analyzed}} & \multicolumn{1}{c|}{\cellcolor[HTML]{C0C0C0}\textbf{Open-source}} & \multicolumn{1}{c|}{\cellcolor[HTML]{C0C0C0}\textbf{Self-declared motivation}}                                                                                                                                                                                                                                                                                                            \\ \hline
\cite{adoit}          & ADOIT: Community Edition                                        & BOC Products \& Services AG                                             & 2003                                                                & 2020                                                        & ADOIT:CE based on ADOIT 12.0   & No          & Enterprise architecture management                                                                                                                                                                                                                                                                                                                                          \\ \hline
\cite{archi}          & Archi                                                       & Beauvoir, P and Sarrodie, JB                                            & 2010                                                                & 2021                                                        & 4.8.1                          & Yes         & Enterprise architecture                                                                                                                                                                                                                                                                                                                                                     \\ \hline
\cite{aris}          & ARIS                                                            & Software AG                                                             & 2009                                                                & 2017                                                        & 2.4d - 7.1.0.1161389           & No          & Business process modeling                                                                                                                                                                                                                                                                                                                                                   \\ \hline
\cite{etas-ascet}          & ASCET Developer                                                 & ETAS                                                                    & 2002                                                                & 2020                                                        & 7.6.0 Build ID 209             & No          & "easily combine texts and graphics suiting your programming needs"                                                                                                                                                                                                                                                                                                          \\ \hline
\cite{atompm}          & AToMPM                                                          & Université de Montréal                                                  & 2013                                                                & 2020                                                        & 0.8.5                          & Yes         & Multi-paradigm modeling on the web                                                                                                                                                                                                                                                                                                                                          \\ \hline
\cite{blended-profile}          & BlendedProfile                                                  & Mälardalen University                                                   & 2018                                                                & 2020                                                        & 0.3                            & Yes         & Blended modelling for UML profiles                                                                                                                                                                                                                                                                                                                                          \\ \hline
\cite{boston}          & Boston                                                          & Viev                                                                    & 2015                                                                & 2020                                                        & 5.0                            & No          & Fact-based modeling via Object-Role Modeling (ORM)                                                                                                                                                                                                                                                                                                                          \\ \hline
\cite{cardanit}          & Cardanit                                                        & ESTECO SpA                                                              & 2013                                                                & 2020                                                        & Online @07.04.2021.            & No          & Modeling BPMN with diagrams and tabular views                                                                                                                                                                                                                                                                                                                               \\ \hline
\cite{certware}          & Certware                                                        & NASA                                                                    & 2013                                                                & 2016                                                        & 2.0                            & Yes         & Safety case modeling                                                                                                                                                                                                                                                                                                                                                        \\ \hline
\cite{dbdiagram}          & DBDiagrams                                                      & Holistics Software                                                      & 2018                                                                & 2021                                                        & Online @07.04.2021.            & No          & Visualize textual DB schema definition                                                                                                                                                                                                                                                                                                                                      \\ \hline
\cite{papyrus}          & Eclipse Papyrus                                                 & The Eclipse Foundation                                                  & 2008                                                                & 2020                                                        & 5.0.0                          & Yes         & Generic-purpose MBSE tool, based on UML and providing support for DSLs via UML Profiles                                                                                                                                                                                                                                                                                     \\ \hline
\cite{epf}          & Eclipse Process Framework Project                               & The Eclipse Foundation                                                  & 2006                                                                & 2018                                                        & 1.5.2                          & Yes         & Software process modeling                                                                                                                                                                                                                                                                                                                                                   \\ \hline
\cite{magicdraw}          & MagicDraw                                                       & CATIA No Magic                                                          & 1998                                                                & 2021                                                        & MagicDraw 2021x LTR Enterprise & No          & Modelling tool that facilitates analysis and design of Object Oriented (OO) systems and databases. It provides code engineering mechanism (with full round-trip support for Java, C++, C\#, CL (MSIL) and CORBA IDL programming languages), as well as database schema modeling, DDL generation and reverse engineering facilities.                                         \\ \hline
\cite{mbeddr}          & mbdeddr                                                         & itemis AG                                                               & 2012                                                                & 2018                                                        & 2018.2.0 based on MPS 2018.2.6 & Yes         & "Boosting productivity and quality by using extensible DSLs, flexible notations and integrated verification tools."                                                                                                                                                                                                                                                         \\ \hline
\cite{memo4ado}          & MEMO4ADO                                                        & OMiLAB                                                                  & 2015                                                                & 2018                                                        & 1.10                           & No          & Multi-Perspective Enterprise Modeling                                                                                                                                                                                                                                                                                                                                       \\ \hline
\cite{modelio}          & Modelio                                                         & Modelisoft                                                              & 2011                                                                & 2020                                                        & 4.1.0 (202001232131)           & Yes         & Generic modeling tool for UML, BPMN, ArchiMate, SysML, etc                                                                                                                                                                                                                                                                                                                  \\ \hline
\cite{osate}          & OSATE                                                           & Carnegie Mellon University                                              & 2004                                                                & 2021                                                        & 2.9.1                          & Yes         & AADL is a language, with different representations. A textual representation provides a comprehensive view of all details of a system, and graphical if one want to hide some details, and allow for a quick navigation in multiple dimensions.                                                                                                                             \\ \hline
\cite{quickdb}          & QuickDataBaseDiagrams                                           & Dovetail Technologies Ltd                                               & 2002                                                                & 2021                                                        & Online @07.04.2021.            & No          & Modeling DB schemas by text and diagram                                                                                                                                                                                                                                                                                                                                     \\ \hline
\cite{sequencediagramorg}          & SequenceDiagramOrg                                              & -                                                                       & 2014                                                                & 2021                                                        & Online - 9.1.1                 & No          & Improve the efficiency when creating and working with sequence diagrams by combining text notation scripting and drawing by clicking and dragging in the same model.                                                                                                                                                                                                        \\ \hline
\cite{somadoxx}          & SOM/ADOxx                                                       & OMiLAB                                                                  & 1996                                                                & 2014                                                        & SOM 3.0 on ADOxx 1.5           & No          & Semantic Object Model. Comprehensive approach for object-oriented and semantic modeling of business systems.                                                                                                                                      \\ \hline
\cite{swimlanes}          & Swimlanes                                                       & -                                                                       & 2014                                                                & 2021                                                        & Online @07.04.2021.            & No          & Visualize sequence diagrams                                                                                                                                                                                                                                                                                                                                                 \\ \hline
\cite{topbraid}          & TopBraid Composer Maestro Edition                                            & TopQuadrant, Inc                                                        & 2006                                                                & 2021                                                        & 7.1.0                          & No          & "TopBraid Composer™ Maestro Edition (TBC-ME) is a comprehensive Knowledge Graph modeling and SPARQL query tool. In use by thousands of commercial customers, Composer offers robust and comprehensive support for building and testing configurations of rich knowledge graphs." \\ \hline
\cite{umlet}          & UMLet                                                           & TU Wien                                                                 & 2002                                                                & 2018                                                        & 14.3 Standalone                & Yes         & Allow textual+visual modeling of UML diagrams                                                                                                                                                                                                                                                                                                                               \\ \hline
\cite{umletino}          & UMLetino                                                        & TU Wien                                                                 & 2013                                                                & 2018                                                        & \multicolumn{1}{r|}{14.3}      & Yes         & Allow textual+visual modeling of UML diagrams                                                                                                                                                                                                                                                                                                                               \\ \hline
\cite{umple}          & Umple                                                           & University of Ottawa                                                    & 2008                                                                & 2020                                                        & Online - 1.30.1.5099 .60569f335 & Yes         & Support the convenient modeling across different formalisms. No particular domain targeted, thus, it's a pretty abstract tool.                                                                                                                                                                                                                                              \\ \hline
\cite{use}          & USE                                                             & Universität Bremen                                                      & 2007                                                                & 2020                                                        & 6.0.0                          & Yes         & System modeling via a subset of UML + OCL                                                                                                                                                                                                                                                                                                                                   \\ \hline
\end{tabular}%
}

\end{table*}

In this section, we elaborate on the findings of this study. 
First, we discuss the general findings in \secref{sec:findings-general}. Then, we elaborate on the two research questions of our study: the user-oriented characteristics (RQ1) and the realization-oriented characteristics (RQ2) of the sampled tools, in Section \ref{sec:rq1} and \ref{sec:rq2}, respectively. In both cases, we contextualize our findings in terms of the three core blended modeling aspects: multi-notation, seamless interaction, and flexibility, as shown in Table~\ref{tab:blended-rqs}.

\subsection{Overview}\label{sec:findings-general}

In this section, we review some of the general findings regarding the analyzed blended modeling tools. The list of the included tools is shown in Table~\ref{tab:selectedtools}.

\begin{figure}[htb]
    \centering
    \includegraphics[width=\linewidth]{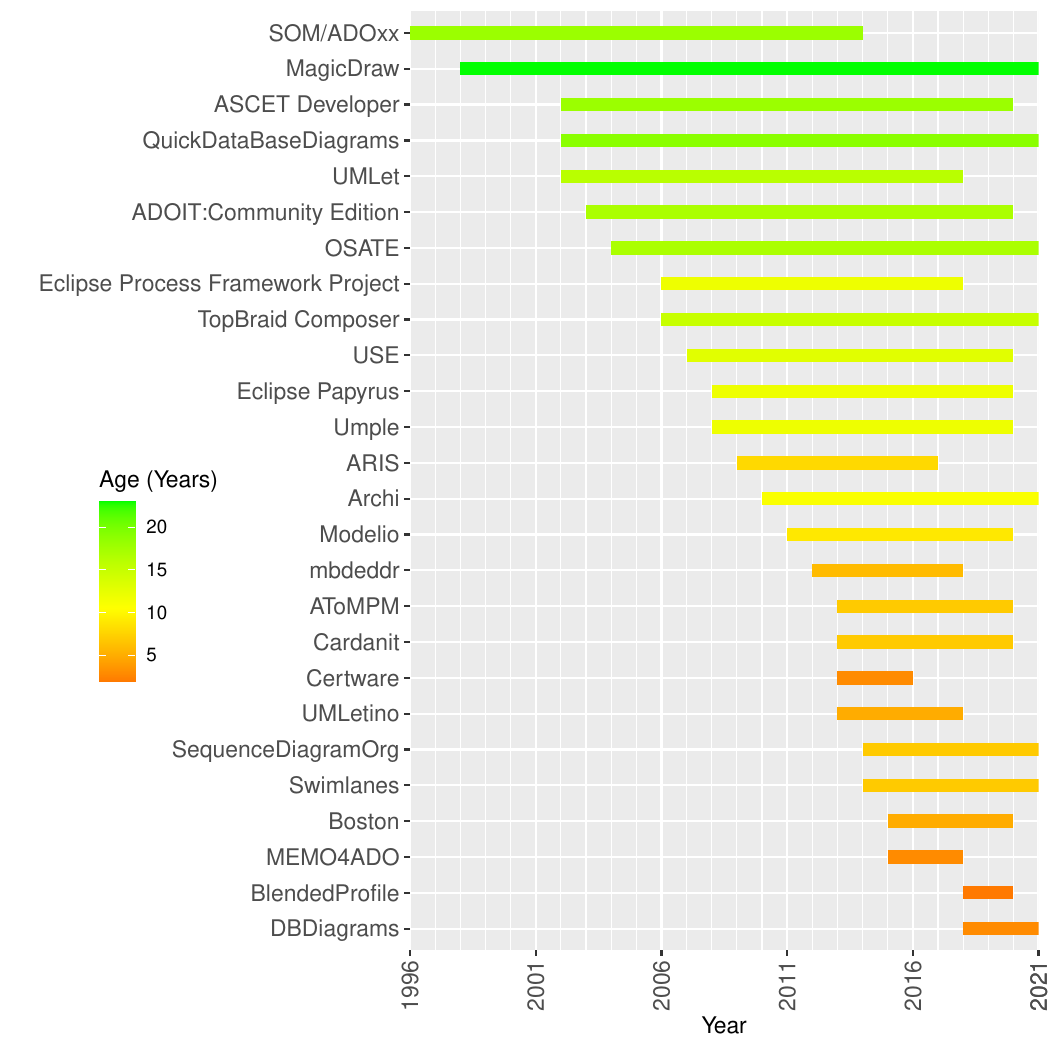}
    \caption{Overview of the age of the tool projects, spanned by their respective first and last releases.}
    \label{fig:releaseYears}
\end{figure}

\paragraph{Project age and timeline.} The tools and their respective projects spread over 25 years, with SOM/ADOxx~\cite{somadoxx} being the oldest tool (first release in 1996) in our sample. On average, the age of the tool projects is 10.6 years ($\sigma=5.9$). The means of the first and last releases are 2008.8 ($\sigma=5.9$) and 2019.4 ($\sigma=1.8$), respectively. These numbers suggest a sample of mature enough tools with sufficient recency in terms of the latest release. \figref{fig:releaseYears} provides a visual overview of the age and timeline of tool projects.

\paragraph{Motivations.}
The self-declared motivations of the tools vary greatly. We have recorded the mission statements of the tools and clustered them.
General-purpose modeling tools are typical in our sample, usually offering multi-notation support for UML-based modeling, e.g. Modelio~\cite{modelio}, USE~\cite{use}, and Papyrus~\cite{papyrus}. Some of these tools are very specific about their intentions to combine or augment the traditional graphical notation of UML with textual elements, such as UMLet~\cite{umlet} and ETAS ASCET Developer~\cite{etas-ascet}. Among the tools with specific modeling purposes are the ones aiming at process modeling (e.g., SOM/ADOxx~\cite{somadoxx}, ARIS~\cite{aris}), database modeling (e.g., DBDiagram~\cite{dbdiagram}, QuickDBD~\cite{quickdb}), and enterprise architecture (e.g., Archi\\~\cite{archi}, ADOIT~\cite{adoit}).

\paragraph{Web-based implementation.} We have found that the majority of the sampled tools, \xofy{17}{\numberToolsIdentified{}}{}, are exclusively desktop-based applications, as shown in Table~\ref{tab:webbased}.

\begin{table}[H]
\centering
\caption{The web-based nature of tools.}
\label{tab:webbased}
\begin{tabular}{ @{}p{1.7cm}p{1.2cm}p{4.3cm}@{}  } 
\toprule
\textbf{Web-based} & \textbf{\#Tools} & \textbf{Tools} \\
\midrule
No & \relativeDatabar{17}{\numberToolsIdentified} & \scriptsize{\cite{archi}, \cite{aris}, \cite{etas-ascet}, \cite{blended-profile}, \cite{boston}, \cite{certware}, \cite{papyrus}, \cite{epf}, \cite{magicdraw}, \cite{mbeddr}, \cite{memo4ado}, \cite{modelio}, \cite{osate}, \cite{somadoxx}, \cite{topbraid}, \cite{umlet}, \cite{use}} \\ 
Yes & \relativeDatabar{9}{\numberToolsIdentified} & \scriptsize{\cite{adoit}, \cite{atompm}, \cite{cardanit}, \cite{dbdiagram}, \cite{quickdb}, \cite{sequencediagramorg}, \cite{swimlanes}, \cite{umletino}}, \cite{umple}\\
\bottomrule
\end{tabular}
\end{table}

\paragraph{Open-source.} Half of the sampled tools are released as open-source software (Table~\ref{tab:opensource}), allowing access to the source code of the tool.

\begin{table}[H]
\centering
\caption{The open-source nature of tools.}
\label{tab:opensource}
\begin{tabular}{ @{}p{1.9cm}p{1.3cm}p{4cm}@{} } 
\toprule
\textbf{Open-source} & 
\textbf{\#Tools} & 
\textbf{Tools} \\
\midrule
No & \relativeDatabar{13}{\numberToolsIdentified} & \scriptsize{\cite{adoit}, \cite{aris}, \cite{etas-ascet}, \cite{boston}, \cite{cardanit}, \cite{dbdiagram}, \cite{magicdraw}, \cite{memo4ado}, \cite{quickdb}, \cite{sequencediagramorg}, \cite{somadoxx}, \cite{swimlanes}, \cite{topbraid}}\\
Yes & \relativeDatabar{13}{\numberToolsIdentified} & \scriptsize{\cite{archi}, \cite{atompm}, \cite{blended-profile}, \cite{certware}, \cite{papyrus}, \cite{epf}, \cite{mbeddr}, \cite{modelio}, \cite{osate}, \cite{umlet}, \cite{umletino}, \cite{umple}, \cite{use}} \\ 
\bottomrule
\end{tabular}
\end{table}

\paragraph{Collaboration.}

Collaborative modeling is the joint creation of a shared representation of a system through means of modeling~\cite{franzago2017collaborative,david2021collaborative}. Collaboration enables an orchestrated interplay among stakeholders of different domains, and thus, very often, collaboration raises the need for multiple different notations.
In real-time collaborative settings, the groupwork of stakeholders happens synchronously. Off-line collaborative settings do not assume synchronicity, but rather stakeholders who work on shared models at different times.
As shown in Table~\ref{tab:collaboration}, the majority of tools, \xofy{15}{\numberToolsIdentified{}}{}, provides some means of collaboration. Specifically, off-line techniques are typical, accounting for \xofy{9}{15}{collaborative tools} or \xofy{9}{\numberToolsIdentified{}}{tools overall}, respectively. Finally, \xofy{11}{\numberToolsIdentified{}}{sampled tools} do not support any means of collaboration.

\begin{table}[H]
\centering
\caption{Support for collaboration.}
\label{tab:collaboration}
\begin{tabular}{ @{}p{2cm}p{1.2cm}p{4cm}@{} } 
\toprule
\textbf{Collaboration} & \textbf{\#Tools} & \textbf{Tools} \\
\midrule
No & \relativeDatabar{11}{\numberToolsIdentified} & \scriptsize{ \cite{aris}, \cite{blended-profile}, \cite{certware}, \cite{papyrus}, \cite{epf}, \cite{memo4ado}, \cite{sequencediagramorg}, \cite{somadoxx}, \cite{swimlanes}, \cite{topbraid} \cite{use}} \\ 
Yes: Off-line & \relativeDatabar{9}{\numberToolsIdentified} & \scriptsize{\cite{archi}, \cite{etas-ascet}, \cite{magicdraw}, \cite{mbeddr}, \cite{modelio}, \cite{osate}, \cite{umlet}, \cite{umletino}}, \cite{umple} \\
Yes: Real-time & \relativeDatabar{6}{\numberToolsIdentified} & \scriptsize{\cite{adoit}, \cite{atompm}, \cite{boston}, \cite{cardanit}, \cite{dbdiagram}, \cite{quickdb}} \\
\bottomrule
\end{tabular}
\end{table}

\subsection{User-oriented characteristics (RQ1)}\label{sec:rq1}

In this section, we discuss the findings related to the user-oriented characteristics of the sampled tools. We contextualize our findings in terms of the three aspects of blended modeling tools: the support for multiple notations (\secref{sec:rq1-blended1}), seamless interaction (\secref{sec:rq1-blended2}), and flexibility (\secref{sec:rq1-blended3}).

\subsubsection{Notations}\label{sec:rq1-blended1}

\begin{figure}
    \centering
    \subfloat[\centering Number of notation types.
    \label{fig:concreteSyntaxTypesCount-a}]{\includegraphics[trim=0 0 0 0,width=0.285\linewidth]{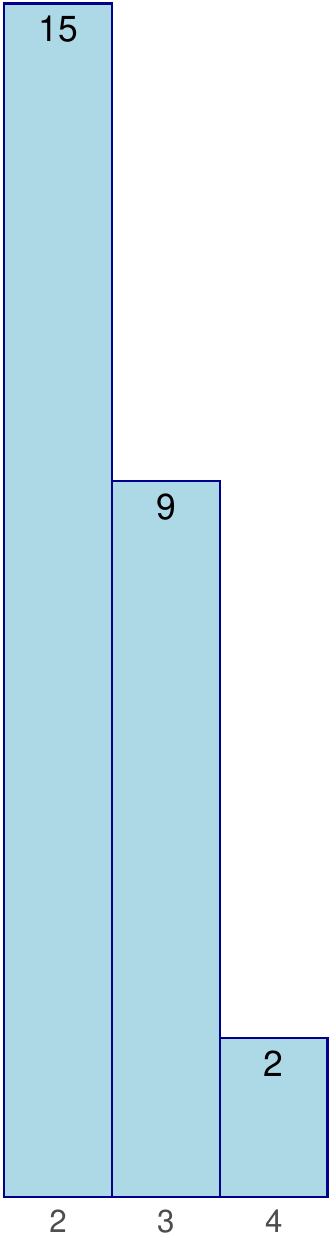}}
    \qquad
    \subfloat[\centering Support for specific notation types.\label{fig:concreteSyntaxTypesCount-b}]{\includegraphics[width=0.62\linewidth]{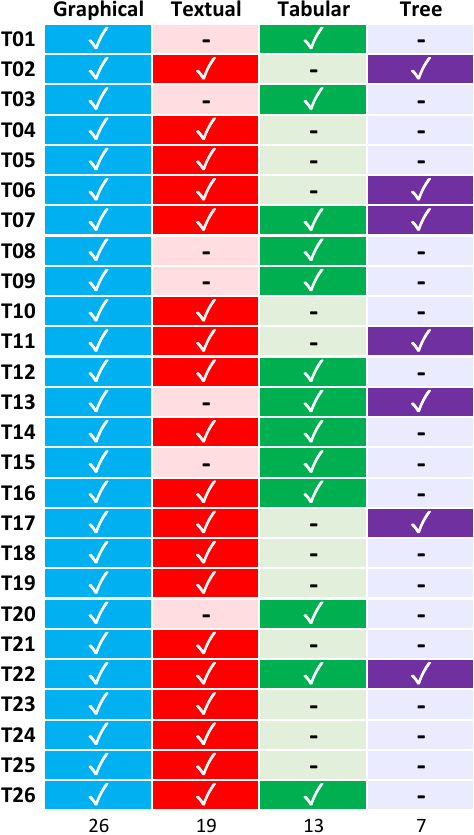}}
    \caption{Number and combinations of notation types.}
    \label{fig:concreteSyntaxTypesCount}
\end{figure}

\paragraph{Notation types.} 
As shown in~\figref{fig:concreteSyntaxTypesCount-a} and Table~\ref{tab:noconcretesyntax}, the majority of tools, \xofy{5}{\numberToolsIdentified{}}{}, support two types of notation, with additional nine tools supporting three types, and two tools supporting four types.

Every tool, \xofy{26}{\numberToolsIdentified{}}{}, features a graphical notation. Textual notations are supported by 19 tools. Additional 13 tools were found with a support for tabular notations, and seven with a support for tree-like notations. This information is detailed in~\figref{fig:concreteSyntaxTypesCount-b} and Table~\ref{tab:notationsupport}.

\begin{table}[H]
\centering
\caption{Number of supported notation types.}
\label{tab:noconcretesyntax}
\begin{tabular}{ @{}p{1.6cm}p{1.3cm}p{4cm}@{} } 
\toprule
\textbf{\#Notation types} & \textbf{\#Tools} & \textbf{Tools} \\
\midrule
2 & \relativeDatabar{15}{\numberToolsIdentified} & \scriptsize{\cite{adoit}, \cite{aris}, \cite{etas-ascet}, \cite{atompm}, \cite{cardanit}, \cite{certware}, \cite{dbdiagram}, \cite{memo4ado}, \cite{quickdb}, \cite{sequencediagramorg}, \cite{somadoxx}, \cite{swimlanes}, \cite{umlet}, \cite{umletino}, \cite{umple}}\\ 
3 & \relativeDatabar{9}{\numberToolsIdentified} & \scriptsize{\cite{archi}, \cite{blended-profile}, \cite{epf}, \cite{papyrus}, \cite{magicdraw}, \cite{mbeddr}, \cite{modelio}, \cite{osate}, \cite{use}}\\ 
4 & \relativeDatabar{2}{\numberToolsIdentified} & \scriptsize{\cite{boston}, \cite{topbraid}} \\ 
\bottomrule
\end{tabular}
\end{table}

\begin{table}[H]
\centering
\caption{Support for specific notation types.}
\label{tab:notationsupport}
\begin{tabular}{ @{}p{1.6cm}p{1.3cm}p{4cm}@{} } 
\toprule
\textbf{Notation type} & \textbf{\#Tools} & \textbf{Tools} \\
\midrule
Graphical & \relativeDatabar{26}{\numberToolsIdentified} & \scriptsize{\cite{adoit}, \cite{archi}, \cite{aris}, \cite{etas-ascet}, \cite{atompm}, \cite{blended-profile}, \cite{boston}, \cite{cardanit}, \cite{certware}, \cite{dbdiagram}, \cite{papyrus}, \cite{epf}, \cite{magicdraw}, \cite{mbeddr}, \cite{memo4ado}, \cite{modelio}, \cite{osate}, \cite{quickdb}, \cite{sequencediagramorg}, \cite{somadoxx}, \cite{swimlanes}, \cite{topbraid}, \cite{umlet}, \cite{umletino}, \cite{umple}, \cite{use}}\\ 
Textual & \relativeDatabar{19}{\numberToolsIdentified} & \scriptsize{\cite{archi}, \cite{etas-ascet}, \cite{atompm}, \cite{blended-profile}, \cite{boston}, \cite{dbdiagram}, \cite{epf}, \cite{papyrus}, \cite{mbeddr}, \cite{modelio}, \cite{osate}, \cite{quickdb}, \cite{sequencediagramorg}, \cite{swimlanes}, \cite{topbraid}, \cite{umlet}, \cite{umletino}, \cite{umple}, \cite{use}}\\ 
Tabular & \relativeDatabar{13}{\numberToolsIdentified} & \scriptsize{\cite{adoit}, \cite{aris}, \cite{boston}, \cite{cardanit}, \cite{certware}, \cite{epf}, \cite{magicdraw}, \cite{mbeddr}, \cite{memo4ado}, \cite{modelio}, \cite{somadoxx}, \cite{topbraid}, \cite{use}} \\ 
Tree & \relativeDatabar{7}{\numberToolsIdentified} & \scriptsize{\cite{archi}, \cite{blended-profile}, \cite{boston}, \cite{papyrus}, \cite{magicdraw}, \cite{osate}, \cite{topbraid}} \\ 
\bottomrule
\end{tabular}
\end{table}

\paragraph{Embedded notations.}

We found a single occurrence of embedded notations, i.e., a host notation being enriched by fragments of another notation (Table~\ref{tab:embeddedlanguages}). While the host notation is prevalent during the entirety of the interaction, the embedded notation is accessible in a specific subset of the host notation. For example, in the \texttt{Statecharts + Class Diagrams} (SCCD) formalism~\cite{vanmierlo2016sccd}, Class Diagram fragments are used to augment the Statecharts formalism, and provide structural information to compose complex systems.

\begin{table}[H]
\centering
\caption{Support for embedded languages.}
\label{tab:embeddedlanguages}
\begin{tabular}{ @{}p{1.6cm}p{1.3cm}p{4cm}@{} } 
\toprule
\textbf{Embedded languages} & \textbf{\#Tools} & \textbf{Tools} \\ \midrule
No & \relativeDatabar{25}{\numberToolsIdentified} & \scriptsize{\cite{adoit}, \cite{archi}, \cite{aris}, \cite{etas-ascet}, \cite{atompm}, \cite{blended-profile}, \cite{boston}, \cite{cardanit}, \cite{certware}, \cite{dbdiagram}, \cite{papyrus}, \cite{epf}, \cite{magicdraw}, \cite{memo4ado}, \cite{modelio}, \cite{osate}, \cite{quickdb}, \cite{sequencediagramorg}, \cite{somadoxx}, \cite{swimlanes}, \cite{topbraid}, \cite{umlet}, \cite{umletino}, \cite{umple}, \cite{use}}\\ 
Yes & \relativeDatabar{1}{\numberToolsIdentified} & \scriptsize{ \cite{mbeddr}} \\ 
\bottomrule
\end{tabular}
\end{table}

\paragraph{Overlap.}
The majority of tools, \xofy{21}{\numberToolsIdentified{}}{}, comes with notations that are not fully overlapping. This means that different notations provide different modeling aspects in these tools.
An example of full overlap is where a graphical state machine language can render a state machine model with every structural feature; whereas a table only shows which states have transitions to which states.

\begin{table}[htp]
\centering
\caption{Overlap between notations.}
\label{tab:overlap}
\begin{tabular}{ @{}p{1.6cm}p{1.3cm}p{4cm}@{} } 
\toprule
\textbf{Overlap} & \textbf{\#Tools} & \textbf{Tools} \\
\midrule
Partial & \relativeDatabar{21}{\numberToolsIdentified} &  \scriptsize{\cite{adoit}, \cite{archi}, \cite{aris}, \cite{etas-ascet}, \cite{atompm}, \cite{blended-profile}, \cite{boston}, \cite{cardanit}, \cite{certware}, \cite{papyrus}, \cite{epf}, \cite{magicdraw}, \cite{memo4ado}, \cite{modelio}, \cite{osate}, \cite{somadoxx}, \cite{topbraid}, \cite{umlet}, \cite{umletino}, \cite{umple}, \cite{use}} \\ 
Complete & \relativeDatabar{5}{\numberToolsIdentified} & \scriptsize{\cite{dbdiagram}, \cite{mbeddr}, \cite{quickdb}, \cite{sequencediagramorg}, \cite{swimlanes}}\\
\bottomrule
\end{tabular}
\end{table}

\subsubsection{Visualization and navigation}\label{sec:rq1-blended2}

Usability aspects in general are hard to measure. To gain reliable results, it is necessary to conduct a complex user study with concrete tasks, a larger number of participants, interviews and/or surveys, and a thorough evaluation of the answers. This is not feasible in the context of this study and therefore, we decided to focus on usability aspects that are i) easily measured objectively and ii) specific to blended modeling. 

We do not consider the usability of modeling languages themselves as discussed in~\cite{moody2009physics} and~\cite{barisic2012usability}. Instead, we focus on the usability of the tools in terms of the topics that are crucial for blended modeling. The idea of blended modeling is to use the notation that is best suited for the current task at hand. This makes it necessary to switch frequently between the available notations. Therefore, for pleasant usability with good support for the user, a tool must offer the possibility to visualize multiple notations side by side and/or provide seamless navigation between notations, or even synchronized navigation. To clarify this more focused view of usability, we use the term ``seamless interaction''.

\paragraph{Visualization of multiple syntaxes.} In general, a blended modeling tool must have the ability to support multiple concrete syntaxes of the same abstract syntax. This parameter, in particular, addresses the possibility of simultaneously viewing multiple notations within a modeling tool, e.g., side-by-side or in an integrated manner such as projectional editors as mbeddr \cite{mbeddr} do. All \numberToolsIdentified{} identified tools support the simultaneous view of two or more notations. 

\paragraph{Synchronized navigation.} In addition to the previous parameter, this parameter investigates whether the navigation across multiple notations in the models' editors is synchronized. For instance, this can be the case in a side-by-side view, if an element in one notation is selected, also its corresponding element in the other notation is selected. Another example of such synchronized navigation is the usage of the double click feature to jump between different views showing corresponding elements but belonging to different notations. As shown in Table~\ref{tab:synchronizednavigation},
more than half of the tools, \xofy{16}{\numberToolsIdentified{}}{}, provides synchronized navigation facilities.

\begin{table}[htp]
\centering
\caption{Support for synchronized navigation.}
\label{tab:synchronizednavigation}
\begin{tabular}{ @{}p{1.6cm}p{1.3cm}p{4cm}@{} } 
\toprule
\textbf{Sync'd navigation} & \textbf{\#Tools} & \textbf{Tools} \\
\midrule
Yes & \relativeDatabar{16}{\numberToolsIdentified} &  \scriptsize{\cite{archi}, \cite{aris}, \cite{atompm}, \cite{blended-profile}, \cite{boston}, \cite{cardanit}, \cite{dbdiagram}, \cite{epf}, \cite{magicdraw}, \cite{mbeddr}, \cite{modelio}, \cite{quickdb}, \cite{sequencediagramorg}, \cite{swimlanes}, \cite{topbraid}, \cite{umple}} \\ 
No & \relativeDatabar{10}{\numberToolsIdentified} & \scriptsize{\cite{adoit}, \cite{etas-ascet}, \cite{certware}, \cite{papyrus}, \cite{memo4ado}, \cite{osate}, \cite{somadoxx}, \cite{umlet}, \cite{umletino}, \cite{use}}\\
\bottomrule
\end{tabular}
\end{table}

\paragraph{Navigation among notations.} Blended modeling tools introduce the benefit that the same model can be viewed and modified using different notations. To enable a fluent modeling experience, the effort required to navigate across notations should be minimal. This binary parameter classifies the effort. It can be either \textit{immediate} (e.g., a click or a keyboard shortcut), or it can involve more complex steps, such as the navigation through multiple (context) menus or wizards. The majority of tools, \xofy{20}{\numberToolsIdentified{}}{}, provide immediate navigation from one notation to the other, suggesting a better user experience in terms of seamless interaction.

\begin{table}[htp]
\centering
\caption{Navigation among notations.}
\label{tab:navamongsyntaxes}
\begin{tabular}{ @{}p{1.6cm}p{1.3cm}p{4cm}@{}} 
\toprule
\textbf{Navigation} & \textbf{\#Tools} & \textbf{Tools} \\
\midrule
Immediate & \relativeDatabar{20}{\numberToolsIdentified}  & \scriptsize{\cite{archi}, \cite{aris}, \cite{etas-ascet}, \cite{atompm}, \cite{blended-profile}, \cite{boston}, \cite{cardanit}, \cite{dbdiagram}, \cite{papyrus}, \cite{epf}, \cite{magicdraw}, \cite{mbeddr}, \cite{memo4ado}, \cite{modelio}, \cite{quickdb}, \cite{sequencediagramorg}, \cite{somadoxx}, \cite{swimlanes}, \cite{topbraid}, \cite{umple}}\\ 
Complex & \relativeDatabar{6}{\numberToolsIdentified} & \scriptsize{\cite{adoit}, \cite{certware}, \cite{osate}, \cite{umlet}, \cite{umletino}, \cite{use}}\\
\bottomrule
\end{tabular}
\end{table}

\subsubsection{Flexibility}\label{sec:rq1-blended3}

Flexibility is the user-related embodiment of tolerating vertical and horizontal inconsistencies~\cite{vanherpen2016ontological} at various levels of abstraction in the modeling stack and various modeling facilities. In this study, we specifically consider three types of flexibility, as follows.

\paragraph{Flexibility -- models.} 
As shown in Table~\ref{tab:flexibilitymodels}, the majority of tools, \xofy{19}{\numberToolsIdentified{}}{}, does not provide flexibility at the model level. This means that there are no inconsistency tolerance mechanisms in place that would allow deviations between different notations describing the same model. However, a small set of six tools support model-level flexibility.

\begin{table}[ht]
\centering
\caption{Support for model-level flexibility.}
\label{tab:flexibilitymodels}
\begin{tabular}{ @{}p{1.6cm}p{1.3cm}p{4cm}@{} } 
\toprule
\textbf{Flexibility: models} & \textbf{\#Tools} & \textbf{Tools} \\
\midrule
No &  \relativeDatabar{19}{\numberToolsIdentified} &  \scriptsize{\cite{adoit}, \cite{archi}, \cite{aris}, \cite{atompm}, \cite{blended-profile}, \cite{cardanit}, \cite{certware}, \cite{papyrus}, \cite{magicdraw}, \cite{mbeddr}, \cite{modelio}, \cite{osate}, \cite{sequencediagramorg}, \cite{somadoxx}, \cite{swimlanes}, \cite{topbraid}, \cite{umlet}, \cite{umletino}, \cite{use}} \\ 
Yes &  \relativeDatabar{7}{\numberToolsIdentified} & \scriptsize{\cite{boston}, \cite{epf}, \cite{etas-ascet}, \cite{dbdiagram}, \cite{memo4ado},  \cite{quickdb}, \cite{umple}}\\
\bottomrule
\end{tabular}
\end{table}

\paragraph{Flexibility -- language.}
The majority of tools, \xofy{22}{\numberToolsIdentified{}}{}, does not provide flexibility at the language-level. (Table~\ref{tab:flexibilitylanguage}) This means that vertical inconsistencies between model and language (e.g., broken conformance or typing relationships) are not tolerated. We found three exceptions, which are, however, different from the ones with support for model-level flexibility discussed above: mbeddr~\cite{mbeddr}, OSATE~\cite{osate}, TopBraid Composer~\cite{topbraid}. Only a single tool, Umple~\cite{umple}, supports both model- and language-level flexibility.

\begin{table}[ht]
\centering
\caption{Support for language-level flexibility.}
\label{tab:flexibilitylanguage}
\begin{tabular}{ @{}p{1.6cm}p{1.3cm}p{4cm}@{} } 
\toprule
\textbf{Flexibility: language} & \textbf{\#Tools} & \textbf{Tools} \\
\midrule
No & \relativeDatabar{22}{\numberToolsIdentified} &  \scriptsize{\cite{adoit}, \cite{archi}, \cite{aris}, \cite{etas-ascet}, \cite{atompm}, \cite{blended-profile}, \cite{boston}, \cite{cardanit}, \cite{certware}, \cite{dbdiagram}, \cite{epf}, \cite{papyrus}, \cite{magicdraw}, \cite{memo4ado}, \cite{modelio}, \cite{quickdb}, \cite{sequencediagramorg}, \cite{somadoxx}, \cite{swimlanes}, \cite{umlet}, \cite{umletino}, \cite{use}} \\ 
Yes & \relativeDatabar{4}{\numberToolsIdentified} & \scriptsize{\cite{mbeddr}, \cite{osate}, \cite{topbraid}, \cite{umple}}\\
\bottomrule
\end{tabular}
\end{table}

\paragraph{Flexibility -- persistence.} 
The majority of tools, \xofy{22}{\numberToolsIdentified{}}{}, does not support persisting inconsistent models. (Table~\ref{tab:flexibilitypersistence}) Out of the ones with support for persistence-level flexibility, ETAS ASCET Developer~\cite{etas-ascet} and\\ Umple~\cite{umple} support model-flexibility and flexibility at both levels, respectively. The other two tools with support for persistence-level flexibility are MagicDraw~\cite{magicdraw} and SequenceDiagramOrg~\cite{sequencediagramorg}.

\begin{table}[ht]
\centering
\caption{Support for persistence flexibility.}
\label{tab:flexibilitypersistence}
\begin{tabular}{  @{}p{1.7cm}p{1.3cm}p{4cm}@{}  } 
\toprule
\textbf{Flexibility: persistence} & \textbf{\#Tools} & \textbf{Tools} \\
\midrule
No &  \relativeDatabar{22}{\numberToolsIdentified} & \scriptsize{\cite{adoit}, \cite{archi}, \cite{aris}, \cite{atompm}, \cite{blended-profile}, \cite{boston}, \cite{cardanit}, \cite{certware}, \cite{dbdiagram}, \cite{papyrus}, \cite{epf}, \cite{mbeddr}, \cite{memo4ado}, \cite{modelio}, \cite{osate}, \cite{quickdb}, \cite{somadoxx}, \cite{swimlanes}, \cite{topbraid}, \cite{umlet}, \cite{umletino}, \cite{use}} \\ 
Yes &  \relativeDatabar{4}{\numberToolsIdentified} & \scriptsize{\cite{etas-ascet}, \cite{magicdraw}, \cite{sequencediagramorg}, \cite{umple}}\\
\bottomrule
\end{tabular}
\end{table}

\subsection{Realization-oriented characteristics (RQ2)}\label{sec:rq2}

In this section, we discuss the findings related to the implementation characteristics of the sampled tools. We contextualize our findings in terms of the three aspects of blended modeling tools: the support for multiple notations (\secref{sec:rq2-blended1}), seamless interaction (\secref{sec:rq2-blended2}), and flexibility (\secref{sec:rq2-blended3}).

\subsubsection{Mapping and platforms}\label{sec:rq2-blended1}

\paragraph{Mapping.} The mapping between abstract syntax and notation is typically implemented either in a parser-based or in a projectional fashion.
In \textit{parser-based} approaches, the user modifies the models via different notations, and a parser produces the abstract syntax tree. In \textit{projectional} approaches, however, the abstract syntax tree is modified directly. Since projectional editors bypass the stages of parser-based editors, they provide support for notations that cannot be easily parsed, but at the same time deliver a different editing experience for textual notations. As shown in Table~\ref{tab:mapping}, the majority of tools, \xofy{22}{\numberToolsIdentified{}}{}, implement a parser-based editor, while four come with projectional facilities.

\begin{table}[ht]
\centering
\caption{Type of mapping.}
\label{tab:mapping}
\begin{tabular}{ @{}p{1.6cm}p{1.3cm}p{4cm}@{} } 
\toprule
\textbf{Mapping} & \textbf{\#Tools} & \textbf{Tools} \\
\midrule
Parser-based & \relativeDatabar{22}{\numberToolsIdentified} & \scriptsize{\cite{archi}, \cite{aris}, \cite{etas-ascet}, \cite{atompm}, \cite{blended-profile}, \cite{boston}, \cite{cardanit}, \cite{certware}, \cite{dbdiagram}, \cite{papyrus}, \cite{epf}, \cite{modelio}, \cite{osate} \cite{quickdb}, \cite{sequencediagramorg}, \cite{somadoxx}, \cite{swimlanes}, \cite{topbraid}, \cite{umlet}, \cite{umletino}, \cite{umple}, \cite{use}}  \\ 
Projectional & \relativeDatabar{4}{\numberToolsIdentified} & \scriptsize{\cite{adoit}, \cite{magicdraw}, \cite{mbeddr}, \cite{memo4ado}} \\
\bottomrule
\end{tabular}
\end{table}

\paragraph{Platforms.}
Eclipse is the only frequently encountered platform in our sample.
As shown in Table~\ref{tab:platforms}, \xofy{10}{\numberToolsIdentified{}}{tools} are built on top of Eclipse, and 18 are built on other, mainly custom platforms. mbdeddr~\cite{mbeddr} is the only MPS-based tool in our sample.
One tool, MagicDraw~\cite{magicdraw}, also supports more than one platform.

\begin{table}[ht]
\centering
\caption{Platforms of implementation.}
\label{tab:platforms}
\begin{tabular}{  @{}p{1.6cm}p{1.3cm}p{4cm}@{} } 
\toprule
\textbf{Platform} & \textbf{\#Tools} & \textbf{Tools} \\
\midrule
Other & \relativeDatabar{17}{\numberToolsIdentified} & \scriptsize{\cite{adoit}, \cite{aris}, \cite{atompm}, \cite{boston}, \cite{cardanit}, \cite{dbdiagram}, \cite{magicdraw}, \cite{memo4ado}, \cite{mbeddr}, \cite{modelio},  \cite{quickdb}, \cite{sequencediagramorg}, \cite{somadoxx}, \cite{swimlanes}, \cite{umletino}, \cite{umple}, \cite{use}} \\ 
Eclipse & \relativeDatabar{10}{\numberToolsIdentified} & \scriptsize{\cite{archi}, \cite{etas-ascet}, \cite{blended-profile}, \cite{certware}, \cite{papyrus}, \cite{epf}, \cite{magicdraw}, \cite{osate}, \cite{topbraid}, \cite{umlet}} \\
\bottomrule
\end{tabular}
\end{table}

\subsubsection{Change propagation and traceability}\label{sec:rq2-blended2}

Change propagation and traceability are the realization-oriented manifestations of the \textit{seamless integration} blended modeling aspect. (See Table~\ref{tab:blended-rqs}.) During the data extraction phase, however, we have failed to obtain any useful information in these two categories. In the vast majority of cases (exception for ADOIT~\cite{adoit}, ARIS~\cite{aris}, the Eclipse Process Framework~\cite{epf}, and MagicDraw~\cite{magicdraw}), we have not found explicit discussions of these concerns, nor any evidence of these concerns being explicit in the tool.

We report these negative results to maintain the symmetry of our classification framework. We suggest replications of this study to be carried out in a conceptual way~\cite{dennis2015replication}, i.e., attempting to answer the research questions using different methods.

\subsubsection{Inconsistency management}\label{sec:rq2-blended3}

\paragraph{Inconsistency visualization.} 
As shown in Table~\ref{tab:inconsistencyvisualization}, the majority of tools, \xofy{15}{\numberToolsIdentified{}}{}, does not provide any visualization for inconsistencies. Out of the remaining 11 tools, eight implement an internal visualization mechanism, and three rely on external services.

\begin{table}[ht]
\centering
\caption{Support for inconsistency visualization.}
\label{tab:inconsistencyvisualization}
\begin{tabular}{ @{}p{1.9cm}p{1.2cm}p{4cm}@{} } 
\toprule
\textbf{Inconsistency visualization} & \textbf{\#Tools} & \textbf{Tools} \\
\midrule
No & \relativeDatabar{15}{\numberToolsIdentified} & \scriptsize{\cite{adoit}, \cite{aris}, \cite{atompm}, \cite{blended-profile},  \cite{cardanit}, \cite{certware}, \cite{epf}, \cite{magicdraw}, \cite{osate}, \cite{sequencediagramorg}, \cite{somadoxx}, \cite{swimlanes}, \cite{umlet}, \cite{umletino}, \cite{use}}\\ 
Internal & \relativeDatabar{8}{\numberToolsIdentified} & \scriptsize{\cite{archi}, \cite{boston}, \cite{etas-ascet}, \cite{dbdiagram}, \cite{mbeddr}, \cite{modelio}, \cite{quickdb}, \cite{umple}}\\ 
External & \databar[3]{11} & \scriptsize{\cite{papyrus}, \cite{memo4ado}, \cite{topbraid}}\\
\bottomrule
\end{tabular}

\end{table}

\paragraph{Inconsistency management type.} 
The two fundamental approaches to manage inconsistencies are prevention, and allow-and-resolve~\cite{david2019foundation}. Preventive techniques effectively prohibit the emergence of inconsistencies, either by serializing user operations (e.g., via locking), or by constructing the underlying data structures in a way that they can never be inconsistent (e.g., in conflict-free replicated data types (CRDT)~\cite{shapiro2011conflict}. Allow-and-resolve approaches embrace the existence of inconsistencies~\cite{finkelstein2000foolish} instead of preventing them. This allows treating inconsistencies with highly sophisticated operations for tolerance~\cite{balzer1991tolerating,david2016towards}, and resolution~\cite{mens2006detecting,nentwich2003consistency,gausemeier2009management}. As shown in Table~\ref{tab:inconsistencymanagementtype}, half of the tools, \xofy{13}{\numberToolsIdentified{}}{}, prevent inconsistencies. The remaining tools either manage inconsistencies on-the-fly (\xofyp{11}{\numberToolsIdentified{}}{}) or on-demand (\xofyp{2}{\numberToolsIdentified{}}{}).

\begin{table}[htp]
\centering
\caption{Support for different inconsistency management types.}
\label{tab:inconsistencymanagementtype}
\begin{tabular}{  @{}p{1.9cm}p{1.2cm}p{3.9cm}@{} } 
\toprule
\textbf{Inconsistency mgmt type} & \textbf{\#Tools} & \textbf{Tools} \\
\midrule
Preventive &  \relativeDatabar{13}{\numberToolsIdentified} & \scriptsize{\cite{adoit}, \cite{aris}, \cite{blended-profile}, \cite{cardanit}, \cite{epf}, \cite{magicdraw}, \cite{memo4ado}, \cite{modelio}, \cite{swimlanes}, \cite{umlet}, \cite{umletino}, \cite{umple}, \cite{use}} \\ 
On-the-fly &  \relativeDatabar{11}{\numberToolsIdentified} & \scriptsize{\cite{archi}, \cite{etas-ascet}, \cite{atompm}, \cite{boston}, \cite{certware}, \cite{dbdiagram}, \cite{mbeddr}, \cite{quickdb}, \cite{sequencediagramorg}, \cite{somadoxx},  \cite{topbraid}}\\ 
On-demand &  \relativeDatabar{2}{\numberToolsIdentified} & \scriptsize{\cite{papyrus}, \cite{osate}}\\
\bottomrule
\end{tabular}

\end{table}

\paragraph{Inconsistency management automation.} 
As shown in Table~\ref{tab:inconsistencymanagementsupport}, \xofy{13}{\numberToolsIdentified{}}{tools} do not provide inconsistency resolution due to their preventive inconsistency management approach. These tools are identical to the ones of the \emph{preventive} category in~Table~\ref{tab:inconsistencymanagementtype}. Out of the remaining 13 tools, 11 provide some level of automation for resolving inconsistencies, while two tools rely on manual resolution.

\begin{table}[htp]
\centering
\caption{Level of automation of inconsistency management.}
\label{tab:inconsistencymanagementsupport}
\begin{tabular}{ @{}p{2.2cm}p{1.4cm}p{3.1cm}@{} } 
\toprule
\textbf{Inconsistency automation} & \textbf{\#Tools} & \textbf{Tools} \\
\midrule
Not applicable &  \relativeDatabar{13}{\numberToolsIdentified} & \scriptsize{\cite{adoit}, \cite{aris}, \cite{blended-profile}, \cite{cardanit}, \cite{epf}, \cite{magicdraw}, \cite{memo4ado}, \cite{modelio}, \cite{swimlanes}, \cite{umlet}, \cite{umletino}, \cite{umple}, \cite{use}}\\
Fully automated &  \relativeDatabar{6}{\numberToolsIdentified} & \scriptsize{\cite{boston}, \cite{osate}, \cite{quickdb}, \cite{sequencediagramorg}, \cite{somadoxx}, \cite{topbraid}} \\ 
Semi-automated &  \relativeDatabar{5}{\numberToolsIdentified} & \scriptsize{\cite{archi}, \cite{certware}, \cite{dbdiagram}, \cite{mbeddr}, \cite{papyrus}}\\ 
Manual &  \relativeDatabar{2}{\numberToolsIdentified} & \scriptsize{\cite{etas-ascet}, \cite{atompm}}\\
\bottomrule
\end{tabular}
\end{table}

%% file: TexFiles/paperA/Orthogonal_Finding.tex
\section{Orthogonal findings}\label{sec:orthogonal}

We have analyzed the extracted data for horizontal findings, orthogonal to the vertical analysis reported in the previous section. Specifically for this purpose, we have generated contingency tables for each pair of categories of the classification framework and looked for relevant emerging correlations.
In this section, we discuss these findings and contextualize them in terms of the aspects of blended modeling: the support for multiple notations (\secref{sec:orthogonal-blended1}), seamless interaction (\secref{sec:orthogonal-blended2}), and flexibility and inconsistency management (\secref{sec:orthogonal-blended3}); and in the additional aspect of technological trends that are independent from the blended aspects (\secref{sec:orthogonal-tech}).

\subsection{Number of notation types and Overlap of notations}\label{sec:orthogonal-blended1}

As shown by the data in \secref{sec:rq1}, the sampled tools support 2.5 types of notation on average. In about 81\% of the cases, the overlap between the specific notations is only partial, thus providing a richer way to build models.

\paragraph{Notation types count vs Web-based nature.}
The number of types of notation tends to be higher in desktop tools. Tools with more than two types of notation are exclusively desktop-based. While every web-based tool in our sample provides a maximum of two types of notations, \xofy{11}{17}{desktop tools} provide three or more types of notations.
We have measured a statistically significant difference at $p=0.0064$.\footnote{For the remainder of the paper, $\alpha=0.05$, unless specifically noted otherwise. Following the directions of Haviland~\cite{haviland1990yates}, we report the p-values of the conventional Chi-square test without Yates's correction for continuity.}

\paragraph{Overlap of notations vs Web-based nature.}
We found significantly more completely overlapping notations in web-based tools than in desktop-based tools. \xofy{4}{9}{web-based tools} come with completely overlapping notations. This ratio is 5.9\% in desktop-based tools ($p=0.0176$). This is in line with the previous observation of web-based tools typically providing fewer types of notation. It is plausible to assume that in desktop tools, the higher number of notation types might result in relevant differences between the notations and, thus, less overlap among them.

\paragraph{Notation types count vs open-source nature.} The number of notation types tends to be higher in open-source tools than in commercial ones. Three or more types of notations are supported in \xofy{8}{13}{open-source tools}, while this number is only \xofy{3}{13}{} in commercial tools. However, a deeper look also reveals that the only two tools supporting four types of notations are commercial ones (\cite{boston}, \cite{topbraid}). While \xofy{2}{13}{commercial tools} provide four types of notations, \xofy{10}{13}{} of them support only two. These differences are significant at $p=0.0105$. It is plausible to assume that while commercial tool vendors have the capabilities to develop sophisticated tools with many types of notations, they still opt for a more streamlined user experience either due to explicit user requirements, or to minimize the technological risks and improve the maintainability of the tools.

\subsection{Seamless interaction}\label{sec:orthogonal-blended2}

In terms of seamless interaction, we have found significant relationships between the navigation among notations, their synchronicity, and the presence of inconsistency visualization.

\paragraph{Navigation among notations vs Synchronous navigation.}
We have observed a statistically significant difference ($p= $4E-4) between the \textit{complexity of navigation among notations}, and the \textit{synchronicity of navigation}. The two features go hand in hand. \xofy{16}{16}{tools} with support for synchronous navigation also support immediate navigation across different notations. In contrast, only \xofy{4}{10}{tools} without synchronous navigation support immediate navigation. That is, in over half of such tools, navigation between notations becomes a complex and tedious task, significantly impacting the user experience in terms of seamless interaction.
Synchronous navigation is more frequently observed in tools with completely overlapping notations. While \xofy{5}{5}{tools} with completely overlapping notations operate with synchronous navigation, this ratio is only \xofy{11}{21}{} in tools with partially overlapping notations.

\paragraph{Navigation among notations vs Inconsistency visualization.}
We observed that \xofy{11}{11}{tools} that support inconsistency visualization operate with immediate navigation; while tools without inconsistency visualization support immediate navigation only in \xofy{9}{15}{cases}. The difference is significant at $p=0.0168$.

\subsection{Flexibility and inconsistency management}\label{sec:orthogonal-blended3}

As discussed in \secref{sec:rq1}, flexibility, in general, is sporadically supported by the tools we have sampled. We have found that the three types of flexibility features (model-level, language-level, persistence-level), often correlate with inconsistency management aspects.

\paragraph{Model-level flexibility vs Inconsistency visualization.}
Inconsistency visualization is significantly better supported in tools with model-level flexibility. We have found that \xofy{7}{7}{tools} with model-level flexibility also support inconsistency visualization, while this ratio drops to \xofy{4}{19}{} in tools without model-level flexibility ($p=3\mathrm{E}{-4}$). It is plausible to assume that inconsistency visualization is an enabler to model-level flexibility. Visualizing inconsistencies certainly helps the stakeholders to keep track of inconsistencies and reason about the most appropriate time and approach to resolving them.

\paragraph{Inconsistency visualization vs collaboration.} Tools with internal inconsistency visualization features are also collaborative tools. This holds for \xofy{8}{26}{tools}. Conversely, the \xofy{11}{26}{tools} without collaborative features do not support internal means of inconsistency visualization. The ratio of collaborative and non-collaborative tools is split almost evenly when inconsistency visualization is not present. \xofy{15}{26}{tools} come without inconsistency visualization, out of which seven (27\%) support collaboration and eight (31\%) lack collaborative features. These relationships are significant at $p=0.0047$. These observations can be explained by the strong relationship between collaboration and inconsistencies: as the lack of collaboration might severely reduce the cases when inconsistencies can appear, tools vendors whose tools do not support collaboration might be less interested in developing internal inconsistency visualization techniques.

\subsection{Technological trends}\label{sec:orthogonal-tech}

We have further identified some purely technological trends, orthogonal to the three facets of blended modeling, mainly related to the web-based nature of tools (Table~\ref{tab:webbased}), their collaborative features (Table~\ref{tab:collaboration}), and their platforms of implementation (Table~\ref{tab:platforms}).

\paragraph{Collaboration on the web.} The type of collaboration tends to correlate with the type of client software. \xofy{5}{6}{tools} that operate with synchronous (real-time) collaboration, are implemented as web-based tools. In contrast, \xofy{7}{9}{tools} that operate with asynchronous (off-line) collaboration, are implemented as desktop tools. The type of client software is nearly evenly split in collaborative tools between web clients (\xofyp{7}{15}) and desktop clients (\xofyp{8}{15}). However, \xofy{9}{11}{non-collaborative tools} are built as desktop applications, and we found only two web-based non-collaborative tools. These differences are significant at $p=0.0164$. These observations are in line with the observations of our previous work~\cite{david2021collaborative}, especially on the apparent mobilization of collaborative modeling.

\paragraph{"Modeling" platforms are primarily desktop-based.} We observed that neither of the web-based tools in our sample is implemented on a platform that explicitly aims to provide \textit{modeling} capabilities. In contrast, \xofy{11}{18}{desktop tools} are implemented on top of a modeling platform, such as Eclipse (\xofyp{10}{18}), JetBrains MPS (\xofyp{1}{18}), and other, custom platforms (\xofyp{7}{18}). While the web-based tools in our sample leverage web frameworks that provide reusable elements to build front-end and back-end functionality, the lack of \textit{modeling} frameworks tailored to the web are apparent. These differences are significant at $p=0.0097$.

%% file: TexFiles/paperA/Discussion.tex
\section{Discussion}\label{sec:discussion}

The corpus of this paper consists of \numberAcademicTools{} academic papers and \numberGreyTools{} entries of grey literature survey, which eventually resulted in \numberToolsIdentified{} identified tools. Based on the rigorously constructed research protocol, we are reasonably confident in the representativeness of our sample for the field under study.

\subsection{Takeaways}

The main takeaway of our investigation is that the state-of-the-art and state-of-the-practice tools only provide \textit{partial and accidental support for blended modeling}. This is not a surprising result, considering the novel and emerging nature of the concept of blended modeling. We have found adequately scaling tools in terms of the number of supported notation types. \xofy{11}{26}{tools} provide more than the minimal two notation types (Table~\ref{tab:noconcretesyntax}). Various aspects related to flexibility, however, pose a potentially serious obstacle for multi-notation tools to become true blended modeling tools. Only \xofy{7}{26}{tools} provide flexibility at the instance model level, i.e., tolerance of horizontal inconsistencies between models (Table~\ref{tab:flexibilitymodels}). \xofy{4}{26}{tools} support flexibility at the language level, i.e., tolerance of vertical inconsistencies, such as conformance or type discrepancies (Table~\ref{tab:flexibilitylanguage}). In terms of user experience (UX), and especially seamless interaction, we noticed encouraging signs in cross-notation navigability and inconsistency management automation. \xofy{16}{26}{tools} support a synchronized navigation across their supported notations (Table~\ref{tab:synchronizednavigation}) and, in \xofy{20}{26}{tools}, immediate navigation is also available (Table~\ref{tab:navamongsyntaxes}). This enables a better concert of notations, allowing using them in a truly complementary fashion. \xofy{11}{13}{tools} that allow inconsistencies to occur treat them with a substantial level of automation; only \xofy{2}{13}{} of such tools (a grand total of 8\% -- 2 of 26) rely on manual resolution of inconsistencies (Table~\ref{tab:inconsistencymanagementsupport}).\\

In terms of \textit{user-oriented characteristics} (RQ1), we observed a strong dominance of graphical notations, supported by \xofy{26}{26}{tools}, followed by textual (\xofyp{19}{26}), tabular (\xofyp{13}{26}), and tree-based ones (\xofyp{7}{26}) (Table~\ref{tab:notationsupport} and \figref{fig:concreteSyntaxTypesCount-b}). Only \xofy{5}{26}{tools} feature a combination of notations that are completely overlapping in terms of modeling language concepts (Table~\ref{tab:overlap}). This means that multi-notation tools tend to leverage the complementary nature of different types of notation. This is a welcome direction as it opens up for opportunities of a richer modeling experience, paramount in approaches such as MVM and MPM and, as such, it motivates the efforts of blended modeling.\\

In terms of \textit{realization-oriented characteristics (RQ2)}, we observed the dominance of parser-based solutions, employed in \xofy{22}{26}{tools} (Table~\ref{tab:mapping}). Evidence suggests that projectional editors align better with multi-view and multi-notation principles~\cite{berger2016efficiency,voelter2011language,voelter2014towards}, which are now the typical modeling settings for complex systems~\cite{persson2013characterization}. The average age of tools in our sample is 10.6 years ($\sigma=5.9$), dating the typical modeling tool earlier than the uptick in research interest in projectional editors.\footnote{A directed search on Google Scholar using the \texttt{(intitle:"projectional editing" OR intitle:"projectional editor" OR intitle:"projectional editors") OR ("projectional editing" OR "projectional editor" OR "projectional editors")} search string suggests an increasing publication output starting from 2013.} We foresee the support for projectional editors to grow as modeling tools are becoming more complex in their denotational and semantic functionalities.
We observed a relatively high support for automation of inconsistency management (Table~\ref{tab:inconsistencymanagementsupport}). Inconsistency management, and tolerance in particular (Tables \ref{tab:flexibilitymodels}-\ref{tab:flexibilitypersistence}), are key enablers to the flexibility of modeling tools. Only \xofy{2}{26}{tools} come without some level of automation in resolving conflicts and these are either research tools, such as~\cite{atompm}, or tools that are explicitly not supporting groupwork, such as~\cite{etas-ascet}.

\subsection{Challenges and opportunities}

By mapping the state-of-the-art and state-of-the-practice, we have identified challenges and opportunities related to the concept of blended modeling in relation to tools.

\paragraph*{Multi-formalism.}~Our study assumed one single underlying abstract syntax and a single underlying formalism, but even with this simplification, the support for multi-notation is sporadic. Multi-formalism, and especially multi-semantics, exacerbates this problem as we anticipate the interest in blended modeling gradually shifting towards more complex domains~\cite{cicchetti2019multi,mosterman2004computer,vanherpen2016ontological}. We see an opportunity for tool builders and integrators in complex engineering domains that inherently work in an MVM/MPM setup, such as mechatronics, automotive, and robotics, to incorporate blendedness as an enabling concept into their existing tool ecosystems.
However, this should be preceded by academic research on extending blended modeling, especially on topics such as coordination between models of different languages~\cite{engelen2010integrating}, and synchronization of abstract and concrete syntax in DSLs~\cite{rath2010synchronization}.
Nevertheless, we expect an early maturation and rapid take-off of blended modeling techniques in an array of applied modeling settings. Therefore, we advise technology transfer entities to closely follow academic and semi-academic advancements to propel the transition of the concept to applied industrial settings.

\paragraph*{Seamless interaction.}~As a primary user experience (UX) concern, seamless interaction can make a substantial difference in user satisfaction~\cite{wixom2005theoretical} towards modeling tools. The user-oriented aspects of our study (\secref{sec:rq1-blended2}) show that current tools are often equipped with related features (e.g., synchronous navigation among notations). Such tools have the opportunity to provide holistic support for blended modeling. The evaluation and comparison of realization-oriented aspects, however, is certainly a challenge, as demonstrated in \secref{sec:rq2-blended2}. The scope of our study did not include the development of methods that would allow extracting information about user experience and seamless integration of the different modeling paradigms in blended modeling tools. In general, the evaluation of such user-facing aspects remains a challenge. We encourage researchers to develop methods suitable for extracting the types of information outlined in \secref{sec:rq2-blended2}; and to further enrich the user-facing aspects, based on \secref{sec:rq1-blended2}. We suggest facilitating dedicated evaluation events, e.g., hands-on workshops at major conferences, where crowdsourcing models for hands-on experimentation and evaluation are feasible because of the volume of the co-located participants and their significant expertise, such as the Hands-on Workshop on Collaborative Modeling (HoWCoM)\footnote{\url{http://howcom2021.github.io/}}, and the workshop on Human Factors in Modeling / Modeling of Human Factors (HuFaMo)\footnote{\url{https://www.monash.edu/it/humanise-lab/hufamo21}} at MODELS\footnote{\url{http://www.modelsconference.org/}}, as well as the Conference on Human Factors in Computing Systems (CHI)\footnote{\url{https://chi2021.acm.org/}}.
Explicitly modeled user interfaces~\cite{syriani2021generation} and API protocols~\cite{vanmierlo2018multi} provide especially good foundations for developing software tools that allow seamless switching between notations. Seamless interaction across textual and graphical notations is especially challenging~\cite{engelen2010integrating} due to the differences between their respective grammar-based and metamodel-based approaches~\cite{gjosaeter2008meta}. Projectional editing~\cite{voelter2014towards} provides appropriate means to overcome these limitations, thus, we advise researchers to investigate seamless interaction from this standpoint as well.

\paragraph*{Flexibility.}~The flexibility of modeling tools in terms of (temporarily) tolerating inconsistencies, such as violations of well-formedness rules and inter-notation/inter-view discrepancies, is best approached by employing state-of-the-art inconsistency models, such as eventual and strong eventual consistency~\cite{shapiro2011conflict}. Although the scope of this study does not entail the particularities of inconsistency management, we have identified traces and patterns of shortcomings in this aspect. While the majority of tools operate in a preventive inconsistency management fashion (\secref{sec:rq2-blended3}), they implement prevention in the traditional way, i.e., by prohibiting consistency-breaking operations. Such approaches stem from the limitations of strict consistency, whereas novel developments in the field offer much better inconsistency management and, by extension, better flexibility. Strong eventual consistency (SEC)~\cite{shapiro2011conflict}, for example, offers a convenient trade-off between the strictness of strong consistency and the guarantees of eventual consistency. As such, SEC is especially well-suited for tools whose developers are more comfortable with preventive inconsistency management models. Such avenues have been explored in multiple collaborative modeling frameworks, such as lowkey\footnote{\url{https://github.com/geodes-sms/lowkey}}, and C-Praxis~\cite{michaux2011semantically}.
We see an opportunity in developing advanced inconsistency tolerance methods that work at the semantic level of models, especially if blended modeling is extended to support multiple abstract syntaxes or multiple semantics.
Recently, inconsistency management between the data and (meta)model level has been investigated, e.g., by Zaher et al.~\cite{zaheri2021towards}. Such directions align well with the persistence flexibility aspect of modeling tools, which is sporadically supported currently.
In general, we encourage tool builders to treat inconsistencies as first-class citizens and, instead of overspending on resources to prevent them, we suggest appropriately managing them~\cite{finkelstein2000foolish,david2019foundation}.

\paragraph*{The many facets of web-based tools.} The interconnected nature of web-based tools and the advanced communication and networking standards of the Internet align well with building collaborative modeling tools. We observed a tendency of tool builders to use web technologies more in collaborative tools (\secref{sec:orthogonal-tech}). However, we also observed that web-based tools come with significantly less types of notations (\secref{sec:orthogonal-blended1}), and that \textit{modeling} platforms and frameworks are built for desktop applications (\secref{sec:orthogonal-tech}). It is possible that the shortage of modeling frameworks and language workbenches with a web-based focus limits the ability of tool vendors to provide rich modeling tools with numerous types of notations and advanced modeling facilities. Modeling platforms such as Eclipse already started providing support for deploying modeling tools onto the web, but this is merely a workaround. We foresee an increasing industrial interest in web-based modeling frameworks, such as WebGME~\cite{maroti2014next}, providing researchers of language engineering and language workbenches with opportunities.

\paragraph*{Tools performance assessment.}~The current generation of modeling tools is facing challenges to manage large-scale complex models~\cite{bucchiarone2020grand,kolovos2013research}. Given the presence of multiple different notations in blended modeling, estimating tool performance when dealing with large-scale and complex models is crucial for the future technical sustainability of blended modeling. However, in our data analysis, we did not observe that tool builders discuss the performance of their blended modeling tools. We conjecture that this lack of communication is mainly because (i) tool performance is still an open problem in MDE~\cite{bucchiarone2020grand}, and (ii) there are still no standard benchmarks for objectively and fairly comparing the performance of different modeling tools. We suggest that researchers investigate a shared and open benchmark for assessing the performance of modeling tools when dealing with models of different levels of size and complexity (\ie from a few up to millions of modeling elements). To avoid bias concerning specific DSMLs or application domains,  populating such benchmark should be a community effort, where researchers and tool builders coming from different domains collaborate and contribute their models, language definitions, and requirements (\eg expected time to open a model with 1M elements, expected time to propagate a model change from a visual syntax to the corresponding textual one, \etc). Having such shared benchmarks will provide practitioners with an evidence-based instrument for comparing similar modeling tools and choosing the best one according to their project and organizational needs. Also, a shared benchmark will help MDE researchers in designing and conducting empirical studies assessing the performance of (blended) modeling tools, thus providing objective and replicable knowledge for addressing the grand challenge of scalability in Model-Driven Engineering~\cite{bucchiarone2020grand}.

%% file: TexFiles/paperA/Threats_to_Validity.tex
\section{Threats to validity}\label{sec:threats}

The study reported in this paper has been carried out based on a carefully designed protocol. To minimize the threats to validity, we have designed our protocol based on well-established guidelines for systematic studies in software engineering~\cite{kitchenham2007guidelines,wohlin2012experimentation,zhang2011identifying} and those for including grey literature by Garousi et al.~\cite{garousi2016highly}.

We have assessed the quality of our study following the guidelines by Petersen et al.~\cite{petersen2015guidelines} and achieved a 63.6\% result. This score is significantly higher than the median and absolute maximum scores (33\% and 48\%, respectively) reported in~\cite{petersen2015guidelines}. This high score can be mainly attributed to the detailed search strategy; the involvement of external senior consultants in the study design phase; and the involvement of multiple authors in the screening phase, minimizing the number of false inclusions and exclusions.

In the following, we discuss the possible threats to the validity of our study and elaborate on how we have mitigated them.

\subsection{External validity}
External validity concerns the generalizability of the results~\cite{wohlin2012experimentation} and it is primarily associated with the sampling method. The most severe threat to external validity is the lack of representativeness of the selected tools to the field of interest in general. We have mitigated this threat by an appropriately constructed protocol with two orthogonal concerns. First, our search strategy included manual and automated search steps, with exhaustively iterative backward and forward snowballing. Second, we have carried out this search both for the academic and the grey literature~\cite{garousi2016highly}.

Another class of threats to external validity can be attributed to the inclusion and exclusion criteria used in the screening. To mitigate these threats, we defined exclusion criteria specific to the type of literature (white or grey) being surveyed. Some threats remain, for example, due to the exclusion of non-peer-reviewed academic material (\texttt{A-E2} in \secref{sec:study_design}), and the exclusion of proprietary tools that do not allow experimentation with at least a trial version (\texttt{GEN-E4}). We consider these threats minimal.

\subsection{Internal validity}

Internal validity is the extent to which claims are supported by data and it is primarily associated with the study design. We have mitigated this risk by the thorough construction and validation of our protocol. The protocol has been developed by multiple authors with relevant expertise on the topics related to blended modeling. Additionally, the protocol has been validated by an external reviewer with significant expertise in empirical research. We have employed rigorous descriptive statistical methods for orthogonal analysis and validation of the data to further mitigate the threats.

\subsection{Construct validity}
Construct validity is concerned with the generalizability of the measures of the study to the investigated concepts, and it is primarily associated with the categories and parameters employed during the data extraction and the subsequent analysis.
We have mitigated the threats by mapping the research questions to typical parameters before constructing our search strategy. Consequently, we are reasonably confident about the construction validity of the search strings used in the automatic search steps. We have further minimized the threats in the screening phase by refining the inclusion and exclusion criteria in multiple iterations, to reach unambiguous definitions. Each study was assigned to two researchers randomly, and a third researcher was involved to oversee the results and make the final decisions on the inclusion.

\subsection{Conclusion validity}
Conclusion validity is the degree of credibility of the conclusions, based on the relationship between cause and effect. Specifically, in our case, conclusion validity is concerned with the relationship between the conclusions communicated in Sections~\ref{sec:rq1}--\ref{sec:discussion} and the extracted data. We mitigated the main threats in two steps. First, considering that different researchers might interpret the same data in different ways, we have documented our research protocol in great detail and made it available along with our datasets and statistical analysis scripts in the publicly available replication package.\footref{fn:replication-package}
Second, we have constructed conclusions based only on the available data. Any hypotheses and conjunctures were explicitly marked as such.

%% file: TexFiles/paperA/Conclusions.tex
\section{Conclusions}\label{sec:conclusions}

In this paper, we have reported the results of our systematic, multi-vocal study on the potential, opportunities, and challenges of the emerging approach of blended modeling.  We have reviewed nearly \numberAcademicScreenedApprox{} academic papers, and nearly \numberGreyScreenedApprox{} entries of grey literature. Based on these, we have identified \numberToolCandidates{} candidate tools, and eventually selected \numberToolsIdentified{} state-of-the-art and state-of-the-practice modeling tools which represent the current spectrum of modeling tools. We defined a classification framework for these tools which we used to map their support for other blended aspects, such as navigation and inconsistency tolerance.

Our findings show that current tooling only provides partial support for the features of blended modeling, in particular for inconsistencies between different notations of the same model. The existing support for automated consistency management is encouraging. We also observe that the overlap between notations is not complete. Projectional editing seems to be a promising avenue for future blended modeling, but most existing tools we reviewed are not projectional.
Concerning the challenges, we observe that support for multi-formalism and multi-semantics is still largely lacking. We also see opportunities for improvements when it comes to the seamless integration of the different modeling notations and the evaluation of the user experience. Finally, we identify incorporating ``softer'' models of consistency that directly use the semantics of the models to achieve eventual consistency as a promising area of future research.

We foresee a new generation of modeling tools that will take blended modeling further by introducing semantic techniques that will allow basing the modeling workflow on multiple different abstract syntaxes.

As for future work, we are working on implementing a generator that produces blended modeling tools for arbitrary domain-specific languages. These tools will be based on the takeaways of this study as well as on a prototype implementation that already embraces the blended principles by Addazi et al.~\cite{addazi2021blended}. We intend to keep our dataset up-to-date and report increments on the efforts made on improving blended modeling. Finally, we plan to develop methods for the evaluation of the user experience of blended modeling tools based on hands-on events and workshops.

%% file: TexFiles/paperA/appendix-tools.tex
\renewcommand\refname{Referred Tools}

\renewcommand\refname{References}

%% file: TexFiles/paperB/abstract.tex
\newpage
\section*{Abstract}
When generating textual editors for large and highly structured meta-models, it is possible to extend Xtext's generator capabilities and the default implementations it provides. These extensions provide additional features such as formatters and more precise scoping for cross-references. 
However, for large metamodels in particular, the realization of such extensions typically is a time-consuming, awkward, and repetitive task.
For some of these tasks, we motivate, present, and discuss in this position paper automatic solutions that exploit the structure of the underlying metamodel.
Furthermore, we demonstrate how we used them in the development of a textual editor for \eatxt, a textual concrete syntax for the automotive architecture description language \eastadl. This work in progress contributes to our larger goal of building a language workbench for blended modelling.
\newpage

%% file: TexFiles/paperB/main.tex
\section{Introduction}
Xtext~\cite{eysholdt2010xtext} is a framework for the development of domain-specific languages (DSLs). It can either take an existing meta-model and derive a grammar from it or allows a language engineer to create a grammar directly which is then translated into a meta-model. Once a grammar exists, Xtext can generate editors that integrate seamlessly into the Eclipse IDE and offer many convenient features such as an outline view of the file which is currently edited. On the other hand, it offers extension mechanisms for more advanced editor features.

In practice, using these extensions mechanisms can pose significant technical challenges. 
For example, auto-formatting or the use of template proposals\,---\,both common features in modern editors\,---\,are not supported for DSLs based on Xtext out-of-the-box. Despite comprehensive documentation of the corresponding extension mechanisms, these challenges re-occur and have to be solved manually. In particular, implementing these features for sufficiently large languages can be cumbersome and involves a lot of repetitive code.

In other cases where Xtext provides support out-of-the-box, the default implementations provided by Xtext are not always suitable for large DSLs since the performance they provide (e.g., for cross-reference auto-completion) is insufficient for practical purposes or for certain use cases.

In the context of a prototype for a textual variant of the automotive systems modeling language \eastadl~
\cite{debruyne2004eastadl,cuenot2007managing}, we implemented several automated solutions for these re-occurring challenges by using custom \emph{Xtext generator fragments} that exploit the structure of the language's meta-model. These fragments are used when Xtext generates the editors for a DSL. We created fragments for formatting, content-assist, and template proposals. In addition, we created a solution for providing the scope of cross-references automatically suggested by the editor which addresses several short-comings of the default solution. In all cases, the structure of the meta-model provides the necessary support to enable the automation.

In this position paper, we discuss these automated solutions and describe how other DSL engineers can adapt them for their needs. Our solutions are particularly suitable for very large, but highly structured DSLs that are available as a metamodel, but should also translate to other situations in which Xtext is used.
Furthermore, we discuss the limitations of the automatic solutions where present.
Whereas these automations are not generic enough to add them to Xtext properly, they are interesting for other language engineers and can be applied to other languages as well.

Our work in progress contributes to a larger vision of a language workbench for blended modelling in which editors for different concrete syntaxes for the same abstract syntax co-exist and enable engineers to seamlessly switch between the representations~\cite{ciccozzi2019blended}. Especially when used in conjunction with evolving languages where the meta-model structure changes regularly, it is necessary to be able to quickly regenerate the editors with as little manual effort as possible. 

\section{Related Work}


Neubauer et al.\ proposed an approach~\cite{neubauer2015xmltext} to automatically create modern editors for XML-based DSLs by bridging the ``technical spaces'' 
\emph{XMLware}, \emph{grammarware}, and \emph{modelware}. Their automation mainly focuses on providing automatic customization of textual concrete syntaxes for the target DSL, such as which symbols to use. In contrast, our automation mainly focuses on the enhancement of the editors, such as automatically providing default names for new elements, improved suggestions for cross-references, etc.

Latifaj et al.\ focused on enabling different stakeholders to perform blended modeling~\cite{david2022blended} in an automated manner, i.e., using different modeling notations to seamlessly handle overlapping parts of the model \cite{latifaj2021towards}. They discuss four challenges to achieving this automation goal in the commercial tool RTist. However, they do not address the challenges of content assistance, template proposals, etc., which are the core challenges discussed in this paper.

Cooper and Kolovos acknowledge some of the challenges we address in this work in their paper on requirements and challenges of blended modelling~\cite{cooper2019engineering}. For instance, they consider scoping across concrete syntaxes to be a challenge. We address this with our custom \fe{ScopeProvider} as discussed in Section~\ref{sec:solutions:scoping}.

\section{Background}\label{sec:background}

In the following, we provide relevant information on Xtext as well as \eatxt, the textual variant of \eastadl which we used as an exemplar and a case study for our work.

\subsection{Xtext}\label{sec:background_Xtext}

Xtext is a framework for the development of textual domain-specific languages (DSL)~\cite{eysholdt2010xtext}. At its core is a grammar language that allows defining the syntax of a DSL. Xtext can either generate a grammar out of an existing meta-model or the user can specify the grammar directly and the meta-model is generated by Xtext. A grammar can be used as the input to generate editors for the Eclipse IDE or for browsers. In addition, Xtext can generate a language server that allows integrating the DSL into VS.Code or Eclipse Theia.

The generation process is controlled by a workflow for the Modeling Workflow Engine (MWE2) \cite{XtextConfiguration}. 
Xtext provides a \emph{language generator} which can be customized and extended with custom \emph{fragments}. A fragment generates code based on the generator's configuration, the grammar and the corresponding meta-model. Xtext out-of-the-box provides a number of such fragments, which can be extended or replaced. 
These fragments add a number of features to the generated editors. For instance, Xtext automatically generates a syntax validator which highlights incorrect syntax directly in the editor and in Eclipse's ``Problems'' view. For editors in the Eclipse IDE, support for the outline view is also generated. 
We use Xtext's ability to change the standard configuration to add custom fragments that provide better formatting, content-assist and template proposals as described below. These custom fragments are written in Xtend \cite{Xtend}.

Xtext splits the generated code into different bundles, distinguishing between the infrastructure for the language itself, the user interface of the editors (bundle name ends with ``.ui'') and the integration into the Eclipse IDE (bundle name ends with ``.ide'').

\subsection{\eastadl and \eatxt}\label{sec:background_Eatxt}

\eastadl is an automotive systems modeling language~\cite{cuenot2007managing} and is based on a large metamodel with more than 200 metaclasses and a hierarchy of nested elements describing different aspects of electronic vehicle systems. It can be edited in \textsc{Eatop}, an Eclipse-based editing environment that provides a hierarchical view of an \eastadl model along with form- and table-based editing capabilities. \eatxt provides a textual syntax for \eastadl with the goal to enable blended modeling~\cite{ciccozzi2019blended}, that is, the ability to switch between the hierarchy-based and the textual representation seamlessly depending on the editing task.

\begin{figure}[tb]
	\centering
		\includegraphics[width=\linewidth,page=2]{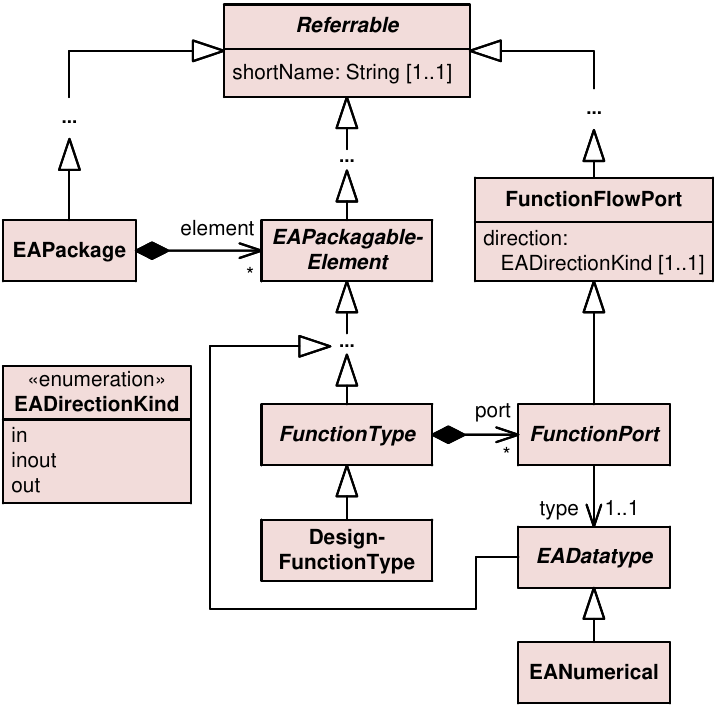}
	\caption{\eastadl Metamodel Excerpt}
	\label{fig:eadl_MM_excerpt}
\end{figure}

Figure~\ref{fig:eadl_MM_excerpt} depicts an \eastadl metamodel excerpt, which we use in this paper as a running example for illustrative purposes.
The excerpt contains a set of metaclasses (partially containing attributes) and relationships (i.e., generalizations, compositions, and cross-references), which we explain in the following.

\begin{figure}[htb]
	\centering
		\includegraphics[width=0.9\linewidth]{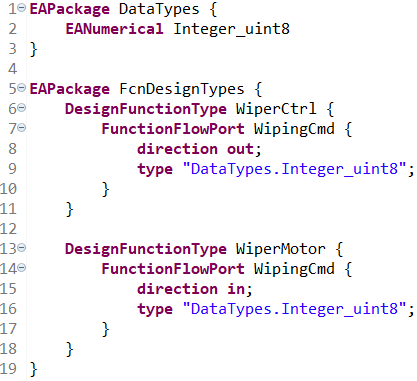}
	\caption{Excerpt from an \eatxt File Specifying a Windshield Wiper Control System}
	\label{fig:eatxt_Screenshot}
\end{figure}

An example of an \eatxt file is shown in Figure~\ref{fig:eatxt_Screenshot}.
The textual concrete syntax follows the hierarchy and structure of the \eastadl metamodel. 
Metaclasses in \eastadl are represented as blocks delimited by curly braces (e.g., the \fe{FunctionFlowPort} \texttt{WipingCmd} in lines 7--10 as part of the \fe{DesignFunctionType} \texttt{WiperCtrl}). 
Attributes are represented as lists with the attribute name and the values (e.g., the attribute \fe{direction} in line 8 as part the \fe{FunctionFlowPort} \texttt{WipingCmd}). 
Cross-references are represented as the reference name followed by the actual path to the reference (e.g., the cross-reference \fe{type} points to ``\texttt{DataTypes.Integer\_uint8}'').
In the case of \eatxt, the Xtext grammar is specified in such a way that the textual keywords representing the metamodel concepts like metaclasses, attributes, and cross-references are named as in the metamodel (e.g., both the keyword and its corresponding metaclass have the same name \fe{FunctionFlowPort}).

\section{Challenges and Solutions}
\begin{figure*}[tb]
	\centering
		\includegraphics[width=0.9\textwidth,page=1]{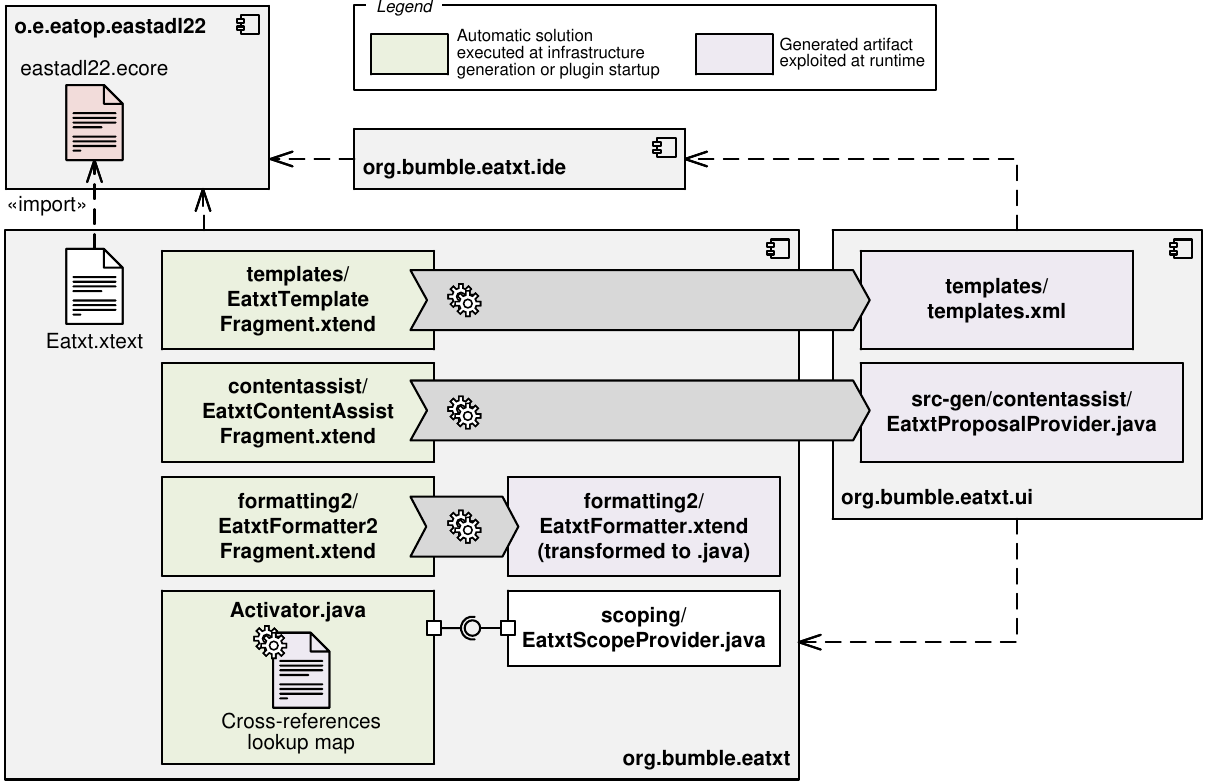}
	\caption{\eatxt Architecture and Automatic Solutions for Re-occurring Challenges}
	\label{fig:eatxt_architecture}
\end{figure*}

In this section, we explain our solutions to re-occurring challenges in the development of Xtext-based language workbenches by referring to Figure~\ref{fig:eatxt_architecture}, which depicts the coarse-grained architecture of our Xtext-based editor for \eatxt.

As described in Section~\ref{sec:background_Eatxt}, \eastadl is a stable language based on an Ecore metamodel (cf. \fe{eastadl22.ecore} in the plugin \fe{o.e.eatop.eastadl22} in the top-left corner of Figure~\ref{fig:eatxt_architecture}).
We conceived its textual syntax by automatically deriving an initial grammar from that metamodel with Xtext and subsequently adapting the grammar to the stakeholders' needs.
Figure~\ref{fig:eatxt_architecture} indicates the resulting grammar as the artifact \fe{Eatxt.xtext} as part of the plugin \fe{org.bumble.eatxt}.
In this grammar, we named the production rules and keywords equally as the language concepts in the metamodel (i.e., metaclasses, attributes, and \textsf{EReference}s).
This enables the exploitation of the metamodel structure and thereby the development of our automatic solutions.

The green Xtend and Java classes that are part of the plugin \fe{org.bumble.eatxt} represent our solutions to the re-occurring challenges described below, exploiting the structure of the metamodel \fe{eastadl22.ecore}.
Three of the solutions are custom fragments that are executed by the Xtext language generator (cf. Section~\ref{sec:background_Xtext}) and generate further artifacts (depicted in lilac) that are used in the \eatxt runtime.
Another solution generates an artifact during the activation of the plugin which can then be exploited at runtime.
We provide the implementation in our \eatxt development repository \cite{EATXT4MODELSWARD23}.
In the following four subsections, we explain and discuss these automatic solutions, the challenges they solve, and the artifacts they generate.

\subsection{Template Proposals}
\label{sec:solutions:templates}
The Eclipse IDE provides the possibility of using \emph{code templates} to enter complex code constructs that contain a significant amount of text and are more complicated than simple keywords. For instance, in the case of the Java programming language, the proposal of complete constructors or different kinds of loops is handled by code templates.
Eclipse ships with a set of pre-defined code templates and allows adding user-defined templates or customize existing ones.
Likewise, Xtext editors also provide the possibility to use and define code templates by supporting \emph{template proposals}, which consist of the actual code template and a context type~\cite{XtextTemplateProposals}. 
During the editor generation, Xtext automatically registers such a context type for each production rule and keyword (e.g., the production rule for textual instances of the \eastadl metaclass \fe{FunctionFlowPort}), and it provides the context for the code template proposal.  

However, the development of template proposals is cumbersome, because the DSL engineer has to define each template manually in an XML file \fe{templates/\_templates.xml} as part of the UI plugin (cf.~top of the plugin \fe{org.bumble.eatxt.ui} on the right-hand side of Figure~\ref{fig:eatxt_architecture}) \cite{XtextTemplateProposals}.
Moreover, it is recommended practice to define them in the runtime workspace of the Xtext editor through a corresponding dialog and to export the resulting \fe{\_templates.xml} to the development workspace~\cite{XtextTemplateProposals}, which impedes rapid prototyping of the template proposals.
Particularly, this manual practice is awkward for large metamodels. 

As an automatic solution to this challenge in \eatxt, we implemented the generator fragment \fe{templates/EatxtTemplateFragment.xtend} as part of the plugin \fe{org.bumble.eatxt} (cf.~fragment of the plugin in Figure~\ref{fig:eatxt_architecture}).
This fragment is executed by the MWE2 workflow and automatically generates the XML file \fe{templates/\_templates.xml} as part of the plugin \fe{org.bumble.eatxt.ui}.
The fragment iterates over the metamodel and generates a template proposal for all metaclasses and all of their mandatory sub-elements (i.e., attributes, containments and their nested structures, as well as cross-references).
We restrict the code templates to contain only sub-elements that are mandatory in the metamodel instead of proposing all potential sub-elements, because it might be much effort for the user to delete all proposed optional, but potentially not required, elements.

\begin{figure*}[tb]
	\centering
		\includegraphics[width=0.8\textwidth,page=3]{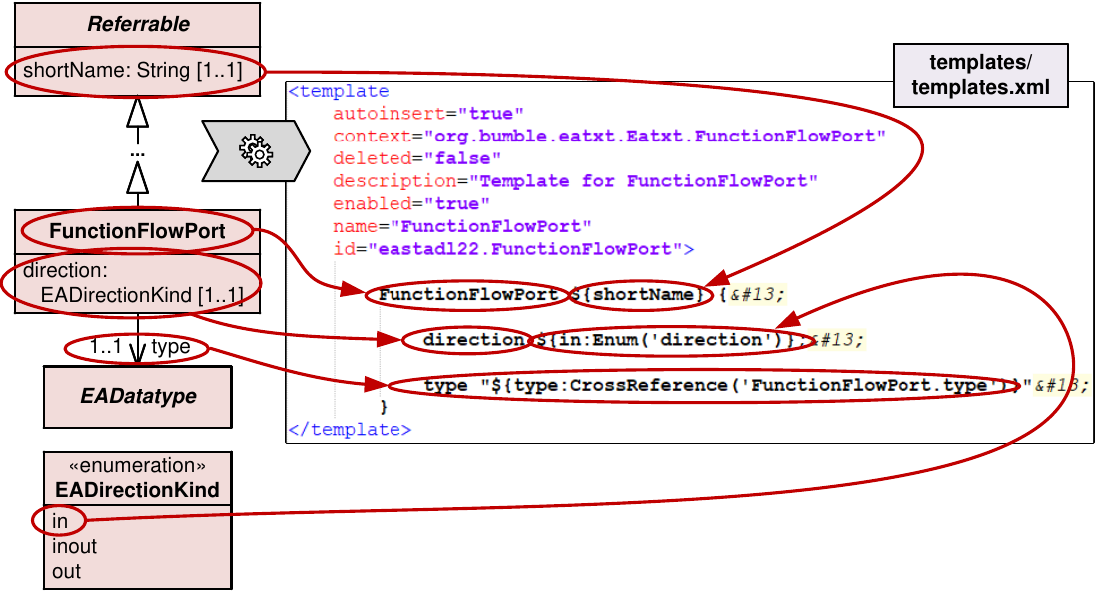}
	\caption{Exemplary Generation of a Template Proposal}
	\label{fig:TemplateProposalGeneration}
\end{figure*}

Figure~\ref{fig:TemplateProposalGeneration} depicts the generation scheme of the fragment using the example of the metaclass \fe{FunctionFlowPort} and its mandatory attribute and mandatory cross-reference.
We specify the context type as the name of the production rule, which in the \eatxt case is always named the same as the corresponding metaclass (e.g., \fe{org.bumble.eatxt.Eatxt.FunctionFlowPort} for the XML template attribute \fe{context}).
The actual code template is then generated as the name of the production rule and metaclass (e.g., \fe{FunctionFlowPort}) followed by a set of further texts as well as opening and closing curly braces.

Beyond proposing simple static texts that would typically not fit to the remainder of the text file and hence would lead to error messages in the editor, the template proposal approach also supports \emph{template variable resolvers} \cite{XtextTemplateProposals}.
For simple attributes, we distinguish the different plain data types of attributes (e.g., String, Integer, Float) and corresponding template variable resolvers.
For example in Figure~\ref{fig:TemplateProposalGeneration}, we translate the String attribute \fe{shortName} of the metaclass \fe{Referrable} to a corresponding template variable resolver \fe{\$\{shortName\}}.
For enumeration attributes, we generate enumeration template variable resolvers.
For example, the value of the attribute \fe{direction} is automatically set to the first literal \fe{in} of the enumeration \fe{EADirectionKind}.
Furthermore, we also support cross-reference template variable resolvers that propose an existing element that fits to the type of the cross-reference.
For example, the cross-reference \fe{type} is translated to the corresponding resolver. 

\begin{figure*}[htb]
	\centering
		\includegraphics[width=0.8\textwidth]{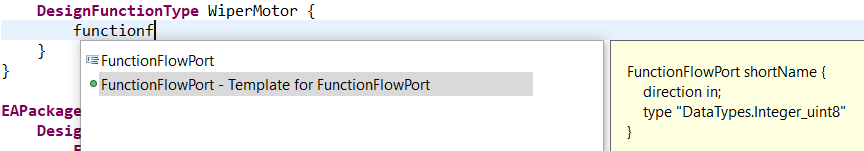}
	\caption{Proposed Code Template}
	\label{fig:TemplateProposalScreenshot}
\end{figure*}

Figure~\ref{fig:TemplateProposalScreenshot} depicts a screenshot for the proposal of a code template for the context type \fe{FunctionFlowPort} with its mandatory two attributes, where the default enumeration value of the \fe{direction} attribute as well as a target candidate for the cross-reference \fe{type} is directly proposed.
In the case of \eatxt, the generated template file contains 194 code templates with a total of more than 1,000 XML lines and covers all metaclasses of the metamodel and their mandatory sub-elements.

As we exploit the concept names of the underlying metamodel for this solution, such a generation of template proposals is restricted to grammars that have the same concept names as the corresponding metamodel (i.e., metaclass names, attribute names, association role names).
Thus, the DSL engineer is not allowed to rename the resulting keywords.

\subsection{Content-assist for new Model Elements with Unique Names}
\label{sec:solutions:content-assist}
Beyond the usual keyword-based content-assist known from IDEs, Xtext provides a customizable approach to provide content-assist for the specification of new model elements by means of \emph{proposal providers} \cite{XtextContentAssist}.
In this approach, Xtext generates an abstract proposal provider class with default functionality with content-assist methods for all metaclasses as well as an empty concrete subclass that can be customized.
To customize the content-assist for a specific model element type (e.g., for the metaclass \fe{FunctionFlowPort} with exemplary instances in lines 7--10 and 14--17 in Figure~\ref{fig:eatxt_Screenshot}), the DSL engineer has to override the corresponding method in a concrete subclass.

In the case of \eatxt, we had the requirement to provide content-assist proposals with unique names in the model namespace for all metaclasses with a mandatory \fe{shortName} attribute.
Since almost all of the more than 200 metaclasses in the \eastadl metamodel have this mandatory attribute due to sub-classing (cf. Figure~\ref{fig:eadl_MM_excerpt}), the implementation of the corresponding particular methods in the concrete proposal provider would have been very repetitive.

In order to provide an automatic solution for this challenge, we implemented a fragment \fe{contentassist/EatxtContentAssistFragment.xtend} (cf.~plugin \fe{org.bumble.eatxt} in Figure~\ref{fig:eatxt_architecture}).
Again, this fragment is executed by the MWE2 workflow and exploits the metamodel structure by iterating over all metaclasses with the mandatory attribute and generating the concrete subclass \fe{src-gen/contentassist/EatextProposalProvider.java} (cf.~plugin \fe{org.bumble.eatxt.ui} on the right-hand side of Figure~\ref{fig:eatxt_architecture}).

\begin{figure*}[htb]
	\centering
		\includegraphics[width=0.8\textwidth,page=5]{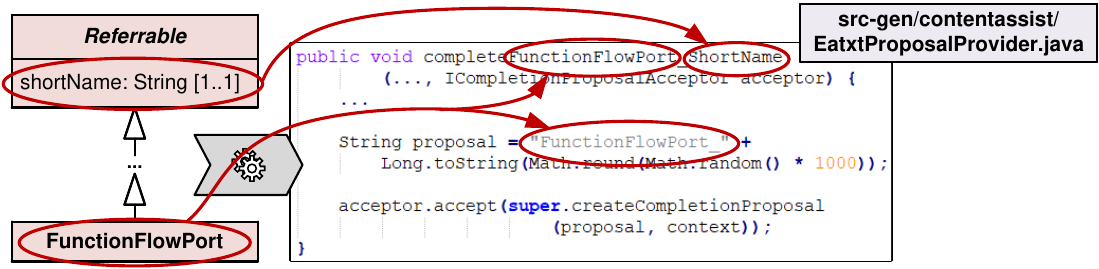}
	\caption{Exemplary Generation of a Content-assist Method for Unique Names of new \fe{FunctionFlowPort} Model Elements}
	\label{fig:ContentAssistGeneration}
\end{figure*}

Figure~\ref{fig:ContentAssistGeneration} depicts the scheme for the generation of the proposal provider methods using the example for the metaclass \fe{FunctionFlowPort}.
For each metaclass with the mandatory attribute, this generated proposal provider includes a corresponding overriding content-assist method that proposes a unique name consisting of the corresponding prefix \fe{$<$metaclassName\_$>$} followed by a randomized number for a new instance of the metaclass.
For example, we generate for the metaclass \fe{FunctionFlowPort} a corresponding proposal provider method that proposes the prefix \fe{``FunctionFlowPort\_''} followed by the randomized number.
Overall, the generated \fe{EatxtProposalProvider} encompasses 188 such methods.

\begin{figure}[htb]
	\centering
		\includegraphics[width=0.6\linewidth]{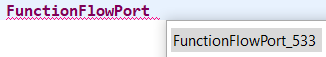}
	\caption{Content-assist for a new Model Element}
	\label{fig:ContentAssistScreenshot}
\end{figure}

Figure~\ref{fig:ContentAssistScreenshot} depicts a screenshot of the content-assist proposal for a new \fe{FunctionFlowPort} instance.

\subsection{Formatters}
\label{sec:solutions:formatters}
\emph{Formatting} text documents (e.g., Java source code) in the Eclipse IDE is the process of rearranging the documents' texts without semantically changing their contents. Xtext in principle supports this process as well, but does not provide out-of-the-box formatters~\cite{XtextFormatters}.

Instead, it requires that the language engineer extends an abstract formatter class and implements a corresponding dispatcher method for any metaclass to be formatted in its text representation.
These methods are called when the formatting process is triggered and apply a series of text replacements for the corresponding model objects.
As formatting should treat the text document and its particular contents in a uniform way, the implementation of these methods is highly repetitive.

In order to automate this tedious task and to provide a way of formatting the particular model elements uniformly, we implemented the generator fragment \fe{formatting2/EatxtFormatter2Fragment.xtend} as part of the plugin \fe{org.bumble.eatxt} (cf.~fragment of the plugin in Figure~\ref{fig:eatxt_architecture}).
This fragment is executed by the MWE2 workflow and automatically generates the formatter class \fe{formatting2/EatxtFormatter.xtend} as part of the same plugin.
The fragment iterates over the metaclasses of the metamodel and generates the corresponding dispatcher method for each container metaclass.
Furthermore, the fragment generates the individual formatting methods for all nested containments or sub-elements of these container classes calls.
The generated formatter class in the \eatxt case encompasses 51 dispatch methods and 141 calls of the formatting methods for the nested sub-elements.


\subsection{Scoping for Cross-references}
\label{sec:solutions:scoping}
Programming languages, DSLs, and metamodels typically provide means to establish links between the semantic concepts defined in the corresponding models (e.g., for specifying the type of a language concept through referencing a different type concept). 
In the Xtext language development framework, such links are called \emph{cross-references}~\cite{XtextCrossReferences}.
The \emph{scoping API} of Xtext provides the means for finding the target of a cross-reference based on its source context~\cite{XtextScoping}.
For example, the \fe{FunctionFlowPort} instances in the lines 7--9 and 14--17 in Figure~\ref{fig:eatxt_Screenshot} are source contexts, and their elements \fe{type} are cross-references pointing to target objects typed by the metaclass \fe{EADatatype} (cf. Figure~\ref{fig:eadl_MM_excerpt}).
If the target concept is nested in a container hierarchy (e.g., hierarchies of packages like the \fe{EAPackage} instance \fe{DataTypes} in lines 1--3 in Figure~\ref{fig:eatxt_Screenshot}), this procedure for finding a cross-reference target particularly includes the computation of the scope within the nested container hierarchy. 

Xtext provides a default out-of-the-box approach for the cross-reference scoping within container hierarchies by means of a \emph{scope provider}~\cite{XtextScoping}, where the scope provider also enables content-assist for the cross-reference targets.
In this default approach, the fully qualified name of the cross-reference target in the container hierarchy is only proposed by the content-assist if the target is part of a different container.
For example in Figure~\ref{fig:eatxt_Screenshot}, the \fe{type} cross-reference in line 9 as part of the container hierarchy \fe{FcnDesignTypes.WiperCtrl.WipingCmd}  points to the model element \fe{Integer\_uint8} as part of the container \fe{DataTypes}). 
In contrast, if the target is part of the same container as the source context, then only the plain name of the target is proposed but not its fully qualified name.

In the case of \eatxt, a requirement is to provide content-assist that always proposes the fully qualified name. 
This requires a custom implementation of the scope provider (see also~\cite{latifaj2021towards} for a different use case requiring such a custom implementation).
In such a custom implementation, the DSL engineers have to compute for any source context metaclass the corresponding cross-reference target candidate metaclasses.
For example, for the source context metaclass \fe{FunctionFlowPort} and its cross-reference \fe{type}, they have to realize that the target candidates are instances of the metaclass \fe{EADatatype} (cf. Figure~\ref{fig:eadl_MM_excerpt}).
In this context, multiple cross-reference target metaclasses are possible if the source context metaclass has multiple cross-references.

Basically, this computation is straightforward, but needs to consider all cross-references of the underlying metamodel, resulting in a large switch-case statement (i.e., if the source context of a cross-reference is an instance of a certain metaclass, return all candidates that are instances of the metaclasses of all target references).
Particularly for large grammars and metamodels, this straightforward custom implementation is very awkward.
Beyond that, both the Xtext default approach and custom implementation suffer from performance issues for large grammars and metamodels with many cross-references. In the runtime editor, the target candidate types are always computed on any content-assist keystroke for the given source context type. 
In the worst case, the lookup needs to traverse the entire switch-case-statement, which consists of all $n$ possible cases for a complexity of $O(n)$.

\begin{figure}[htb]
	\centering
		\includegraphics[width=0.8\linewidth]{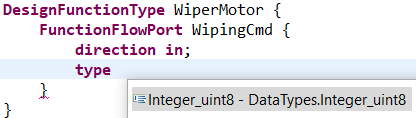}
	\caption{Cross-referencing an Existing Model Element}
	\label{fig:CrossReferenceScreenshot}
\end{figure}

As an automatic solution for this challenge in \eatxt, we generate a \fe{cross-references lookup map} in the activator of the plugin (cf.~\fe{Activator.java} in the plugin \fe{org.bumble.eatxt} in Figure~\ref{fig:eatxt_architecture}). This generation traverses the metamodel exactly once with a complexity of $O(n)$.
This lookup map contains the corresponding type of the cross-reference target for any source context type, and we compute it by iterating over all cross-references in the metamodel.
We generate the lookup map during the first activation of the plugin. 
After that, the \fe{scoping/EatxtScopeProvider.java} accesses it via an interface but does not need to perform the same computation on every cross-reference content-assist keystroke. 
The lookup map is implemented as a Java \fe{HashMap} whose \fe{get()} method has a complexity of $O(1)$ in most cases.
In the \eatxt case, the map encompasses the source context metaclasses and corresponding target metaclasses for 261 cross-references of the \eastadl metamodel.
Figure~\ref{fig:CrossReferenceScreenshot} depicts a screenshot of the content-assist for cross-referencing an existing model element in a different container hierarchy.

The solution has two advantages, in particular for large grammars and metamodels.
First, we automate the awkward custom implementation of a scope provider with a generic solution that exploits the underlying metamodel cross-references structure.
Second, it significantly improves the performance of the Xtext scope provider approach by computing a cross-reference lookup table once at plugin activation, instead of computing the target types on every content-assist keystroke for a source context. Thus, the expensive operation takes place only once rather than repeatedly at runtime.

\section{Conclusion and Outlook}


Our solutions to the four challenges of template proposals, formatters, content-assist, and scoping have been developed and evaluated in the context of \eatxt, but are not only useful there. Instead, we argue that similar approaches are helpful for any highly structured abstract syntax for which Xtext is used to generate editors for a textual concrete syntax. A specific language will benefit from adaptations, especially in the area of scoping. As such, our solutions are not generic enough to contribute to Xtext directly, but will hopefully serve as a blueprint that is useful to other engineers with similar challenges.

We view our contributions in the light of a future language workbench that supports the blended modeling for large DSLs that evolve. 
As described in our previous work~\cite{holtmann2022migrating}, \eastadl is such a language and support for its evolution within Eclipse is a relatively new capability. Our work further supports this approach by providing a streamlined way to generate the editors whenever the language changes. When coupled with the generation of a graphical editor (see, e.g.,~\cite{cooper2019engineering}), this brings us closer to the ideal of a \emph{blended modeling language workbench}.

\section*{\uppercase{Acknowledgements}}
Parts of this research were sponsored by Vinnova under grant agreement nr. 2019-02382 as part of the ITEA4 project BUMBLE.

%% file: TexFiles/paperC/abstract.tex
\newpage
\section*{Abstract}

Xtext is a well-known domain-specific language design framework and technology. It automatically generates a textual grammar for a language, given a meta-model specified in Ecore. These generated textual grammars are typically not user-friendly. Python-style languages are popular among developers for their usability and conciseness. We aim to propose a systematic approach to transform a DSL with a generated grammar into a Python-style DSL. To achieve this, we analyze the problems of grammars generated with Xtext, based on a lightweight architecture description language. In response to these problems, we propose a general semi-automated grammar adaptation approach. We apply the approach to two other DSLs to validate the generalization of the approach. We also discuss the limitations of this approach and prospects for the future.

\newpage

%% file: TexFiles/paperC/main.tex
\section{Introduction}
\label{sec:introduction}
In contrast to general-purpose programming languages (GPLs) like Java, domain-specific languages (DSLs) are computer languages tailored for a particular application domain~\cite{kosar2016domain}. 
DSLs come in a variety of forms and are employed in a wide range of domains\cite{do2012systematic}. For example, DSLs are used to improve the level of abstraction and automation in the application development in the robotic domain~\cite{nordmann2014survey}. But it is important to take into account the workload that goes into creating the DSL~\cite{mernik2005and}. 
The good news is that there are many tools available for designing and developing DSLs, like JetBrains MPS, MontiCore, Xtext, Racket, etc.~\cite{iung2020systematic}. 
One of them is Xtext~\cite{XtextHomepage}, an open-source software framework that simplifies the development of DSLs. 
Xtext can generate a complete DSL infrastructure, including parsers, linkers, compilers, etc. Developers utilize Ecore meta-models to represent domain ideas and their relationships when creating DSLs based on Xtext. From such a meta-model, Xtext generates grammar that specifies the concrete syntax. The editor and its different components are automatically generated from the meta-model of this language as Xtext artifacts. 

The grammar generated by Xtext tends to specify languages that are not user-friendly to use.
For example, the generated DSLs include default requirements that developers type in all keywords, use  braces to an extensive amount, etc. This leads to a high effort in writing when programming in these DSLs.
In contrast, the Python~\cite{PythonHomepage} language provides a very clean coding style, is renowned for increasing programmer productivity, and is considered easy to learn~\cite{gmys2020comparative}.
Both are properties are also relevant and desirable for DSLs. 
However, there is currently no systematic approach for converting DSLs to Python-like languages.

In this paper, we introduce an approach to semi-automatically convert textual DSL grammars generated by Xtext to a Python-like DSL.
This means that the resulting languages will be easy to code, easy to read, user-friendly, concise, and use whitespace and indents instead of braces~\cite{gmys2020comparative}.
We developed the approach by developing an architectural description language (DemoAdl) as an exemplar for automatically transforming a DSL to a Python-like language. Once our transformation worked for DemoAdl, we evaluated the approach with additional DSLs, \emph{Xenia} and \emph{ACME}.
This way we validate the generalizability of our approach and explore its limitations.


\section{Background}

\begin{figure}[tb]
  \centering
  \includegraphics[width=\linewidth]{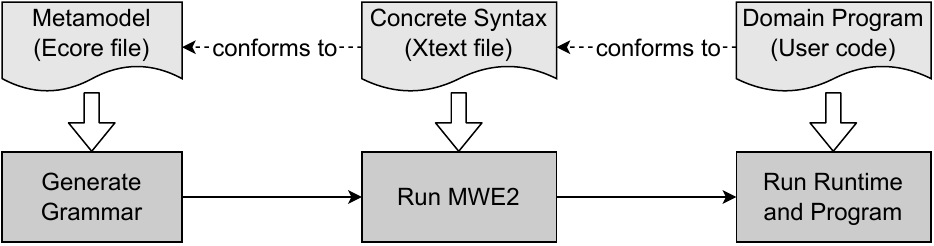}
  \caption{The relationship between different artifacts in the Xtext-based MDSE solution.}
  \label{fig:relationship-between-different-xtext-artifacts}
\end{figure}

As outlined in the introduction, Xtext is a framework for developing DSLs, and Xtext-based Model-Driven Software Engineering (MDSE) is a common solution for developing DSLs~\cite{behrens2008xtext}. In this solution, the DSL developer uses an Ecore meta-model to describe the concepts of the domain and the relationships between the concepts~\cite{steinberg2008emf}. For example, ``port'' becomes a class in the meta-model when using a meta-model to describe concepts in the field of system architecture. Next, the concrete syntax is generated from the meta-model by using the Xtext framework. Xtext provides Modeling Workflow Engine 2 (MWE2)~\cite{MWE2}, a declarative, externally configurable generator engine. After having a concrete syntax, all Xtext artifacts can be generated by running the MWE2 file. These artifacts actually make up the editor for the DSL, including linker, parser, type-checker, etc., common components of editors, and editing support for Eclipse. After the editor is generated, code written in the DSL could be resolved and supported at runtime. Figure~\ref{fig:relationship-between-different-xtext-artifacts} shows the relationship between the above concepts. When Xtext generates a textual grammar, it will generate a corresponding grammar rule for each class or enumeration in the meta-model and create an attribute in the textual grammar for each attribute or association (i.e., reference or containment) in the meta-model. A grammar rule corresponding to a class may contain multiple attributes, and each attribute occupies a line of text. A grammar rule corresponding to an enumeration contains multiple alternative values, usually on the same line of text.

\section{Methodology}
As mentioned above, we developed the approach by developing an architectural description language (i.e., DemoADL) for an example embedded system. The DSL can be used to describe simple embedded system structures, which include hardware components and software components. There are different types of hardware components, and each software component contains a number of sub-components. There are direct links between hardware components and software components, which together make up the system.

We created an Ecore meta-model that describes the domain concepts from this system and generated textual grammar from the meta-model by using Xtext. This constitutes the default grammar for the concrete syntax for DemoAdl. We used this DSL to code an example program with a default style (we called this program a ``Default-style program''). We analyzed the shortcomings of the grammar and the program conforms to it. We also created a draft of the program to illustrate what it might look like in a Python-style grammar. We call this program a ``Python-like draft''.

Based on that analysis, we developed a systematic approach to adapt grammars that were generated with Xtext.
The approach is based on grammar adaptation rules in response to the problems analyzed from the default-style DSL. We applied these rules to the grammar Xtext generated for DemoAdl. To semi-automate the grammar adaptations, we developed a script in the form of an eclipse Java plug-in. After the adaptation of the grammar, we generated an editor and wrote the same program according to the newly adapted grammar (we called this code a ``Python-like program''). This was to test whether we could successfully parse such a program in the new editor. 

To evaluate the generalizability of our approach, we chose two other DSLs. Our selection criteria are that the language 1) has an accessible homepage and 2) with examples on the homepage, and 3) has a downloadable Ecore meta-model. We apply the proposed approach to these two languages to determine whether the script and approach can be applied to create Python-style languages. 

\section{Results}
We present our analysis, our semi-automated approach, and an evaluation of our approach.

\begin{figure}[H]
  \centering
  \includegraphics[scale=0.5]{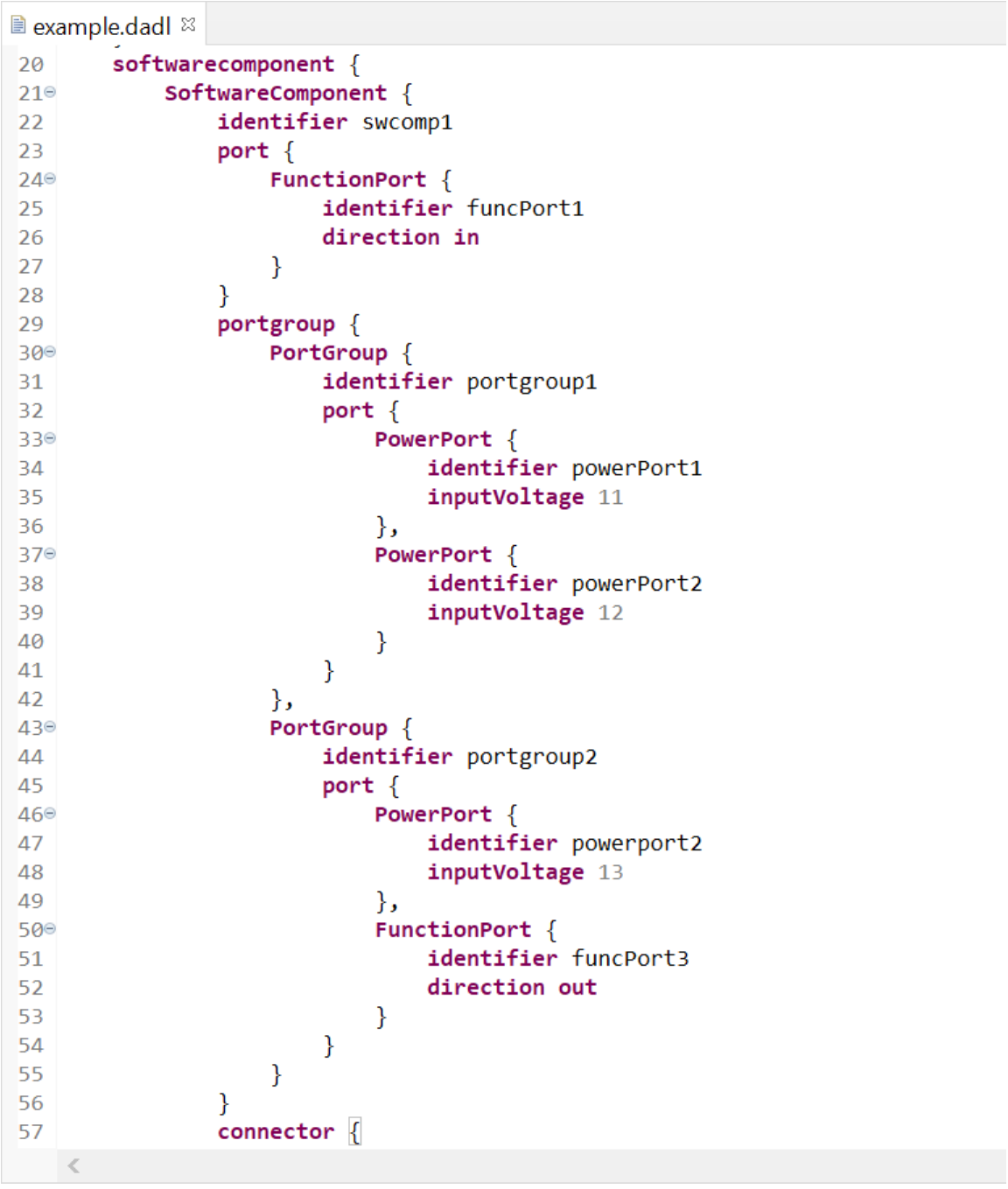}
  \caption{A screenshot of part of the domain program with a default Xtext style.}
  \label{fig:example-program-in-the-default-style}
\end{figure}

\subsection{Analysis}
\begin{figure}[tb]
 \centering
 \includegraphics[width=\linewidth]{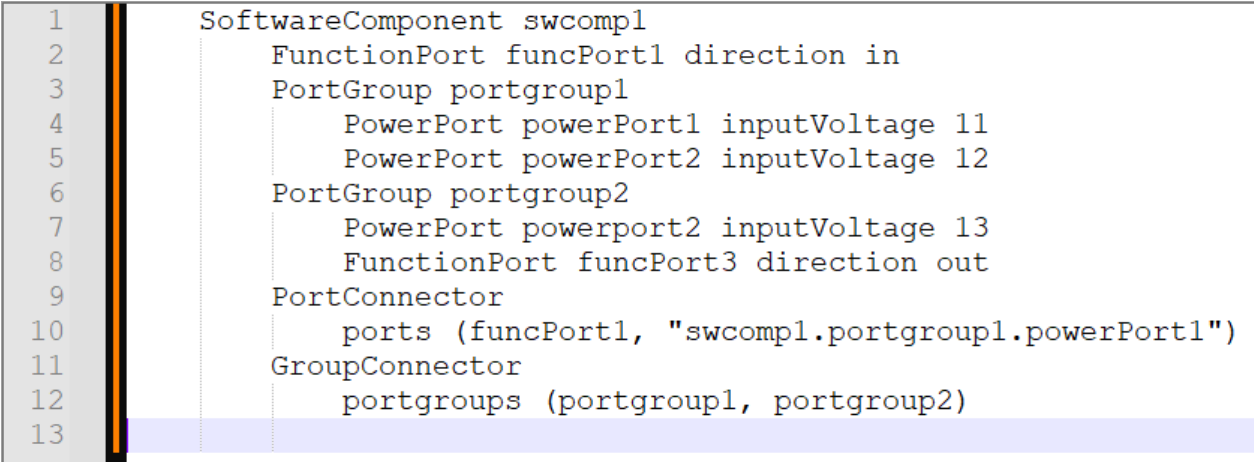}
 \caption{Snippet of \texttt{SoftwareComponent} in the system architecture description example, in Python style}
 \label{fig:Envisioned_code_for_DemoAdl_in_Python_style}
\end{figure}

As mentioned in the methodology section, we first analyzed the shortcomings of the sample DSL (i.e., DemoADL). 
We illustrate our findings with the help of a snippet of the default style program in 
Figure~\ref{fig:example-program-in-the-default-style}. We made the following observations about the grammar generated by Xtext:
 
\begin{itemize}
  \item \textbf{Inappropriate position of identifier.} In Xtext, `name' is the default name for an element's identifier. However, when we use an attribute named \emph{identifier} to identify an element (e.g., a software component in our case language DemoAdl), then it will default to a normal attribute and be placed within braces. This means that, for example, when we type in keyword \emph{SoftwareComponent}, we have to type in the left brace first and then the attribute identifier. For example, in Figure~\ref{fig:example-program-in-the-default-style} (line 22), attribute \emph{identifier} with value `swcomp1' is used to identify an element \emph{SoftwareComponent}, however, the attribute \emph{identifier} is existing inside of braces in the second line which reduces the brevity of the code. If the same content is expressed in Python, e.g., in Figure~\ref{fig:Envisioned_code_for_DemoAdl_in_Python_style}, the identifier value `swcomp1' (line 1) will follow the keyword \emph{SoftwareComponent} to identify the element.
  \item \textbf{Heavy Separation of Code Blocks} The default-style program uses both  braces and commas to separate and distinguish code blocks. For example, \emph{PortGroup} `portgroup1' and `portgroup2' are on the same level while they are separated by a comma symbol. Obviously, these commas are redundant, because  braces have already been able to separate and distinguish them. In Python, for example in Figure~\ref{fig:Envisioned_code_for_DemoAdl_in_Python_style}, `portgourp1' and `portgroup2' are separated by being on different lines though there is no  brace or comma for them. This is because there are indents to express hierarchy, and `portgourp1' and `portgroup2' are on the same hierarchy level. 
  \item \textbf{Repetitive Keyword.} There are different keywords with the same functional implication. For example, there are two keywords \emph{softwarecomponent} and \emph{SoftwareComponent}, which are highly redundant in function. This also reduced the usability of the language.
  \item \textbf{Nested braces.} There are many nested  braces, for example in Figure~\ref{fig:example-program-in-the-default-style}, after typing in a keyword, it's necessary to open a  brace. Next, we should type in the keyword \emph{PortGroup} and go into a brace again. These nesting braces are unnecessary and redundant, which increases the complexity of the code. In Python, braces are avoided by having a whitespace-sensitive syntax.
\end{itemize}

The draft shown in Figure~\ref{fig:Envisioned_code_for_DemoAdl_in_Python_style} shows how the program from Figure~\ref{fig:example-program-in-the-default-style} could appear in a Python-style language.

The envisioned code in Figure~\ref{fig:Envisioned_code_for_DemoAdl_in_Python_style} uses whitespace and indentations to define and separate code blocks, which greatly improves the conciseness of the code. Identifier of e.g. \emph{SoftwareComponent} directly follows after the keyword `SoftwareComponent' (cf. line 1 in Figure~\ref{fig:Envisioned_code_for_DemoAdl_in_Python_style}), which makes the code more readable. There are no more braces and the number of keywords and symbols is greatly reduced. 

\subsection{The Semi-automated Approach}
As mentioned earlier, we developed a semi-automated approach to adapt a generated grammar to become more like Python. On the one hand, we develop a script that applies some changes automatically. On the other hand, we require the language engineer to take some specific decisions that must be done by hand.
The adaptations that are automatically completed by the script are removing  braces, repositioning attributes, and removing commas. Hand-crafted adaptations address which keywords need to be refined. In the following we describe these adaptations in more detail:

\paragraph{Remove Braces and Introduce White-Space Awareness.}
Removing all braces directly from the grammar definition may cause errors, such as left recursion errors\cite{bettini2019type}, which will prevent the creation of a textual editor.

\begin{figure*}[tb]
 \centering
 \includegraphics[width=\linewidth]{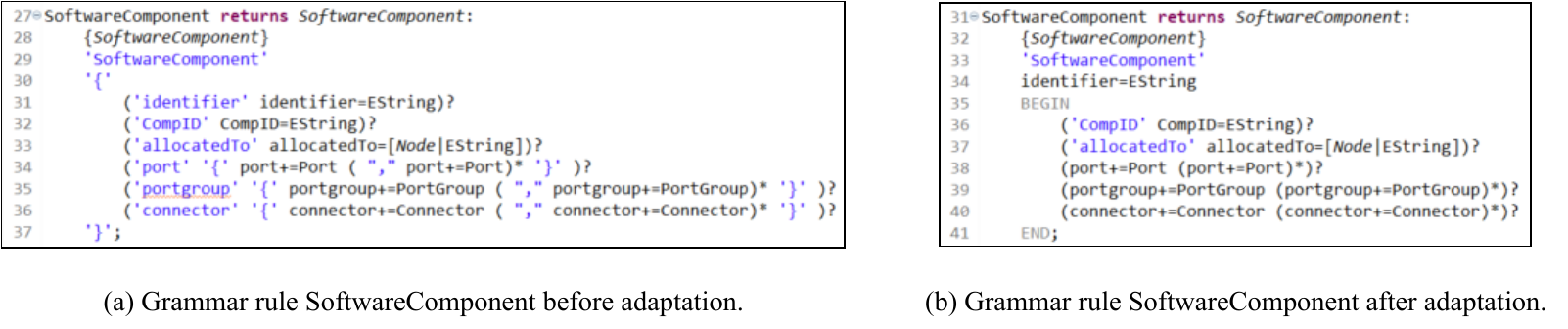}
 \caption{Comparison of grammar rules SoftwareComponent before and after adaptation.}
 \label{fig:Grammar_rule_SoftwareComponent_before_and_after_the_adaptation}
\end{figure*}

Therefore, the functionality of the braces needs to be substituted with the use of whitespace and indents to define code blocks. Thus, we need the language to be whitespace-aware~\cite{Whitespace-awareness}. To this end, the script introduces the following changes:

\begin{description}
\item{\textbf{Import required features.}} Change the reference in the grammar definition file from \emph{org.eclipse.xtext.common.Terminals} to \emph{org.eclipse.xtext.xbase.Xbase}. In addition, import \emph{Xbase} to refer to \emph{EClassifiers} from that model. This gives access to features that allow white-space-aware grammars in Xtext.
\item{\textbf{Create BEGIN/END terminals.}} Include whitespace-aware blocks in your language by using synthetic tokens in the grammar of the form `\texttt{synthetic:<terminal name>}'. An example using \emph{BEGIN}/\emph{END} is shown in Figure~\ref{fig:Grammar_rule_SoftwareComponent_before_and_after_the_adaptation}.
\item{\textbf{Redefine expression.}} Inherits expressions from \emph{Xbase} and redefines the syntax of block expressions by overriding the definition of \emph{xbase::XExpression}.
\end{description}

\paragraph{Reposition attribute identifier}
With the help of the script, we moved the attribute \emph{identifier} from its original position to after the keyword with the same name as the grammar rule. For example, \emph{SoftwareComponent} would be exactly followed by the attribute \emph{identifier}. Here, the script uses regular expressions to find the attribute identifier and move it. If the identifier is called \emph{name}, we do not need to do anything with it.

\paragraph{Remove Commas}
In this step, we removed commas that separate code blocks. For example, if there are multiple \emph{port}s under the same \emph{PortGroup} (cf. line 11 \& 12 in Figure~\ref{fig:example-program-in-the-python-style}), we do not need to separate these ports with commas because whitespace and indentations are used instead. 

\paragraph{Refine Keywords}
In this step, functionally redundant keywords are manually removed. The aforementioned keywords \emph{SoftwareComponent} and \emph{softwarecomponent}, e.g., were functionally redundant and we removed one of them (in our case, we kept the upper case one). We did not implement this removal in the script, because the keywords are language-specific and people make different decisions about what to keep. At the same time, we also did not implement the removal of the \emph{BEGIN}/\emph{END} related to them in the script. We will address these two automated operations in our future work by configurable rules with a finite set of options. For manually removing functionally redundant keywords, we recommend defining a rule, e.g., to remove the keyword which is written entirely in lowercase. However, our script automatically removes the keyword `identifier', because, in our envisioned draft, the value of the \emph{identifier} should directly follow the keyword (e.g., \emph{SoftwareComponent}) without the existence of the keyword `identifier', which would make the code more concise. 

When the script adapts the grammar, it uses regular expressions to search for the target texts in the entire grammar text, and then performs operations on it, including deletion, modification, etc.





\subsection{Evaluation}
\begin{figure}[tb]
  \centering
  \includegraphics[width=\linewidth]{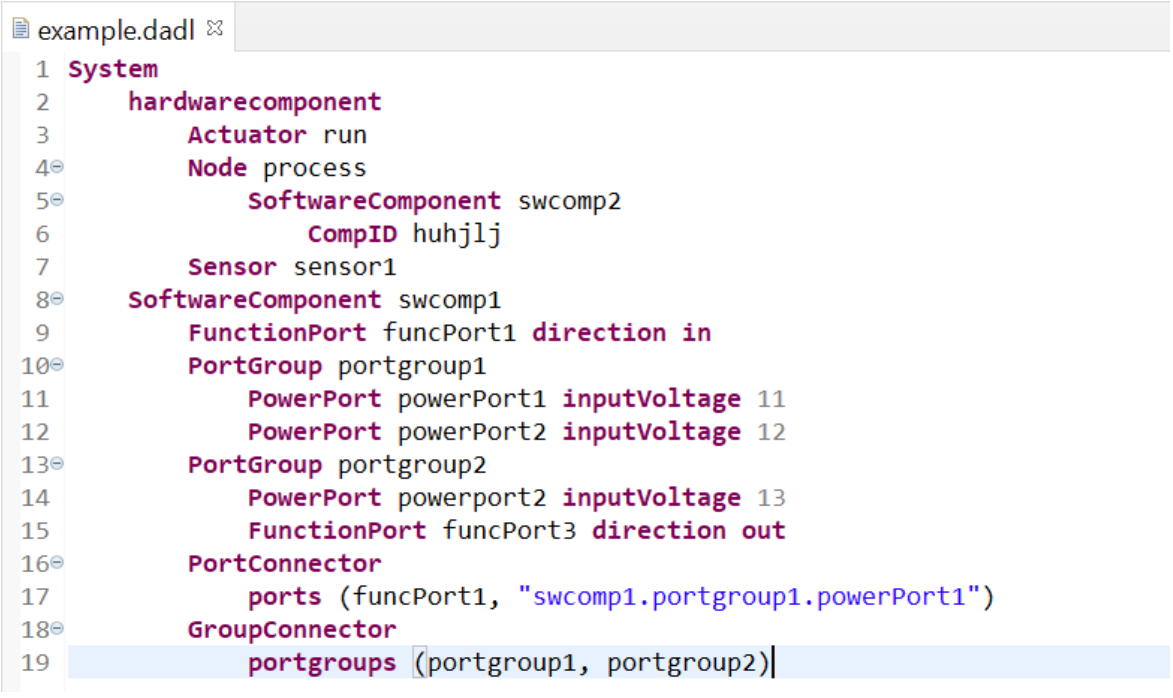}
  \caption{Screenshot of part of the example program for the DemoADL language with a Python-like style in eclipse.}
  \label{fig:example-program-in-the-python-style}
\end{figure}

With the above modifications, the grammar of DemoADL was changed. We generated the Xtext artifacts for the language by running the MWE2 workflow file. We wrote a program (cf. Figure~\ref{fig:example-program-in-the-python-style}) that conforms to the newly adapted textual grammar, which was parsed successfully by the generated editor. 

To evaluate the usability and generalizability of the proposed approach, we applied it to two additional DSLs, \emph{Xenia} and \emph{ACME}, which we selected based on the criteria described above. The basic information of the two DSLs is shown in Table~\ref{tab:Two-chosen-DSLs}.

\begin{table*}[tb]
\scriptsize
\centering
\caption{The two languages chosen for the evaluation.}
\label{tab:Two-chosen-DSLs}
\begin{tabular}{@{}llllll@{}}
\toprule
\textbf{Language}   & \textbf{Domain}     & \textbf{Language} & \textbf{meta-model} & \textbf{Homepage} \\
                    &                     & \textbf{elements} & \textbf{Source}    &                   \\
\midrule
    Xenia   & Generate web pages          & 14    & \cite{XeniaMM} & \cite{XeniaHomePage} \\
    ACME    & SW Architecture description & 16    & \cite{Atlanticzoo} & \cite{AcmeHomePage} \\
\bottomrule
\end{tabular}
\end{table*}

\paragraph{Grammar generation with Xtext}
The Ecore meta-model of ACME is from the Atlantic Zoo~\cite{Atlanticzoo}. To fulfill all technical constraints necessary for the generation of model code and grammar with Xtext, some small adaptations were necessary:
\begin{itemize}
  \item We filled in the namespace values (i.e., `Ns Prefix' and 'Ns URI') for all packages in the meta-model. 
  \item We filled in the value (i.e., `Instance Type Name' for all the types under the package \emph{primitivetypes}.
  \item We changed the lower-bound value of two attributes under the type \emph{Link} from 1 to 0.
\end{itemize}
We then generated a textual grammar from both DSLs' meta-models using the Xtext framework. 

\paragraph{Applying the approach}
With the proposed approach, we adapted both DSLs to a Python-like DSL with the help of our script. 
All manual steps were performed by the first author of this paper. 
For both \emph{ACME} and \emph{Xenia}, executing these manual steps took less than 30 minutes each.

\begin{figure*}[tb]
  \centering
  \includegraphics[width=\linewidth]{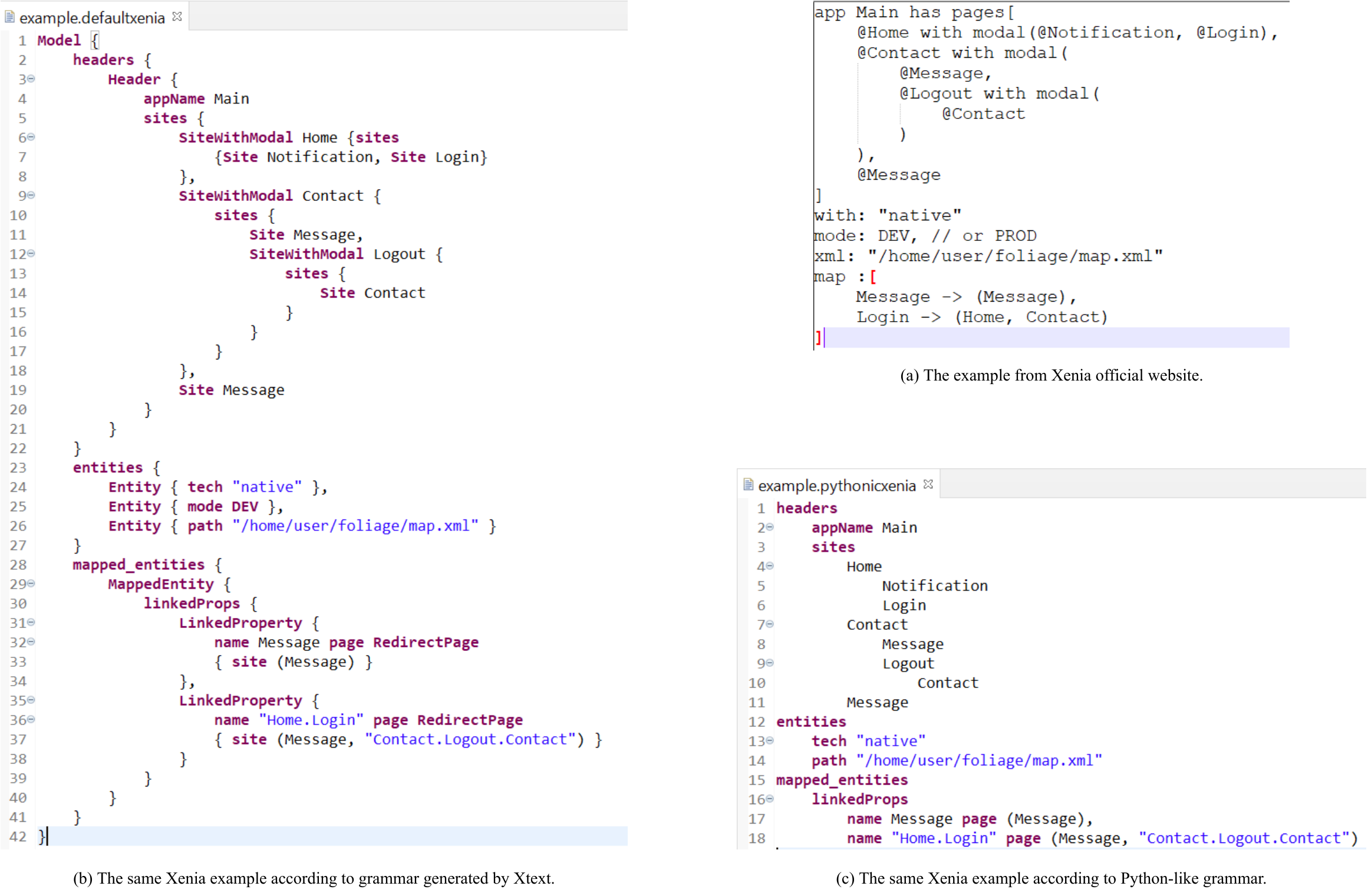}
  \caption{Comparison of Xenia programs in different styles (The original example is from \url{https://github.com/rodchenk/xenia}).}
  \label{fig:Comparison_of_Xenia_examples_in_different_styles}
\end{figure*}

\paragraph{Outcome}
An example program for the grammar that Xtext generated for \emph{Xenia} is shown in Figure~\ref{fig:Comparison_of_Xenia_examples_in_different_styles}-(b) while the program for the resulting Python-like grammar of \emph{Xenia} is shown in Figure~\ref{fig:Comparison_of_Xenia_examples_in_different_styles}-(c). For comparison, two programs correspond to the ``app Main'' example from the \emph{Xenia} home page~\cite{XeniaHomePage} (shown in Figure~\ref{fig:Comparison_of_Xenia_examples_in_different_styles}-(a)). 
The comparison shows that the program in (c) is much more concise than the one in (b) because there are fewer keywords and nesting. Like Python, the program in (c) uses whitespace and indents to express hierarchy. 
The comparison to the original program (Figure~\ref{fig:Comparison_of_Xenia_examples_in_different_styles}-(a)) also shows that the resulting Python-like grammar is much closer in terms of compactness to the actual, intended grammar of Xenia.

\begin{figure}[tb]
  \centering
  \includegraphics[width=\linewidth]{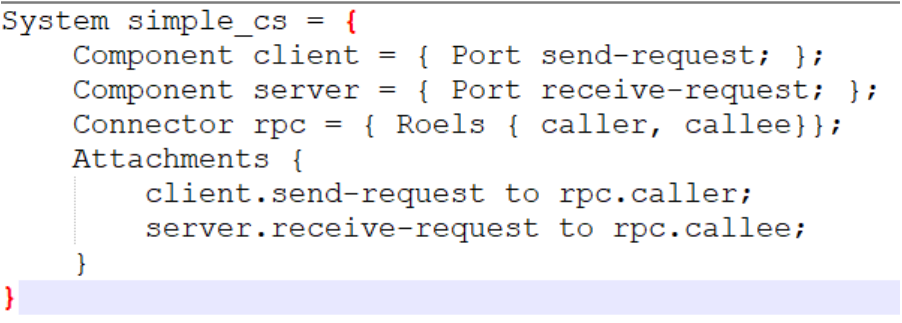}
  \caption{ACME original syntax from \url{https://www.cs.cmu.edu/~acme/docs/language_overview.html}.}
  \label{fig:acme:original}
\end{figure}

\begin{figure}[tb]
  \centering
  \includegraphics[width=\linewidth]{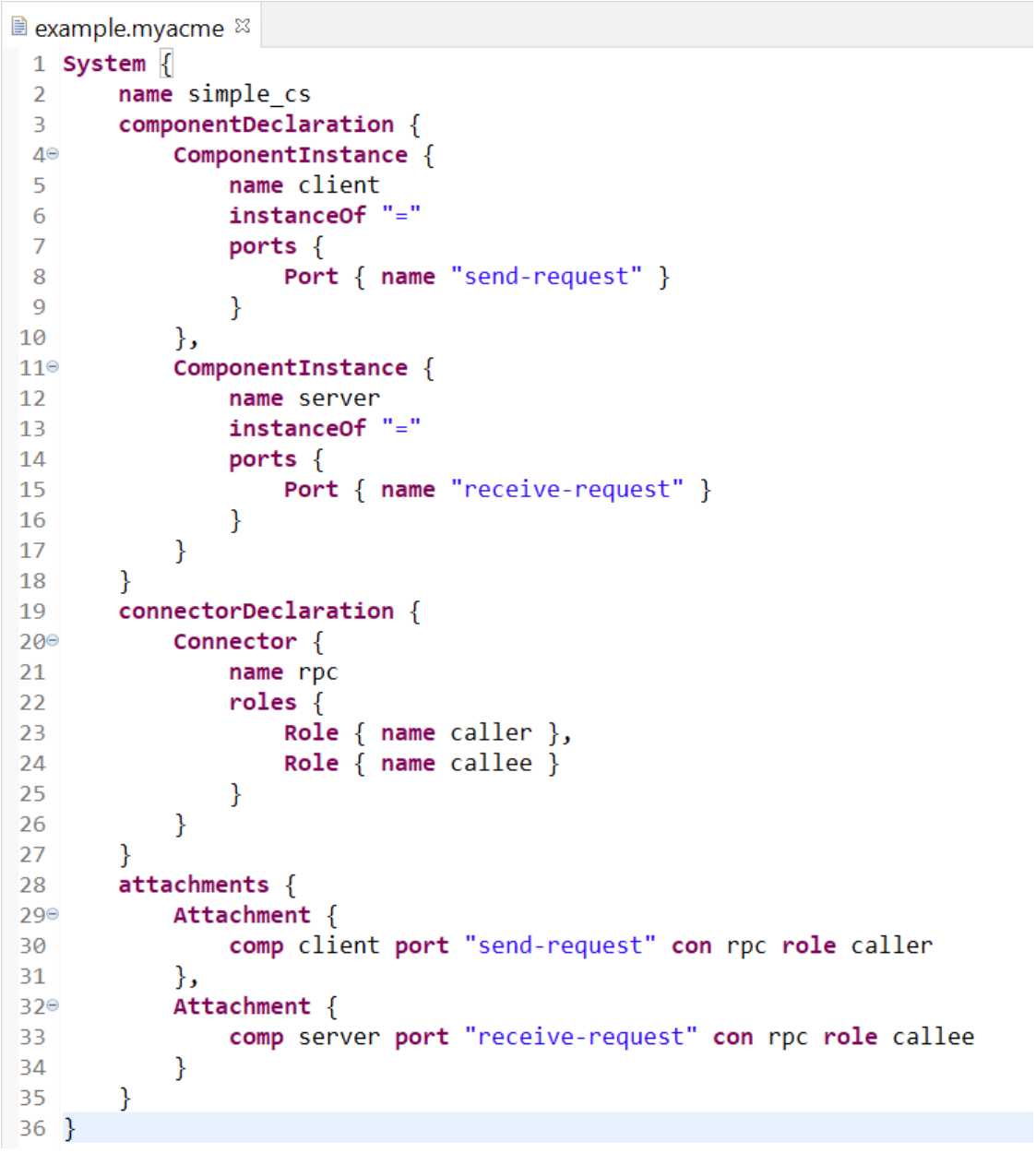}
  \caption{The example from Figure~\ref{fig:acme:original} in the syntax automatically generated by Xtext.}
  \label{fig:acme:xtext}
\end{figure}

\begin{figure}[tb]
  \centering
  \includegraphics[width=\linewidth]{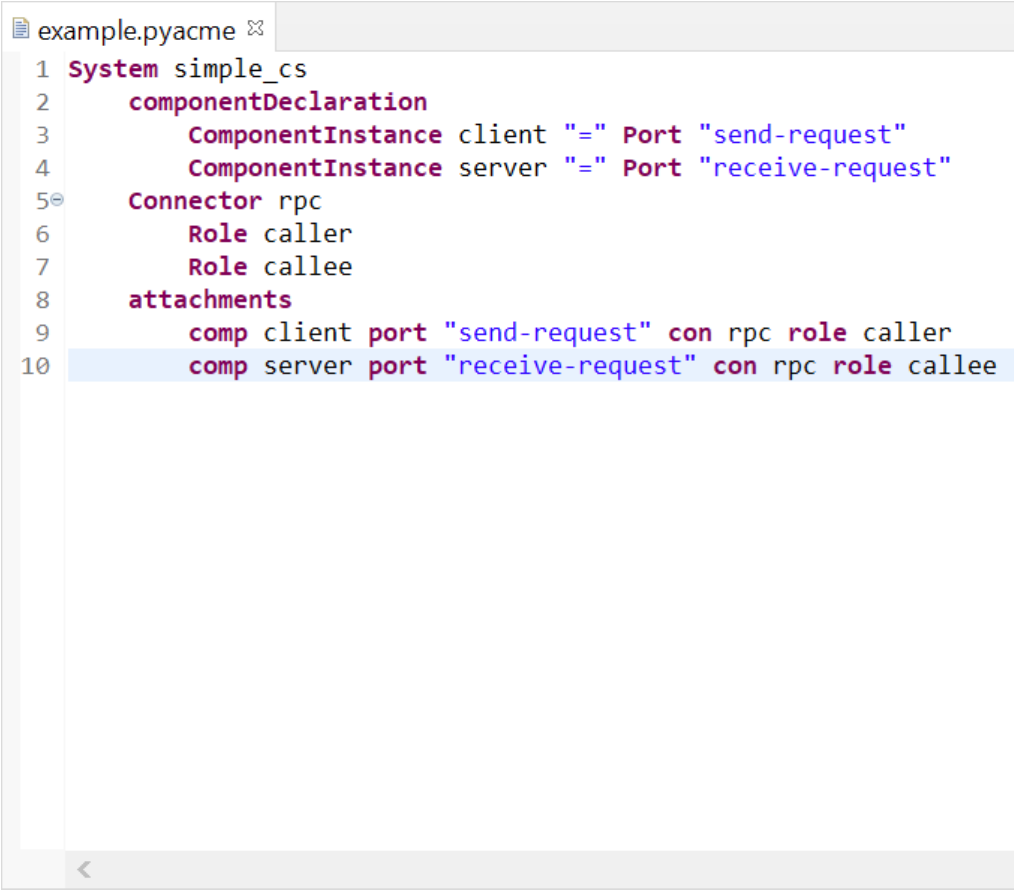}
  \caption{The example from Figure~\ref{fig:acme:original} in the Python-like syntax after our tool modified the generated grammar.}
  \label{fig:acme:python}
\end{figure}

An example program for the grammar that Xtext generated for \emph{ACME} is shown in Figure~\ref{fig:acme:xtext} while the program for the resulting Python-like grammar of \emph{ACME} is shown in Figure~\ref{fig:acme:python}. Similarly, the two programs conform to the example ``simple\_cs'' from the subpage ``An Overview Of Acme'' of the \emph{ACME} homepage \cite{AcmeHomePage} (shown in Figure~\ref{fig:acme:original}). 
Again we can see that the Python-like program in (c))contains much fewer lines of code, and there is no need to input braces in the program. Every identifier (here e.g., the name) follows the main keyword to identify a certain structure. Also, the comparison to the original grammar shows that the Python-like style is much closer to the original and intended grammar of \emph{ACME} (Figure~\ref{fig:acme:original}) in terms of compactness.

Note that the original syntaxes of both \emph{Xenia} and \emph{ACME} are quite different in style and neither is originally white-space sensitive.
This shows that our approach and the script are applicable to diverse DSLs and can be used by DSL developers to quickly reach a Python-like grammar, which could then be used as a basis for further refinements of the grammar.



\section{Discussion}
\subsection{Threats to Validity and Limitations}
\paragraph{Threats to Validity}
As discussed in the evaluation part, we applied the proposed approach to the two DSLs Xenia and ACME. The results show that the approach could be successfully used for these two DSLs. However, to ensure generalizability, we plan to apply the approach to additional languages.
Although we could show that our method provides a quick way to adapt diverse Xtext-generated grammars to Python-style languages, there are still some limitations.


\paragraph{Expressiveness of language.}
Even after adapting a grammar with our approach, there is often room for further modifications to the grammar to improve the expressiveness of the language itself. In the Xenia example, we can change the keyword \emph{page} to the arrow \emph{->}, indicating a ``progressive'' or ``link'' relationship as it was done in the original grammar. Such special keywords are typical for DSLs.
Our proposed approach does not include such operations, since they are highly language-specific. 
We focus currently only on adapting a DSL with a default Xtext-generated grammar to Python style. 
However, in future work, it might be interesting to support such operations with automation tools. 

\paragraph{Automation of Grammar Modification}
We implemented a small script in the form of an Eclipse plugin to simplify grammar modification. However, the functionality of the script is limited. For example, adding a colon \emph{:} after a certain type of keyword is not supported by the script at present. A feasible solution will be to extend the functionality to support various modifications of keywords.

Also, while adapting the grammar, one may encounter problems such as \emph{left-recursion} errors. Other than the focused change to white-space awareness, we currently provide no additional support to help developers avoid changes that lead to these errors. 

\subsection{Future Work}
As we mentioned above, in the future we will apply our approach to more languages to evaluate generalizability. In addition, we plan to apply the approach to larger DSLs, for example, EAST-ADL.
We hope that this will also help us refine our approach and extend the capabilities of the script. Namely, we hope that we can address the previously mentioned limited functionality of our script and the semi-automated nature of our approach. 

Converting a DSL generated from a meta-model into a Python-like language can improve the language's simplicity, friendliness, and usability. However, these are all issues on the language appearance level. In the future, we will discuss and explore how to design a good DSL, including defining the standards of what a good DSL is. 

Another interesting and related topic is blended modeling~\cite{addazi2021blended}, which refers to modeling in different notations (such as textual, graphical, etc.) at the same time. When creating a textual language for a DSL that already has a non-textual notation, the approach in this paper can help create a concise, user-friendly textual language in the first stage, thereby improving the user experience of modeling activities~\cite{david2022blended}. In future work, we plan to evaluate our approach in this context.

\section{Related Work}
Since the first version of Xtext was released in 2006 \cite{efftinge2006oaw}, the German company \emph{itemis} released several versions of Xtext. Eysholdt and Behrens from the company briefly introduced the motivation and capabilities of Xtext in \cite{eysholdt2010xtext}. In addition, \emph{itemis} also officially released the Xtext User Guide, which introduces in detail what Xtext is, how to use it and some of its internal principles \cite{behrens2008xtext}. This is the technical manual that is the main reference for this paper. Bettini's book \cite{bettini2016implementing} goes further and shows how to develop DSLs using Xtext and Xtend. Moreover, the book has a dedicated and independent chapter to introduce Xbase, which is not included in \cite{behrens2008xtext} and it supplements the knowledge about Xbase for us.

Mernik et al.\ in \cite{mernik2005and} provide empirical data and valuable experience on when and how to develop DSLs, this study refers to these experiences when designing the DSL. Neubauer et al.\ customized the grammar generated from the Ecore meta-model \cite{neubauer2015xmltext}. However, their purpose was to solve the problem that the Xtext grammar generator did not create rules for meta-model data types. In contrast, the modification in this paper aims to bring the grammar closer to Python. Manual changes were made by Sredojevi to the DSL's grammar definition file in \cite{sredojevic2015domain}. By including keywords and identifiers, this change primarily addresses the issue that the graphical representation of the meta-model does not fully define the grammar structure. In contrast, our approach replaces the language's overall style by modifying the entire grammar definition file.

\section{Conclusion}
\label{sec:conclusion}
Xtext is one of the most commonly used methods for developing DSL. It can be used to generate a textual grammar from an Ecore meta-model. However, this auto-generated grammar has a format that is often considered cumbersome and difficult to work with. 
Python is a language known for its simplicity, which helps programmers become more productive. 
We aim to make it easy for DSL developers to create their languages with a Python-style syntax, in the hope that these DSLs would benefit from the advantages of Python's syntax.
In this paper, we analyze the primary inadequacies of an auto-generated DSL (i.e., DSL generated by Xtext) which is to describe the architecture of an embedded system. 
Based on this analysis, this paper presents an approach to semi-automatically change a grammar that was generated with Xtext, so that the DSL becomes a Python-like language. We applied this approach to the design of the aforementioned lightweight example language and obtained an intermediate result, a Python-like structure description language. 
We applied the approach to two additional DSLs, \emph{Xenia} and \emph{ACME}, to validate the generalizability of our approach. The contribution of this paper is a generalized approach for transforming generated DSLs into Python-like DSLs.

We intend to investigate the practical applications of this approach in the future, e.g., the adaptation of highly sophisticated and large DSLs in industrial settings. 
In our future efforts, we intend to develop a more complete system for the adaptation of automatically generated grammar. We envision a rule-based system that allows an engineer to configure a number of modifications of the grammar that, together, change the concrete syntax significantly. 
We believe such a system will contribute to making Xtext more attractive to language engineers who want to combine work on an evolving and rapidly changing language with a user-friendly and compact concrete syntax, a combination that Xtext currently does not support.

%% file: TexFiles/paperD/Abstract.tex
\newpage
\section*{Abstract}

In model-driven engineering, developing a textual domain-specific language (DSL) involves constructing a meta-model, which defines an underlying abstract syntax, and a grammar, which defines the concrete syntax for the DSL.
Language workbenches such as Xtext allow the grammar to be automatically generated from the meta-model, yet
the generated grammar usually needs to be manually optimized to improve its usability. When the meta-model changes during rapid prototyping or language evolution, it can become necessary to re-generate the grammar and optimize it again, causing repeated effort and potential for errors.

In this paper, we present \grammaroptimizer, an approach for optimizing generated grammars in the context of meta-model-based language evolution. To reduce the effort for language engineers during rapid prototyping and language evolution, it offers a catalog of configurable \textit{grammar optimization rules}.
Once configured, these rules can be automatically applied and re-applied after future evolution steps, greatly reducing redundant manual effort.
In addition, some of the supported optimizations can globally change the style of concrete syntax elements, further significantly reducing the effort for manual optimizations.
The grammar optimization rules were extracted from a comparison of generated and existing, expert-created grammars, based on seven available DSLs.
An evaluation based on the seven languages shows \grammaroptimizer's ability to modify Xtext-generated grammars in a way that agrees with manual changes performed by an expert and to support language evolution in an efficient way, with only a minimal need to change existing configurations over time.

\newpage

%% file: TexFiles/paperD/main.tex
\section{Introduction}

Domain-Specific Languages (DSLs) are a common way to describe certain application domains and to specify the relevant concepts and their relationships~\cite{van2000domain}. 
They are, among many other things, used to describe model transformations  (the Operational transformation language of the MOF Query, View, and Transformation\,---\,QVTo~\cite{qvt} and the \mbox{ATLAS} Transformation Language\,---\,ATL~\cite{ATLSyntax}), bibliographies (BibTeX~\cite{Bibtex}), graph models (DOT~\cite{Dot}), formal requirements (the Scenario Modeling Language\,---\,SML~\cite{smlRepo} and Spectra~\cite{Spectra}), meta-models (Xcore~\cite{EclipseXcore}), or web-sites (Xenia~\cite{Xenia}).

In many cases, the syntax of the language that engineers and developers work with is textual. 
For example, DOT is based on a clearly defined and well-documented grammar so that a parser can be constructed to translate the input in the respective language into an abstract syntax tree which can then be interpreted. 

A different way to go about constructing DSLs is proposed by model-driven engineering. There, the concepts that are relevant in the domain are  captured in a meta-model which defines the \emph{abstract syntax} (see, e.g., \cite{roy2019methodology,frank2013domain,mernik2005and}). Different \emph{concrete syntaxes}, e.g., graphical, textual, or form-based, can  be defined to describe actual models that adhere to the abstract syntax. 

In this paper, we consider the Eclipse ecosystem and Xtext~\cite{Xtext} as its de-facto standard framework for developing textual DSLs.
Xtext relies on the Eclipse Modeling Framework (EMF)~\cite{EMF} and uses its Ecore (meta-)modeling facilities as basis.
Developing a textual DSL in Xtext involves two main artifacts: a grammar, which defines the concrete syntax of the language, and a meta-model, which defines the abstract syntax.
Xtext allows either the grammar or the meta-model to be created first, and then automatically generating the one from the other (or alternatively, writing both manually and aligning them).

Software languages change over time.
This is due to \textit{language evolution}, which entails that languages change over time to address new and changed requirements, and due to \textit{rapid prototyping}, which involves many quick iterations on an initial design.
In the case of an Xtext-based language, grammar and  meta-model  need to be modified to stay consistent with each other.
We consider two options for evolving a language in Xtext:
First, the developers can change the grammar and then use Xtext to automatically create an updated version of the meta-model from it. 
Second, the developers can change the meta-model then use Xtext to derive an updated version of the grammar from it.
We call the first approach \textit{grammar-based evolution}, and the second approach \textit{meta-model-based evolution}.

In this paper, we focus on meta-model-based evolution, for the following rationale: While grammar-based evolution is a common way of developing languages in Xtext,
it is not geared for three scenarios that we  encountered in the real world, including collaborations with an industrial partner. In particular: 
1. Several concrete syntaxes (e.g., visual, textual, tabular) for the same underlying metamodel co-exist and evolve at the same time. This is particularly common in the context of blended modeling \cite{Ciccozzi.2019}, a timely modeling paradigm.
2. The metamodel comes from some external source (such as a third-party supplier or a standardization committee), which prohibits independent modification.
3. The metamodel is the central artifact of a larger ecosystem of available tools, including. e.g., automated analyses, transformations and visualizations. As such, the language engineers might prefer to evolve it directly, instead of relying on the, potentially sub-optimal, output of automatically co-evolving it after grammar changes.
The real-world case that inspired this paper has aspects of the first two scenarios: we work on a language from an industry partner for which there already exists an evolving metamodel and graphical editor available.

Compared to grammar-based evolution, meta-model-based evolution has one major disadvantage: Co-evolving the grammar after meta-model changes is more complicated than vice versa, as it involves dealing with both abstract and concrete syntax aspects, whereas updating the meta-model after grammar changes only involves abstract syntax aspects.
The goal of this paper is to substantially mitigate this disadvantage, as we will now explain.




One problem that prohibits using a grammar generated from the meta-model directly is that the grammars Xtext automatically generates are not particularly user-friendly. At the same time, the grammars themselves are hard to understand and the languages defined by them are verbose, use many braces, and enforce very strict rules about the presence of keywords and certain constructs. While the usability of DSLs is largely dependent on the right choice of concept names (see, e.g., \cite{albuquerque2015quantifying}), the syntax also plays a significant role in how easily a language can be learned. For example, \cite{stefik2013empirical} find that
\texttt{if}-statements are used by novices more accurately if they are written without parentheses and braces.
We also find that Xtext tends to add a number of keywords that are not strictly necessary and that make the generated language more verbose without adding clarity. 

These issues can be addressed by improving the Xtext-generated grammar.
In the state of the art, this is a manual optimization process, in which the user tweaks the grammar to improve its usability, e.g., changing and removing keywords, parentheses, and order of rule elements.
However, manual optimization has a significant drawback: Once that the meta-model evolves and the grammar is re-generated, the same optimizations have to be performed again on the generated grammar.  This is a time-consuming and error-prone task, even more so when done after any meta-model change. 
Furthermore, for certain changes that address a larger scope within the grammar (e.g., removing inner parentheses around attributes in every grammar rule), it is a tedious and error-prone manual process even before evolution takes place.
Alternatively, instead of auto-generating the grammar when the meta-model evolves, the existing grammar could be manually evolved by new grammar rules and by modifying existing ones. This process is, again, time-consuming and error-prone and can easily lead to inconsistencies.

We propose a different approach: Automated optimization of the generated grammar based on simple rules, which we call \textit{grammar optimization rules}. Instead of modifying the grammar directly, the language engineer creates a set of simple optimization rule applications that modify the grammar file to make the resulting language easier to use and less verbose. Whenever the meta-model changes and the grammar is regenerated, the same or a slightly modified set of optimization rules can be used to update the new grammar to have the same properties as the previous version. 

In meta-model-based evolution scenarios, our approach can considerably reduce the manual effort for optimizations and, consequently, enable faster turnaround times.
This is due to two factors that we demonstrate in our evaluation:
First, the potential to reuse existing configurations across successive evolution steps.
For example, we considered four evolution steps from the history of QVTo.
Initially, we created a configuration that fully optimized the generated grammar to be consistent with the expert-created grammar for that evolution step. 
For the following three iterations, we only needed to modify 2, 0, and 1 configuration lines, respectively, to automatically optimize the generated grammar. Without our approach, language engineers would need to manually modify 228 lines of 66 grammar rules in each evolution step.
Second, the availability of powerful rules that enforce a large-scope change affects many grammar rules at the same time.
 For example, for the EAST-ADL case, modifying the Xtext-generated towards the expert-created grammar required curly braces for all attributes to be removed, while keeping the outer surrounding curly braces for each rule. 
 Performing this change manually entails manually revising 303  rules, whereas it took only one line of configuration in GrammarOptimizer.

While our approach clearly unfolds these benefits in the case of evolving languages and complex changes, it does not come for free. For locally-scoped changes, creating a configuration generally leads to more effort than a   manual grammar edit and hence, presents an upfront investment that pays off only when the language evolves over time.
In a different paper \cite{zhang2023automated}, we present an approach for automating the extraction of configurations from user-provided manual edits, thus reducing the initial manual effort to be the same as in the traditional process while keeping the long-term benefits. Together with the present paper, for the supported kinds of changes, it supports a fully automated process for aligning the grammar after changes to the meta-model.

The contribution of this paper is  \grammaroptimizer, an approach that modifies a generated grammar by applying a set of configurable, modular, simple optimization rules. It integrates into the workflow of language engineers working with Eclipse, EMF, and Xtext technologies and is able to apply rules to reproduce the textual syntaxes of common, textual DSLs.

We demonstrate its applicability on seven domain-specific languages from different application areas. We also show its support for language evolution in two cases:
1), we recreate the textual model transformation language QVTo in all four versions of the official standard~\cite{qvt} with only small changes to the configuration of optimization rule applications and with high consistency of the syntax between versions; and  
2), we conceived for the automotive systems modeling language EAST-ADL~\cite{eadl} together with an industrial partner a textual concrete syntax~\cite{EATXT}, where we initially started with a grammar for a subset of the EAST-ADL meta-model (i.e., textual language version 1) and subsequently evolved the grammar to encompass the full meta-model (i.e., textual language version 2).

The remainder of this paper is structured as follows. First, in Section~\ref{sec:background}, we provide an overview of the background of this paper, in particular, on metamodel-based textual DSL engineering. In Section~\ref{sec:related-work}, we review related research. In Section~\ref{sec:methodology}, we define the methodology of this paper. Subsequently, in Section~\ref{sec:identified-optimization-rules}, we describe the identified optimization rules, which are the main technical contribution of this paper. Following that, in Section~\ref{sec:solution}, we present our solution of the \grammaroptimizer, which implements the identified optimization rules. In Section~\ref{sec:eval}, we present our evaluation.  Section~\ref{sec:discussion} is devoted to our discussion, where we address threats to validity, the effort required to use \grammaroptimizer, implications for practitioners and researchers, and future work. Finally, in the last section, we conclude.

\section{Background: Textual DSL Engineering based on Meta-models}
\label{sec:background}


As outlined in the introduction, the engineering of textual DSLs can be conducted through the traditional approach of specifying grammars, but also by means of meta-models.
Both approaches have commonalities, but also differences \cite{Paige.2014}.
Like grammars specified by means of the Extended Backus Naur Form (EBNF)~\cite{iso1996iec}, meta-models enable formally specifying how the terms and structures of DSLs are composed.
In contrast to grammar specifications, however, meta-models describe DSLs as graph structures and are often used as the basis for graphical or non-textual DSLs.
Particularly, the focus in meta-model engineering is on specifying the abstract syntax. The definition of concrete syntaxes is often considered a subsequent DSL engineering step. However, the focus in grammar engineering is directly on the concrete syntax~\cite{Kleppe.2007a} and leaves the definition of the abstract syntax to the compiler.

\paragraph{Meta-model-based textual DSLs}
There are also examples of textual DSLs that are built with meta-model technology.
For example, the Object Management Group (OMG) defines textual DSLs that hook into their meta-model-based Meta Object Facility (MOF) and Unified Modeling Language ecosystems, for example, the Object Constraint Language (OCL)~\cite{OCLVersions} and the Operational transformation language of the MOF Query, View, and Transformation (QVTo)~\cite{qvt}.
However, this is done in a cumbersome way:
Both the specifications for OCL and QVTo define a meta-model specifying the abstract syntax and a grammar in EBNF specifying the concrete syntax of the DSL.
This grammar, in turn, defines a different set of concepts and, therefore, a meta-model for the concrete syntax that is different from the meta-model for the abstract syntax.
As Willink~\cite{willink2020reflections} points out, this leads to the awkward fact that the corresponding tool implementations such as Eclipse OCL \cite{EclipseOCL} and Eclipse QVTo~\cite{qvto-eclipse} also apply this distinction. 		
That is, both tool implementations each require an abstract syntax and a concrete syntax meta-model and, due to their structural divergences, a dedicated transformation between them.
Additionally, both tool implementations provide a hand-crafted concrete syntax parser, which implements the actual EBNF grammar.
Maintaining these different parts and updating the manually created ones incurs significant effort whenever the language should be evolved.

\paragraph{Grammar generation and Xtext}
A much more streamlined approach to language engineering would, instead, use a single meta-model and use this in a model-driven approach to derive the concrete syntax directly from it. 
With the exception of EMFText~\cite{heidenreich2009derivation} and the Grasland toolkit~\cite{Kleppe.2007b} that are both not maintained anymore, Xtext is currently the only textual DSL framework that allows generating a grammar from a meta-model. 
Using an EBNF-based Xtext grammar, Xtext applies the ANTLR parser generator framework~\cite{antlr} to derive the actual parser and all its required inputs. It also generates editors along with syntax highlighting, code validation, and other useful tools.

\begin{figure}[tb]
  \centering
  \includegraphics[width=\linewidth]{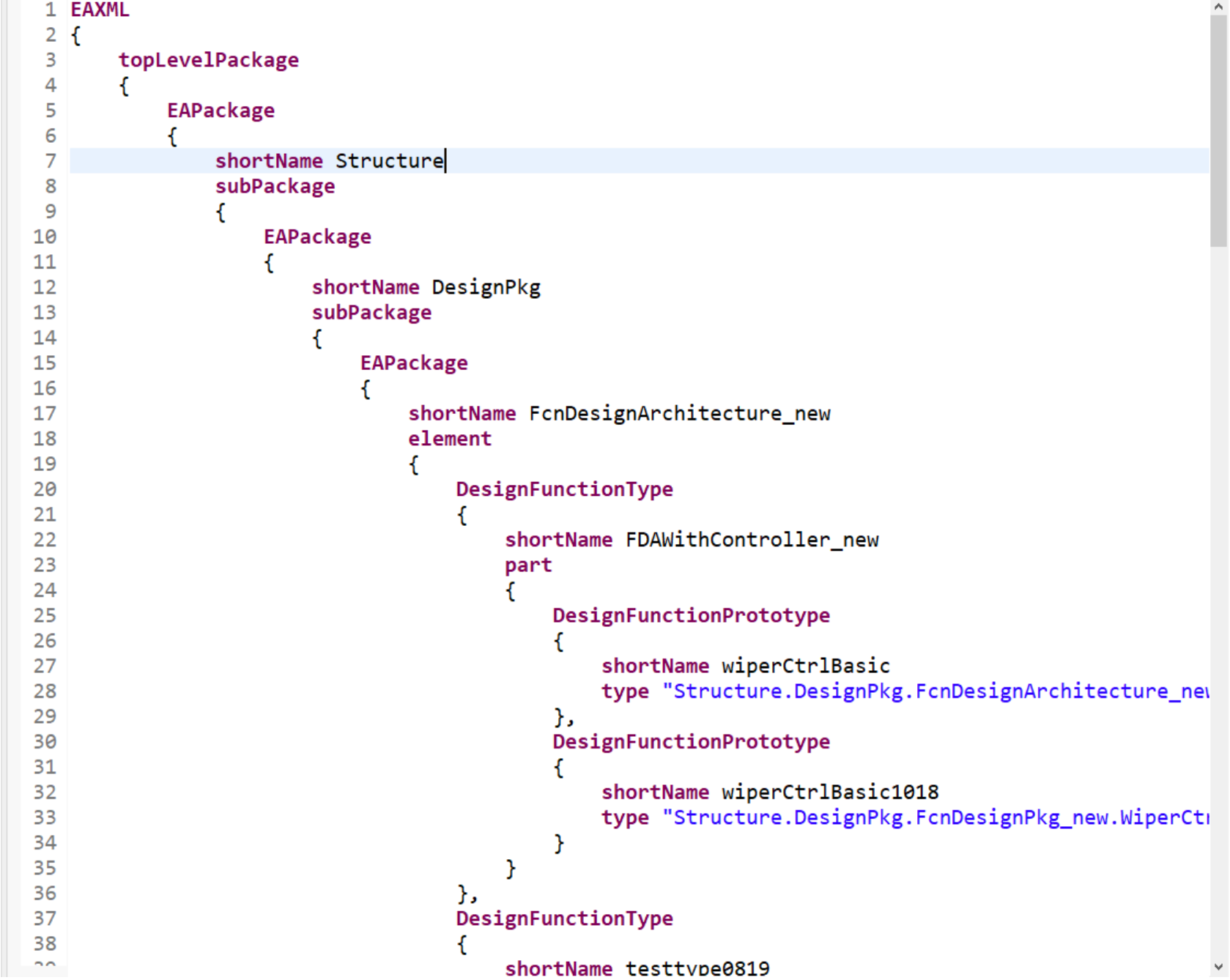}
  \caption{Instance of the generated grammar for EAST-ADL.}
  \label{fig:example_snippet_east-adl}
\end{figure}

A language engineer has two options when constructing a new language from a meta-model in Xtext:
\begin{itemize}
  \item \textbf{Hand-craft a grammar} that maps syntactical elements of the textual concrete syntax to the concepts of the abstract syntax. This is the way many DSLs have been built in Xtext (e.g., Xcore~\cite{EclipseXcore}, Spectra~\cite{Spectra}, and Xenia~\cite{Xenia}). However, this approach is not very robust when the meta-model changes since the grammar needs to be adapted manually to that meta-model change. 
  \item \textbf{Generate a grammar} from the meta-model using Xtext's built-in functionality (we call this grammar \emph{generated grammar} in this paper). This creates a grammar that contains grammar rules for all meta-model elements that are contained in a common root node and resolves references, etc., to a degree (see Section~\ref{sec:methodology_analysis_MMPrepsAndGrammarGen} for details). This approach deals very well with meta-model changes and only requires the re-generation of the grammar which is very fast and can be automated.
  However, the grammar is going to be very verbose, structured extensively using braces, and uses a lot of keywords. Such a situation is shown in Figure~\ref{fig:example_snippet_east-adl}, depicting an instance of the generated grammar for EAST-ADL. This makes it difficult to use such a generated grammar in practice. 
\end{itemize}
In this paper, we focus on making the second option more usable to give language engineers the ability to quickly re-generate their grammars when the meta-model changes, e.g., for rapid prototyping or for language evolution. Thus, we provide the ability to optimize the automatically generated grammars to improve their usability and make them similar in this regard to hand-crafted grammars. We show that this optimization can be re-applied to evolving versions of the language. 
Our contribution, \grammaroptimizer, therefore combines the advantages of both approaches while mitigating their respective disadvantages.

\section{Related Work}
\label{sec:related-work}
In the following, we discuss approaches for grammar optimization, approaches that are concerned with the design and evolution of DSLs, and other approaches.

\paragraph{Grammar Optimization}

There are a few works that aim at optimizing grammar rules with a focus on XML-based languages. For example, \cite{neubauer2015xmltext,neubauer2017xmltextToolPaper}  also mention optimization of grammar rules in Xtext. Their approach XMLText and the scope of their optimization focus only on XML-based languages. 
They convert an XML schema definition to a meta-model using the built-in capabilities of EMF.
Based on that meta-model, they then use an adapted Xtext grammar generator for XML-based languages to provide more human-friendly notations for editing XML files.
XMLText thereby acts as a sort of compiler add-on to enable editing in a different notation and to automatically translate to XML and vice versa.
In contrast, we develop a post-processing approach that enables the optimization of any Xtext grammar, not only XML-based ones,  
cf.~also our discussion in Section~\ref{sec:discussion}). 

The approach of \cite{chodarev2016development} shares the same goal and a similar functional principle as XMLText, but uses other technological frameworks.
In contrast to XMLText, Chodarev supports more straightforward customization of the target XML language by directly annotating the meta-model that is generated from the XML schema.
The same distinction applies here as well: \grammaroptimizer enables the optimization of any Xtext grammar and is not restricted to XML-based languages.

Grammar optimization for DSLs in general is addressed by
\cite{Jouault.2006}. They propose an approach to specify a syntax for textual, meta-model-based DSLs with a dedicated DSL called Textual Concrete Syntax, which is based on a meta-model. 
From such a syntax specification, a concrete grammar and a parser are generated.
The approach is similar to a template language restricting the language engineer and thereby, as the authors state, lacks the freedom of grammar specifications in terms of syntax customization options. 
In contrast, we argue that the \grammaroptimizer provides more syntax customization options to achieve a well-accepted textual DSL.

Finally, \cite{novotny2013model} designed a model-driven Xtext pretty printer, which is used for improving the readability of the DSL by means of improved, language-specific, and configurable code formatting and syntax highlighting. 
In contrast, our \grammaroptimizer is not about improving code readability but focused on how to design the DSL itself to be easy to use and user-friendly.

\paragraph{Designing and Evolving Meta-model-based DSLs}
Many papers about the design of DSLs focus solely on the construction of the abstract syntax and ignore the concrete syntaxes (e.g.,~\cite{roy2019methodology,frank2011some}), or focus exclusively on graphical notations (e.g.,\cite{frank2013domain,tolvanen2018effort}).
In contrast, the guidelines proposed by \cite{karsai2014design} contain specific ideas about concrete syntax design, e.g., to ``balance compactness and comprehensibility''. 
Arguably, the languages automatically generated by Xtext are neither compact nor comprehensible and therefore require manual changes. 

\cite{mernik2005and} acknowledge that DSL design is not a sequential process. 
The paper also mentions the importance of textual concrete syntaxes to support common editing operations as well as the reuse of existing languages. 
Likewise, \cite{van2010exercise} describe DSL development as an iterative process and use EMF and Xtext for the textual syntax of the DSL. They also discuss the evolution of the language, and that ``it is hard to predict which language features will improve understandability and modifiability without actually using the language''. Again, this is an argument for the need to do prototyping when developing a language. 
\cite{karaila2009evolution} broadens the scope and also argues for the need for evolving DSLs along with the ``engineering environment'' they are situated in, including editors and code generators. 
\cite{pizka2007tool} also acknowledge the ``constant need for evolution'' of DSLs. 

There is a lot of research supporting different aspects of language change and evolution.
Existing approaches focus on how diverse artifacts can be co-evolved with evolving meta-models, namely the models that are instances of the meta-models~\cite{hebig2016approaches}, OCL constraints that are used to specify static semantics of the language~\cite{khelladi2017semi, khelladi2016metamodel}, graphical editors of the language~\cite{ruscio2010automated,di2011needed}, and model transformations that consume or produce programs of the language~\cite{garcia2012model}.
Specifically, the evolution of language instances with evolving meta-models is well supported by research approaches. For example, Di Ruscio et al.~\cite{di2011needed} support language evolution by using model transformations to simultaneously migrate the meta-model as well as model instances.

Thus, while these approaches cover a lot of requirements, there is still a need to address the evolution of textual grammars with the change of the meta-model as it happens during rapid prototyping or normal language evolution.
This is a challenge, especially since fully generated grammars are usually not suitable for use in practice. This implies that upon changing a meta-model, it is necessary to co-evolve a manually created grammar or a grammar that has been generated and then manually changed. 
\grammaroptimizer has been created to support prototyping and evolution of DSLs and is, therefore, able to support and largely automate these activities.

\paragraph{Other Approaches}
As we mentioned above, besides Xtext, there are two more approaches that support the generation of EBNF-based grammars and from these the generation of the actual parsers. These are EMFText~\cite{heidenreich2009derivation} and the Grasland toolkit~\cite{Kleppe.2007b}, which are both not maintained anymore.

Whereas our work focuses on the Eclipse technology stack based on EMF and Xtext, there are a number of other language workbenches and supporting tools that support the design of DS(M)Ls and their evolution. 
However, none of these approaches are able to derive grammars directly from meta-models, a prerequisite for the approach to language engineering we propose and the basis of our contribution, \grammaroptimizer. 
Instead, tools like textX~\cite{dejanovic2017textx} go the other way around and derive the meta-model from a grammar. Langium~\cite{langium} is the self-proclaimed Xtext successor without the strong binding to Eclipse, but does not support this particular use case just yet and instead focuses on language construction based on grammars. 
MetaEdit+~\cite{kelly2018collaborative} does not offer a textual syntax for the languages, but instead a generator to create text out of diagrams that are modeled using either tables, matrices, or diagrams. 
JetBrains MPS~\cite{mps} is based on projectional editing where concrete syntaxes are projections of the abstract syntax. 
However, these projections are manually defined and not automatically derived from the meta-model as it is the case with Xtext.
Finally, \cite{pizka2007tool} propose 
an approach to evolve DSLs including their concrete syntaxes and instances. For that, they present ``evolution languages'' that evolve the concrete syntax separately. However, they focus on DSLs that are built with classical compilers and not with meta-models.

\section{Methodology}\label{sec:methodology}

\begin{figure*}[tb]
  \centering
  \includegraphics[width=\linewidth]{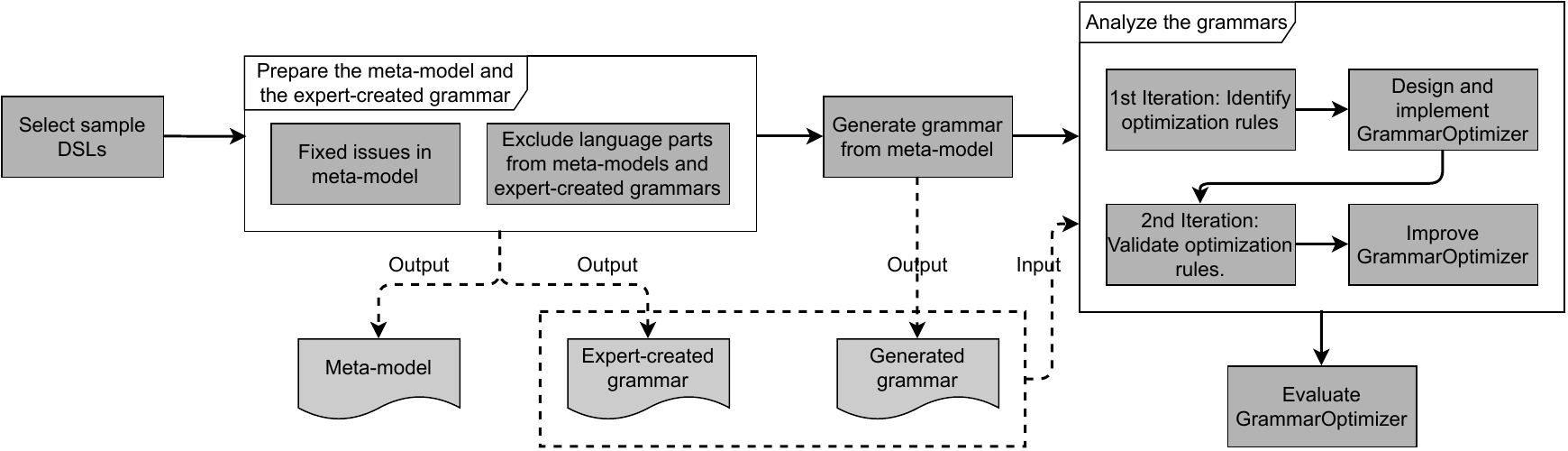}
  \caption{Overview of our methodology.}
  \label{fig:methodology}
\end{figure*}

In this section, we describe our research methodology, shown in an overview in Fig.~\ref{fig:methodology}.
Our methodology consists of a number of sequential steps, in particular: selecting the case languages,  preparing  metamodels and grammars (including the exclusion of certain parts of the language), and  two iterations of analysis, including extraction of grammar optimization rules and tool development.
We now describe all of these steps in detail.

\subsection{Selection of Sample DSLs}
We selected a number of DSLs for which both an expert-created grammar and a meta-model were available. Our key idea was that the expert-created grammar serves as a \textit{ground truth}, specifying what a desirable target of an optimization process would look like.
As the starting point for this optimization process, we considered the Xtext-generated grammars for the available meta-models.
The goal of our grammar optimization rules was to support an automated transformation to turn the Xtext-generated grammar into the expert-created grammar. By selecting a number of DSLs with a grammar or precise syntax definition from which we could derive such a ground truth, we aimed to generalize the grammar optimization rules so that new languages can be optimized based on rules that we include in \grammaroptimizer.

\paragraph{Sources}
To find language candidates, we collected well-known languages, such as DOT, and used language collections, such as the Atlantic Zoo \cite{Atlanticzoo}, a list of robotics DSLs~\cite{robotic2020dsls}, and similar collections~\cite{dslwikipedia,zooofdsls,dslsemanticdesign,financialdsl,van2000domain}.
However, it turned out that the search for suitable examples was not trivial despite these resources.
The quality of the meta-models in these collections was often insufficient for our purposes. In many cases, the meta-model structures were too different from the grammars or there was no grammar in either Xtext or in EBNF publicly available as well as no clear syntax definition by other means. 
We therefore extended our search to also use Github's search feature to find projects in which meta-models and Xtext grammars were present and manually searched the Eclipse Foundation's Git repositories for suitable candidates. Grammars were either taken from the language specifications or from the repositories directly.

\paragraph{Concrete Grammar Reconstruction for BibTeX}
In some cases, the syntax of a language is described in detail online, but no EBNF or Xtext grammar can be found. In our case, this is the language BibTeX. It is a well-known language to describe bibliographic data mostly used in the context of typesetting with LaTeX that is notable for its distinct syntax. In this case, we utilized the available detailed descriptions~\cite{Bibtex} to reconstruct the grammar. To validate the grammar we created, we used a number of examples of bibliographies from~\cite{Bibtex} and from our own collection to check that we covered all relevant cases.

\paragraph{Meta-model Reconstruction for DOT}
DOT is a well-known language for the specification of graph models that are input to the graph visualization and layouting tool Graphviz.
Since it is an often used language with a relatively simple, but powerful syntax, we decided to include it, even if we could not find a complete meta-model that contains both the graph structures and formatting primitives. The repository that also contains the grammar we ended up using~\cite{DotXtext}, e.g., only contains meta-models for font and graph model styles. 

Therefore, we used the Xtext grammar that parses the same language as DOT's expert-created grammar to derive a meta-model~\cite{DotXtext}.
Xtext grammars include more information than an EBNF grammar, such as information about references between concepts of the language. Thus, the fact that the DOT grammar was already formulated in Xtext allowed us to directly generate DOT's Ecore meta-model from this Xtext grammar. 
This meta-model acquisition method is an exception in this paper. Since this paper focuses on how to optimize the generated grammar, we consider this way of obtaining the meta-model acceptable for this one case.

\paragraph{Selected Cases}
As a result, we identified a sample of seven DSLs (cf.~Table~\ref{tab:analysed_DSLs}), which has a mix of different sources for meta-models and grammars. This convenience sampling consists of a mix of well-known DSLs with lesser-known, but well-developed ones. We believe this breadth of domains and language styles is broad enough to extract a generically applicable set of candidate optimization rules for \grammaroptimizer. 
%
We analyzed these selected languages in two iterations, the 1st analyzing four of them and the 2nd analyzing the remaining three.
%
%
%
In Table~\ref{tab:analysed_DSLs}, we list all seven languages, including
information about the meta-model (source and the number of classes in the meta-model) and the expert-created grammar (source and the number of grammar rules).

\lipsum[1][1-4]

\begin{table*}
\scriptsize
    \centering
    \rotatebox{-90}{
        \begin{minipage}[t]{1.5\textwidth}
            \centering
            \caption{DSLs used in this paper, the sources of the meta-model and the grammar used, as well as the size of the meta-model and grammar. The first set of DSLs was analyzed to derive necessary optimization rules, and the second set to validate the candidate optimization rules and extend them if necessary.}
            \label{tab:analysed_DSLs}
            \begin{threeparttable}
            \begin{tabular}{ll l lr r r r r}
              \toprule					
               ~ & ~ 	& \multicolumn{2}{c}{\textbf{Meta-model}} & \multicolumn{2}{c}{\textbf{Expert-created Grammar}} &	\multicolumn{3}{c}{\textbf{Generated Grammar}}\\
               \cmidrule(lr){3-4}\cmidrule(lr){5-6}\cmidrule(lr){7-9}
               Iteration & DSL  	& Source 	& Classes\tnote{1} & Source & Rules & lines & {rules} & {calls}\\
                \midrule
                ~ & ATL\tnote{2} 	& Atlantic Zoo 				& 30	&  ATL Syntax 			& 28 	& 275 & 30 	& 232 \\ 
        				~ & ~ 						& \cite{Atlanticzoo} 	& ~		&  \cite{ATLSyntax} & ~ 	& ~ 	& ~ 	& ~ \\
                ~ & BibTex 		& Grammarware			& 48 	& Self-built      & 46   & 293	& 48  & 188 \\ 
        				1st 	& ~					& \cite{BibtexMM} & ~ 	& Based on \cite{Bibtex}								& ~ 	 & ~	& ~  & ~ \\ 
                ~ & DOT 		& Generated & 19 &   Dot          & 32 & 125	& 23  & 51 \\ 
        				~ & ~ 		& ~						& ~ &   \cite{Dot}          & ~ & ~	& ~  & ~ \\ 
                ~ & SML\tnote{3} & SML repository & 48 & SML repository & 45 & 658 & 96 & 377 \\ 
        				~ & ~ & \cite{smlRepo} & ~ & \cite{smlRepo} & ~ & ~ & ~ & ~ \\ 
                \midrule
                ~ & Spectra & GitHub Repository & 54 	& GitHub Repository & 58 & 442	& 62  & 243 \\
        				~ & ~ 			& \cite{SpectraMM} & ~ 	& \cite{Spectra} & ~ & ~	& ~  & ~ \\
                2nd   & Xcore 	& Eclipse & 22	& Eclipse   & 26 & 243 & 33 & 149 \\ 
        				~   & ~ & \cite{XcoreMM} & ~	& \cite{EclipseXcore}   & ~ & ~ & ~ & ~ \\ 
                ~ & Xenia	    & Github Repository & 13 & Github Repository  & 13 & 84 & 15  & 36 \\ 
        				~ & ~	    & \cite{XeniaMM} & ~ & \cite{Xenia}  & ~ & ~ & ~  & ~ \\ 
              \bottomrule
            \end{tabular}
            \begin{tablenotes}		
                \item[1] After adaptations, containing both classes and enumerations.
                \item[2] Excluding embedded OCL rules.
                \item[3] Excluding embedded SML expressions rules.
            \end{tablenotes}
            \end{threeparttable}
        \end{minipage}}
\end{table*}

\lipsum[2][2-4]

\subsection{Exclusion of Language Parts for Low-level Expressions} \label{subsubsect:Meth:exlusionParts}
Two of the analyzed languages encompass language parts for expressions, which describe low-level concepts like binary expressions (e.g., addition).
We excluded such language parts in ATL and in SML due to several aspects.
Both languages distinguish the actual language part and the expression language part already on the meta-model level and thereby treat the expression language part differently.
The respective expression parts are similarly large than the actual languages (i.e., 56 classes for the embedded OCL part of ATL and 36 classes for the SML scenario expressions meta-model), which implies a high analysis effort. 
Finally, although having a significantly large meta-model, the embedded OCL part of ATL does not specify the expressions to a sufficient level of detail (e.g., it does not allow to specify binary expressions).
Therefore, we excluded such language parts by introducing a fake class \texttt{OCLDummy}. The details for the exclusion is described in the supplemental material~\cite{datasource2023go}\footnote{See folder ``Section\_4\_Methodology''}.


\paragraph{Exclusion from the Grammar}
In addition, we need to ensure that we can compare the language without the excluded parts to the expert-created grammar. To do so, we derive versions of the expert-created grammars in which these respective language parts are substituted by a dummy grammar rule, e.g., \texttt{OCLDummy} in the case of ATL. This dummy grammar rule is then called everywhere where a rule of the excluded language part would have been called.

\subsection{Meta-model Preparations and Generating an Xtext Grammar}\label{sec:methodology_analysis_MMPrepsAndGrammarGen}
The first step of the analysis of any of the languages is to generate an Xtext grammar based on the language's meta-model. 
This is done by using the Xtext project wizard within Eclipse.

Note that it is sometimes necessary to slightly change the meta-model to enable the generation of the Xtext grammar or to ensure that the compatibility with the expert-created grammar can be reached. 
These changes are necessary in case the meta-model is already ill-formed for EMF itself (e.g., purely descriptive Ecore files that are not intended for instantiating runtime models) or if it does not adhere to certain assumptions that Xtext makes (e.g., no bidirectional references).
The method of metamodel modification is described in detail in our  supplementary material~\cite{datasource2023go}\footnote{See directory ``Section\_4\_Methodology'.}.

In Table~\ref{tab:analysed_DSLs}, we list how many lines, rules, and calls between rules the generated grammars included for the seven languages.

\subsection{Comparing EBNF and Xtext grammars}
\label{sec:methodology:imitation}

As a prerequisite for our analysis of grammars, we present a strategy for dealing with a noteworthy aspect of our methodology: in several cases, we dealt with languages where the expert-created grammar was available in EBNF, whereas our contribution targets Xtext, which augments EBNF with additional technicalities, such as cross-references and datatypes. 
Hence, to validate whether our approach indeed produces grammars that are equivalent to expert-created ones, 
we needed a concept that allows comparing EBNF to Xtext grammars.

To this end, we introduce the concept of \emph{imitation}.
Imitation is a form of semantic equivalence of grammars that abstracts from Xtext-specific technicalities.
Specifically, we consider a set of EBNF rules 
$\{rr_{x} | 1 \leq x \leq n\} $ to be \emph{imitated} by a set of Xtext rules  $\{ro_{y} | 1 \leq y \leq m\}$ if both produce the exact same language, modulo Xtext-specific details.
Note that the cardinalities $n$ and $m$ may differ due to situations in which one expert-created rule is replaced by several optimized rules in concert, explained below.

Like semantic equivalence of context-free grammars, in general, \cite{hopcroft1969equivalence}, imitation is undecidable if two arbitrary grammars are considered.
However, in the scope of our analysis, we deal with specific cases that come from our evaluation subjects. These are generally of the following form:
1. Two syntactically identical---and thus, inherently semantically equivalent---grammar rules
2. Situations in which a larger rule from the first grammar is, in a controlled way, split up into several rules in the second grammar. For these, we consider them as equivalent based on a careful manual analysis, explained later.

\begin{lstlisting}[caption=EBNF rule \texttt{edge\_stmt} from the expert-created grammar for DOT, label=DOTRealEdgeStmt,float=tb, basicstyle=\footnotesize, breaklines=true]
edge_stmt  :  (node_id | subgraph) edgeRHS [ attr_list ]
\end{lstlisting}

\begin{lstlisting}[caption=Xtext rules \texttt{EdgeStmtNode} and \texttt{EdgeStmtSubgraph} from the optimized generated grammar, label=DOTOptimizedEdgeStmt,float=tb, basicstyle=\footnotesize, breaklines=true]
EdgeStmtNode returns EdgeStmtNode:
    {EdgeStmtNode}
         node=NodeId
          (edgeRHS+=EdgeRhs)+  
          (attrLists+=AttrList)*  
    ;

EdgeStmtSubgraph returns EdgeStmtSubgraph:
    {EdgeStmtSubgraph}
         subgraph=Subgraph
          (edgeRHS+=EdgeRhs)+  
          (attrLists+=AttrList)*  
    ;
\end{lstlisting}

For example, the rule \texttt{edge\_stmt} shown in Listing~\ref{DOTRealEdgeStmt} is imitated by the combination of the rules \texttt{EdgeStmtNode} and \texttt{EdgeStmtSubgraph} shown in Listing~\ref{DOTOptimizedEdgeStmt}.
Merging the Xtext rules to form one rule, like the EBNF counterpart, was not possible in this case, due to the necessity of specifying a distinct return type in Xtext, which is not required in EBNF.
In addition, the Xtext rules contain Xtext-specific information for dealing with references and attribute types, which is not present in the EBNF rule.

\subsection{Analysis of Grammars}
\label{sec:methodology_analysis}
We performed the analysis of existing languages in two iterations. The first iteration was purely exploratory. Here we analyzed four of the languages with the aim of finding as many candidate grammar optimization rules as possible. In the second iteration, we selected three additional languages to validate the candidate rules collected from the first iteration, add new rules if necessary, and generalise the existing rules when applicable. 

Our general approach was similar in both iterations. 
Once we had generated a grammar for a meta-model, we created a mapping between that generated grammar and the expert-created grammar of the language. The goal of this mapping was to identify which grammar rules in the generated grammar correspond to which grammar rules in the expert-created grammar.
Note that a grammar rule in the generated grammar may be mapped to multiple grammar rules in the expert-created grammar and vice versa.
From there, we inspected the generated and expert-created grammars to identify how they differed and which changes would be required to adjust the generated grammar so that it produces the same language as the expert-created grammar, i.e., \emph{imitates} the expert-created grammar rules.
We documented these changes per language and summarized them as optimization rule candidates in a spreadsheet.

For example, the expert-created grammar rule \texttt{node\_stmt} in DOT (see Listing~\ref{lst:dot-node_stmt-ebnf}) maps to the generated grammar rule \texttt{NodeStmt} in Listing~\ref{lst:dot-node_stmt-generated-xtext}.
Multiple changes are necessary to adjust the generated Xtext grammar rule:
\begin{itemize}
    \item Remove all the braces in the grammar rule \texttt{NodeStmt}.
    \item Remove all the keywords in the grammar rule \texttt{NodeStmt}.
    \item Remove the optionality from all the attributes in the grammar rule \texttt{NodeStmt}.
    \item Change the multiplicity of the attribute \texttt{attrLists} from 1..* to 0..*.
\end{itemize}

Note that in most cases the expert-created grammar was written in EBNF instead of Xtext. For example, the \texttt{returns} statement in line 1 of Listing~\ref{lst:dot-node_stmt-generated-xtext} is required for parsing in Xtext. We took that into account when comparing both grammars.

\subsubsection{First Iteration: Identify Optimization Rules}
\label{sec:methodology:analysis:first_iteration}
The analysis of the grammars of the four selected DSLs in the first iteration had two concrete purposes:
\begin{itemize}
    \item identify the differences between the expert-created grammar and generated grammar of the language;
    \item derive grammar optimization rules that can be applied to change the generated grammar so that the optimized grammar parses the same language as the expert-created grammar.
\end{itemize}
Please note that it is not our aim to ensure that the optimized grammar itself is identical to the expert-created grammar. 
Instead, our goal is that the optimized grammar is an \emph{imitation} of the expert-created grammar 
and therefore is able to parse the same language as the original, usually hand-crafted grammar of the DSL.
Each language was assigned to one author who performed the analysis. 

\begin{lstlisting}[caption={Non-terminal \texttt{node\_stmt} in the expert-created grammar of DOT, in EBNF}, label={lst:dot-node_stmt-ebnf},float=tb, basicstyle=\footnotesize, breaklines=true]
node_stmt : node_id [ attr_list ]
\end{lstlisting}

\begin{lstlisting}[caption={Grammar rule \texttt{NodeStmt} in the generated grammar of DOT, in Xtext}, label={lst:dot-node_stmt-generated-xtext},float=tb, basicstyle=\footnotesize, breaklines=true]
NodeStmt returns NodeStmt:
        {NodeStmt}
        'NodeStmt'
        '{'
                ('node' node=NodeId)?
                ('attrLists' '{' attrLists+=AttrList ( "," attrLists+=AttrList)* '}' )?
        '}';
\end{lstlisting}

\begin{lstlisting}[caption={Grammar rule \texttt{NodeStmt} in the optimized grammar of DOT, in Xtext}, label={lst:dot-node_stmt-optimized-xtext},float=tb, basicstyle=\footnotesize, breaklines=true]
NodeStmt returns NodeStmt:
    {NodeStmt}

         node=NodeId
          (attrLists+=AttrList)*  
    ;
\end{lstlisting}

As a result of the analysis, we obtained an initial set of grammar optimization rules, which contained a total of 58 candidate optimization rules. 
Table~\ref{tab:rules-and-iterations} summarizes in the second column the number of identified rule candidates and in the second row the number for the first iteration.
Since the initial set of grammar optimization rules was a result of an analysis done by multiple authors, it included rules that were partially overlapping and rules that turned out to only affect the grammar's formatting, but not the language specified by the grammar. Thus, we filtered rules that belong to the latter case. For rule candidates that overlapped with each other, we selected a subset of the rules as a basis for the next step. This filtering led to a selection of 46 optimization rules (cf.~third column in Table~\ref{tab:rules-and-iterations}).

\begin{table}
\scriptsize
\centering
\caption{Summary of identified rules their rule variants and their sources}
\label{tab:rules-and-iterations}
\begin{tabular}{@{}lp{1.5cm}p{2cm}p{1.2cm}@{}}
\toprule
\textbf{Iteration} & \textbf{Rule \newline Candidates} & \textbf{Selected Rules}  & \textbf{Rule \newline Variants} \\
\midrule
Iteration 1 & 58 & 46 & 57 \\
Iteration 2 & 10 & 10 & 10 \\
\toprule
Intermediate sum & 68 & 56 & 67\\
\bottomrule
Evaluation & 4 & 4 & 4 \\
\midrule
Overall sum & 72 & 60 & 71  \\
\bottomrule
\end{tabular}
\end{table}

We processed these 46 selected optimization rules to identify required \emph{rule variants} that could be implemented directly by means of one Java class each, which we describe more technically as part of our design and implementation elaboration in Section~\ref{sec:solution_ruleDesign}. 
For identifying the rule variants, we focused on the following aspects:

\begin{itemize}
    \item \textbf{Specification of scope.} Small changes in the meta-model might lead to a different order of the lines in the generated grammar rules or even a different order of the grammar rules. Therefore, the first step was to define a suitable concept to identify the parts of the generated grammar that can function as the \emph{scope} of an optimization rule, i.e., where it applies. We identified different suitable scopes, e.g., single lines only, specific attributes, specific grammar rules, or even the whole grammar. Initially, we identified separate rule variants for each scope. Note that this also increased the number of rule variants, as for some rule candidates multiple scopes are possible.
    \item \textbf{Allowing multiple scopes.} In many cases, selecting only one specific scope for a rule is too limiting. In the example above (Listing~\ref{lst:dot-node_stmt-generated-xtext}), pairs of braces in different scopes are removed: in the scope of the attribute \texttt{attrLists} in line 6 and in the scope of the containing grammar rule in lines 4 and 7. This illustrates that changes might be applied at multiple places in the grammar at once. 
    When formulating rule variants, we analyzed the rule candidates for their potential to be applied in different scopes. When suitable, we made the scope configurable. This means that only one optimization rule variant is necessary for both cases in the example. Depending on the provided parameters, it will either replace the braces for the rule or for specific attributes.
    \item \textbf{Composite optimization rules.} We decided to avoid optimization rule variants that can be replaced or composed out of other rule variants, especially when such compositions were only motivated by very few cases. However, such rules might be added again later if it turns out they are needed more often. 
\end{itemize}

While we identified exactly one rule variant for most of the selected optimization rules, we added more than one rule variant for several of the rules. We did this when slight variations of the results were required. 
For example, we split up the optimization rule \texttt{SubstituteBrace} into the three variants, i.e., 1) \texttt{ChangeBracesToParentheses}, 2) \texttt{ChangeBracesToSquare}, and 3) \texttt{ChangeBracesToAngle}.
Note that this split-up into variants is a design choice and not an inherent property of the optimization rule, as, e.g., the type of target bracket could be seen as nothing more than a parameter of the rule. 
As a result, we settled on 57 rule variants for the 46 identified rules (cf.~fourth column of second row in Table~\ref{tab:rules-and-iterations}).


\subsubsection{Second iteration: Validate Optimization Rules}
\label{sec:methodology:analysis:second_iteration}
The last step left us with 46 selected optimization rules from the first iteration (cf.~second row in Table~\ref{tab:rules-and-iterations}). 
We developed a preliminary implementation of \grammaroptimizer by implementing the 57 rules variants belonging to these 46 optimization rules (we will describe the implementation in the \emph{Solution} section). 
To validate this set of optimization rules, we performed a second iteration. In the second iteration, we selected the three DSLs Spectra, Xenia, and Xcore. As in the first iteration, we generated a grammar from the meta-model, analyzed the differences between the generated grammar and the expert-created grammar, and identified optimization rules that need to be applied to the generated grammar to accommodate these differences. 
In contrast to the first iteration, we aimed at utilizing as many existing optimization rules as possible and only added new rule candidates when necessary.

We configured the preliminary \grammaroptimizer for the new languages by specifying which optimization rules to apply on the generated grammar.
The execution results showed that the existing optimization rules were sufficient to change the generated grammar of Xenia to imitate the expert-created grammar used as the ground truth. However, we could not fully transform the generated grammar of Xcore and Spectra with the preliminary set of 46 optimization rules from the first iteration. For example, Listing~\ref{lst:xcore-xoperation-two-attributes-in-generated-grammar} shows two attributes \texttt{unordered} and \texttt{unique} in the grammar rule \texttt{XOperation} in the generated grammar for Xcore. However, in the expert-created grammar, the rule portions for the two attributes each refer to the other attribute in a way that allows using the keywords in several possible orders, as shown in Listing~\ref{lst:xcore-xoperation-two-attributes-in-original-grammar}. This optimization could not be performed with the optimization rules from the first iteration.

\begin{lstlisting}[caption={Two attributes in the grammar rule \texttt{XOperation} in the generated grammar of Xcore}, label={lst:xcore-xoperation-two-attributes-in-generated-grammar},float=tb, basicstyle=\footnotesize, breaklines=true]
...
        (unordered?='unordered')?
        (unique?='unique')?
...
\end{lstlisting}

\begin{lstlisting}[caption={Two attributes in the grammar rule \texttt{XOperation} in the expert-created grammar of Xcore}, label={lst:xcore-xoperation-two-attributes-in-original-grammar},float=tb, basicstyle=\footnotesize, breaklines=true]
...
          unordered?='unordered' unique?='unique'? |
          unique?='unique' unordered?='unordered'?
...
\end{lstlisting}

Based on the non-optimized parts of the grammars of Xcore and Spectra, we identified another ten optimization rules for the \grammaroptimizer. 
These ten newly identified optimization rules optimize all the non-optimized parts of the grammar of Xcore, including, e.g., optimizing the grammar in Listing~\ref{lst:xcore-xoperation-two-attributes-in-generated-grammar} to Listing~\ref{lst:xcore-xoperation-two-attributes-in-original-grammar}. These new optimization rules also optimize part of the non-optimized parts of the grammar of Spectra. We will interpret the remaining non-optimized parts in the \emph{Evaluation} section.
In the end, after two iterations, we identified a total of 56 optimization rules (which will be implemented by a total of 67 rule variants) (cf.~fourth row in Table~\ref{tab:rules-and-iterations}).

\section{Identified Optimization Rules}
\label{sec:identified-optimization-rules}

In total, we identified 56 distinct optimization rules for the grammar optimization after the 2nd iteration, which we further refined into 67 rule variants (cf.~fourth row in Table~\ref{tab:rules-and-iterations}). 
Note that 4 additional rules were identified during the evaluation (this will be interpreted in the \emph{Evaluation} section), increasing the final number of identified optimization rules to 60 (cf.~bottom row in Table~\ref{tab:rules-and-iterations}) and the final number of rule variants to 71.

Table~\ref*{tab:optimizer-rules} shows some examples of the optimization rules. The rules we implemented can be categorized by the primitives they manipulate: 
grammar rules,
attributes
keywords,
braces,
multiplicities,
optionality (a special form of multiplicities),
grammar rule calls,
import statements,
symbols,
primitive types, and
lines.
They either `add' things (e.g., \emph{AddKeywordToRule}), `remove' things (e.g., \emph{RemoveOptionality}), or `change' things (e.g., \emph{ChangeCalledRule}). 
All optimization rules ensure that the resulting changed grammar is still valid and syntactically correct Xtext. 

Most optimization rules are `scoped' which means that they only apply to a specific grammar rule or attribute. In other cases, the scope is configurable, depending on the parameters of the optimization rule. For instance, the \emph{RenameKeyword} rule takes a grammar rule and an attribute as a parameter. If both are set, the scope is the given attribute in the given rule. If no attribute is set, the scope is the given grammar rule. If none of the parameters is set, the scope is the entire grammar (``Global''). All occurrences of the given keyword are then renamed inside the respective scope.

Changes to optionality are used when the generated grammar defines an element as mandatory, but the element should be optional according to the expert-created grammar. This can apply to symbols (such as commas), attributes, or keywords. Additionally, when all attributes in a grammar rule are optional, we have an optimization rule that makes the container braces and all attributes between them optional. This optimization rule allows the user of the language to enter only the grammar rule name and nothing else, e.g., ``\texttt{EAPackage DataTypes;}''.

Likewise, \grammaroptimizer contains rules to manipulate the multiplicities in the generated grammars. The meta-models and the expert-created grammars we used as inputs do not always agree about the multiplicity of elements. We provide optimization rules that can address this
within the constraints allowed by EMF and Xtext.


\begin{table}
\scriptsize
\centering
\caption{Excerpt of implemented grammar optimization rules. A configurable scope (``Config.'') means that, depending on provided parameters, the rule either applies globally to a specific grammar rule or to a specific attribute.}
\label{tab:optimizer-rules}
\begin{tabular}{@{}llll@{}}
\toprule
\textbf{Subject} & \textbf{Op.} & \textbf{Rule} & \textbf{Scope} \\
\midrule
Keyword & Add    & \emph{AddKeywordToAttr}        & Attribute \\
        &        & \emph{AddKeywordToRule}        & Rule \\
        &        & \emph{AddKeywordToLine}        & Line \\
        & Change & \emph{RenameKeyword}           & Config. \\
        &        & \emph{AddAlternativeKeyword}   & Rule \\
\midrule
Rule    & Remove & \emph{RemoveRule}              & Global \\
        & Change & \emph{RenameRule}              & Rule \\
        &        & \emph{AddSymbolToRule}         & Rule \\
\midrule
Optionality & Add & \emph{AddOptionalityToAttr}   & Attribute \\
            &     & \emph{AddOptionalityToKeyword} & Config. \\
\midrule
Import & Add    & \emph{AddImport}               & Global \\
       & Remove & \emph{RemoveImport}            & Global \\
\midrule
Brace  & Change & \emph{ChangeBracesToSquare}   & Attribute \\
       & Remove & \emph{RemoveBraces}           & Config.\\
\bottomrule
\end{tabular}
\end{table}

For the example in Listing~\ref{lst:dot-node_stmt-generated-xtext}, this means that the necessary changes to reach the same language defined in Listing~\ref{lst:dot-node_stmt-ebnf} can be implemented using the following \grammaroptimizer rules:
\begin{itemize}
    \item \emph{RemoveBraces} is applied to the grammar rule \texttt{NodeStmt} and all of its attributes. This removes all the curly braces (`\{' and `\}' in lines 4, 6, and 7) within the grammar rule.
    \item \emph{RemoveKeyword} is applied to the grammar rule \texttt{NodeStmt} and all of its attributes. This removes the keywords \texttt{`NodeStmt'}, \texttt{`node'} and \texttt{`attrLists'} (lines 3, 5, and 6) from this grammar rule.
    \item \emph{RemoveOptionality} is applied to both attributes. This removes the question marks (`?') in lines 5 and 6. 
    \item \emph{convert1toStarToStar} is applied to the attribute \texttt{attrLists}. This rule changes line 6. Before this change, this line is ``\texttt{attrLists+=AttrList ( "," attrLists+=AttrList)*}'' (the braces, keyword `\texttt{attrLists}' and the optionality `?' have been removed by previous optimization rules). After this change, it becomes \texttt{(attrLists+=AttrList)*}.
    Note that the DOT grammar is specified using a syntax that is slightly different from standard EBNF. In that syntax, square brackets ([ and ]) enclose optional items~\cite{Dot}.
\end{itemize}
Note that line 2 in Listing~\ref{lst:dot-node_stmt-generated-xtext} has no effect on the syntax of the grammar but is required by and specific to Xtext, so that we do not adapt such constructs. 
After the above steps, the grammar rule \texttt{NodeStmt} is adapted from Listing~\ref{lst:dot-node_stmt-generated-xtext} to Listing~\ref{lst:dot-node_stmt-optimized-xtext}.

\section{Solution: Design and Implementation}\label{sec:solution}
The \grammaroptimizer is a Java library that offers a simple API to configure optimization rule applications and execute them on Xtext grammars. 
The language engineer can use that API to create a small program that executes \grammaroptimizer, which in turn will produce the optimized grammar.


\subsection{Grammar Representation}\label{sec:solution_grammarRepresentation}
\begin{figure}[tb]
  \centering
  \includegraphics[width=\linewidth]{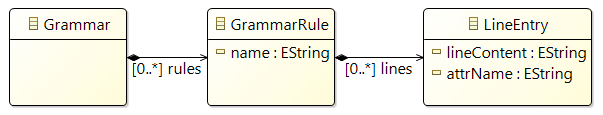}
  \caption{The class design for representing grammar rules.}
  \label{fig:class-diagram-grammar}
\end{figure}

We designed \grammaroptimizer to parse an Xtext grammar into an internal data structure which is then modified and written out again. 
%
This internal representation of the grammar follows the structure depicted in Figure~\ref{fig:class-diagram-grammar}. A \texttt{Grammar} contains a number of \texttt{GrammarRule}s that can be identified by their names. In turn, a \texttt{GrammarRule} consists of a sorted list of \texttt{LineEntry}s with their textual \texttt{lineContent} and an optional \texttt{attrName} that contains the name of the attribute defined in the line. Note that we utilize the fact that Xtext generates a new line for each attribute.

\subsection{Optimization Rule Design}\label{sec:solution_ruleDesign}
\begin{figure}[tb]
  \centering
  \includegraphics[width=\linewidth]{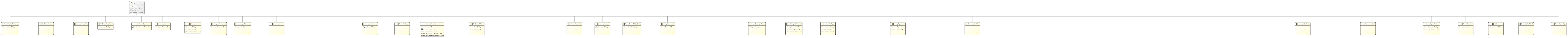}
  \caption{Excerpt of the class diagram for optimization rules.}
  \label{fig:class-diagram-optimization-rule}
\end{figure}

Internally, all optimization rules derive from the abstract class \texttt{OptimizationRule} as shown in Figure~\ref{fig:class-diagram-optimization-rule}. Derived classes overwrite the \texttt{apply()}-method to perform the specific text modifications for this rule. By doing so, the specific rule can access the necessary information through the class members: \texttt{grammar} (i.e., the entire grammar representation as explained in Section~\ref{sec:solution_grammarRepresentation} and depicted in Figure~\ref{fig:class-diagram-grammar}), \texttt{grammarRuleName} (i.e., the name of the specified grammar rule that a user wants to optimize exclusively), and \texttt{attrName} (i.e., the name of an attribute that a user wants to optimize exclusively). Sub-classes can also add additional members if necessary.
This architecture makes the \grammaroptimizer extensible, as new optimization rules can easily be defined in the future. 

We built the optimization rules in a model-based manner by first creating the meta-model shown in Figure~\ref{fig:class-diagram-optimization-rule} and then using EMF to automatically generate the class bodies of the optimization rules. This way we only needed to overwrite the \texttt{apply()}-method for the concrete rules. 
Internally, the \texttt{apply()}-methods of our optimization rules are implemented using regular expressions.
Each optimization rule takes a number of parameters, e.g., the name of the grammar rule to work on or an attribute name to identify the line to work on. In addition, some optimization rules take a list of exceptions to the scope. For example, the optimization rule to remove braces can be applied to a global scope (i.e., all grammar rules) while excluding a list of specific grammar rules from the processing. This allows to configure optimization rule applications in a more efficient way.
We implemented all identified optimization rules.\footnote{See folder `1\_Source\_Code/org.bumble.xtext.grammaroptimizer' in our supplemental material~\cite{datasource2023go}, which contains the `optimizationrule' project with the full implementation.}
For testing, we built a comprehensive test suite, based on the optimized grammars considered in our design methodology.
We created one test case per scenario, to ensure that the grammar produced by our implementation after applying a full given configuration to an Xtext-generated grammar exactly matches an expected ground-truth grammar, for which we previously manually established that it agrees (in the sense of \textit{imitation}) with an expert-created one).

\subsection{Configuration}

The language engineer has to configure what optimization rules the \grammaroptimizer should apply and how. This is supported by the API offered by \grammaroptimizer. 
Listing~\ref{lst:dot-config-example-rule} shows an example of how to configure the optimization rule applications in a method \texttt{executeOptimization()}, where the configuration revisits the DOT grammar optimization example transforming Listing~\ref{lst:dot-node_stmt-generated-xtext} into Listing~\ref{lst:dot-node_stmt-ebnf}. The lines 3 to 6 configure optimization rule applications. For example, line 3 removes all curly braces in the grammar rule \emph{NodeStmt}. 
The value of the first parameter is set to ``NodeStmt'', which means that the operation of removing curly braces will occur in the grammar rule \emph{NodeStmt}. If this first parameter is set to ``null'', the operation would be executed for all grammar rules in the grammar.
The second parameter is used to indicate the target attribute. Since it is set to ``null'', all lines in the targeted grammar rule will be affected. However, if the parameter is set to a name of an attribute, only curly braces in the line containing that attribute will be removed.
Finally, the third parameter can be used to indicate names of attributes for which the braces should not be removed. This can be used in case the second parameter is set to ``null''. 


\begin{lstlisting}[caption=Excerpt of the configuration of \grammaroptimizer{} for the QVTo 1.0 language.), label=lst:dot-config-example-rule,language=Java,float=tb, basicstyle=\footnotesize, breaklines=true]
public static boolean executeOptimization(GrammarOptimizer go) {
    ...
    go.removeBraces("NodeStmt", null, null);
    go.removeKeyword("NodeStmt", null, null, null);
    go.removeOptionality("NodeStmt", null);
    go.convert1toStarToStar("NodeStmt", "attrLists");
    ...
}
\end{lstlisting}

Similarly, the optimization rule application in line 4 is used to remove all keywords in the grammar rule \emph{NodeStmt}. 
Again, the second parameter can be used to specify which lines should be affected using an attribute. 
The third parameter is used to indicate the target keyword. 
Since it is set to ``null'', all keywords in the targeted lines will be removed. However, if the keyword is set, only that keyword will be removed.
The last parameter can be used to indicate names of attributes for which the keyword should not be removed. This can be used in case the second parameter is set to “null”.

Line 5 is used to remove the optionality from all lines in the grammar rule \emph{NodeStmt}.
If the second parameter gets an argument that carries the name of an attribute, the optionality is removed exclusively from the grammar line specifying the syntax for this attribute.

Finally, line 6 changes the multiplicity of the attribute \texttt{attrLists} in the grammar rule \texttt{NodeStmt} from 1..* to 0..*.
If the second parameter would get the argument ``null'', this adaptation would have been executed to all lines representing the respective attributes.

\subsection{Execution}
Once the language engineer has configured \grammaroptimizer, they can invoke the tool using \texttt{GrammarOptimizerRunner} on the command line and providing the paths to the input and output grammars there.
Alternatively, instead of invoking \grammaroptimizer via the command line and modifying \texttt{executeOptimization()}, it is also possible to use JUnit test cases to access the API and optimize grammars in known locations. This is the approach we have followed in order to generate the results presented in this paper.

\begin{figure*}[ht]
  \centering
  \includegraphics[page=1, width=\linewidth]{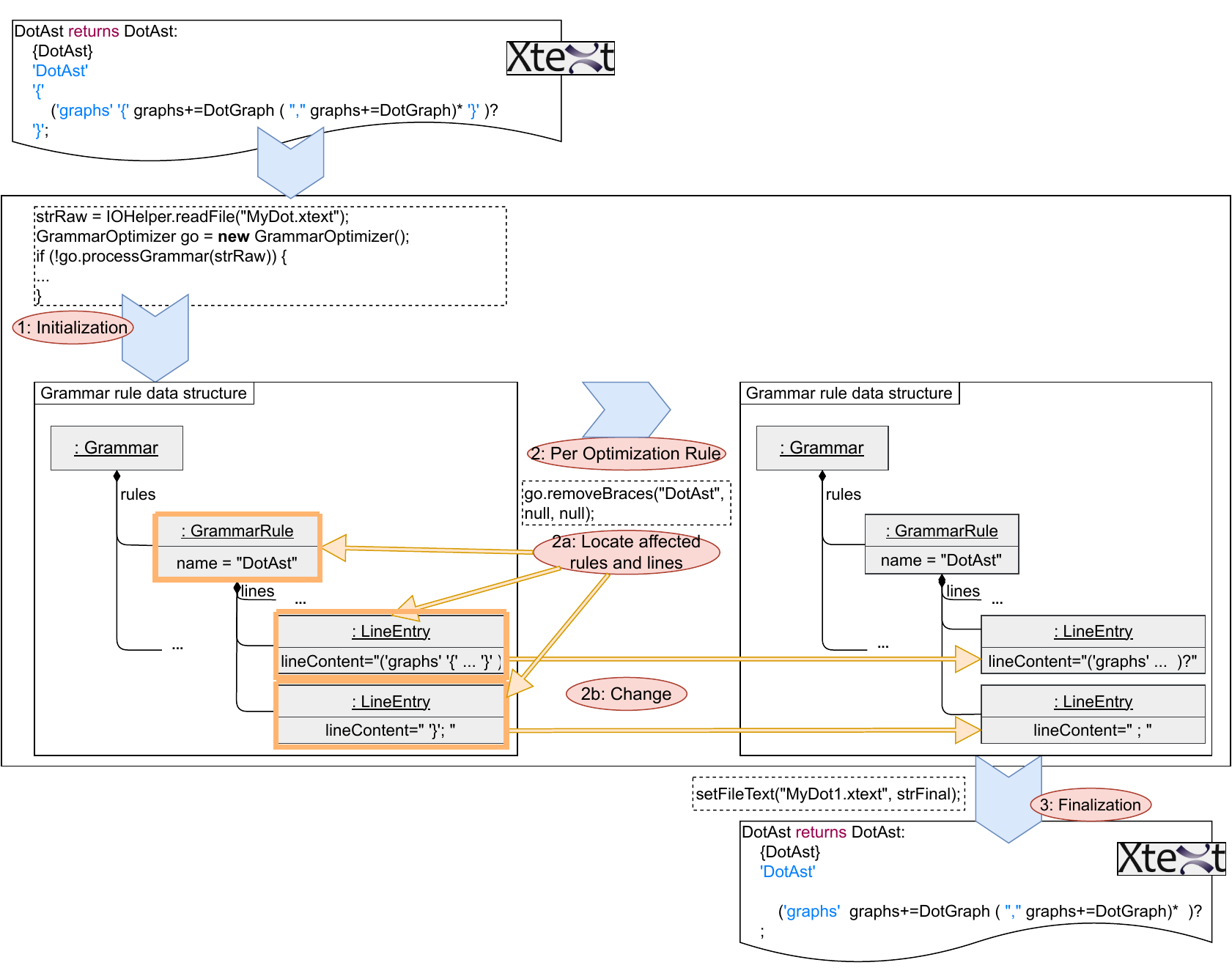}
  \caption{Exemplary Interplay of the Building Blocks of the \grammaroptimizer}
  \label{fig:GrammarOptimizer-Internal-processing-flow}
\end{figure*}

Figure~\ref{fig:GrammarOptimizer-Internal-processing-flow} uses the first optimization operation from  Listing~\ref{lst:dot-config-example-rule} removing curly braces as an example to depict how \grammaroptimizer works internally when optimizing grammars.
The top of the figure shows an example input, which is the grammar rule \texttt{NodeStmt} generated from the meta-model of DOT (cf. Listing~\ref{lst:dot-node_stmt-generated-xtext}). In the lower right corner, the resulting optimized Xtext grammar rule is illustrated. In both illustrated grammar rule excerpt, blue fonts are the keywords and symbols (braces and commas).

In \textbf{Step 1 (initialization)}, \grammaroptimizer builds a data structure out of the grammar initially generated by Xtext.
That is, it builds a \texttt{:Grammar} object containing multiple \texttt{:GrammarRule} objects, with each of them containing several \texttt{:LineEntry} objects in an ordered list.
For example, the \texttt{:Grammar} object contains a \texttt{:GrammarRule} object with the name \texttt{"NodeStmt"}.
This \texttt{:GrammarRule} object contains seven \texttt{:LineEntry} objects, which represent the seven lines of the grammar rule in Listing~\ref{lst:dot-node_stmt-generated-xtext}. Three of these \texttt{:LineEntry} objects contain at least one curly brace (\texttt{`` `\{' ''} or \texttt{`` `\}' ''}).
These lines are explicitly represented in order to later map relevant optimization rules to them.
Figure \ref{fig:GrammarOptimizer-Internal-processing-flow} shows an excerpt of the object structure created for the example with the three line objects for the \texttt{NodeStmt} rule.

In \textbf{Step 2 (per Optimization Rule)} each optimization rule application is processed by executing the \texttt{apply()}-method.
For our example, the optimization rule \texttt{removeBraces} is applied via the \grammaroptimizer API as configured in line 3 of Listing~\ref{lst:dot-config-example-rule}.

In \textbf{Step 2a (localization of affected grammar rules and lines)}, 
the grammar rule and lines that need to be changed are located, based on the configuration of the optimization rule application.
In the case of our example, the grammar rule \texttt{NodeStmt} (cf.~line 1 in Listing~\ref{lst:dot-node_stmt-generated-xtext}) is identified. 
Then, all lines of that grammar rule are identified that include a curly brace. For example, the lines represented by \texttt{:LineEntry} objects as shown in Figure \ref{fig:GrammarOptimizer-Internal-processing-flow} are identified.

In \textbf{Step 2b (change)}, the code uses regular expressions for character-level matching and searching. If it finds curly braces surrounded by single quotes (i.e., \texttt{`` `\{' ''} and \texttt{`` `\}' ''}), it removes them.

Finally, in \textbf{Step 3 (finalization)}, the \grammaroptimizer writes the complete data structure containing the optimized grammar rules to a new file by means of the call \textsf{setFileText(...)}.

After the execution of these steps, the optimized versions of the grammar is ready for use.
The typical next step is to re-generate the parser, textual editor and other artifacts for the grammar via Xtext.
We recommend that the language engineer should systematically test the resulting grammar to check whether it matches their expectations, based on the generated artifacts and a test suite with diverse  language instances.
After evolution steps, previously developed tests can act as regression tests.

\subsection{Post-Processing vs.\ Changing Grammar Generation}
\grammaroptimizer is designed to modify grammars that Xtext generated out of meta-models.
An alternative to this post-processing approach is to directly modify the Xtext grammar generator as, e.g., in XMLText \cite{neubauer2015xmltext,neubauer2017xmltextToolPaper}.
However, we deliberately chose a post-processing approach, because the application of conventional regular expressions enables the transferability to other recent language development frameworks like Langium~\cite{langium} or textX~\cite{dejanovic2017textx}, if they support the grammar generation from a meta-model in a future point in time. 
While the optimization rules implemented in grammar optimizer are currently tailored to the structure of Xtext grammars, \grammaroptimizer does not technically depend on Xtext and the rules could easily be adapted to a different grammar language.
Furthermore, as the implementation of an Xtext grammar generator necessarily depends on many version-specific internal aspects of Xtext, the post-processing approach using regular expressions is considerably more maintainable.

\subsection{Limitations and Caveats}
Our solution has the following limitations and caveats.

First, \grammaroptimizer works on the generated grammar, which is generated from a meta-model. This means that the meta-model must contain all the concepts that the expert-created grammar has. Otherwise, the generated grammar will lack the necessary classes or attributes. This would result in the inability to imitate the expert-created grammar. 
A feasible solution would be to expand the working scope of the \grammaroptimizer, e.g., to provide a feature to detect whether all the concepts contained in the expert-created grammar corresponding elements can be found in the meta-model. 
However, we decided against implementing such a feature for now, as we see the main use case of the \grammaroptimizer not in imitating existing grammars, but in building and maintaining new DSLs.

Second, we were not able to completely imitate one of the seven languages.
In order to do so, we would have had to provide an optimization rule that would require the \grammaroptimizer user to input a multitude of parameter options.
This would have strongly increased the effort and reduced the usability to use this one optimization rule, and the rule is only required for this one language.
Thus, we argue that a manual post-adaptation is more meaningful for this one case.
However, the inherent extensibility of the \grammaroptimizer allows to add such an optimization rule if desired.
We describe the issue in a more detailed manner in Section~\ref{subsubsection:eval:imtatio:results}, which summarizes the evaluation results for the grammar adaptions of the seven analyzed languages.

Third, our solution is non-commutative, that is, applying the same rules with the same parametrization, but in a different order might lead to different results. For example, if \texttt{ChangeBracesToAngle} and \texttt{ChangeBracesToSquare} are successively applied to the same grammar rule, the outcome is ``last write wins'', i.e., the rule obtains square braces.  Users should be aware of this property to ensure that the achieved outcome is consistent with their intended outcome.

Fourth, our solution does not strive to maintain backwards comparability to previous grammar versions---in general, after rule applications, instances of the previous, un-optimized grammar can no longer be parsed. 
This lack of backwards compatibility is generally desirable, as the alternative would be support for a mixing of old and new grammar elements (e.g., changed keywords and parantheses styles) in the same instance, which would generally be confusing to the user, and lead to issues with parsing and other tool support.
However, to reduce manual effort in cases where legacy grammar instances exist, automated co-evolution of grammar instances after grammar changes is generally possible and leads to  a promising future work direction (discussed in Section~\ref{sec:discussion_coevolution_instances}).

\section{Evaluation}\label{sec:eval} 
In this evaluation, we focus on two research questions:
\begin{itemize}
\item \emph{RQ1: Can our solution be used to adapt generated grammars so that they produce the same language as available expert-created grammars?} \\
The goal of this question is to validate the claim that our approach can automatically perform the changes that an expert would need to do manually. To this end, we consider languages for which an expert-created grammar exists, and validate the capability of our approach to re-create an equivalent grammar. 

\item \emph{RQ2: Can our solution support the co-evolution of generated grammars when the meta-model evolves? } \\
Our original motivation for the work was to enable evolution and rapid prototyping for textual languages build with a meta-model. The aim here is to evaluate whether our approach is suitable for supporting these evolution scenarios.
\end{itemize}

In the following, we address both questions. Our supplemental material~\cite{datasource2023go} contains the source code of the implementation as well as all experiments.

\subsection{Grammar Adaptation (RQ1)}
\label{sec:eval-grammar-adaptation}
To address the first question, we evaluate the \grammaroptimizer by transforming the generated grammars of the seven DSLs, so that they parse the same syntax as the expert-created grammars. 

\subsubsection{Cases}
Our goal is to evaluate whether the \grammaroptimizer can be used to optimize the generated grammars so that their rules imitate the rules of the expert-created grammars.
%
We reused the meta-model adaptations and generated grammars from Section~\ref{sec:methodology_analysis_MMPrepsAndGrammarGen}. 
%
Furthermore, we continued working with the versions of ATL and SML in which parts of their languages were excluded as described in Section~\ref{subsubsect:Meth:exlusionParts}.

\subsubsection{Method}
For each DSL, we wrote a configuration for the final version of \grammaroptimizer which was the result of the work described in Section~\ref{sec:methodology,sec:identified-optimization-rules,sec:solution}. The goal was to transform the generated grammar so as to `imitate' as many grammar rules as possible from the expert-created grammar of the DSL.
Note that this was an iterative process in which we incrementally added new optimization rule applications to the \grammaroptimizer's configuration, using the expert-created grammar as a ground truth and using our notion of `imitation' (cf.~Section~\ref{sec:methodology:imitation}) as the gold standard. Essentially, we updated the \grammaroptimizer configuration and then ran the tool before analysing the optimized grammar for imitation of the original.
We repeated the process and adjusted the \grammaroptimizer configuration until the test grammar's rules `imitated' the expert-created grammar. 
Note that in the case of \textit{Spectra}, we did not reach that point. We explain this in more detail in Section~\ref{subsubsection:eval:imtatio:results}.
For all experiments, we used the set of 56 optimization rules that were identified after the two iterations described in Section~\ref{sec:methodology} and as summarized in Section~\ref{sec:identified-optimization-rules}.

To verify whether the optimized grammar imitates the expert-created grammar, we adopted a manual verification method, in which we systematically compared the grammar rules in the optimized grammar with the grammar rules in the expert-created grammar. An expert-created grammar is imitated by an optimized grammar if every grammar rule in it is imitated by one (or several) grammar rules from the optimized grammar.
The procedure and results of this step are documented in our supplementary materials \cite{datasource2023go}.\footnote{
See directory `2\_Supplemental\_Material/Section\_7\_Evaluation`.}

\subsubsection{Metrics}\label{sec:eval-grammar-adaptation_metrics}
To evaluate the optimization results of the \grammaroptimizer on the case DSLs, we assessed the following metrics.

\begin{description}
    \item[$\#$\textit{GORA}] Number of \grammaroptimizer rule applications used for the configuration. 
    \item[Grammar rules] The changes in grammar rules performed by the \grammaroptimizer when adapting the generated grammar towards the expert-created grammar. We measure these changes in terms of 
		\begin{itemize}			
			\item mod: Number of modified grammar rules 
			\item add: Number of added grammar rules 
			\item del: Number of deleted grammar rules 
		\end{itemize}
    \item[Grammar lines] The changes in the lines of the grammar performed by the \grammaroptimizer when adapting the generated grammar towards the expert-created grammar. We measure these changes in terms of
		\begin{itemize}			
			\item mod: Number of modified lines 
			\item add: Number of added lines 
			\item del: Number of deleted lines 
		\end{itemize}	

    \item[Optimized grammar] Metrics about the resulting optimized grammar. We assess
		\begin{itemize}		
			\item lines: Number of overall lines 
			\item rules: Number of grammar rules 
			\item calls: Number of calls between grammar rules 
		\end{itemize}		
	\item[$\#$\textit{iGR}] Number of grammar rules in the expert-created grammar that were successfully \emph{imitated} by the optimized grammar. 

    \item[$\#$\textit{niGR}] Number of grammar rules in the expert-created grammar that were not \emph{imitated} by the optimized grammar. 
\end{description}


\subsubsection{Results} \label{subsubsection:eval:imtatio:results}

\lipsum[1][1-4]
\begin{table*}
\scriptsize
    \centering
    \rotatebox{-90}{
        \begin{minipage}[t]{1.5\textwidth}
            \centering
            \caption{Result of applying the GrammarOptimizer to different DSLs (RQ1)}
            \label{tab:result-of-generalizing-to-different-DSLs}
            \begin{threeparttable}
            \begin{tabular}{@{}llrrrrrrrrrrrr@{}}
                \toprule
        					~&\textbf{Optimization} 	&~    & \multicolumn{3}{c}{\textbf{Grammar Rules}}	 & \multicolumn{3}{c}{\textbf{Lines in Grammar}}	& \multicolumn{3}{c}{\textbf{Optimized Grammar}}  & ~     &~	\\
        					\textbf{DSL}                &\textbf{degree}   &\textbf{$\#GORA$} 	 & \textbf{mod}  & \textbf{add}                 & \textbf{del}  & \textbf{mod}	& \textbf{add}	& \textbf{del}	& \textbf{lines}	& \textbf{rules} 	& \textbf{calls}~\tnote{1}	& \textbf{$\#iGR$} 	&\textbf{$\#niGR$} \\
                \midrule
                ATL        & Complete  & 178   & 30    & 0     & 0     & 187   & 0     & 23    & 187   & 30    & 76    & 28 & 0 \\ 
                BibTeX     & Complete  & 14    & 47    & 0     & 1     & 291   & 0     & 0     & 291   & 47    & 188   & 46 & 0 \\
                DOT        & Complete  & 79    & 24    & 1     & 3     & 112   & 2     & 0     & 114   & 25    & 41    & 13 & 0 \\
                SML        & Complete  & 421   & 40    & 5     & 56    & 267   & 18    & 2     & 285   & 45    & 121   & 44 & 0 \\
                \midrule
                Spectra    & Close     & 585   & 54    & 3     & 8     & 190   & 9     & 13    & 414   & 57    & 223   & 54 & 2 \\
                Xcore      & Complete  & 307   & 20    & 7     & 14    & 179   & 35    & 10    & 214   & 27    & 100   & 25 & 0 \\
                Xenia      & Complete  & 74    & 13    & 0     & 2     & 74    & 0     & 0     & 74    & 13    & 28    & 13 & 0 \\ 
                \bottomrule
            \end{tabular}
            \begin{tablenotes}
                \item[1] The number includes the calls to dummy OCL and dummy SML expressions. 
            \end{tablenotes}
            \end{threeparttable}
        \end{minipage}}
\end{table*}

\lipsum[2][2-4]

Table~\ref{tab:result-of-generalizing-to-different-DSLs} shows the results of applying the \grammaroptimizer to the seven DSLs. See Table~\ref{tab:analysed_DSLs} for the corresponding metrics of the initially generated grammars.

\paragraph{Imitation}
For all case DSLs in the first two iterations except \textit{Spectra}, we were able to achieve a complete adaptation, i.e., we were able to modify the grammar by using \grammaroptimizer so that the grammar rules of the optimized grammar \emph{imitate} all grammar rules of the expert-created grammar. 

\paragraph{Limitation regarding Spectra}
\begin{lstlisting}[caption={Example\,---\,grammar rule \texttt{TemporalPrimaryExpr} in the generated grammar of Spectra}, label={lst:excerpt-of-generated-grammar-in-Spectra},float=tb, basicstyle=\footnotesize, breaklines=true]
TemporalPrimaryExpr returns TemporalPrimaryExpr:
	{TemporalPrimaryExpr}
	'TemporalPrimaryExpr'
	'{'
		('operator' operator=EString)?
		('predPatt' predPatt=[PredicateOrPatternReferrable|EString])?
		('pointer' pointer=[Referrable|EString])?
		('regexpPointer' regexpPointer=[DefineRegExpDecl|EString])?
		('predPattParams' '{' predPattParams+=TemporalExpression ( "," predPattParams+=TemporalExpression)* '}' )?
		('tpe' tpe=TemporalExpression)?
		('index' '{' index+=TemporalExpression ( "," index+=TemporalExpression)* '}' )?
		('temporalExpression' temporalExpression=TemporalExpression)?
		('regexp' regexp=RegExp)?
	'}';
\end{lstlisting}

\begin{lstlisting}[caption={Example\,---\,grammar rule \texttt{TemporalPrimaryExpr} in the expert-created grammar of Spectra}, label={lst:excerpt-of-original-grammar-in-Spectra},float=tb, basicstyle=\footnotesize, breaklines=true]
TemporalPrimaryExpr returns TemporalExpression:
    Constant | '(' QuantifierExpr ')' | {TemporalPrimaryExpr}
    (predPatt=[PredicateOrPatternReferrable] 
    ('(' predPattParams+=TemporalInExpr (',' predPattParams+=TemporalInExpr)* ')' | '()') | 
    operator=('-'|'!') tpe=TemporalPrimaryExpr | 
	pointer=[Referrable]('[' index+=TemporalInExpr ']')*  | 
	operator='next' '(' temporalExpression=TemporalInExpr ')' | 
	operator='regexp' '(' (regexp=RegExp | regexpPointer=[DefineRegExpDecl]) ')' | 
	pointer=[Referrable] operator='.all' | 
	pointer=[Referrable] operator='.any' | 
	pointer=[Referrable] operator='.prod' | 
	pointer=[Referrable] operator='.sum' | 
	pointer=[Referrable] operator='.min' | 
	pointer=[Referrable] operator='.max');
\end{lstlisting}

For one of the languages, Spectra, we were able to come very close to the expert-created grammar. 
Many grammar rules of Spectra could be nearly imitated. However, we did not implement all grammar rules that would have been necessary to allow the full optimization of Spectra.
Listing~\ref{lst:excerpt-of-generated-grammar-in-Spectra} shows the grammar rule \texttt{TemporalPrimaryExpr} in Spectra's generated grammar, while Listing~\ref{lst:excerpt-of-original-grammar-in-Spectra} shows what that grammar rule looks like in the expert-created grammar. In order to optimize the grammar rule \texttt{TemporalPrimaryExpr} from Listing~\ref{lst:excerpt-of-generated-grammar-in-Spectra} to Listing~\ref{lst:excerpt-of-original-grammar-in-Spectra}, we need to configure the \grammaroptimizer so that it combines the attribute \texttt{pointer} and \texttt{operator} multiple times, and the default value of the attribute \texttt{operator} is different each time. The language engineers using the \grammaroptimizer need to input multiple parameters to ensure that the \grammaroptimizer gets enough information, and this complex optimization requirement only appears in Spectra. Therefore we did not do such an optimization.

\paragraph{Size of the Changes}
It is worth noting that the number of optimization rule applications is significantly larger than the number of grammar rules for all cases but BibTeX. This indicates that the effort required to describe the optimizations once is significant. 
However, the actual changes to the grammar, e.g., in terms of modified lines in the grammar are in most cases comparable to the number of optimization rule applications (e.g., for ATL with 178 optimization rule applications and 187 changed lines in the grammar) or even much larger (e.g., for BibTeX with 14 optimization rule applications and 291 modified lines).
Note that the number of changed, added, and deleted lines is also an underestimation of the number of necessary changes, as many lines will be changed in multiple ways, e.g., by changing keywords and braces in the same line. This explains why for some languages the number of optimization rule applications is bigger than the number of changed lines (e.g., for SML we specified 421 optimization rule applications which changed, added, and deleted 287 lines in the grammar).

\paragraph{Effort for the Language Engineer}
We acknowledge that the number of optimization rule applications that are necessary to adapt a generated grammar to imitate the expert-created grammar indicates that it is more effort to configure \grammaroptimizer than to apply the desired change in the grammar manually once. However, even with that assumption, we argue that the effort of configuring \grammaroptimizer is in the same order of magnitude as the effort of applying the changes manually to the grammar.

Furthermore, we argue that it is more efficient to configure \grammaroptimizer once than to manually rewrite grammar rules every time the language changes -- under the assumption that the configuration can be reused for new versions of the grammar.
In that case, the effort invested in configuring \grammaroptimizer would quickly pay off when a language is going through changes, e.g., while rapidly prototyping modifications or when the language is evolving. 
In the next section (Section \ref{sec:support_eval}), we evaluate this assumption.


In terms of reusability of the configurable optimization rules, we observe that most of the languages we cover require at least one \emph{unique} optimization rule that is not needed by any other language. This applies to DOT, BibTeX, and ATL with one unique optimization rule, each. Spectra was our most complicated case with six unique rules, whereas Xcore requires four and SML requires five unique rules. This indicates that using \grammaroptimizer for a new language might require effort by implementing a few new optimization rules. However, we argue that this effort will be reduced as more optimization rules are added to \grammaroptimizer and that, in particular for evolving languages, the small investment to create a new optimization rule will pay off quickly.

\subsection{Supporting Evolution (RQ2)}\label{sec:support_eval}
To address the second question, we evaluate the \grammaroptimizer on two languages' evolution histories: The industrial case of EAST-ADL and the evolution of the DSL QVTo. We focus on the question to what degree a configuration of the \grammaroptimizer that was made for one language version can be applied to a new version of the language.

\subsubsection{Cases}
The two cases we are using to evaluate how \grammaroptimizer supports the evolution of a DSL are a textual variant of EAST-ADL~\cite{eadl} and QVT Operational (QVTo)~\cite{qvt}.

\paragraph{EAST-ADL}
EAST-ADL is an architecture description language used in the automotive domain~\cite{eadl}. 
Together with an industrial language engineer for EAST-ADL, we are currently developing a textual notation for version 2.2 of the language~\cite{EATXT}. We started this work with a simplified version of the meta-model to limit the complexity of the resulting grammar. In a later step, we switched to the full meta-model. We treat this switch as an evolution step here. The meta-model of EAST-ADL is taken from the EATOP repository~\cite{eatop-bitbucket}. The meta-model of the simplified version contains 91 classes and enumerations, and the meta-model of the full version contains 291 classes and enumerations.

\paragraph{QVTo}
QVTo is one of the languages in the OMG QVT standard~\cite{qvt}. We use the original meta-models available in Ecore format on the OMG website~\cite{qvt}. The baseline version is QVTo 1.0~\cite{qvt10spec} and we simulate evolution to version 1.1~\cite{qvt11spec}, 1.2~\cite{qvt12spec} and 1.3~\cite{qvt13spec}.
Our original intention was to use the Eclipse reference implementation of QVTo~\cite{qvto-eclipse}, but due to the differences in abstract syntax and concrete syntax (see Section~\ref{sec:background}), 
we chose to use the official meta-models instead. 
We analyzed four versions of QVTo's OMG official Ecore meta-model. There are 50 differences between the meta-models of version 1.0 and 1.1, 29 of which are parts that do not contain OCL (as for ATL as described in Section~\ref{subsubsect:Meth:exlusionParts}, we exclude OCL in our solution for QVTo). These 29 differences include different types, for example, 1) the same set of attributes has different arrangement orders in the same class in different versions of the meta-model; 2) the same class has different superclasses in different versions; 3) the same attribute has different multiplicities in different versions, etc. There are 3 differences between versions 1.1 and 1.2, all of which are from the OCL part. There is only one difference between versions 1.2 and 1.3, and it is about the same attribute having a different lower bound for the multiplicity in the same class in the two versions.
Altogether we observed 54 meta-model differences in QVTo between the different versions (cf. the file ``Comparison of QVTo metamodel versions'' in the folder ``Section\_7\_Evaluation/Subsection\_7.2\_Support'' lists all the metamodel differences).

The OMG website provides an EBNF grammar for each version of QVTo, which is the basis for our imitations of the QVTo languages. Among them, versions 1.0, 1.1, and 1.2 share the same EBNF grammar for the QVTo part except for the OCL parts, despite the differences in the meta-model. The EBNF grammar of QVTo in version 1.3 is different from the other three versions.


\subsubsection{Preparation of the QVTo Case}\label{sec:support_eval_QVTo_Preparation}
In contrast to the EAST-ADL case, we needed to perform some preparations of the grammar and the meta-model to study the QVTo case. All adaptations were done the same way on all versions of QVTo.

\paragraph{Exclusion of OCL}
As described in detail in Section~\ref{subsubsect:Meth:exlusionParts}, we excluded the embedded OCL language part from QVTo. 
For the meta-model, we introduced a dummy class for OCL, changed all calls to OCL types into calls to that dummy class, and removed the OCL metaclasses from the meta-model.

As described in Section~\ref{subsubsect:Meth:exlusionParts}, excluding a language part such as the embedded OCL from the scope of the investigation also implies that we need to exclude this language part when it comes to judging whether a grammar is imitated. 
Therefore, we substituted all grammar rules from the excluded OCL part with a placeholder grammar rule called \texttt{ExpressionGO} where an OCL grammar rule would have been called.
This change allows us to compare the expert-created grammar of the different QVTo versions to the optimized grammar versions.

\paragraph{QVTo Meta-model Adaptations}
We found that some non-terminals of QVTo's EBNF grammar are missing in the QVTo meta-model provided by OMG. For example, there is a non-terminal \texttt{<top\_level>} in the EBNF grammar, but there is no counterpart for it in the meta-model. Therefore, we need to adapt the meta-model to ensure that it contains all the non-terminals in the EBNF grammar. 
To ensure that the adaptation of the meta-model is done systematically, we defined seven general adaptation rules that we followed when adapting the meta-models of the different versions. We list these adaptation rules in the supplemental material~\cite{datasource2023go}. 

As a result, we added 62 classes and enumerations with their corresponding references to each version of the meta-model. Note that this number is high compared to the original number of classes in the meta-model (24 classes). 
This massive change was necessary because the available Ecore meta-models were too abstract to cover all elements of the language. The original meta-model did contain most key concepts, but would not allow to actually specify a complete QVTo transformation. For example, with the original meta-model, it was not possible to represent the scope of a mapping or helper. 

These changes enable us to imitate the QVTo grammar. 
However, they do not bias the results concerning the effects of the observed meta-model evolution as, with the exception of a single case, these evolutionary differences are neither erased nor increased by the changes we performed to the meta-model. The exception is a meta-model evolution change between version 1.0 and 1.1 where the class \texttt{MappingOperation} has super types \texttt{Operation} and \texttt{NamedElement}, while the same class in V1.1 does not. 
The meta-model change performed by us removes the superclass \texttt{Operation} from \texttt{MappingOperation} in version 1.0. We did this change to prevent conflicts as the attribute \emph{name} would have been inherited multiple times by \texttt{MappingOperation}. This in turn would cause problems in the generation process. Thus, only two of the 54 meta-model evolutionary differences could not be studied. The differences and their analysis can be found in the supplemental material~\cite{datasource2023go}.

\subsubsection{Method}
To evaluate how \grammaroptimizer supports the evolution of meta-models we look at the effort that is required to update the optimization rule applications after an update of the meta-models of EAST-ADL and QVTo. 

\paragraph{Baseline \grammaroptimizer Configuration}
First, we generated the grammar for the initial version of a language's meta-model (i.e., the simple version for EAST-ADL and version 1.0 for QVTo). Then we defined the configuration of optimization rule applications that allows the \grammaroptimizer to modify the generated grammar so that its grammar rules \emph{imitate} the expert-created grammar for each case. 
Doing so confirmed the observation from the first part of the evaluation that a new language of sufficient complexity requires at least some new optimization rules (see Section \ref{subsubsection:eval:imtatio:results}). Consequently, we identified the need for four additional optimization rules for QVTo, which we implemented accordingly as part of the \grammaroptimizer (this is also summarized in Section~\ref{sec:identified-optimization-rules} in Table~\ref{tab:rules-and-iterations}).  
This step provided us with a baseline configuration for the \grammaroptimizer.

\paragraph{Evolution}
For the following language versions, i.e., the full version of EAST-ADL and QVTo 1.1, we then generated the grammar from the corresponding version of the meta-model and applied the \grammaroptimizer with the configuration of the previous version (i.e., simple EAST-ADL and QVTo 1.0). 
We then identified whether this was already sufficient to \emph{imitate} the language's grammar or whether changes and additions to the optimization rule applications were required. We continued adjusting the optimization rule applications accordingly to gain a \grammaroptimizer configuration valid for the new version (full EAST-ADL and QVTo 1.1, respectively). 
For QVTo, we repeated that process two more times: For QVTo 1.2, we took the configuration of QVTo 1.1 as a baseline, and for QVTo 1.3, we took the configuration of QVTo 1.2 as a baseline.

\subsubsection{Metrics}
We documented the metrics used in Section~\ref{sec:eval-grammar-adaptation_metrics} for EAST-ADL and QVTo in their different versions. In addition, we also documented the following metric:
\begin{description}
    \item [$\#$\textit{cORA}] The number of changed, added, and deleted optimization rule applications compared to the previous language version.
\end{description}

\subsubsection{Results}
Table~\ref{tab:result-of-supporting-evolution} shows the results of the evolution cases. 

\paragraph{EAST-ADL}
Compared with the simplified version of EAST-ADL, the full version is much larger. It contains 291 metaclasses, i.e., 200 metaclasses more than the simple version of EAST-ADL, which leads to a generated grammar with 291 grammar rules and 2,839 non-blank lines in the generated grammar file (cf. Table~\ref{tab:result-of-supporting-evolution}).

The 22 optimization rule applications for the simple version of EAST-ADL already change the grammar significantly, causing modifications of all 91 grammar rules and changes in nearly every line of the grammar. This also illustrates how massive the changes to the generated grammar are to reach the desired grammar. The number of changes is even larger with the full version of EAST-ADL.

We only needed to change and add a total of 10 grammar optimization rule applications to complete the optimization of the grammar of full EAST-ADL. For example, we excluded the primary type \texttt{String0} from the full version of the EAST-ADL grammar, which led us to add a line of configuration \texttt{go.removeRule(String0)}. While this is increasing the \grammaroptimizer configuration from the simple EAST-ADL version quite a bit (from 22 optimization rule applications to 31 optimization rule applications), the increase is fairly small given that the meta-model increased massively (with 200 additional metaclasses).


The reason is that our grammar optimization requirements for the simplified version and the full version of EAST-ADL are almost the same. This optimization requirement is mainly based on the look and feel of the language and is provided by an industrial partner. These optimization rule applications have been configured for the simplified version. When we applied them to the generated grammar of the full version of EAST-ADL, we found that we can reuse all of these optimization rule applications. Furthermore, we benefit from the fact that many optimization rule applications are formulated for the scope of the whole grammar and thus can also influence grammar rules added during the evolution step. We do not list a number of grammar rules in a expert-created grammar of EAST-ADL in Table~\ref{tab:result-of-supporting-evolution}, because there is no ``original'' text grammar of EAST-ADL. Instead, we optimize the generated grammar of EAST-ADL according to our industrial partner's requirements for EAST-ADL's textual concrete syntax.


\lipsum[1][1-4]

\begin{table*}
\scriptsize
    \centering
    \rotatebox{-90}{
        \begin{minipage}[t]{1.5\textwidth}
            \centering
            \caption{Result of supporting evolution (RQ2)}
            \label{tab:result-of-supporting-evolution}
            \begin{threeparttable}
            \begin{tabular}{@{}llrrrrrrrrrrrrrrrr@{}}
                \toprule
        		~ & \textbf{Meta-m.} & \multicolumn{3}{c}{\textbf{Generated grammar}} & \multicolumn{3}{c}{\textbf{Optimized grammar}} & \multicolumn{3}{c}{\textbf{Grammar rules}} &
                \multicolumn{3}{c}{\textbf{Lines in Grammar}} \\
        		\textbf{DSL} & \textbf{Classes}~\tnote{1} & \textbf{lines}	& \textbf{rules}	& \textbf{calls} & \textbf{lines} & \textbf{rules} & \textbf{calls}~\tnote{2} & \textbf{mod} & \textbf{add} & \textbf{del} & \textbf{mod} & \textbf{add} & \textbf{del} & \textbf{$\#GORA$} 	&\textbf{$\#cORA$} \\
                \midrule
                EAST-ADL & 91 & 755 & 91  & 735   & 767   & 103   & 782   & 70    & 12    & 0     & 517   & 14    & 2   & 22  & / \\
        				(simple) & ~ & ~ & ~  & ~ & ~			& ~			& ~ 		& ~			& ~			&	~ 		& ~			& ~			& ~   	& ~		& ~ \\
                EAST-ADL  & 291 & 2,839 & 291 & 3,062 & 2,851 & 303   & 3,074 & 233  & 12    & 1     & 2,046 & 16    & 4   & 31  & 10\\
        				(full)  & ~ & ~ & ~  & ~ & ~			& ~			& ~ 		& ~			& ~			&	~ 		& ~			& ~			& ~   	& ~		& ~ \\
                \midrule
                QVTo 1.0     & 85  & 1,026 & 109   & 910   & 444   & 77    & 181   & 66    & 1     & 33    & 228   & 2     & 580 & 733 & / \\
                QVTo 1.1     & 85  & 992   & 110   & 836   & 444   & 77    & 181   & 66    & 1     & 34    & 228   & 2     & 546 & 733 & 2 \\
                QVTo 1.2     & 85  & 992   & 110   & 836   & 444   & 77    & 181   & 66    & 1     & 34    & 228   & 2     & 546 & 733 & 0 \\
                QVTo 1.3     & 85  & 991   & 110   & 835   & 443   & 77    & 180   & 66    & 1     & 34    & 228   & 2     & 546 & 733 & 1 \\
                \bottomrule
            \end{tabular}
            \begin{tablenotes}
                \item[1] The number is after adaptation, and it contains both classes and enumerations.
                \item[2] The number includes the calls to dummy OCL and dummy SML expressions. 
            \end{tablenotes}
            \end{threeparttable}
        \end{minipage}}
\end{table*}

\lipsum[2][2-4]

\paragraph{QVTo}
The baseline configuration of the \grammaroptimizer for QVTo includes 733 optimization rule applications, which is a lot given that the expert-created grammar of QVTo 1.0 has 115 non-terminals. 
Note that the optimized grammar has even fewer grammar rules (77) as some of the rules in the optimized grammar \emph{imitate} multiple rules from the expert-created grammar at once. This again is a testament to how different the expert-created grammar is from the generated one (over 228 lines in the grammar are modified, 2 lines are added, and 580 lines are deleted by these 733 optimization rule applications).

However, if we look at the evolution towards versions 1.1, 1.2, and 1.3 we witness that very few changes to the \grammaroptimizer configuration are required. 
In fact, only between 0 and 2 out of the 733 optimization rule applications needed adjustments. 
This significantly reduces the effort required compared to manually modifying a grammar generated from a new version of the QVTo metamodel, which would require modifying hundreds of lines.
The reason is that, even though there are many differences between different versions of the QVTo meta-model, there are only 0 to 2 differences that affect the optimization rule applications. For example, version 1.0 of the QVTo meta-model has an attribute called \texttt{bindParameter} in the class \texttt{VarParameter}, whereas it is called \texttt{representedParameter} in version 1.1. 
This attribute is not needed according to the expert-created grammars, so the \grammaroptimizer configuration includes a call to the optimization rule \emph{RemoveAttribute} to remove the grammar line that was generated based on that attribute.
The second parameter of the optimization rule \emph{RemoveAttribute} needs to specify the name of the attribute. As a consequence of the evolution, we had to change that name in the optimization rule application. 
Another example concerns the class \texttt{TypeDef}, which contains an attribute \texttt{typedef\_condition} in version 1.2 of the QVTo meta-model. We added square brackets to it by applying the optimization rule \emph{AddSquareBracketsToAttr} in the grammar optimization. However, in version 1.3 of the QVTo meta-model, the class \texttt{TypeDef} does not contain such an attribute, so the optimization rule application \emph{AddSquareBracketsToAttr} was unnecessary.

Most of the differences between different versions of the meta-model do not lead to changes in the optimization rule applications. For example, the multiplicity of the attribute \texttt{when} in the class \texttt{MappingOperation} is different in version 1.0 and 1.1. We used \emph{RemoveAttribute} to remove the attribute during the optimization of grammar version 1.0. The same command can still be used in version 1.1, as the removal operation does not need to consider the multiplicity of an attribute. Therefore, this difference does not affect the configuration of optimization rule applications.

\section{Discussion}
\label{sec:discussion}
In the following, we discuss the threats to the validity of the evaluation, different aspects of the \grammaroptimizer, and future work implied by the current limitations.

\subsection{Threats to Validity}\label{sec:threats}
The threats to validity structured according to the taxonomy of Runeson et al. \cite{Runeson.2008,Runeson.2012} are as follows.

\subsubsection{Construct Validity}\label{sec:threats_construct}
We limited our analysis to languages for which we could find meta-models in the Ecore format. Some of these meta-models were not ``official'', in the sense that they had been reconstructed from a language in order to include them in one of the ``zoos''. An example of that is the meta-model for BibTeX we used in our study. 
In the case of the DOT language, we reconstructed the meta-model from an Xtext grammar we found online. 
We adopted a reverse-engineering strategy where we generated the meta-model from the expert-created grammar and then generated a new grammar out of this meta-model.
This poses a threat to validity since many of the languages we looked at can be considered ``artificial'' in the sense that they were not developed based on meta-models. 
However, we do not think this affects the construct validity of our analysis since our purpose is to analyze what changes need to be made from an Xtext grammar file that has been generated.
In addition, we address this threat to validity by also including a number of languages (e.g., Xenia and Xcore) that are based on meta-models and using the meta-models provided by the developers of the language.

Furthermore, we had to make some changes to some of the meta-models to be able to generate Xtext grammars out of them at all (cf. Section~\ref{sec:methodology_analysis_MMPrepsAndGrammarGen}) or to introduce certain language constructs required by the textual concrete syntax (cf. Section~\ref{sec:support_eval_QVTo_Preparation}).
These meta-model adaptations might have introduced biased changes and thereby imposed a threat to construct validity.
However, we reduced these adaptations to a minimum as far as possible to mitigate this threat and documented all of them in our supplemental material~\cite{datasource2023go} to ensure their reproducibility.

\subsubsection{Internal Validity}\label{sec:threats_internal}
In the evaluation (cf. Section~\ref{sec:eval}), we set up and quantitatively evaluate size and complexity metrics regarding the considered meta-models and grammars as well as regarding the \grammaroptimizer configurations for the use cases of one-time grammar adaptations and language evolution.
Based on that, we conclude and argue in 
Section~\ref{subsubsection:eval:imtatio:results} and Section~\ref{sec:discussion_effort}
about the effort required for creating and evolving languages as well as the effort to create and re-use \grammaroptimizer configurations.
These relations might be incorrect.
However, the applied metrics provide objective and obvious indications about the particular sizes and complexities and thereby the associated engineering efforts. 

\subsubsection{External Validity}\label{sec:threats_external}
As discussed in the analysis part, we analyzed a total of seven DSLs to identify generic optimization rules. Whereas we believe that we have achieved significant coverage by selecting languages from different domains and with very different grammar structures, we cannot deny that analysis of further languages could have led to more
optimization rules. However, due to the extensible nature of \grammaroptimizer{}, the practical impact of this threat to generalisability is low since it is easy to add additional generic optimization rules once more languages are analyzed.

\subsubsection{Reliability}\label{sec:threats_reliability}
Our overall procedure to conceive and develop the \grammaroptimizer encompassed multiple steps.
That is, we first determined the differences between the particular initially generated Xtext grammars and the grammars of the actual languages in two iterations as described in Section~\ref{sec:methodology}.
This analysis yielded the corresponding identified conceptual grammar optimization rules summarized in Section~\ref{sec:identified-optimization-rules}.
Based on these identified conceptual grammar optimization rules, we then implemented them as described in Section~\ref{sec:solution}.
This procedure imposes multiple threats to reliability.
For example, analyzing a different set of languages could have led to a different set of identified optimization rules, which then would have led to a different implementation.
Furthermore, analyzing the languages in a different order or as part of different iterations could have led to a different abstraction level of the rules and thereby a different number of rules.
Finally, the design decisions that we made during the identification of the conceptual optimization rules and during their implementation could also have led to different kinds of rules or implementation.
However, we discussed all of these aspects repeatedly amongst all authors to mitigate this threat and documented the results as part of our supplemental material~\cite{datasource2023go} to ensure their reproducibility.

\subsection{The Effort of Creating and Evolving a Language with the \grammaroptimizer }\label{sec:discussion_effort}
The results of our evaluation show three things.
%
First, the expert-created grammars of all studied languages differ greatly in appearance from the generated grammars.
Thus, in most cases, creating a DSL with Xtext will require the language engineer to perform big changes to the generated grammar.
Second, depending on the language, using the \grammaroptimizer for a single version of the language may or may not be more effort for the language engineer, compared to manually adapting the grammar.
Third, there seems to be a large potential for the reuse of \grammaroptimizer configurations between different versions of a language, thus supporting the evolution of textual languages.

These observations can be combined with the experience that most languages evolve with time and that especially DSLs go through a rapid prototyping phase at the beginning where language versions are built for practical evaluation~\cite{wang2005rapidly}. 
Therefore, we conclude that the \grammaroptimizer has big potential to save manual effort when it comes to developing DSLs.

Additionally, a topic worth mentioning is how the involvement of different people and their skill sets affect the effort when creating and reusing optimization rule configurations. For example, in case updates to an existing configuration are needed after an evolution step, the maintainers need to understand the optimization rule configuration of the previous version, which could take a new contributor more time than the original contributor. Assessing the impact of this aspect is a subject for future work.

\subsection{Implications for Practitioners and Researchers}
Our results have several implications for language engineers and researchers. 

\paragraph{Impact on Textual Language Engineering}	
Our work might have an impact on the way DSL engineers create textual DSLs nowadays. 
That is, instead of specifying grammars and thereby having to be EBNF experts, the \grammaroptimizer also enables engineers familiar with meta-modeling to conceive well-engineered meta-models and to semi-automatically generate user-friendly grammars from them.
Furthermore, Kleppe \cite{Kleppe.2007b} compiles a list of advantages of approaches like the \grammaroptimizer, among them two that apply especially to our solution: 1) the \grammaroptimizer provides flexibility for the DSL engineering process, as it is no longer necessary to define the kind of notation used for the DSL at the very beginning
as well as 2) the \grammaroptimizer enables rapid prototyping of textual DSLs based on meta-models.

\paragraph{Blended Modeling}
Ciccozzi et al.~\cite{Ciccozzi.2019} coin the term \emph{blended modeling} for the activity of interacting with one model through multiple notations (e.g., both textual and graphical notations), which would increase the usability and flexibility for different kinds of model stakeholders. 
However, enabling blended modeling shifts more effort to language engineers.
This is due to the fact that the realization of the different editors for the different notations requires many manual steps when using conventional modeling frameworks.	
In this context, Cicozzi and colleagues particularly stress the issue of the manual customization of grammars in the case of meta-model evolution.
Thus, as one research direction to enable blended modeling, Ciccozzi et al.\ formulate the need to automatically generate the different editors from a given meta-model.	
Our work serves as one building block toward realizing this research direction and opens up the possibility of developing and evolving blended modeling languages that include textual versions.

A relevant question is to which extent our approach enables cost savings in a  larger context, as the cost for evolving the existing tools and applications working with existing languages might be higher than the cost for evolving the languages themselves.
We benefit from the extensive tool support offered by Xtext, which can automatically re-generate large parts of the available textual editor after changes in the underlying grammar, including features such as, e.g., auto-formatting, auto-completion, and syntax highlighting. In consequence, by supporting automated grammar changes (in particular, after evolution steps), we also save effort for the overall adaptation of the textual editor. However, in MDE contexts, other applications and tools typically refer to the metamodel, instead of the grammar, and hence, are outside our scope.

\paragraph{Prevention of Language Flaws}
Willink~\cite{willink2020reflections} reflects on the version history of the Object Constraint Language (OCL) and the flaws that were introduced during the development of the different OCL 2.x specifications by the Object Management Group~\cite{OCLVersions}.
Particularly, he points out that the lack of a parser for the proposed grammar led to several grammar inaccuracies and thereby to ambiguities in the concrete textual syntax.
This, in turn, led to the fact that the concrete syntax and the abstract syntax in the Eclipse OCL implementation \cite{EclipseOCL} are so divergent that two distinct meta-models with a dedicated transformation between both are required, which also holds for the QVTo specification and its Eclipse implementation \cite{willink2020reflections} (cf. Section~\ref{sec:background}).
The \grammaroptimizer will help to prevent and bridge such flaws in language engineering in the future. 
Xtext already enables the generation of the complete infrastructure for a textual concrete syntax from an abstract syntax represented by a meta-model. Our approach adds the ability to optimize the grammar (i.e., the concrete syntax), as we show in the evaluation by deriving an applicable parser with an optimized grammar from the QVTo Specification meta-models.	

\subsection{Future Work}
The \grammaroptimizer is a first step in the direction of supporting the evolution of textual grammars for DSLs. However, there are, of course, still open questions and challenges that we discuss in the following.

\paragraph{Name Changes to Meta-model Elements}
In the \grammaroptimizer configurations, we currently reference the grammar concepts derived from the meta-model classes and attributes by means of the class and attribute names (cf. Listing~\ref{lst:dot-config-example-rule}).
Thus, if a meta-model evolution involves many name changes, likewise many changes to optimization rule applications are required. 
Consequently, we plan as future work to improve the \grammaroptimizer with a more flexible concept, in which we more closely align the grammar optimization rule applications with the meta-model based on name-independent references.

\paragraph{More Efficient Rules and Libraries}
We think that there is a lot of potential to make the available set of optimization rules more efficient. 
This could for example be done by providing libraries of more complex, recurring changes that can be reused. 
Such a library can contain a default set of optimization rule configurations to make the generated grammar follow a particular style (e.g., mimicking an existing language, to be appealing to users of that language). Language engineers can use it as a basis and with minimal effort define optimization rule configurations that perform DSL-specific changes.
Such a change might make the application of the \grammaroptimizer attractive even in those cases where no evolution of the language is expected.
While this use-case still requires effort for defining configurations, the overall effort compared to manual editing can be reduced especially in cases with applicable large-scoped rules that, e.g., globally change the parenthesis style in the grammar.

In addition, the API of \grammaroptimizer could be changed to a fluent version where the optimization rule application is configured via method calls before they are executed instead of using the current API that contains many \texttt{null} parameters.
This could also lead to a reduction of the number of grammar optimization rule applications that need to be executed since some executions could be performed at the same time.

Another interesting idea would be to use artificial intelligence to learn existing examples of grammar optimizations in existing languages to provide optimization suggestions for new languages and even automatically create configurations for the \grammaroptimizer. 


\paragraph{Expression Languages}
In this paper, we excluded the expression language parts (e.g., OCL) of two of the example languages (cf. Section~\ref{subsubsect:Meth:exlusionParts}). 
However, expression languages define low-level concepts and have different kinds of grammars and underlying meta-models than conventional languages. 
In future work, we want to further explore expression languages specifically, in order to ensure that the \grammaroptimizer can be used for these types of syntaxes as well.

\paragraph{Visualization of Configuration}
Currently, we configure the \grammaroptimizer by calling the methods of optimization rules, which is a code-based way of working. In the future, 
we intend to improve the tooling for \grammaroptimizer and embed the current library into a more sophisticated workbench that allows the language engineer to select and parameterize optimization rule applications either using a DSL or a graphical user interface and provides previews of the modified grammar as well as a view of what valid instances of the language look like.

\paragraph{Co-evolving Model Instances}\label{sec:discussion_coevolution_instances}
We also intend to couple \grammaroptimizer with an approach for language evolution that also addresses the model instances. 
In principle, a model instance represented by a textual grammar instance can be read using the old grammar and parsed into an instance of the old meta-model. It can then be transformed, e.g., using QVTo to conform to the new meta-model, and then be serialized again using the new grammar. However, following this approach means that formatting and comments can be lost. Instead, we intend to derive a textual transformation from the differences in the grammars and the optimization rule applications that can be applied to the model instances and maintain formatting and comments as much as possible.

\paragraph{Alternative implementation strategy}\label{sec:discussion_alternative_impl}
Our implementation strategy relies on the format of textual grammars produced by Xtext, which is stable across recent versions of Xtext.
This implementation strategy was suitable for positively answering our evaluation questions and thus, substantiating the scientific contribution of our paper.
An alternative, arguably more elegant implementation strategy would be to use Xtext's abstract syntax tree representation of the grammar.
A benefit of such an implementation would be that it would be more robust in case that the output format of Xtext changes, rendering it a desirable direction for future work.




\section{Conclusion}
In this paper, we have presented \grammaroptimizer, a tool that supports language engineers in the rapid prototyping and evolution of textual domain-specific languages that are based on meta-models. \grammaroptimizer uses a number of optimization rules to modify a grammar generated by Xtext from a meta-model. These optimization rules have been derived from an analysis of the difference between the actual and the generated grammars of seven DSLs.

We have shown how \grammaroptimizer can be used to modify grammars generated by Xtext based on these optimization rules. This automation is particularly useful while a language is being developed to allow for rapid prototyping without cumbersome manual configuration of grammars and when the language evolves. We have evaluated \grammaroptimizer on seven grammars to gauge the feasibility and effort required for defining the optimization rules. We have also shown how \grammaroptimizer supports evolution with the examples of EAST-ADL and QVTo.

Overall, our tool enables language engineers to use a meta-model-based language engineering workflow and still produce high-quality grammars that are very close in quality to hand-crafted ones. We believe that this will reduce the development time and effort for domain-specific languages and will allow language engineers and users to leverage the advantages of using meta-models, e.g., in terms of modifiability and documentation.

In future work, we plan to extend \grammaroptimizer into a more full-fledged language workbench that supports advanced features like refactoring of meta-models, a ``what you see is what you get'' view of the optimization of the grammar, and the ability to co-evolve model instances alongside the underlying language. We will also explore the integration into workflows that generate graphical editors in order to enable blended modeling.

\section*{Acknowledgements}
This work has been sponsored by Vinnova under grant number 2019-02382 as part of the ITEA 4 project \emph{BUMBLE}.